\documentclass[useAMS,usenatbib,usegraphicx]{mn2e}
\usepackage{amssymb}
\usepackage{amsmath}
\usepackage{multirow}
\usepackage{booktabs}
\usepackage{subfig}
\usepackage{color}
\usepackage{fixltx2e}
\usepackage{lscape}
\usepackage{enumitem}
\usepackage{rotating}

\newcommand{\eg}{e.\,g. }
\newcommand{\ie}{i.\,e., }
\newcommand{\etc}{etc.\ }
\newcommand{\etal}{{et al.~}}
\newcommand{\chinu}{$\chi^2_\nu$}

\newcommand{\galcount}{\textsc{galnum} }
\newcommand{\Dev}{\texttt{deV}}
\newcommand{\Exp}{\texttt{Exp}}
\newcommand{\Ser}{\texttt{Ser}}
\newcommand{\DevExp}{\texttt{deV-Exp}}
\newcommand{\SerExp}{\texttt{Ser-Exp}}
\newcommand{\pymorph}{\textsc{PyMorph}}
\newcommand{\galfit}{\textsc{Galfit}}
\newcommand{\sextractor}{\textsc{SExtractor}}
\newcommand{\gimd}{\textsc{gim2d}}
\newcommand{\kcorrect}{\textsc{kcorrect}}
\newcommand{\readatlasimage}{\textsc{readAtlasImages-v5\_4\_11}}
\newcommand{\fits}{\textsc{FITS}}

\usepackage{appendix}
\usepackage{longtable}
\usepackage{url}
\usepackage{float}
\usepackage[T1]{fontenc}
\usepackage{aecompl} 
\usepackage{hyperref}

\title[2D Fit Catalogue]{A catalogue of 2D photometric decompositions 
in the SDSS-DR7 spectroscopic main galaxy sample: preferred models and systematics}
\author[Meert \etal{}]{Alan Meert,$^{1}$\thanks{E-mail:
ameert@physics.upenn.edu} 
Vinu Vikram,$^{1}$\thanks{E-mail: vvinuv@gmail.com} 
and Mariangela Bernardi$^{1}$\thanks{E-mail: bernardm@sas.upenn.edu} \\
$^{1}$Department of Physics and Astronomy, University of Pennsylvania, 
Philadelphia, PA 19104, USA\\}

\begin{document}

\date{Accepted 2014 October 31.  Received 2014 October 31; in original form 2014 June 15}

\maketitle

\label{firstpage}

\begin{abstract}
 
We present a catalogue of two-dimensional, point spread function-corrected de Vacouleurs, S\'{e}rsic,
de Vacouleurs+Exponential, and S\'{e}rsic+Exponential fits of $\sim7\times10^5$
spectroscopically selected galaxies drawn from the Sloan Digital Sky Survey (SDSS) Data Release 7. 
Fits are performed for the SDSS \emph{r} band utilizing the fitting routine \galfit{} and analysis
pipeline \pymorph{}. We compare these fits to prior catalogues. 
Fits are analysed using a physically motivated flagging system. 
The flags suggest that more than 90 per cent of
two-component fits can be used for analysis. We show that the fits follow the expected behaviour 
for early and late galaxy types. The catalogues provide a robust set of structural and photometric 
parameters for future galaxy studies. We show that some biases remain in the measurements, \eg
the presence of bars significantly affect the bulge measurements although the bulge ellipticity
may be used to separate barred and 
non-barred galaxies, and about 15 percent of bulges 
of two-component fits are also affected by resolution.
The catalogues are available in electronic format.
We also provide an interface for generating postage stamp
images of the 2D model and residual as well as the 1D profile.
These images can be generated for a user-uploaded list of galaxies on
demand.
\end{abstract}


\section{Introduction}

The study of the structural components of galaxies has contributed substantially to the understanding
of the formation and evolution of galaxies. The discovery of many scaling relations including the Faber Jackson
\citep{FaberJackson}, Kormendy \citep{Kormendy},  Tully Fisher \citep{TullyFisher}, the Fundamental Plane
\citep{FundamentalPlane} and the morphology density relation \citep{DresslerMDR} refined
models of galaxy formation and evolution. In addition,
the structural components of galaxies in the local Universe 
trace morphological galaxy type and many other galaxy parameters related
to both assembly and evolution of galaxies: colour, metallicity, gas fraction, 
central velocity dispersion \citep[\eg][]{bernardi03b,bernardi03c,bernardi03d, Kauffman_SFR, tremonti04}. 
Properties may also trace halo
size and galaxy environment and place constraints on $\Lambda$ cold dark matter cosmology
\citep[\eg][]{blanton2005,Bernardi2009,shankar10a,shankar10b, kravtsov2014}. 
However, careful estimation of structural parameters 
for large numbers of galaxies is required to test different 
formation and evolution models.

The Sloan Digital Sky Survey \citep[SDSS;][]{sdss_tech,DR7} has already provided a sample of many millions of nearby
galaxies. Future surveys like the Dark Energy Survey \citep{des_whitepaper} and Large
Synoptic Survey Telescope \citep[LSST;][]{lsst2009} will produce larger data sets, both
increasing the number and quality of galaxies available for analysis. At the
same time, the growth of computing power makes it possible to analyse these
data sets at a reasonable rate, making it possible to perform
time-intensive analysis, like galaxy decompositions, on large data sets.

There has been much recent work on improving photometric decomposition of galaxies 
\citep[\eg][]{gadotti09, simard11, Kelvin2011, Lackner2012, Haussler2012}. 
However, the accuracy of such fits is often questioned, particularly when 
multiple components are fitted (\ie bulge+disc+bar \etc{}) and when fits are 
automated without individual inspection. 

This paper presents a catalogue of 2D, point spread funtion (PSF)-corrected de~Vacouleurs, S\'{e}rsic,
de~Vacouleurs+Exponential, and S\'{e}rsic+Exponential fits of $\sim7\times10^5$
spectroscopically selected galaxies drawn from the SDSS. Fits are presented for
the SDSS {\em r} band utilizing the fitting routine \galfit{} \citep[][]{galfit} and analysis
pipeline \pymorph{} \citep[][]{pymorph}. This catalogue is one of the largest galaxy samples
for which structural decompositions have been performed on SDSS galaxies. 

The simulations presented in \citet[][hereafter M13]{meert2013} are used as a benchmark for these fits.
M13 used simulated galaxies drawn from galaxies in this work to test the accuracy of the fitting process. It established 
uncertainties on fitting parameters and showed that the choice of cutout size and background estimation were appropriate 
for the galaxies in this sample.

2D decompositions of SDSS galaxies that overlap with our catalogue have also been carried out by other groups.
\citet[][hereafter S11]{simard11} presented decompositions of the photometric sample 
of SDSS ($\approx 1.4$ million galaxies). \cite{Kelvin2011} limited to fitting a single S\'{e}rsic model
to a subset of our catalogue.
\citet[][hereafter LG12]{Lackner2012} fitted several models to galaxies at lower redshifts . 
We compare to S11, LG12 and other works in Sections~\ref{sec:analysis}~and~\ref{sec:other_cmp}.

This paper focuses on the spectroscopic sample, a subsample of SDSS and S11.
We also present a method for identifying good and bad fits different from the 
statistical approach of S11 and the statistical and qualitative combination of LG12.
Using a combination of comparisons between this work, S11, and LG12 (presented in
Sections~\ref{sec:analysis}~and~\ref{sec:other_cmp}), 
we show that our catalogue improves on previous fits in many respects.

Several studies have already utilized this catalogue. 
\cite{Shankar} tested semi-analytical
modelling of hierarchical formation. \cite{huertas12}
examined environmental effects on the size of galaxies.
\cite{bernardi2013} analysed the uncertainty in the bright end
of the mass and luminosity functions (LF). 
\cite{bernardi2014} also examined the biases automated decompositions impose 
on the size luminosity relation. Finally,
\cite{kravtsov2014} performed detailed fits of approximately 10 brightest cluster galaxies (BCGs) and found that their measurements agree more with our measurements than with the measurements of S11.

The paper is organized as follows: Section~\ref{sec:data} describes the
selection of the data, Section~\ref{sec:fitting} briefly describes the \pymorph{}
fitting routine we used and the specifics of our fitting procedure 
including our choice of cutout size, background fitting, and neighbour fitting. 
Section~\ref{sec:validation} describes the flagging system used to identify poor fits 
and interpret fits as either bulge, disc, or two-component galaxies.
Section~\ref{sec:best_model} describes internal comparisons and consistency checks among the models we fit.
Section~\ref{sec:analysis} describes comparisons to measurements made by other groups including SDSS 
\citep{DR7}, S11, LG12, and \cite{mendel2014}. 
Section~\ref{sec:other_cmp} describes the comparisons incorporating morphological information 
from previous visual or automated classifications.
Section~\ref{sec:discussion} presents the catalogue and describes how to use it. In addition, we describe the webpage content
associated with the catalogue.
Finally, Section~\ref{sec:conclusion} concludes the
paper with a summary of results and final remarks. The fits discussed in this paper and further 
recommendations for their use are available in electronic format as a public release.

\section{The Data} \label{sec:data}

\begin{figure*}
\begin{minipage}{\linewidth}
\begin{center}
\subfloat[ ]{
\includegraphics[width=0.3\columnwidth]{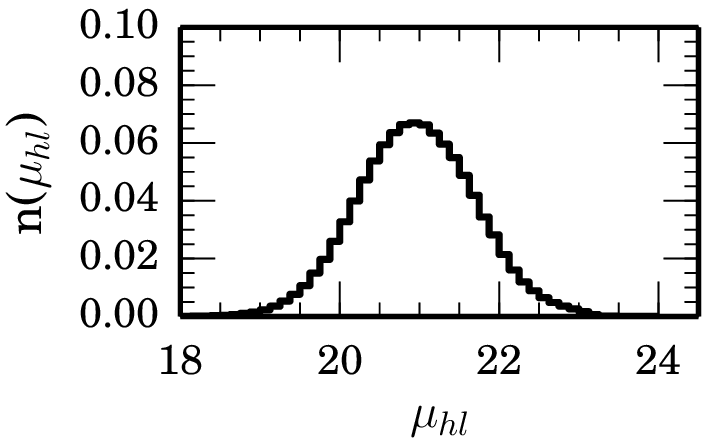}
\label{subfig:orig_dist:surfbright}}
\subfloat[ ]{
\includegraphics[width=0.3\columnwidth]{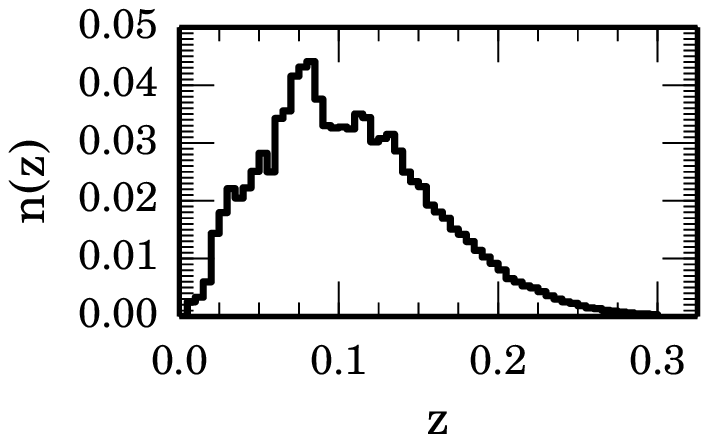}
\label{subfig:orig_dist:z}}
\subfloat[ ]{
\includegraphics[width=0.3\columnwidth]{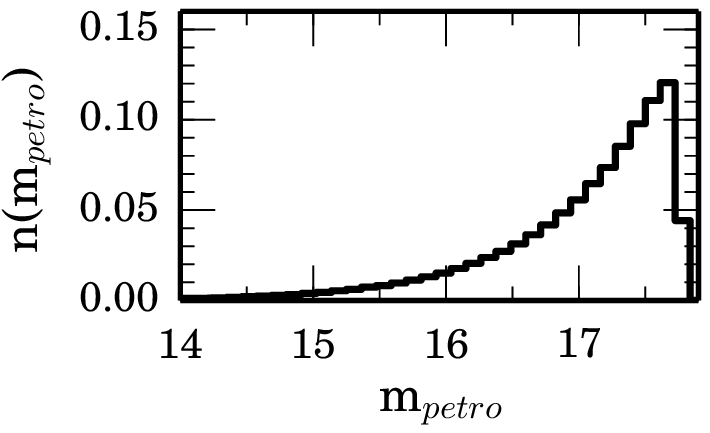}
\label{subfig:orig_dist:appmag}}
\end{center}
\end{minipage}

\begin{minipage}{\linewidth}
\begin{center}
\subfloat[ ]{
\includegraphics[width=0.3\columnwidth]{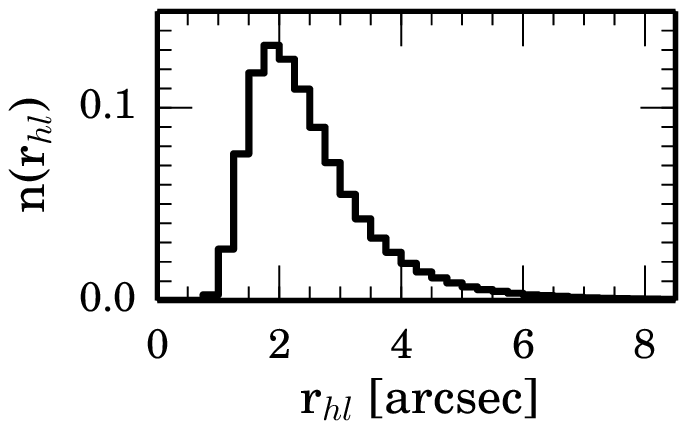}
\label{subfig:orig_dist:rad}}
\subfloat[ ]{
\includegraphics[width=0.3\columnwidth]{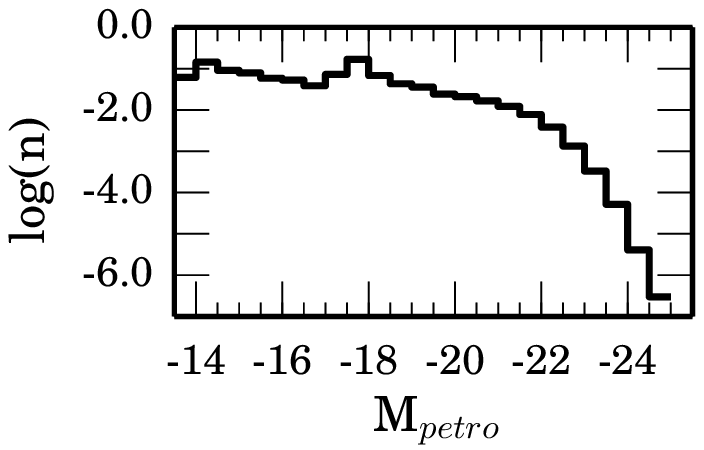}
\label{subfig:orig_dist:lum_func}}
\subfloat[ ]{
\includegraphics[width=0.3\columnwidth]{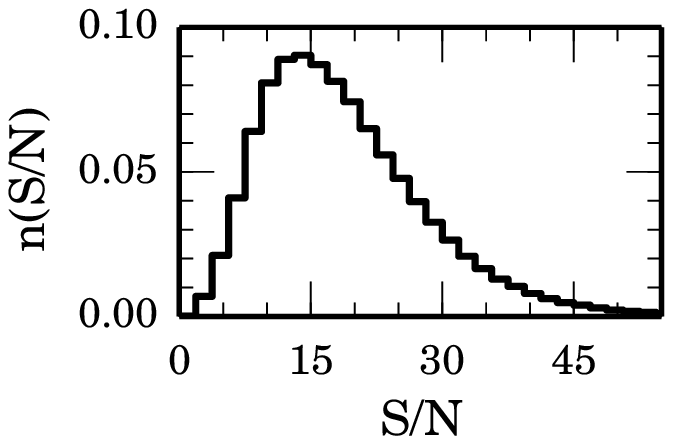}
\label{fig:SN}}
\end{center}
\end{minipage}

\caption{\textbf{\protect\subref{subfig:orig_dist:surfbright}}~The surface brightness distribution,\textbf{\protect\subref{subfig:orig_dist:z}}~redshift distribution, \textbf{\protect\subref{subfig:orig_dist:appmag}}~extinction-corrected {\em r} band Petrosian magnitude, \textbf{\protect\subref{subfig:orig_dist:rad}}~{\em r} band Petrosian half-light radius, \textbf{\protect\subref{subfig:orig_dist:lum_func}}~ V$_{\mathrm{max}}$-weighted LF, and \textbf{\protect\subref{fig:SN}} S/N distribution using the measurement of sky provided by the SDSS photometric pipeline of the sample used in this paper drawn from the DR7 SDSS spectroscopic galaxy sample. Bin counts are normalized {so that each distribution} integrates to 1.}
\label{fig:orig_dists}
\end{figure*}

\subsection{SDSS CasJobs data} \label{sec:data:CasJobs}

The data used in this analysis were drawn from the spectroscopic sample
of the Legacy area\footnote{A list of fields in the Legacy survey is provided at
\url{http://www.sdss.org/dr7/coverage/allrunsdr7db.par}} of the Sloan Digital
Sky Survey Data Release 7
\citep[DR7;][]{DR7}. The spectroscopic sample provides a
well-established sample with well-defined and tested selection
criteria. The criteria are presented in \cite{Strauss2002}. 

Galaxies listed in both the \texttt{PhotoObj} and \texttt{SpecObj} tables of the CasJobs DR7 data base that
satisfy three main selection criteria were selected. Those were: (1) the extinction-corrected {\em r} band Petrosian magnitude between
magnitude 14 and 17.77; (2) the \texttt{Photo} pipeline identified the object as a
galaxy (\texttt{Type = 3}); and (3) the spectrum was also identified as a galaxy
(\texttt{SpecClass = 2}). We place a limit at the faint end of 17.77 mag in the {\em r} band
because this is the lower limit for completeness of the SDSS Spectroscopic
Survey \citep{Strauss2002}. The limit of 14 mag at the bright end is used to exclude
large, nearby galaxies that are typically either too well resolved to be fitted with a
standard smooth light profile or shredded into multiple smaller objects in the
SDSS catalogue. These brightest galaxies may also be segmented over
multiple images or so large that it prevents robust estimation of
the background flux. Such galaxies require additional work to properly combine
neighbouring images \citep[see][for details]{blanton2011}.

The query used, omitting the names of selected data columns, is:\newline
\texttt{SELECT p.objid \ldots
 FROM photoobj as p \ldots LEFT OUTER JOIN SpecObj as s on p.objID $=$
s.BestObjID, segment g, field f,  chunk c WHERE
g.segmentID $=$ f.segmentID and f.fieldID $=$ p.fieldID and  c.chunkID $=$ g.chunkID 
and  (p.petroMag\_r - p.extinction\_r) between 14.0 and 17.77 and p.type $=$ 3 and
s.specclass $=$ 2 ORDER BY p.objid }, which produces 676\,010 matches. 

We apply additional cuts similar to \cite{Shen2003} and S11 to the data prior to fitting.
We remove all galaxies with redshift $<$ 0.005 (1647 galaxies). These galaxies have large apparent sizes 
and resolved structure that make decomposition difficult. We also remove 20 galaxies with redshift $>$ 1.0. 
Visual inspection reveals that these galaxies likely
represent catastrophic failures in the redshift code. 

Galaxies with saturated pixels as indicated by the \texttt{Photo} flags are also 
removed from the sample (3207 galaxies). 
In addition, as discussed in \cite{Strauss2002}, we apply a
surface brightness cut on the mean surface brightness within the Petrosian halflight radius of 
$\mu_{\textrm{50, r}}<23.0$ mag, where
\begin{equation}\label{eq:avg_sb}
\begin{aligned}
\mu_{\textrm{50, r}}&=m_{petro,\ r} + 2.5\log{(2\pi r_{petro,\ 50}^2)}\\
\end{aligned}
\end{equation} because there is
incomplete spectroscopic target selection at brightnesses below this threshold. 
After applying all the cuts, 5529 galaxies (approximately 0.8 per cent of the sample) are removed and
a sample of 670\,722 galaxies remains. We identify this as
our sample used throughout the paper. We consider the cuts described here in the 
completeness given in Section~\ref{sec:discussion}.

Figure~\ref{fig:orig_dists} shows the surface brightness distribution,
redshift distribution, extinction-corrected {\em r} band Petrosian magnitude,
{\em r} band Petrosian half-light radius, V$_{\mathrm{max}}$-weighted
LF, and signal-to-noise ratio (S/N) distribution of the sample used in
this paper. We define the S/N as the mean pixel flux within the half-light radius
divided by the noise associated with that pixel, or
\begin{equation}
\begin{aligned}
\left(\dfrac{\textrm{S}}{\textrm{N}}\right) &\equiv \dfrac{I_{\mu_{50}}}{N_{avg}}\\
N_{avg} &\equiv \sqrt{\dfrac{I_{\mu_{50}}+I_{sky}}{\textrm{gain}} + \textrm{dark variance}}\\
I_{\mu_{50}} &= 10^{-0.4(\mu_{50}-zp)} \times platescale^{-2}
\end{aligned}
\end{equation}
where $I_{\mu_{50}}$ is the source DN (`data numbers' or, equivalently, counts)\footnote{The counts
are related to the number of photo electrons collected by the detector through the gain of the detector amplifier. This distinction
is important since the photo electrons obey Poisson statistics.} of the average surface brightness defined in 
Equation~\ref{eq:avg_sb}. 
The zeropoint, $zp$, is calculated from the SDSS zeropoint, extinction, and airmass terms associated with each image.
The $platescale$ is used to convert the surface brightness from counts per square arcsecond to counts per pixel.
$N_{avg}$ is the noise in a pixel using the SDSS background
measurement as an estimate of the background flux and the average flux per pixel inside the Petrosian
half-light radius as the galaxy flux.

We collect all of the identifying data as well as photometric measurements obtained from the SDSS CasJobs server
into the table named the CasJobs Table distributed with the catalogue. 
For convenience, ID numbers were assigned to all galaxies 
contained in the catalogue to be used in place of
the SDSS ObjID. These ID numbers are referred to as \galcount{}
and used throughout the available data products presented here. 
The \galcount{} are used throughout the analysis as the unique identifier for each galaxy. 

Using the information obtained from the SDSS CasJobs server, we download all
necessary fpC images and PsField files from the SDSS. The PsField files provide the 2D reconstruction of
the point spread function (PSF) necessary for fitting. The fpC images contain the galaxy and surrounding neighbourhood.
We extract our own postage stamps from the fpC images for fitting rather than using the stamps provided
through the atlas images produced by SDSS. This process is described in more detail in Section~\ref{sec:fit:preprocess}.  

\subsection{Additional parameters} \label{sec:data:additional}

In addition to the data provided by CasJobs, we collect parameters from a number
of other studies. The Morphology catalogue of 
\citet[][hereafter H2011]{huertas10} is an automated morphological classification that used
a Bayesian SVM algorithm to classify all galaxies in the spectroscopic sample based on data 
available as part of the SDSS DR7. Objects are matched to our catalogue based on the SDSS DR7
 \texttt{objID}. The morphological parameters from the matching catalogue (H2011)
for our sample is described in the electronic catalogue.

We also calculate {\em K}-corrections, distance modulus, angular diameter distance, and V$_{\textrm{Max}}$ 
correction for each galaxy. {\em K}-corrections are calculated using version 4.2 of the {\em K}-correction code
\kcorrect{} described in \cite{blantonKcorr}. To calculate the
K-correction, the SDSS \texttt{modelmag} and \texttt{modelmag\_err} are used and
data for all band passes ({\em u,g,r,i,z}) are provided to the program. These terms are collected and provided with the data.
We assume a cosmology with ($H_0$,$\Omega_\Lambda$,$\Omega_\mathrm{m}$,$\Omega_k$) =(70 km s$^{-1}$Mpc$^{-1}$, 0.7, 0.3, 0.0) when necessary.

\section{The fitting process} \label{sec:fitting}

\begin{figure*}
\centering
\includegraphics[width=0.95\linewidth]{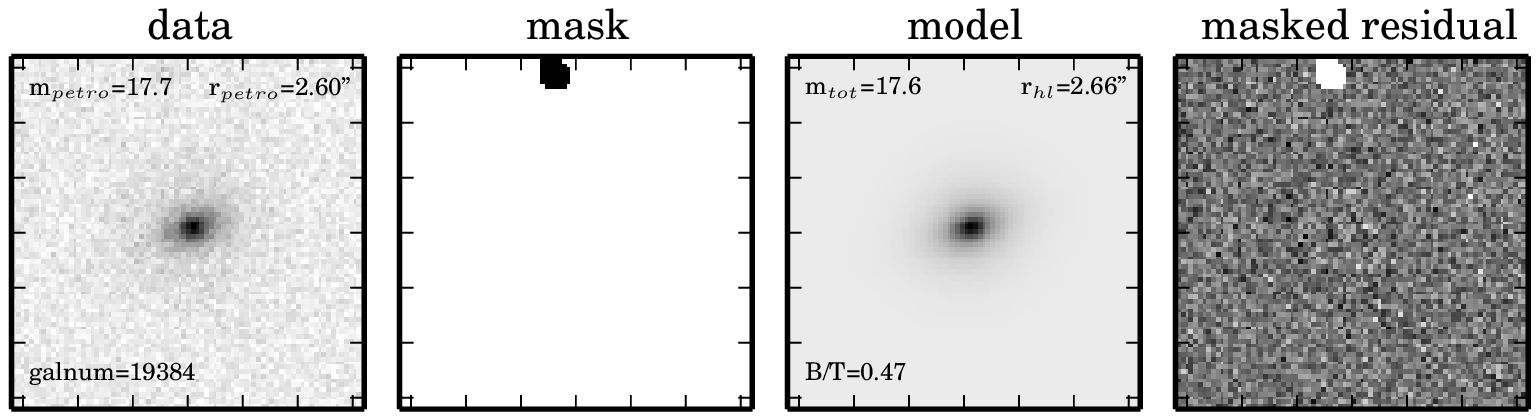}
\includegraphics[width=0.95\linewidth]{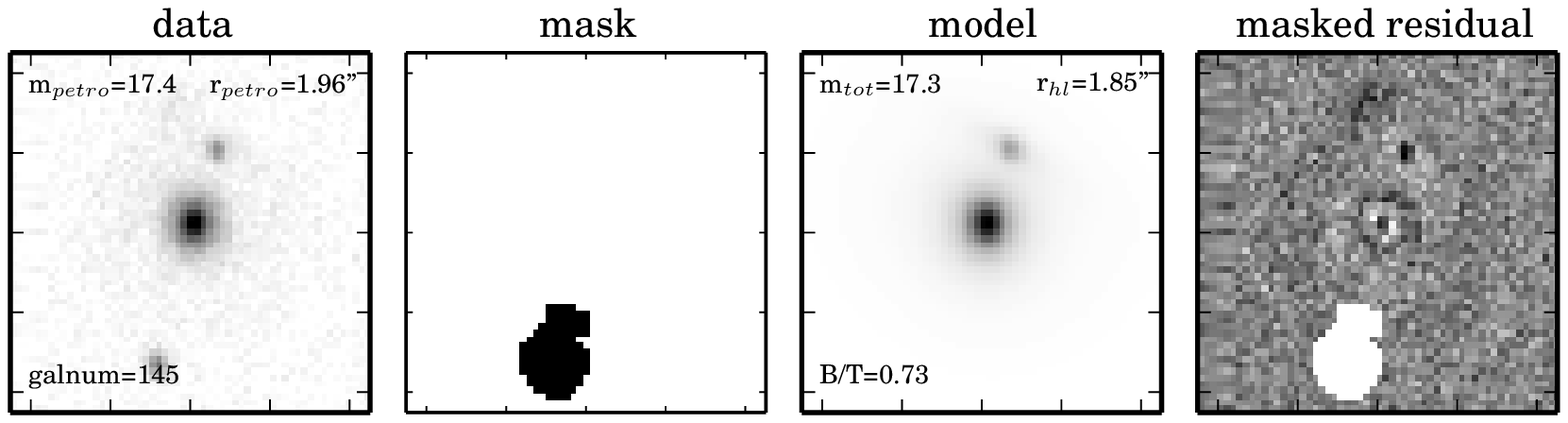}

\caption{Two example fits output by the \pymorph{} pipeline. \SerExp{} fits are shown. Each
column contains the input image (far left), mask (centre left), fitted galaxy (centre right), and the residual image (far right).
The first row shows a late-type galaxy. Some spiral structure is clearly evident in the residual. 
The second row shows an early-type galaxy with a smoother profile and a fitted neighbour.}
\label{fig:successful_fits}
\end{figure*}

In this section we describe the fitting process. Section~\ref{sec:fit:preprocess} describes our choice of cutout size and the 
data used for fitting. Section~\ref{sec:fitting:basic} describes the profiles used during fitting. Section~\ref{sec:fit:pipe} 
briefly describes the \pymorph{} pipeline used to fit the catalogue. Finally, Section~\ref{sec:mask_and_deblend} describes our masking and neighbour identification and focuses on how we verified the masking and simultaneous fitting in crowded fields where fitting is complicated and potentially biased by neighbouring objects.

\subsection{Pre-processing of SDSS images} \label{sec:fit:preprocess}

FpC images and psField files are the primary data used in the fitting procedure. 
The data were downloaded from the SDSS Data Archive Server. We used postage stamp
images of galaxies during fitting. Postage stamp images of each source
were extracted from the fpC image such that the stamp was 40 Petrosian
half-light radii on a side (20$\times$\texttt{petroR50\_r} from the centre
of the image to the edge) and centred on the target source. 
The decision to cut at 20 half-light radii is justified 
in M13 based upon simulations and provides a large number of background pixels
(about 30\,000-40\,000 pixels for an average-sized image).

In addition, a minimum size of 80 pixels on each side was set to ensure that enough
pixels were retained to properly determine the background. In
reality, with such a large postage stamp size, this minimum of 80 pixels is
rarely required.

A smaller cutout size could potentially be used when fitting the galaxy and
would reduce the time needed to fit each galaxy. Section~3.5 and Figure~12 of M13
show that the effect of further reducing the cutout size is insignificant. 
The main driver of postage
stamp size selection is to ensure that there are a sufficient number of
background pixels for sky estimation. Section~3.6 of M13 shows that the
estimate of background sky brightness is accurate to $\sim0.1$ per cent with a slight bias
towards underestimating the background level using this stamp size. This sky bias 
does not noticeably bias the other fitted parameters.

We also extract a PSF from the PsField files using the \readatlasimage{} program distributed on the SDSS 
website. \footnote{The use of \readatlasimage{} for PSF extraction is
described at \url{http://www.sdss.org/dr7/products/images/read_psf.html}}
The PSF provided by SDSS using the \readatlasimage{} program has a
standard image size of 51 pixels on each side.  

In addition, prior to fitting we remove the 1000 DN soft bias from the images
and PSF. We create sigma images from the SDSS image cutout following the
standard deviation calculation \footnote{See the SDSS DR7 online documentation at 
\url{http://www.sdss.org/dr7/algorithms/fluxcal.html} for further discussion.}
\begin{equation} \label{eq:weight}
W_{i,j} = \sqrt{\dfrac{F_{i,j}}{\textrm{gain}} + \textrm{dark variance}}
\end{equation}
where $W_{i,j}$ is the pixel sigma in DN, $F_{i,j}$ is the pixel flux (again in DN), $gain$ for the image as specified in SDSS CasJobs and used to account for the fact that
the photo electrons (rather than the DN) obey Poisson statistics, and `dark variance'
is the term used by SDSS to describe the contribution of the read noise and
dark current to the image noise.
Finally, we normalized the postage stamp and sigma images to a 1 s exposure prior to fitting.

\subsection{The fitted profiles}\label{sec:fitting:basic}

The S\'{e}rsic model has been used extensively in galaxy studies since first being 
proposed by \cite{Sersic1963}:

\begin{equation} \label{eq:sersic}
\begin{aligned}
& I(r)  = I_{e} \exp\left(-b_n\left[\left( \dfrac{r}{  R_{e}}
\right)^{\frac{1}{n}} - 1 \right]\right) \\
& b_n   = 1.9992n - 0.3271, 
\end{aligned}
\end{equation}
where  S\'{e}rsic index  ($n$), half-light radius ($R_{e}$), and surface brightness at
$R_e$ ($I_{e}$) are the parameters used to define the profile. $b_n$ uses 
the approximation from \cite{capaccioli1989} which is valid for $0.5<n<10$. When $n=4$, the
S\'{e}rsic model reduces to the de Vacouleurs model \citep{DeVacouleurs1948}. 
For the fitting presented here, the S\'{e}rsic index is restricted to values  less than or equal to 8.0.
Higher values of the S\'{e}rsic index are not allowed. Although such galaxies may exist in nature 
\citep[\eg][]{graham2007,kormendy2009}, 
we find that higher values of the S\'{e}rsic index are often associated with fitting problems in this sample. 

For two-component models, a de Vacouleurs or S\'{e}rsic model
is used to model the light in the central part  of the galaxy (often associated with the bulge) 
and an exponential disc is added to model the portion of the galaxy farther from the centre (often associated with the rotational disc). The exponential
model is defined by the scale radius ($R_{d}$) and central
surface brightness ($I_{d}$). The disc is modelled using the function
\begin{equation} \label{eq:expdisk}
I_{Exp}(r) = I_{d} \exp\left(\dfrac{-r}{  R_{d}}\right)
\end{equation}

The profiles defined in equations \ref{eq:sersic} and \ref{eq:expdisk} are 
1D profiles. The 1D profiles are used to generate 
2D models by also fitting a centre on the image, position angle ($\phi$), 
and axis ratio ($b/a$) to each component.

Equations~\ref{eq:sersic} and \ref{eq:expdisk} are often interpreted as representing a bulge and a disc, respectively. 
However, fitting two-component models to galaxies does not guarantee that the two components measured are truly present. 
Many early-type galaxies show no signs of disc-like structures. Similarly, many late-type galaxies show little or no sign of
a bulge in the central part of the galaxy. Also, when fitting multiple components, a significant second component may only 
indicate substantial departure from a single component profile rather than the presence of a physically meaningful second component. For example, \cite{gonzalez2005},  \cite{Donzelli2011}, and \cite{huang13} fit multiple components to Ellipticals (Ell) and BCGs without necessarily claiming the existence of additional physically distinct components. 
Also, the presence of a bar will affect fitting,
changing the ellipticity and  S\'{e}rsic index  of the bulge component in the two-component models. We examine the effects of a
bar component in Section~\ref{sec:mod_bar}.

We reserve judgement on the interpretation of the components until after discussion of the flagging system in Section~\ref{sec:validation}. We merely comment here that there are many cases in which the components 
should not be interpreted as a physically meaningful bulge and disc. 

We also note that the S\'{e}rsic model intended to represent the central region of two-component galaxies has broad wings 
and can have a large fraction of its total light at large radii. This is especially true at higher values of 
the S\'{e}rsic index and often causes the S\'{e}rsic model to dominate any exponential disc at large radii. We account for cases where the S\'{e}rsic component  dominates at large radii during the flagging procedure described in Section~\ref{sec:validation} and Appendix~\ref{app:autoflag} by focusing on the relative brightness of the two components only out to the radius where 90 per cent of the total light is enclosed. However, the reader should be aware that the 
S\'{e}rsic component may often dominate at radii beyond this point.

\subsection{The fitting pipeline}\label{sec:fit:pipe}

We performed both one- and two-component fits to the sample described in
Section~\ref{sec:data} using \pymorph{} \citep{pymorph}. \pymorph{} is a Python based
automated software pipeline built on \sextractor{} \citep[][]{sex} and the 2D fitting routine \galfit{} \citep{galfit}. Both \pymorph{} and \galfit{} have been extensively tested
\citep[see][for more tests of \galfit{}]{galfit,pymorph}. 

M13 also tested fitting in SDSS conditions of S/N, platescale, and seeing 
using simulated data. The test presented in M13 showed that estimates of the sky 
are expected to be accurate at the level of 0.05 per cent, and that fitted parameters are not 
significantly affected by sky errors at this level. In addition, fitting a \Ser{} profile to 
a galaxy with two components will cause a bias in size and magnitude at the 5-10 per cent level. When the appropriate
profile (\ie one versus two components) is chosen, magnitude and halflight radius are expected to be accurate
at the 5 per cent level (using 1 $\sigma$ errorbars). The flagging procedure presented in Section~\ref{sec:validation} is intended 
(among other goals) to reliably distinguish between one- and two-component fits, reducing any fitting bias resulting from fitting
a one-component profile to a two-component galaxy (see M13 for more information on the simulations and results).

We use the \pymorph{} pipeline to fit single S\'{e}rsic and single de Vaucouleurs
fits (\Ser{} and \Dev) as well as S\'{e}rsic + exponential disc (\SerExp) and 
de Vaucouleurs + exponential disc (\DevExp) models to each galaxy in the SDSS {\em r} band. 
In addition to photometric decomposition, \pymorph{} also supports measurement of
several non-parametric structural parameters. The final catalogue reports the auto magnitude (a Kron-like magnitude) and 
half-light radius measured by \sextractor{}. We also directly measure a total half-light radius and total axis ratio
for all models using the image of the fitted model. 
These measurements allow direct comparison of the one- and two-component model half-light radii 
(See Section~\ref{sec:ser_serexp_cmp} and Figure~\ref{fig:ser_serexp_meert}).

Two example fits and the corresponding residuals are presented in Figure~\ref{fig:successful_fits}.
Each row of the figure represents a different galaxy with the data, mask, fitted model, and residual presented
in columns 1, 2, 3, and 4, respectively. 

\subsection{Masking and neighbour identification}\label{sec:mask_and_deblend}

\begin{figure*}
\centering
\includegraphics[width=0.95\textwidth]{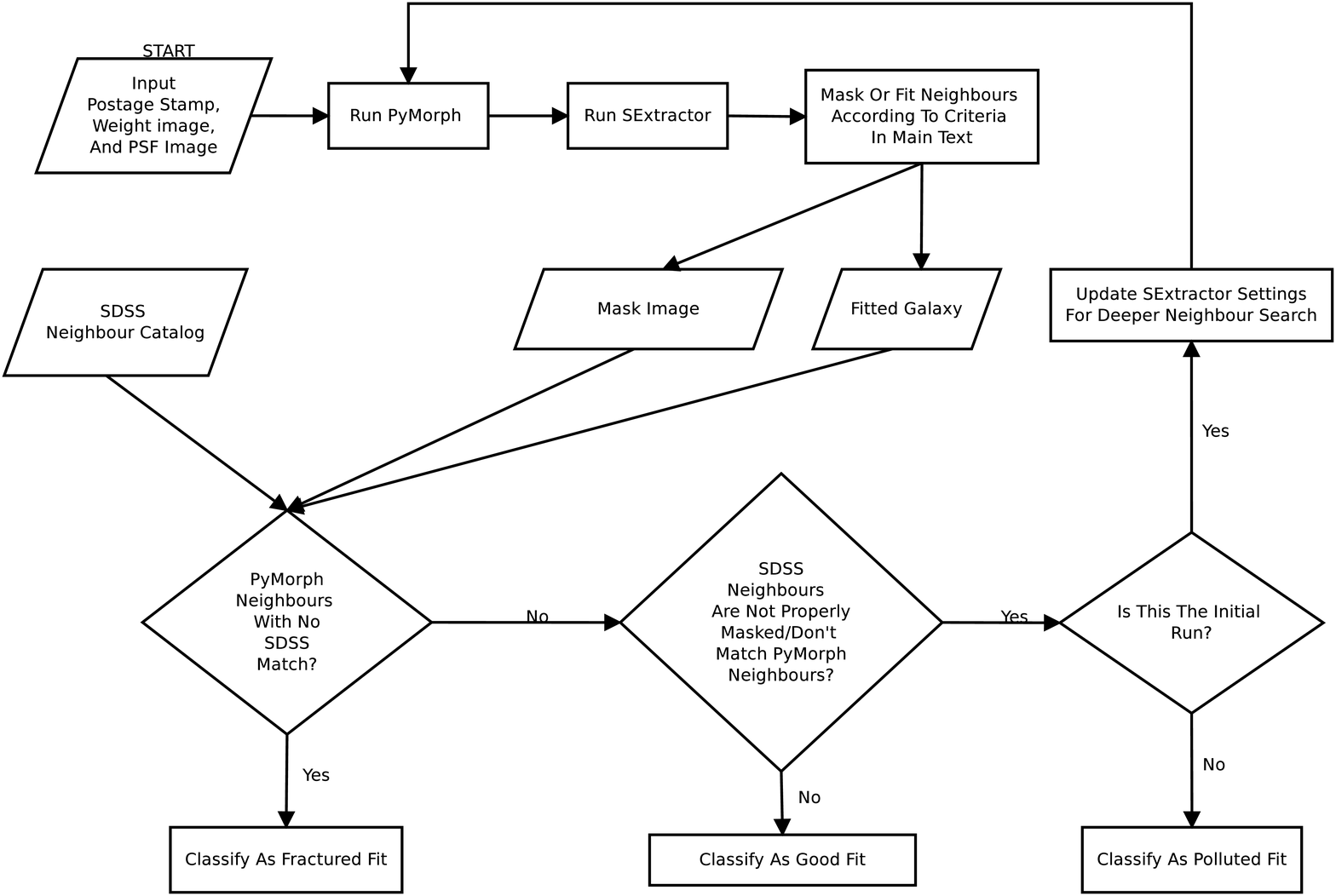}	
\caption{A flowchart showing the process of masking and simultaneous 
fitting described in Section~\ref{sec:mask_and_deblend}. The flowchart also 
includes the decision process used to identify `fractured' 
or `polluted' fits described in 
Section~\ref{subsec:mask_and_deblend_verify}. This entire process is carried
out for each galaxy individually.}
\label{fig:mask_flowchart}
\end{figure*}

\begin{figure*}
\centering
\includegraphics[width=0.95\textwidth]{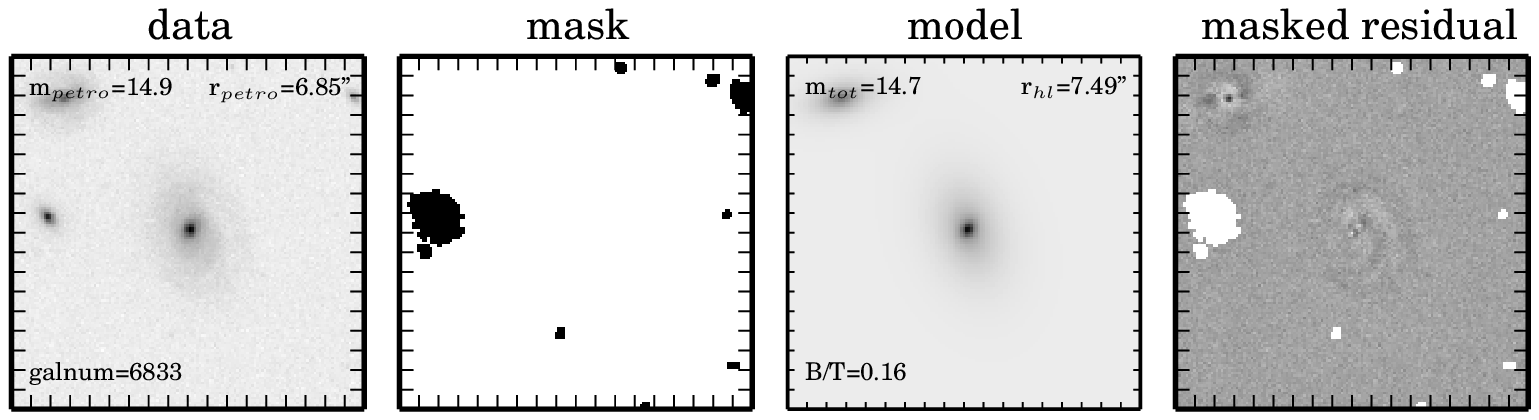}	
\includegraphics[width=0.95\textwidth]{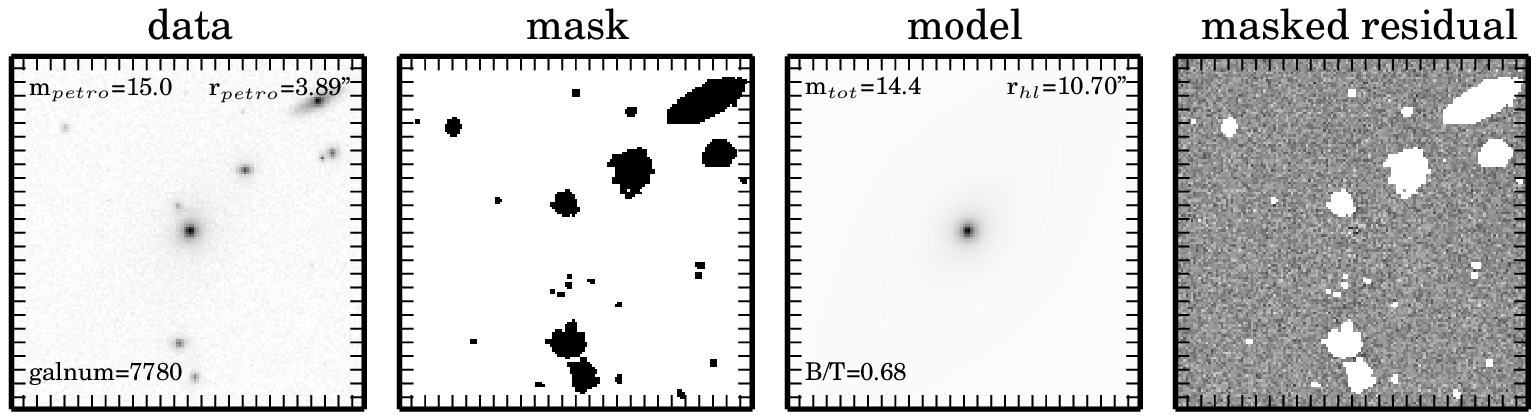}	

\caption{Examples of two galaxies with neighbours fitted simultaneously (top
row) or masked out (bottom row). From left to right, each row shows the input
image, mask created after initial analysis of the image, fitted image, and
residual. Regions that are masked out during fitting are shown in the mask
image as the black areas.}
\label{fig:mask_fig}
\end{figure*}

In this section, we focus on the effects of neighbours on the fitting 
process. We discuss our choices for masking and simultaneously 
fitting neighbours. We justify these choices and describe how galaxies with 
fitting problems caused by neighbours are identified in the catalogue.

Figure~\ref{fig:mask_flowchart} summarizes the process we used to identify neighbour sources, 
decide whether to mask or simultaneously fit the neighbour, and verify that the choice is appropriate. 
This process is described throughout Section~\ref{sec:mask_and_deblend}.

\subsubsection{The masking and neighbour identification process}\label{subsec:mask_neighbour_id}

\pymorph{} performs image masking using the \sextractor{} program \citep{sex}. Any sources
identified by \sextractor{} are masked out for fitting unless the extra source is
too close to the target galaxy to be properly masked. (The threshold for this
case is set by the user.) Neighbours that are not simultaneously fit
are masked according to the \sextractor{} segmentation image.
Simultaneous fitting of the target source and
neighbour source is performed in cases where the proximity
of the sources makes masking ineffective. In this case, the extra source is
simultaneously fitted with a single S\'{e}rsic profile while fitting the target galaxy.

\pymorph{} detects all neighbouring sources in the fitted frame using \sextractor{} with settings 
\texttt{BACK\_SIZE$=64$, ANALYSIS\_THRESHOLD$=1.5$} and \texttt{DETECT\_MINAREA$=6$}. 
We tested several values of these settings but found no effect on the final fits. 
Object detection is largely unaffected by varying the parameters.  

After searching the frame with \sextractor{}, the \sextractor{} catalogue is used to decide whether to mask or 
fit neighbouring sources according to requirements set on the minimum fractional size of a neighbour relative to the target
and the maximum separation between the neighbour and target in multiples of the sum of the half-light radii. 
When both conditions are satisfied simultaneously by a specific neighbour-target combination, the neighbour 
is simultaneously fitted with a \Ser{} profile. Otherwise, the area occupied by the source as 
defined by the \sextractor{} segmentation image is masked out during fitting.

For the fits presented in this paper, neighbouring
sources are simultaneously fit when the separation of the target and neighbour
source is less than three times the sum of the two objects' semi-major half-light
radii as measured by \sextractor{}. The neighbour source must be larger than 20 per cent of the 
area defined by the \sextractor{} radius of the target galaxy 
(\ie $r^2_{sextractor, neighbor}/r^2_{sextractor, target}>0.2$).

We tested several masking methods, using circular and elliptical masks with sizes 2, 4, and 6 times the size 
of the neighbour sources. These masking configurations provided no reduction in the scatter of the recovered 
parameters when tested on simulations. Since no improvement was evident, 
we used the default configuration for analysis (\ie masks are drawn according to the \sextractor{} segmentation 
image produced during fitting). While this likely leaves unmasked light from neighbouring sources, 
it has no effect on our determination of sky due to the large number of sky pixels.
Different masking and handling of neighbouring sources was carried out by S11 as well as in \cite{barden2012}. 
Our masking generally covers a smaller area around each source when compared to
 these two works. The insensitivity of the sky estimation to this choice of masking is likely the result of the large cutout sizes used in our 
fitting, which reduces sensitivity of sky estimation to stray light from masked neighbours.

Figure~\ref{fig:mask_fig} shows some examples of galaxies fitted with \Ser{}
profiles. Each galaxy has masked or simultaneously fit neighbours. The decision
of whether to mask or fit the neighbour was carried out as described in this section. 
Each row shows the input image, mask used during fitting,
the final fitted image, and the residual of the fit. The half-light radius of
the target galaxy and the neighbour jointly determine whether the neighbour is
masked or simultaneously fit.

\subsubsection{Verifying masking and deblending conditions}\label{subsec:mask_and_deblend_verify} 

After fitting, we verify that neighbours are properly identified and masked or 
simultaneously fitted by comparing the catalogue containing each targeted galaxy and any fitted neighbours 
against the five nearest Primary Photometric targets from the DR7 CasJobs PhotoPrimary table. We require that the 
CasJobs objects have Petrosian {\em r} band magnitude brighter than 20 and be within 9 arcsec of the original galaxy fitted by \pymorph{}. 
Objects farther or dimmer than these cuts are unlikely to cause fitting problems because they are more than 
$\sim5$ times dimmer and over 4 half-light radii away from the average galaxy in our catalogue.  

There are $\approx78\,000$ potential neighbours in SDSS and $\approx 100\,000$ simultaneously fitted \pymorph{} neighbours 
(about 79\,000 galaxies have one neighbour, about 8000 galaxies have two neighbours, and about 1500 galaxies 
have three or more simultaneously fitted neighbours, adding up to produce the $\approx 100\,000$ neighbours).

Since the average number of \pymorph{} neighbours per galaxy is small, there should be 
good agreement between the \pymorph{} neighbours and the brightest neighbours
found in SDSS. Therefore, we match our \pymorph{} neighbours to SDSS neighbours by 
cross-matching the two catalogues with a search radius equal to the {\em r} band 
Petrosian half-light radius of the SDSS neighbour (which is usually a few arcseconds). 
If a \pymorph{} neighbour and SDSS neighbour are separated by less than this radius, we consider the two matched.

About 40\,000 of the 100\,000 fitted \pymorph{} neighbours match SDSS objects. The remaining 60\,000 unmatched \pymorph{} neighbours
and $\approx38\,000$ unmatched SDSS neighbours may cause problems during fitting. Following the procedure described in this
section, we show that the number of galaxies with problems due to neighbours is only $\approx$4000 galaxies in the final sample.
 
The two groups, unmatched SDSS neighbours and unmatched \pymorph{} neighbours, possibly represent two different 
failures of the pipeline. Unmatched \pymorph{} neighbours may be spurious neighbours fitted by \pymorph{} 
after \sextractor{} improperly separates the target galaxy into several smaller fragments. Unmatched SDSS objects may be nearby 
neighbours that fail to be deblended from the target galaxy by \sextractor{}.

The vast majority of unmatched \pymorph{} neighbours do not negatively affect fitting. 
These are objects that are dimmer than 20 mag in the {\em r} band or farther than 9 arcsec from the target galaxy, 
and do not have SDSS matches as a result of our original SDSS neighbour catalogue selection. Only 
a small number of unmatched \pymorph{} neighbours are close to the target galaxy. In the largest and nearest galaxies, detailed galaxy structure (\eg spiral arms or dust lanes) triggers improper separation and deblending.
We call these cases `fractured' galaxies and identify them by searching for the fitted galaxies where the {\em r} band 
Petrosian magnitude is at least 0.5 mag brighter than the fitted magnitude and the unmatched fitted neighbour is less than 3.0 Petrosian half-light radii from the target galaxy. These cases are quite rare, representing only about 150 `fractured' galaxies. 
The remainder of unmatched \pymorph{} neighbours (almost all of the 60\,000 \pymorph{} neighbours) do not affect the fit because of their separation from the target and small magnitude.

Unmatched SDSS neighbours tend to happen 
when neighbours are superimposed on the target galaxy or very nearby. In principle, these situations
can happen independent of the true physical separation of the neighbour and target (\ie a star may be superimposed on a galaxy, the two of which should not be correlated in any physically meaningful way). The incidence of nearby neighbours may also be enhanced in 
dense environments (\ie within clusters). This is a potentially important effect as it can bias our measurements of galaxies in 
clusters. 

Many SDSS galaxies may be masked rather than simultaneously fit, causing it to appear that \pymorph{} misses many 
neighbours. In order to examine the SDSS neighbours for contamination of our target galaxies, we examined the 
\pymorph{} fitting masks to verify that these neighbours are not masked out. Any masked neighbours (for which at least 80 per cent of the pixels inside their half-light radii are masked) are removed from our set of potentially problematic SDSS neighbours. This removes 
approximately 4000 SDSS neighbours. We also remove SDSS neighbours that possess the \texttt{DEBLEND\_NOPEAK} {\em r} band flag in the SDSS data (this indicates that no peak was found in the  deblended source by the SDSS photo pipeline, and the source is likely to be a spurious source). 

Any SDSS neighbours that pass these cuts potentially corrupt our fits. We find empirically that a safe cut for considering these 
unmasked, unfit neighbours to be problems is if they are less than 3 half-light radii away and 
\begin{equation}\label{eq:neighbour_cut}
 m_n-m_t < 4.0-r_d/r_{hl}
\end{equation}
where $r_d$ is the radial distance from the target galaxy to the neighbour, r$_{hl}$ is the target galaxy half-light radius, $m_n$ 
is the neighbour magnitude and $m_t$ is the target magnitude. For example, a neighbour galaxy 4 arcsec away from an average
galaxy in our sample (with halflight radius of 2 arcsec) may be no more than 2 mag dimmer than the target in order for
us to be concerned that it was not chosen for simultaneous fitting. This cut naturally tapers, allowing brighter neighbours to 
be considered problematic out to larger radii relative to dimmer neighbour galaxies. After this cut, only about 7500 target galaxies potentially have contamination from unmatched SDSS neighbours.

Once the images with unfitted neighbours are identified, we perform an additional run of \pymorph{} on the smaller sample. 
This run uses a different, deeper set of deblending settings for \sextractor{} as well as more generous settings of \pymorph{} neighbour fitting. The \sextractor{} setting are changed to \texttt{DEBLEND\_NTHRESH  64} rather than the original \texttt{DEBLEND\_NTHRESH  32} and \texttt{DEBLEND\_MINCONT  0.001 } rather than the original \texttt{DEBLEND\_MINCONT  0.005 }. This produces more fragmented sources with less contrast required to determine that a source is actually two blended objects. Changing these
settings not only increases the likelihood of detecting the SDSS neighbours, but also of incorrectly `fracturing' larger sources.
We also reduce the neighbour source area requirement in \pymorph{} for this fitting run. This makes \pymorph{} more likely to perform a neighbour fit, requiring the neighbour to be only 10 per cent of the area of the target rather than the original setting of 20 per cent area.

After refitting we again test for unfit or unmasked neighbours. Galaxies still having neighbour problems at this point are flagged in
the final catalogue as potentially being polluted by neighbour objects. This process improves the fits of $\sim3500$ galaxies, 
or 0.5 per cent of the total sample, for which deeper fitting identifies previously unmasked neighbours.

\begin{figure}
\includegraphics[width=0.97\columnwidth]{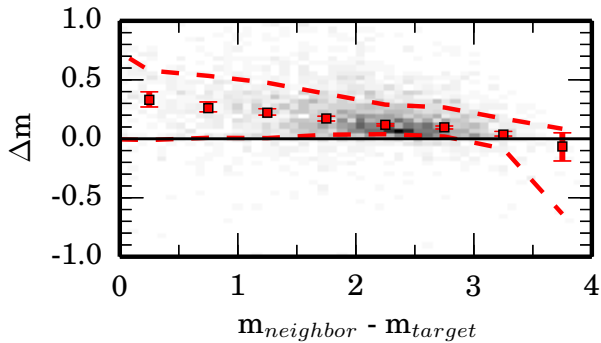}
\caption{The change in magnitude (the fitted magnitude from the second pass minus the original fitted magnitude) of galaxies for which the fitting is rerun with deeper deblending and all the expected neighbours are detected. Individual galaxies are shown as grey scatter points. The median and 68 per cent contour are over-plotted in red. The {\em x}-axis shows the difference in target and neighbour magnitudes. Galaxies tend towards dimmer magnitudes on the second fit which is consistent with a reduction in the contamination of 
the target galaxy by light from the neighbour source. Note that by a difference of 3 mag the median change in the parameters is consistent with zero. There is no need to look at unmatched SDSS sources dimmer than this.}\label{fig:deblend_refit}
\end{figure}

Figure~\ref{fig:deblend_refit} shows the change in fitted \Ser{} magnitude (the fitted magnitude from the second pass minus the original fitted magnitude) for galaxies where all neighbours were detected on a second pass of fitting using deeper deblending. Individual galaxies are shown as grey scatter points. The median and 68 per cent contour are overplotted in red. The {\em x}-axis shows the difference in target and neighbour magnitudes. 

The plot shows that on average, the target galaxies get dimmer on a second pass as light from the neighbour is now properly associated with a different source (the neighbour) rather than being fitted as part of the target galaxy.  A difference between target and neighbour of more than 3 mag causes little change in the fitted magnitude. 

According to the 68 per cent contours on the plot, approximately 16 per cent of the refit galaxies  become brighter during refitting. Since the original fits may include unmasked SDSS neighbours, the fits are usually poor. In some cases, the fit is not just contaminated by the neighbour, but the fit fails catastrophically, producing unreliable magnitudes and other fitting parameters. In these cases, the original fit would have been classified as a failed fit using our flagging system. Such fits may produce a brighter magnitude when refit with proper identification of the neighbour target.

Similarly (although not shown here), separations of the source and target of more than 3 target half-light radii show little-to-no change in magnitude. The other fit parameters (\ie  S\'{e}rsic index , axis ratio, and radius) exhibit similar behaviours. Based on these observations, we believe our cutting criteria in Equation~\ref{eq:neighbour_cut} to be generous enough to capture the
majority of cases where a true SDSS neighbour is causing fitting contamination.

We find about 4000 `polluted' target galaxies for which the second pass with deeper deblending still fails to find the neighbour 
identified in SDSS. The `polluted' and `fractured' galaxies jointly comprise about 0.6 per cent of the sample. 
We mark galaxies that are suspected to have substantial fitting problems with a quality flag discussed 
in Section~\ref{sec:validation} and shown in Table~\ref{tab:finalflag_percents}. We also report the number of simultaneous fits for each galaxy in our catalogue since 
such fits substantially increase the number of free parameters during fitting and make
it much more likely that the resulting fit has problems. 

\section{Flagging and classification of good and bad fits}\label{sec:validation}

\begin{figure*}
\centering
\includegraphics[width=0.9\linewidth]{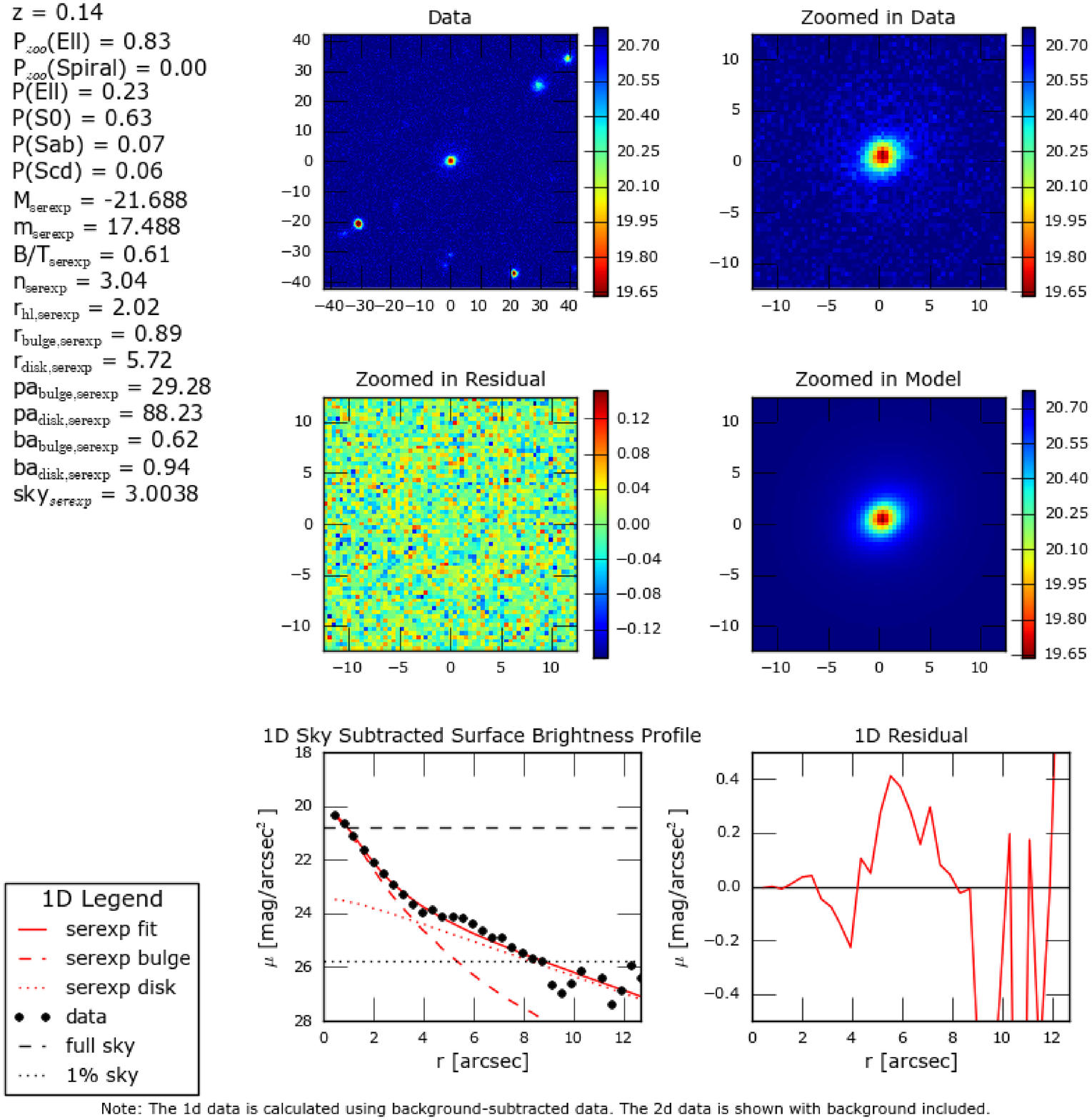}
\caption{An example panel used during visual classification showing the original stamp image, zoomed in image of the galaxy, fitted galaxy, 2D residual, 1D galaxy profile, and 1D residual. In addition, many of the fitting parameters are displayed on the left-hand side of the image for comparison. When classifying the galaxy visually, we look for \Ser{} components that dominate at large radii (which indicates a non-physical bulge), as well as mis-aligned components, large residuals, and other indications of bad fits.} 
\label{fig:vis_panel}
\end{figure*}

\begin{table*}
\begin{tabular}{l l l l c  c  c}
\textbf{Flag bit} & \multicolumn{3}{l}{\textbf{Descriptive category}} &  \textbf{Per cent \SerExp{}} & \textbf{Example galaxy} \\ \hline \hline
-- & \multicolumn{3}{l}{\textbf{Good total and component magnitudes and sizes}}&  39.055 &  \\ \hline
10 & & \multicolumn{2}{l}{\textbf{Two-component galaxies}} &  39.055 & Figure~\ref{fig:finalflag_good}, \ref{fig:some_flag}, \ref{fig:flip_com}\\
11 & & & No flags &  18.095 &  Figure~\ref{fig:finalflag_good} \\
12 & & & Good \Ser{}, good \Exp\ (some flags) &  19.417 &  Figure~\ref{fig:some_flag} \\
13 & & & Flip components &  1.543 &  Figure~\ref{fig:flip_com} \\ \hline
-- & \multicolumn{3}{l}{\textbf{Good total magnitudes and sizes only}} &  54.945 &  \\ \hline
1 & & \multicolumn{2}{l}{\textbf{Bulge galaxies}} &  18.964 & Figure~\ref{fig:finalflag_flags_bulge:no_exp}, \ref{fig:finalflag_flags_bulge:ser_dom} \\
2 & & &No \Exp{} component, n$_{\Ser}>$2&  7.074 & Figure~\ref{fig:finalflag_flags_bulge:no_exp} \\
3 & & &\Ser{} dominates always &  11.889 & Figure~\ref{fig:finalflag_flags_bulge:ser_dom} \\
4 & & \multicolumn{2}{l}{\textbf{Disk galaxies}} &  25.146 & Figure~\ref{fig:finalflag_flags_disk:no_ser}, \ref{fig:finalflag_flags_disk:no_exp_flip}, \ref{fig:finalflag_flags_disk:ser_dom_flip}, \ref{fig:finalflag_flags_disk:exp_dom}, \ref{fig:finalflag_flags_disk:parallel_com} \\
5 & & & No \Ser{} component &  16.876 & Figure~\ref{fig:finalflag_flags_disk:no_ser} \\
6 & & & No \Exp{}, n$_{Ser}<$2, Flip Components &  0.551 & Figure~\ref{fig:finalflag_flags_disk:no_exp_flip}\\
7 & & & \Ser{} dominates always, n$_{\Ser}<$2 & 0.103 &  Figure~\ref{fig:finalflag_flags_disk:ser_dom_flip}\\
8 & & & \Exp{} dominates always &  2.872 & Figure~\ref{fig:finalflag_flags_disk:exp_dom}\\
9 & & & Parallel components &  4.745 &  Figure~\ref{fig:finalflag_flags_disk:parallel_com}\\
14 & & \multicolumn{2}{l}{\textbf{Problematic two-component galaxies}} & 10.835 & Figure~\ref{fig:finalflag_flags_2comprob:ser_outer}, \ref{fig:finalflag_flags_2comprob:exp_inner}, \ref{fig:finalflag_flags_2comprob:good_ser_bad_exp}, \ref{fig:finalflag_flags_2comprob:bad_ser_good_exp} \\
15 & & & \Ser{} outer only &  7.504 & Figure~ \ref{fig:finalflag_flags_2comprob:ser_outer} \\
16 & & & \Exp{} inner only &  0.425 & Figure~ \ref{fig:finalflag_flags_2comprob:exp_inner} \\
17 & & & Good \Ser{}, bad \Exp{}, B/T$>=$0.5 &  0.017 & Figure~\ref{fig:finalflag_flags_2comprob:good_ser_bad_exp} \\
18 & & & Bad \Ser{}, good \Exp{}, B/T$<$0.5 & 0.625 &  Figure~\ref{fig:finalflag_flags_2comprob:bad_ser_good_exp} \\ 
19 & & & Bulge is point & 2.237 &  Figure~\ref{fig:finalflag_flags_2comprob:tinybulge} \\ \hline \hline
20 & \multicolumn{3}{l}{\textbf{Bad total magnitudes and sizes}} & 6.000 &  \\
\end{tabular}
\caption{A breakdown of the automated flags characterized into categories useful for analysis. The first two groups can be used for analysis of total fits. When examining the sub-components, consideration should be given as to exactly which groups of fits should be included. }
\label{tab:useful_flag_breakdown}
\end{table*}

While the fitting process is fairly straightforward, automated fitting
routines tend to produce many poor fits. Non-physical
ellipticity or sizes can easily be produced, especially if the sky is not
properly estimated or neighbours are not properly accommodated. 
Inverted two-component profiles (where the component
intended to fit the bulge fits the disc of a galaxy) are possible when the
bulge has low S\'{e}rsic index, or when the S/N of the bulge is too low. Extreme cases of face-on or edge-on galaxies can also present difficulty.
Overfitting (\ie fitting components not truly present in the galaxy just to improve the \chinu) 
is easy to do, producing meaningless results for the components. 

In this section we discuss the flagging method used to separate good fits from poor fits in our catalogue.
We attempt to provide a straight-forward way of determining which fits to use and an interpretation of the fits that we
claim to be good. Section~\ref{sec:flag_motivation} gives the motivation for our flagging system. 
Section~\ref{sec:flag_vis} briefly describes the visually classified galaxies used to design and tune our flags. 
Section~\ref{sec:flag_auto} describes the automated flags resulting from 
our visual inspection and gives a breakdown of the flags for the \SerExp{} catalogue (a more detailed description of 
the flags are available in the supplementary online material provided in 
Appendix~\ref{app:autoflag} and example cases for each flag are presented in Appendix~\ref{app:example_gals}).

\subsection{The motivation for the flagging system}\label{sec:flag_motivation}

Independent inspection of the $\chi^2$ values from each fitted model
(\Dev{}, \Ser{}, \DevExp{}, and \SerExp) is a poor indicator of properly fitted models in our 
catalogue for some reasons. The  most important reason is that while the $\chi^2$ measures 
the ability of the model to appropriately fit the data, it makes no distinction between 
physical and unphysical models. Also, the number of degrees of freedom (DOF) is not well 
defined for non-linear models \citep[see][for a more in-depth discussion]{Andrae2010}, 
and the distribution of residual values do not approximate a normal distribution. 
Although the number of DOF's and the resulting probability are not well defined, 
minimization of the $\chi^2$ can still produce the `best' fit by minimizing the $\chi^2$ value. 
However, the statistical likelihood associated with the fitted $\chi^2$ is not the best measure 
for determining physically meaningful fits. 

The focus of this work is not only on producing good fits, but on separating the cases where
a second component is needed from those where the second component is not needed. 
We would also like to separate cases where the best fit is physically meaningful from those fits that are unphysical. 
For this, analysis beyond the $\chi^2$ value is required.  
In place of the $\chi^2$ test, we devise a series of physically motivated flags outlined in Table~\ref{tab:useful_flag_breakdown}.
We use these flags to determine the reliability of the various fits and the individual subcomponents.
We also use the flags to mark poorly fitted galaxies. 

The final goal of our flagging system is to identify galaxies as being in one of the following categories: 
\begin{enumerate}[leftmargin=0.25in]
 \item bulge-like galaxies,
 \item disk-like galaxies, 
 \item two-component galaxies, and
 \item unknown type/poorly fitted/failed fits.
\end{enumerate}
In the \Ser{} and \Dev{} fits, these conditions simplify somewhat as we do not have to evaluate the appropriateness of the subcomponent parameters. Rather than focus on these broad (and perhaps vaguely defined) categories for classification, we develop a series of quantitative indicators based on the fitted parameters and assess the indicators to determine the quality of the fits.

Several methods of quality assessment have been used in previous works. \cite{Allen2006} separated galaxies based on the 1D radial profile, comparing the \Ser{} and \Exp{} components to separate galaxies into one of seven categories. More recently, \cite{mendel2014} applied a similar criteria to distinguish one-component galaxies from galaxies that are better fit by two-component \DevExp{} models. A different approach is to apply a statistical test similar to S11 and LG12, which perform an {\em F}-test to separate meaningful fits from galaxies that are likely overfit by the more complex models. Our approach is similar to the former groups rather than the latter. We separate the galaxies using a categorical 
description of the fits and show that these categories match the expected distributions of properties when compared to other 
observables like magnitude and radius. We compare the results of our method to the results of S11, LG12, and \cite{mendel2014} in Sections~\ref{sec:analysis}~and~\ref{sec:other_cmp}.

\subsection{The visual classification}\label{sec:flag_vis}

No catalogue of $10^5$ galaxies can be visually inspected in a reasonable amount of time without employing a large number of observers
following a procedure similar to Galaxy Zoo \citep[GZ;]{galzoo01}. We constructed a small training set of visually classified fits 
that were used to design our flagging criteria. For purposes of training and validation, we randomly selected a 1000 galaxy 
sample from the fitted galaxies for visual classification. The visual classification system has several categories 
including general fit problems and image characteristics. Each of the authors classified the sample by examining the 
fitted parameters, the 2D image, 2D model, 2D residual, the 1D profile and, the 1D residual. 

Figure~\ref{fig:vis_panel} shows an example of the plots used to manually inspect the fits. Visual classification allowed us to enumerate the types of problems commonly found in the fits (\ie effects of contamination or poorly defined bulges) and understand the relative importance of each common problem. The visual sample was selected randomly from the catalogue and fairly represents the full catalogue in photometric parameters (\eg magnitude, radius, redshift, \etc). The incidence of problems associated with fitting in the visually classified sample is expected to be similar to that of the full catalogue. However, rarer problems are potentially missed in such a small sample.

After visual classification, the sample was randomly divided into two samples of approximately equal size, a training sample 
and a test sample. The training sample was used to define the automated flags described in Section~\ref{sec:flag_auto}. The test 
sample was set aside and used to test the reliability of the automated flags once they were defined.

\subsection{The automated flagging system}\label{sec:flag_auto}

\begin{figure*}
\begin{minipage}{\linewidth}
\begin{center}
 \subfloat[Galaxy 35521]{\includegraphics[width=0.45\linewidth]{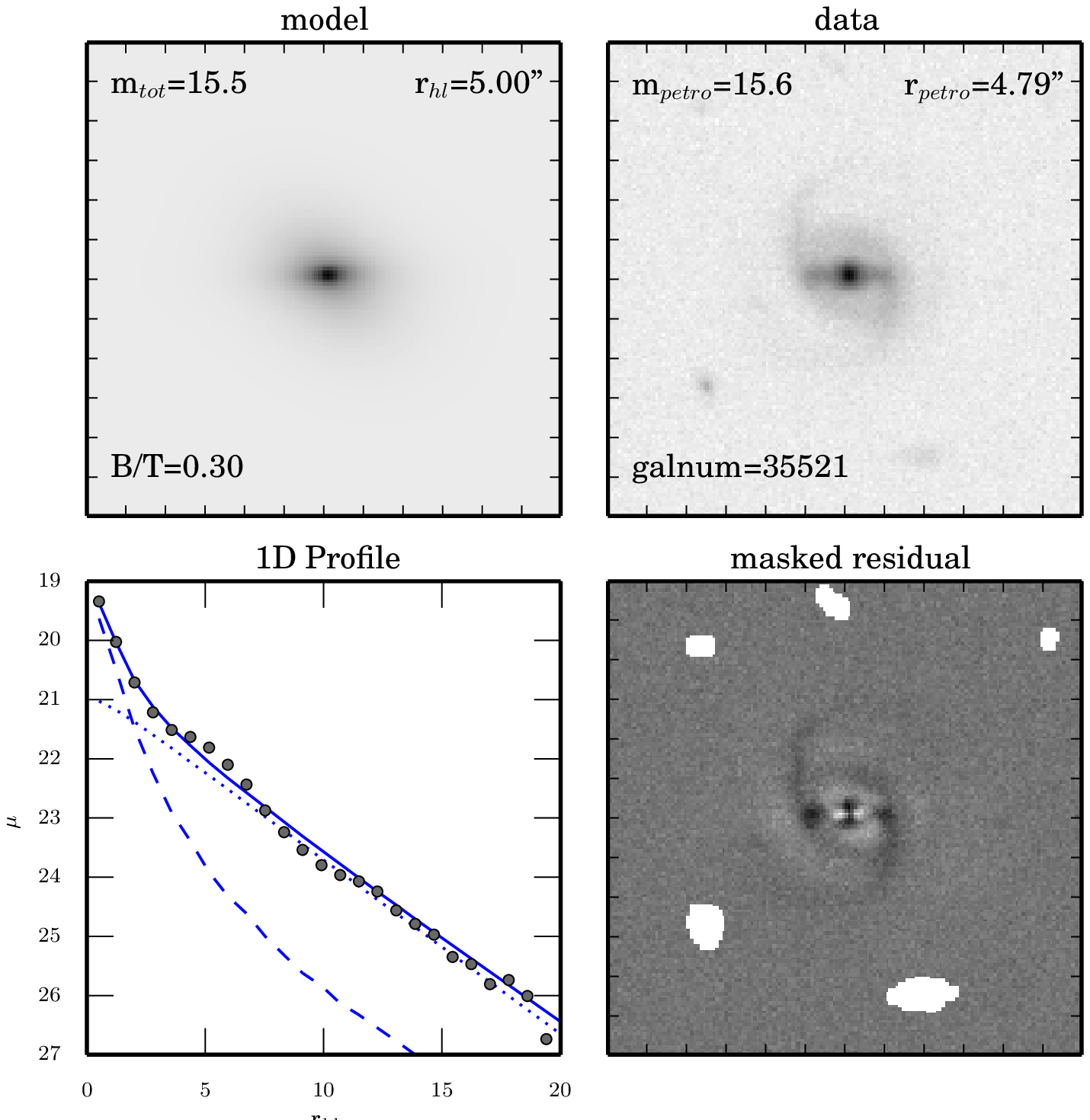}} \,
 \subfloat[Galaxy 408774]{\includegraphics[width=0.45\linewidth]{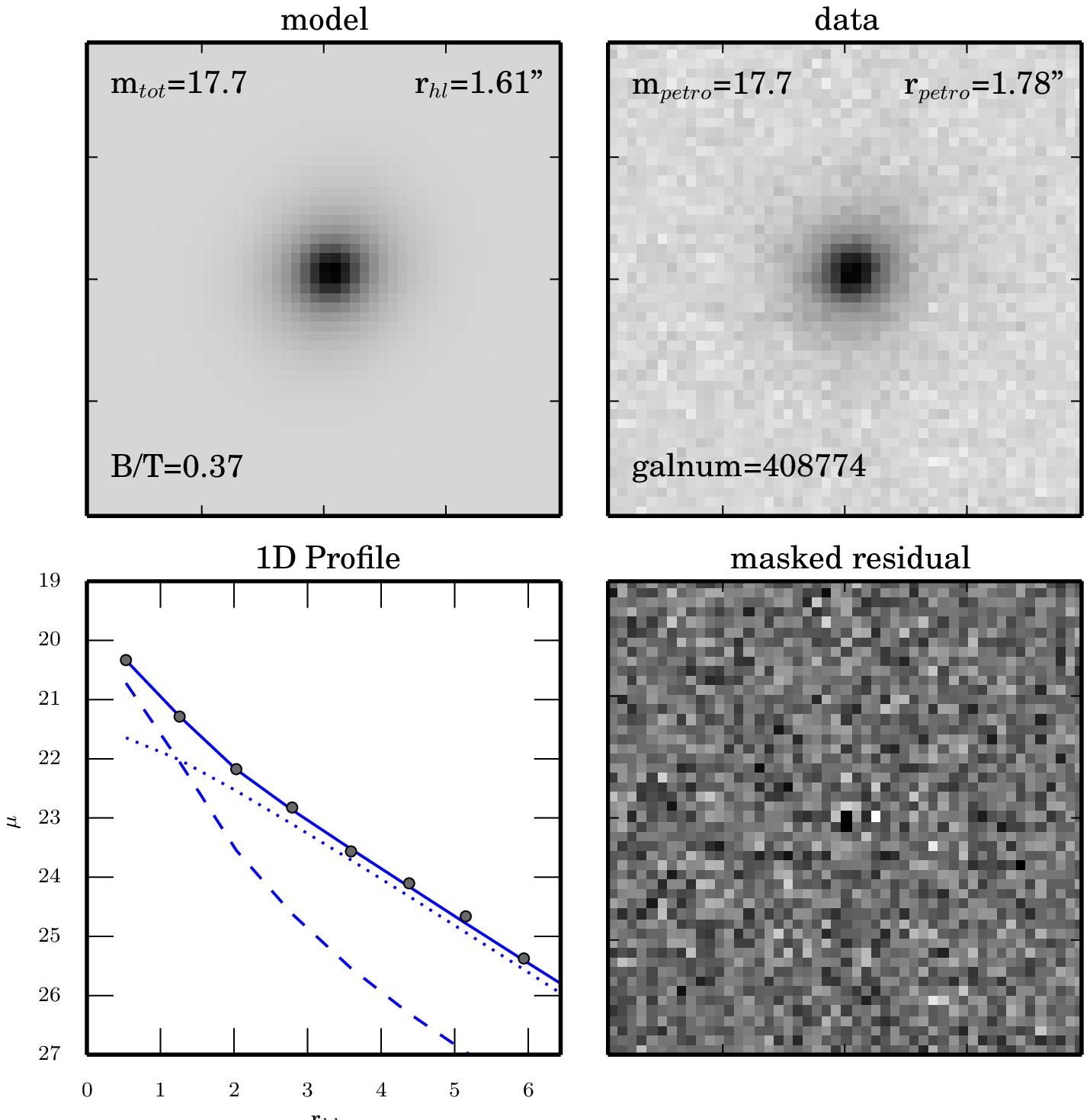}} \,  
\end{center}
\end{minipage}
 \caption{Examples of  \SerExp{} fits which we flag as good; \ie the two components of the fit are real and without fitting problems. These galaxies are in category `No Flags' in Table~\ref{tab:useful_flag_breakdown} and have flag bits 10 and 11 set.}\label{fig:flag_example_text}
\end{figure*}

The automated bit flags are designed to accurately identify problems commonly observed during visual examination of the fits.
Each automated flag has a tunable cutoff that is optimized using the training set of manually classified galaxies. 
Then the efficacy of the flag is evaluated using the test set of galaxies (which are also visually classified). 
Appendix~\ref{app:autoflag} (available online) describes the flags in more detail including the decision criteria for the flags (see Table~\ref{tab:useful_flag_definitions}) and the percent of galaxies possessing each flag (see Table~\ref{tab:finalflag_percents}).  

Table~\ref{tab:useful_flag_breakdown} shows the flagging categories, the final percentage of \SerExp{} galaxies in each category, and references to example galaxies for each flag. Example galaxies selected to characterize each flag are presented in Appendix~\ref{app:example_gals} (available online). Figure~\ref{fig:flag_example_text} shows an example of galaxies with bits 10 and 11 set. These panels and additional examples of other flag categories are available in
Appendix~\ref{app:example_gals}. We show two example galaxies for each flagging category. For each fit, we show the 2D data, 
fitted model, and residual. We also show the 1D radial data profile, bulge and disc component profiles, and the total fitted profile. 

We use a tiered structure to describe the fits. The most general description is the accuracy of the total magnitude and radius. About 94 per cent of the \SerExp{} sample is classified as having an accurate measurement of the total magnitude and half-light radius. Within the first tier assessing the accuracy of the total magnitude and radius, we then separate galaxies into 
single-component galaxies (flag bits 1 and 4), two-component galaxies (flag bit 10), and problematic two-component galaxies (flag bit 14) that have fit parameters that are difficult to interpret.

We recommend using the categories presented in Table~\ref{tab:useful_flag_breakdown} to select samples from the catalogue. 
In particular, we highlight the major categories of our flagging as follows.
\begin{itemize}[leftmargin=0.25in]
\setlength{\itemindent}{-.15in}
 \item {\bf Good Two-Component Fits (bit 10 set)}--\\
 These are the galaxies we find to have two fitted components with intermediate 
 bulge-to-total light ratio (B/T) and reasonably well behaved subcomponents. We recommend using both the subcomponents and the total magnitude and radius in 
 any analysis. Bits 11-13, only one of which will be set, give a more specific description of the galaxy. 
 \item {\bf Good Bulge Fits (bit 1 set)}--\\
 These are the galaxies we find to have little or no evidence of an \Exp{} 
 second component. The B/T can be as low as 0.8, however, the \Exp{} component is not trustworthy. We recommend using the \Ser{} fit for these galaxies, and they should be treated as having B/T$=$1.0 regardless of the fitted B/T.
 Bits 2 and 3, only one of which will be set, give a more specific description of the galaxy.
 \item {\bf Good Disk Fits (bit 4 set)}--\\
 These are the galaxies we find to have little or no evidence of a \Ser{} 
 second component. The B/T can be as high as 0.2, however, the \Ser{} bulge component of the \SerExp{} fit is not trustworthy. We recommend using the \Ser{} fit total magnitude and radius in any analysis. The \Ser{} fit should be used for galaxies, and they
 should be treated as having B/T$=$0 regardless of the fitted B/T. 
 Bits 5-9, only one of which will be set, give a more specific description of the galaxy.
 \item {\bf Problematic Two-Component Fits (bit 14 set)}--\\
 These are the galaxies we find to have two fitted components 
 with intermediate B/T, but at least one sub-component has strange behaviour. 
 These galaxies likely require additional investigation prior to including them in any analysis.
Bits 15-19, only one of which will be set, give a more specific description of the galaxy.
 \item {\bf Bad Fits (bit 20 set)}--\\
 These are the galaxies we find to have severe problems with the fit. They should not be 
 included in any analysis without close examination. Even the total magnitude and total radius are believed to have significant
 errors.
 Bits 21-26, only one of which will be set, give a more specific description of the galaxy.
\end{itemize}

For the single-component \Ser{} and \Dev{} fits, the two-component categories listed above have no galaxies in them, but we retain the same flagging structure for all fits to ensure consistency and ease of use.

The flagging here focuses on identifying \Ser{} versus \SerExp{} fits. While the \Ser{} fit can reproduce the \Dev{} fit in cases 
where the S\'{e}rsic index is 4, it will be different for other values. We choose to carry out and report \Dev{} fits for comparison purposes. Similarly, the \DevExp{} fit is included as a separate fit because the 
\SerExp{} fit is not necessarily equivalent to the \DevExp{} fit.

\section{Internal comparisons and consistency checks}\label{sec:best_model}

\begin{figure*}
 \centering
\includegraphics[width = 0.95\linewidth]{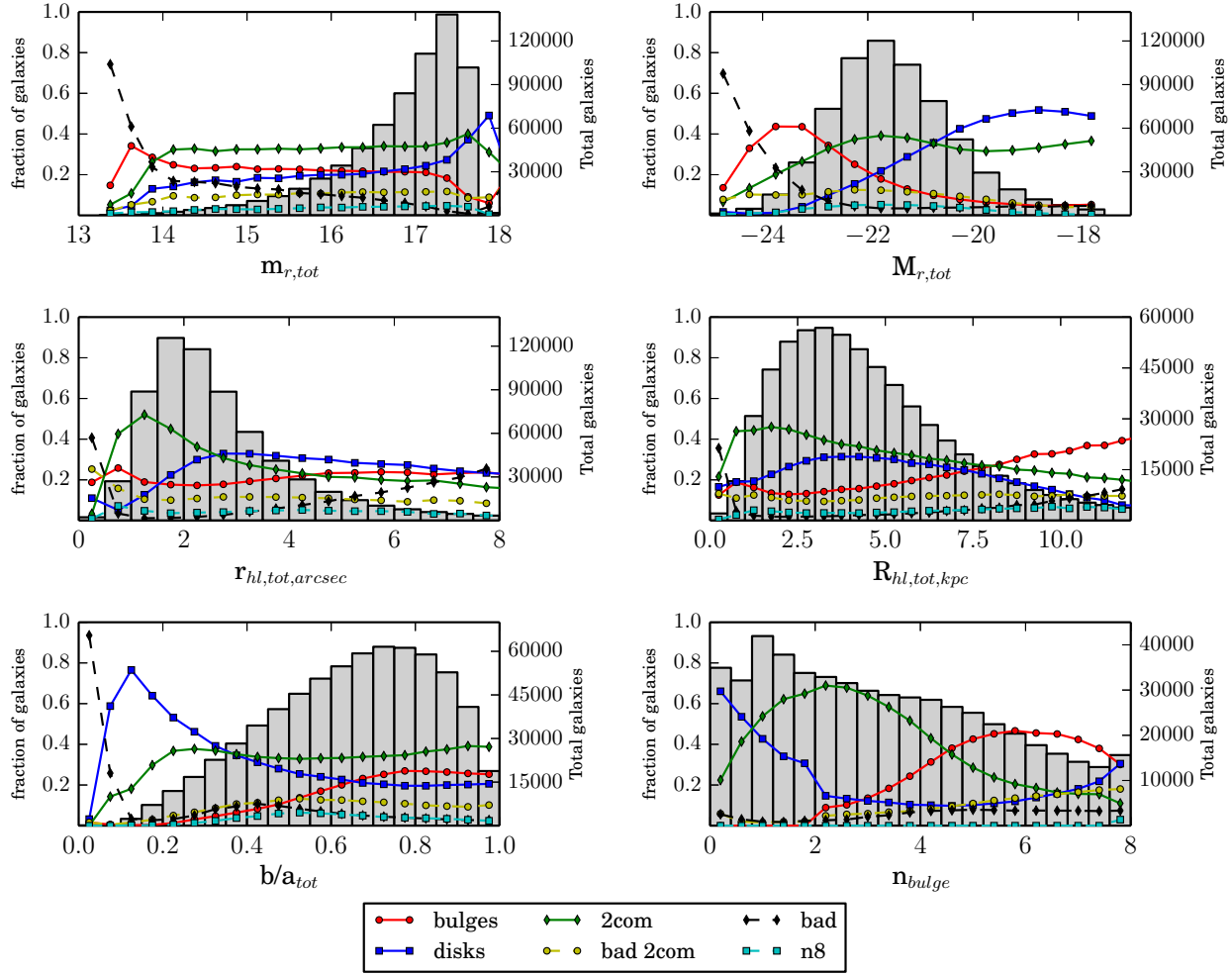}
\caption{The bin-by-bin percent of galaxies of each type: disk (blue points, labelled as `discs'), bulge (red points, labelled as `bulges'), two-component (green points, labelled as `2com'), problematic two-component (yellow points, labelled as `bad 2com'), and failed galaxies (black points, labelled as `bad'), according to our categorical flags. We have also separated the good two-component models into those with S\'{e}rsic indices below 8 and those galaxies with acceptable fits but the  S\'{e}rsic index  of the bulge hits the $n=8$ boundary of the parameter space (cyan points, labelled as `n8') to check for any bias resulting from the restriction on the fitted S\'{e}rsic index. In the background of the plot, we plot the total distribution of galaxies with respect to the parameter used to bin the data. For example, there are approximately 60\,000 galaxies in our catalogue with apparent magnitudes between 16.5 and 16.75 (from the top-left panel). Of these 60\,000 galaxies, about 37 per cent are good two-component fits, 20 per cent are bulges, 20 per cent are discs, and the remaining $\sim20$ per cent are a mixture of the remaining classes.  When summing the percentages over all model types, each bin sums to 100 per cent. The percentage of two-component galaxies is mostly stable with respect to apparent size and magnitude. However, the data favour more two-component models at small half-light radii. This effect is examined in the text.}
\label{fig:flag_by_type}
\end{figure*}

\begin{figure*}
 \centering
\includegraphics[width = 0.95\linewidth]{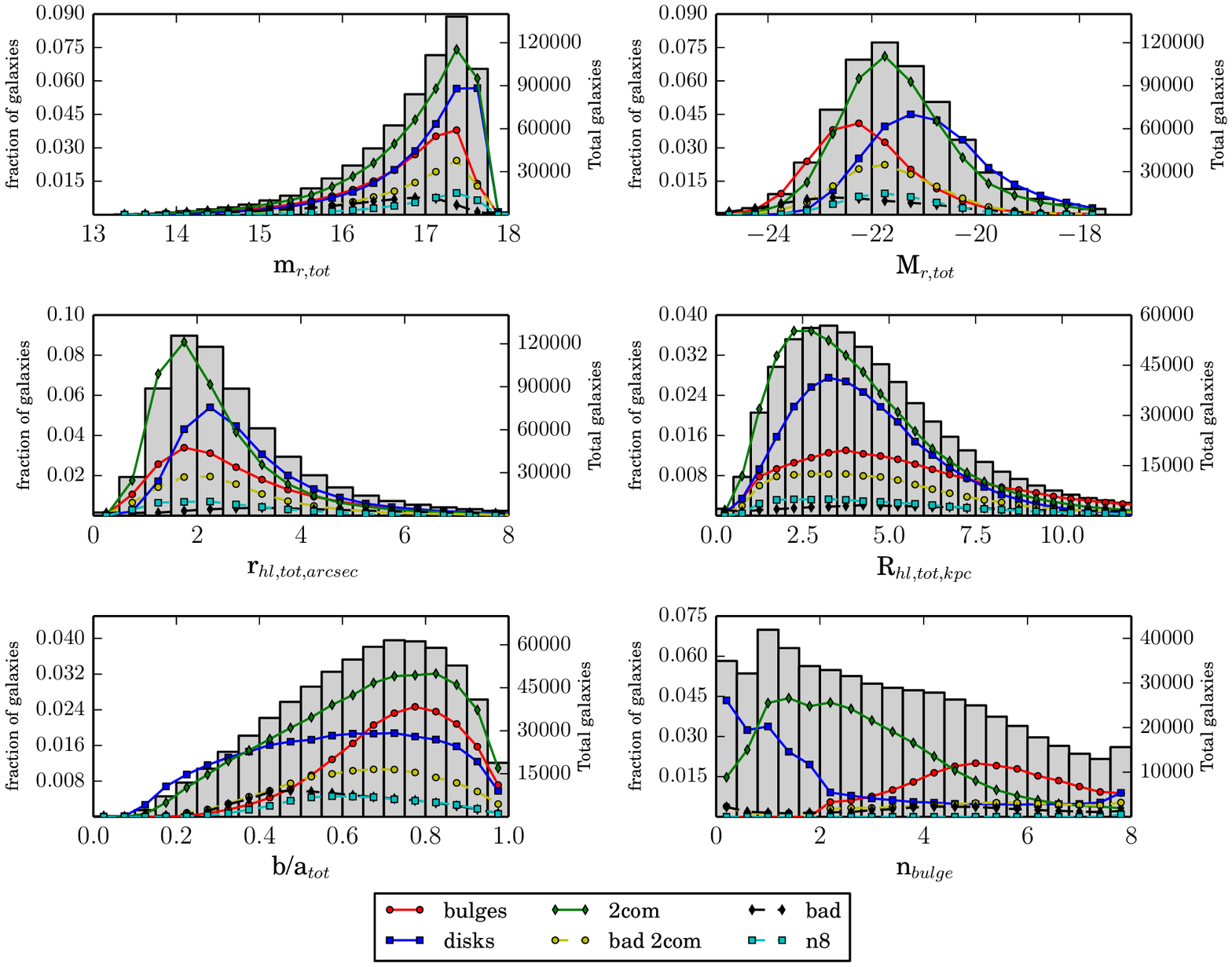}
\caption{The percent of galaxies of each type, given as a percentage of the total sample. The colours and plot design are the same as Figure~\protect\ref{fig:flag_by_type}. For example, approximately 3 per cent of all galaxies are `2com' galaxies fitted with  apparent magnitudes between 16.5 and 16.75 (from the top-left panel)}
\label{fig:flag_by_type_total_per}
\end{figure*}

Before comparing the fits, in particular the \SerExp{} fits, to other groups, we perform some internal consistency checks.
Section~\ref{sec:flag_obs} examines the distribution of general classifications described in Section~\ref{sec:flag_auto}
for the \SerExp{} catalogue as a function of basic observables. 
Section~\ref{sec:bulge_rad_psf} examines the bulge radii with respect to the PSF.
Section~\ref{sec:ser_serexp_cmp} compares some of the fitted values of the \Ser{} fit to the \SerExp{} fit.

\subsection{Examination of flags with basic observables}\label{sec:flag_obs}

Using the flags from Section~\ref{sec:flag_auto}, we should be able to reproduce sensible distributions of 
magnitude, radius, and ellipticity provided that our chosen flags can reliably separate good and bad fits within our sample. 
We also include the bulge S\'{e}rsic index, which is used during classification.
In Figure~\ref{fig:flag_by_type} and Figure~\ref{fig:flag_by_type_total_per}, we show the percentage of two-component (green points, labelled as `2com', flag bit 10), bulge (red points, labelled as `bulges', flag bit 1), 
disc (blue points, labelled as `discs', flag bit 4), 
problematic two-component (yellow points, labelled as `bad 2com', flag bit 14), 
and failed galaxies (black points, labelled as `bad', flag bit 20) in the \SerExp{} catalogue as a function of several parameters.
The figures show the same distribution of the categories using different normalizations. 
Figure~\ref{fig:flag_by_type} is normalized such that the sum of the fractions of all categories sums to 1 in each bin (\ie the
reported fractions for the blue line with square points show the fraction of galaxies at a given magnitude, size \etc{} that are discs). Figure~\ref{fig:flag_by_type_total_per} is normalized by the total number of galaxies in the sample (\ie the
reported fraction for the blue line with square points show the fraction of all galaxies in the sample galaxies that are discs and have a given magnitude, size, or other measured parameter). 

In both figures, the total distribution of galaxies is plotted in the background in order to give the reader a sense of the number of galaxies in each bin. The left-hand axis denotes the fraction of galaxies in given category (\eg bulge, disc, or other category) using the relevant normalization. The right-hand axis denotes the raw number of galaxies plotted in the background histogram relative to 670\,722, the total number of galaxies in our sample.

We have also separated the good two-component models into those with \SerExp{} bulge S\'{e}rsic indices below 8 and those galaxies for which our flags indicate the fitted profiles are acceptable but the \SerExp{} bulge S\'{e}rsic index hits the $n=8$ boundary of the parameter space (cyan points, labelled as `n8'). Galaxies with the \SerExp{} bulge S\'{e}rsic index approaching the boundary of the parameter space may exhibit problems due to the inability of the code to reach the true minimum of the fit. We separate out these fits to check for any strange behaviour that would suggest substantial biases in the fitting due to this effect.

\subsubsection{Behaviour with apparent magnitude}

The original sample selected from SDSS is defined by a cut in extinction-corrected {\em r} band apparent Petrosian magnitude 
removing all galaxies brighter than 14 or dimmer than the spectroscopic sample cut of 17.77.
Examination of the type distribution with respect to total fitted apparent magnitude derived from the \SerExp{} fits by summing both components of the fit (top-left panel) reveals that the percentage of good two-component galaxies is independent of apparent magnitude across magnitudes 14--17.77 where the majority of our sample is located. Independence with apparent magnitude is expected unless there are observational biases in the flags. However, the endpoints (\ie below 14 and above 18) show a large increase in the percentage of failed galaxies. 

Since the Petrosian magnitudes are known to be dimmer at the brighter end of the magnitude distribution \citep[][]{blanton2001, bernardi2013} and dimmer for galaxies with higher S\'{e}rsic index (or greater concentration) \citep[see Figure 7 of][]{Graham2005}, galaxies with fitted magnitudes brighter than 14 (where our original selection cut is made; see Section~\ref{sec:data}) are not immediately identifiable as failed fits. Indeed, Petrosian magnitude is dimmer than fitted magnitude by approximately 0.2 for a \Ser{} galaxy with S\'{e}rsic index of 4 (however this increases to about 0.55 mag for n$=8$ galaxies). Larger concentration (or equivalently, S\'{e}rsic index) is required to account for larger differences between Petrosian and fitted magnitude. 
Similarly, as concentration increases, the proportion of galaxies possessing at least 
that level of concentration decreases. So, a reasonable expectation is that the failure rate should increase brighter than 14. In addition, fitted magnitudes should rarely be dimmer than the Petrosian magnitude unless there was a failure in deblending or some other photometry problem. Therefore, the majority of the galaxies outside the range of 14--17.77 mag should be flagged as failed cases because their magnitudes vary greatly from what is expected. 

The increasingly high failure rate at the bright end of the apparent magnitude plotted above 14 and the increase in the failure rate at the dim end below 17.77 are an indicator of the ability of the flags to identify poorly fit galaxies. Indeed, the majority of the failed galaxies ($\sim800$ out of the $\sim1000$) in the magnitude bins brighter than 14 have either flags 21 or 22 set. These flags identify galaxies with large, extended components that are due to underestimating sky brightness or contamination from nearby neighbours. This makes sense as we expect galaxies with these bright magnitudes to have their brightness substantially overestimated.

Classification of bulge, disc, or 2com galaxies is also increasingly unreliable in galaxies outside of the 14--17.77 mag range. Since these galaxies  are not selected in our original sample, we may not have an appropriate training sample for these galaxies. Therefore, careful consideration of any galaxies dimmer than 17.77 or brighter than 14 should be taken before including them in any analysis.

\subsubsection{Behaviour with absolute magnitude}

Figures~\ref{fig:flag_by_type} and \ref{fig:flag_by_type_total_per} also show the behaviour of our component categories with respect to total fitted absolute magnitude derived from the \SerExp{} fits by summing both components of the fit (top-right panel). Bulge galaxies (\ie elliptical) galaxies dominate at the brightest magnitudes while disc galaxies dominate at the dimmest magnitudes and two-component models dominate at the intermediate magnitudes. There is an increase in failed cases near the bright end of the distribution. Further inspection of this end of the distribution shows that it is a pile-up effect due to the preference of \pymorph{} to over estimate the brightness of a galaxy when a fitting failure occurs. Large components (either \Ser{} or \Exp{} components; both occur at similar rates) can be wrongly used by \pymorph{} to fit sky or neighbours. These large components will contribute to the brightness of the source, making it appear brighter than the true brightness. Therefore, the failed galaxies will tend to shift up the magnitude distribution to brighter magnitudes. 

To test for this problem, we also examined the distribution of galaxy types using the model-independent Petrosian magnitudes. 
This shifts many of the failed cases back to the dimmer Petrosian magnitude bins and smooths out the distribution of failed galaxies at the bright end. As a result, the failure rate at the bright end is substantially lower when viewed in Petrosian magnitudes. The failure rate increases roughly linearly between --23 and --25 mag from about 10 to 30 per cent rather than increasing to 80 per cent as shown in Figure~\ref{fig:flag_by_type}. The percentage of failed cases in the magnitude range --23 to --22 also increases. This shows that a small failure rate in the dimmer bins (magnitudes between --23 and --22) is causing a substantial contribution to the number of galaxies in the brighter bins (brighter than --23). However, these failed cases far outnumber the legitimate galaxies in the brightest bins causing the apparent failure rate to approach or exceed 50 per cent. 

The occurrence of two-component models is also moderately higher in the magnitude bins brighter than --23 when viewed in Petrosian magnitudes. The recent works of \cite{mosleh13} and \cite{Davari14} report improved fitting and recovery of magnitude and radius when using two-component models for nearby elliptical galaxies (as opposed to galaxies at $z\sim1$). \cite{huang13} also provide evidence for using three components when fitting well-resolved elliptical galaxies (requiring resolution substantially better than 1 kpc).
The question of exactly how to interpret such an additional component is beyond the scope of this work, so we only comment on the trend here and caution against a simple bulge+disc interpretation of these galaxies. We will return to this issue in Section~\ref{sec:other_cmp}.

\subsubsection{Behaviour with apparent and absolute half-light radius }\label{subsec:app_abs_rad}

Figure~\ref{fig:flag_by_type} and \ref{fig:flag_by_type_total_per} also show the behaviour of the percentage of our component categories with respect to the total absolute half-light radius derived from the \SerExp{} fits by measuring the halflight radius of the total profile (second row, right). Here, we observe expected trends in physical size (larger physical size should be dominated by bulge galaxies). When observing in apparent size (second row, left), the incidence of two-component galaxies increases with smaller apparent size and peaks at 1.5 arcsec, above the half-width at half-maximum (HWHM) 
of the PSF (which is about 0.7 arcsec in the {\em r} band). At these sizes, the percentage of discs drops substantially while the percentage of bulges remains constant.

The shift in the percentage of galaxy types at small half-light radii is consistent with an interpretation that the observed shift to two-component galaxies is due to observational effects of the magnitude limit on the distribution of galaxies rather than systematics in the fitting. For example, Figure~\ref{fig:flag_by_type} shows that the peak of the pure disc sample is near 5 kpc in size. This is near the typical size of late-type discs for galaxies at --21 {\em r} band Petrosian magnitude in the SDSS sample \citep[see Figures 5 and 6 of][]{Shen2003}. The --21 magnitude also corresponds to the peak of the disc galaxy distribution in Figure~\ref{fig:flag_by_type_total_per}.  When this size (5 kpc) is translated to an apparent size at $z\approx0.05$ the expected size of these disc galaxies would be nearly 4 arcsec, well above the sizes where this effect occurs. Even at $z\approx0.15$, which is higher redshift than approximately 80 per cent of our sample, the expected size of these disc galaxies would be about 2 arcsec. Therefore, it is reasonable not to expect many disc galaxies below 2 arcsec where the drop in pure disc systems occurs. 

Even if such a drop in disc galaxies was believed to be fitting bias, PSF effects have been shown to set in near the HWHM of the PSF (see Section~\ref{sec:bulge_rad_psf} for justification and further discussion). The drop between 1 and 2 arcsec is above the HWHM of the PSF where sizes are potentially biased by the PSF. 

A similar examination of bulge galaxies versus two-component galaxies is less conclusive. Both groups (bulge and two-component) should be present at the smaller radii (below 2 arcsec). 

\begin{figure}
 \centering
\includegraphics[width = 0.95\linewidth]{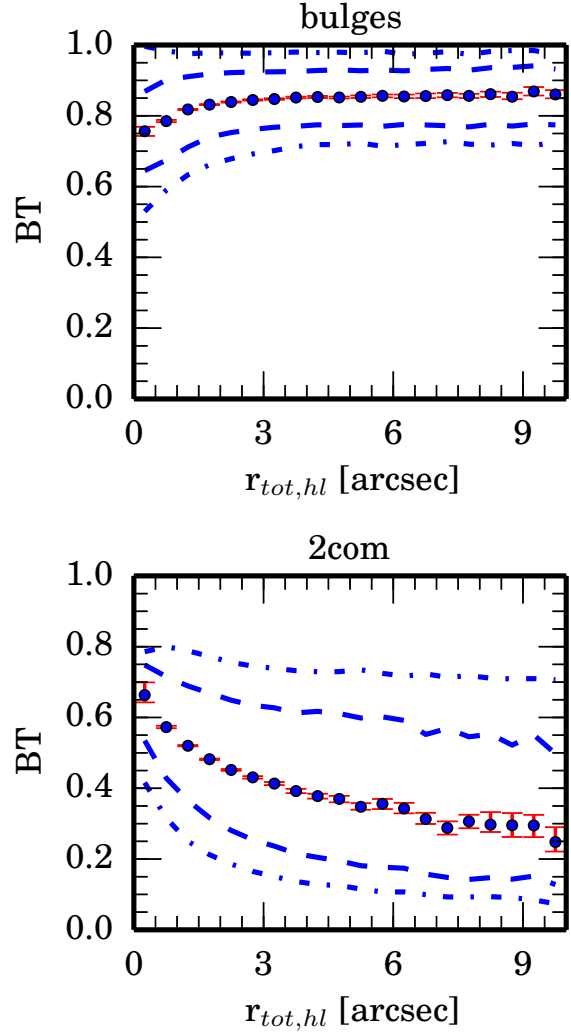}
\caption{Symbols show median B/T as a function of the fitted total half-light radius for pure bulge galaxies (left) and two-component galaxies (right). 68 and 95 per cent contours are plotted as dashed and dot--dashed lines. Errorbars on the median values represent 95 per cent CI obtained from bootstrap resampling. The B/T of pure bulges drops as the apparent size of the galaxy decreases. At the same time, the B/T of the two-component galaxies increases, suggesting the ability to distinguish one- and two-component galaxies is reduced at small apparent size.}
\label{fig:rtot_bt}
\end{figure}

Further examination of the B/T of the two-component and pure bulge galaxies is shown in Figure~\ref{fig:rtot_bt}. The median B/T as a function of the fitted total half-light radius for pure bulge galaxies (left) and two-component galaxies (right) is shown. Symbols show median values. 68 and 95 per cent contours are plotted as dashed and dot--dashed lines. Errorbars on the median values represent 95 per cent confidence intervals (CI) obtained from bootstrap resampling. 

B/T increases to a median value of 0.6 at small radii versus median values of 0.4 in the larger apparent radii bins. The fitted B/T of the \SerExp{} model for pure bulge galaxies also drops from $\approx0.85$ to $\approx0.75$ in the smallest radii bins.  This indicates a possible bias in our catalogue of preferentially classifying apparently small galaxies as two-component galaxies with moderately high (0.6--0.8) B/T due to effects of the PSF. We examine bias due to the PSF further in Section~\ref{sec:bulge_rad_psf}. 
Here, we caution the reader that there may be some classification bias preferring two-component fits at small sizes.

\subsubsection{Behaviour with axis ratio}

Figure~\ref{fig:flag_by_type} and \ref{fig:flag_by_type_total_per} also show the distribution of total axis ratios 
derived from the \SerExp{} fits by measuring the axis ratio of the total profile for the fitted galaxies in our catalogue (bottom left). If the flagging
properly identifies pure bulge systems, we would expect to see pure bulge systems concentrated near axis ratios approaching 1
since early-type galaxies are ellipsoidal. Lower values of axis ratio should be dominated by pure discs and two component systems if these categories properly identify late-type galaxies. Figure~\ref{fig:flag_by_type} shows this behaviour. Figure~\ref{fig:flag_by_type_total_per} also shows that the peak of the bulge galaxy distribution is near 0.8 with few galaxies 
at small axis ratios.

\subsubsection{Behaviour with bulge S\'{e}rsic index}

Figure~\ref{fig:flag_by_type} and \ref{fig:flag_by_type_total_per} also show the distribution of bulge S\'{e}rsic index
derived from the \SerExp{} fits to galaxies in our catalogue (bottom right).
Figure~\ref{fig:flag_by_type} shows that two-component galaxies dominate the objects fitted with bulge indexes between 1 and 4. Galaxies with lower(higher) bulge indexes tend to be classified as pure disc(bulge) galaxies. Figure~\ref{fig:flag_by_type_total_per} also shows similar behaviour. 

However, the reader should be cautioned that interpretation of these plots is complicated by the fact that the bulge  S\'{e}rsic index is used during classification. Also, the B/T of these galaxies is not considered in this plot. Many galaxies, especially with low bulge index, have low B/T, making the observed bulge index unreliable.

\subsubsection{Interpretation of the two-component galaxies}

The distribution of bulge and disc galaxies appear to make sense when the distributions are examined with respect to basic observables like magnitude, size, and axis ratio. Therefore, we consider it appropriate to associate our bulge and disc classes with
early and late-type galaxies, respectively. Further evidence for this claim is discussed in Section~\ref{sec:other_cmp}.
In contrast, we have not yet explored the behaviour of the \SerExp{} bulge and disc components for galaxies that we claim to be two-component galaxies. 

M13 showed that accurate measurement of the total size and magnitude can be accomplished without the components necessarily having physical interpretation. It also showed that \pymorph{} can reliably recover \Ser{} and \Exp{} components of two-component 
models down to component magnitudes of roughly 19 in the {\em r} band. 
However, accuracy of the fitted components does not guarantee that the components represent physical bulges or discs. 
Indeed interpretation of the \SerExp{} subcomponents as true bulge and disc components is complicated. 
We continue to refer to the \Ser{} and \Exp{} components of two-component \SerExp{} fits as `bulge' and `disk' 
throughout the paper. However, we caution the reader that there are many cases where this simple interpretation does not make sense. 
We will return to this issue in Section~\ref{sec:other_cmp}.

\subsection{PSF effects on bulge radius}\label{sec:bulge_rad_psf}

\begin{figure}
\includegraphics[width=0.95\linewidth]{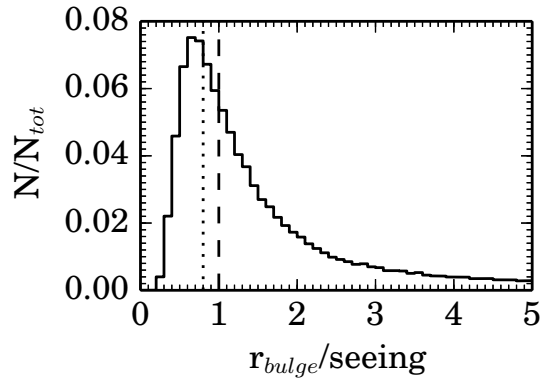}
\caption{The comparison of \protect\SerExp{} bulge circularized effective radius to the PSF HWHM. 15 per cent(24 per cent) of bulges lie below the cut of 0.8$\times$HWHM(1.0$\times$HWHM).}
\label{fig:psf_size}
\end{figure}

The PSF limits our ability to accurately recover component radii and S\'{e}rsic index when the PSF is larger than the component in question. \cite{gadotti08} examined the effects of low resolution on galaxy parameters by performing detailed fitting of 17 nearby ($z\sim0.005$) nearly face-on (b/a$>$0.9) SDSS disc galaxies. The galaxies were then redshifted to $z\sim0.05$ and refit. Figure~13 in \cite{gadotti08} and the accompanying discussion shows that bulge sizes smaller than 80 per cent of the seeing radius (or equivalently 80 per cent of the PSF HWHM) can be biased high (by as much as 50 per cent). Bulge S\'{e}rsic indices can also be suppressed for values greater than 2 (by as much as 1), and B/T can be biased high (by as much as 0.1). However, the authors caution against extending these expectations to higher redshifts as the physical scale of the PSF grows substantially with redshift.

\cite{gadotti09} studied a larger sample of galaxies compared to \cite{gadotti08} ($\sim3000$ compared to 17 galaxies) with a wider range of types (including ellipticals). The decision of whether to fit a second component was based on individual inspection of the radial light profile. Figure~7 in \cite{gadotti09} shows the distribution of the quantity (bulge radius/PSF HWHM) for galaxies with a detected bulge component. The authors find that 3 per cent of their sample are bulges that are smaller than 80 per cent HWHM and 10 percent are smaller than the HWHM. \cite{bernardi2014} found similar results. However, excluding galaxies below the 80 per cent level or even the more conservative HWHM do not change the measurements of physical bulge sizes made by either group.

Figure~\ref{fig:psf_size} shows the comparison of the bulge radius to PSF size for all \SerExp{} bulges in our catalogue. 15 per cent (24 per cent) of bulges lie below $0.8\times$HWHM($1.0\times$HWHM). The percentage of poorly resolved bulges is somewhat higher than either \cite{gadotti09} or \cite{bernardi2014}. For the pure bulge galaxies (\ie flag bit 1 set), only 2 per cent (3 per cent) of pure bulge galaxies have radii smaller than $0.8\times$HWHM ($1.0\times$HWHM). Including pure bulge galaxies and two-component galaxies with B/T$>$0.5 (\ie \SerExp{} fits with flag bit 1 set or with B/T$>$0.5 and flag bit 10 or 14 set), 7 per cent (12 per cent) of the bulges are smaller than $0.8\times$HWHM ($1.0\times$HWHM). These numbers are still higher, but in closer agreement with both \cite{gadotti08} and the sample used in \cite{bernardi2014}. In contrast, 28 per cent (41 per cent) of two-component galaxies with B/T$\leq$0.5 have bulges smaller than $0.8\times$HWHM ($1.0\times$HWHM). The work of \cite{gadotti09} suggests that a significant proportion of our \SerExp{} bulge components with B/T$\leq0.5$ are potentially biased to larger sizes and brighter magnitudes by poor resolution. This must be considered when looking at the bulge components  of galaxies with significant discs. We include the PSF size in the catalogue 
so that this consideration can be made during future analysis.

M13 also showed that the effective bulge radius for bulges smaller 
than 1 arcsec is overestimated by $\sim5$ per cent $\pm20$ per cent. The simulations reflect a tendency to
overestimate bulge radius as reported in \cite{gadotti08}. We do not correct for this effect here, but caution the user that 
small bulges (smaller than the HWHM) are likely biased larger and brighter. This effect likely contributes to the increase in 
two-component galaxies at small apparent sizes as discussed in Section~\ref{subsec:app_abs_rad}.

\subsection{Comparison of the \Ser{} and \SerExp{} models}\label{sec:ser_serexp_cmp}
 
\begin{figure*}
\includegraphics[width=0.329\linewidth]{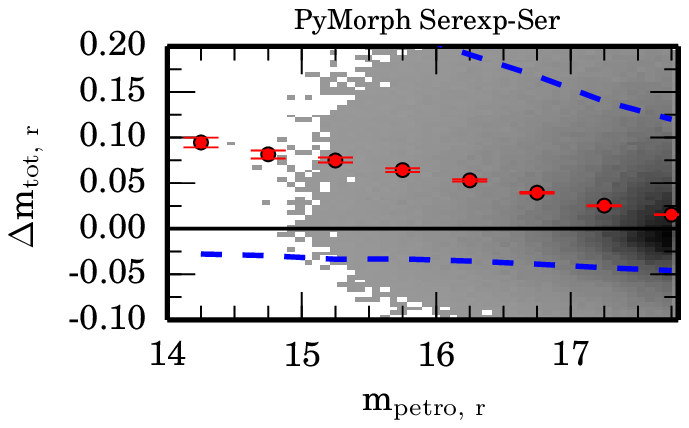}
\includegraphics[width=0.329\linewidth]{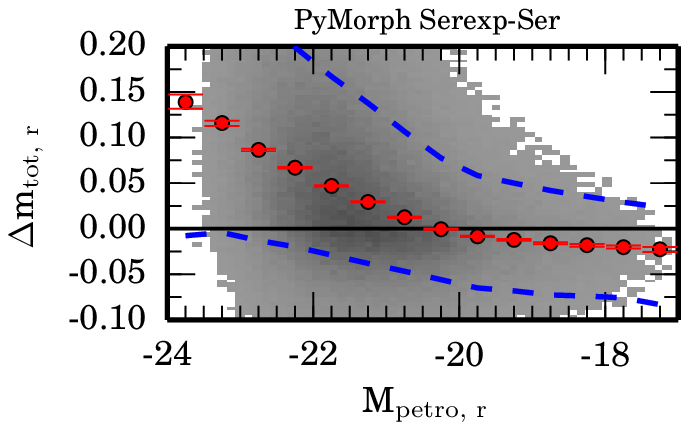}
\includegraphics[width=0.329\linewidth]{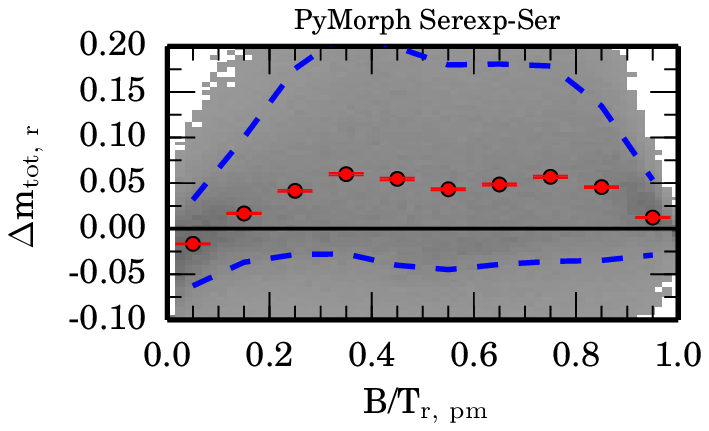}
\includegraphics[width=0.329\linewidth]{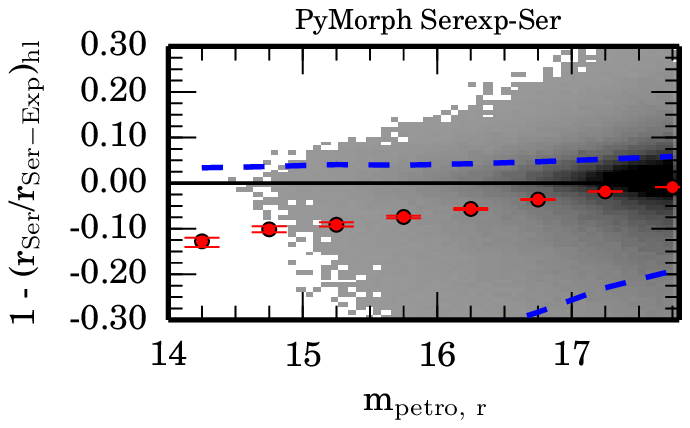}
\includegraphics[width=0.329\linewidth]{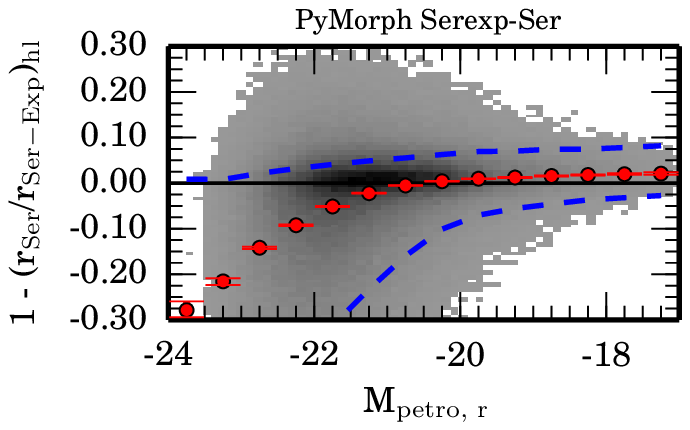}
\includegraphics[width=0.329\linewidth]{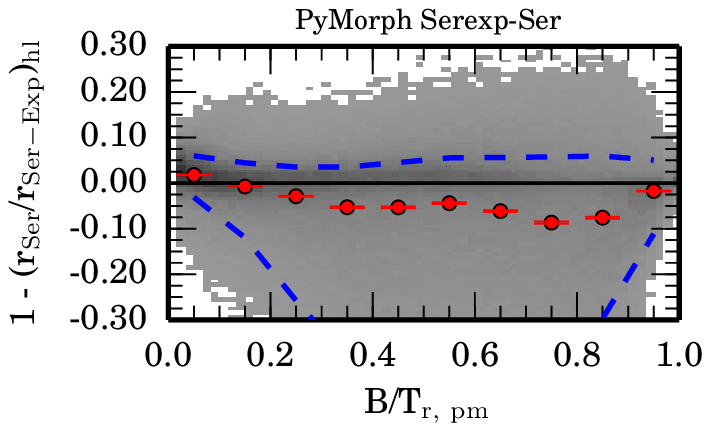}

\caption{The comparison of \Ser{} fit to \SerExp{} fit for galaxies in this work. The top row shows comparisons of magnitude ( \SerExp{} -- \Ser{} )  as a function of apparent Petrosian magnitude, absolute Petrosian magnitude, and B/T.
The bottom row shows comparisons of half-light radius as a function of apparent Petrosian magnitude, 
absolute Petrosian magnitude and the fitted \SerExp{} B/T. At the low and high B/T, 
the bias is nearly 0 with smaller scatter. In the intermediate B/T values where a single-component model would 
be expected to fit poorly, the bias and scatter are larger.}
\label{fig:ser_serexp_meert}
\end{figure*}

\begin{figure}
\includegraphics[width=0.95\linewidth]{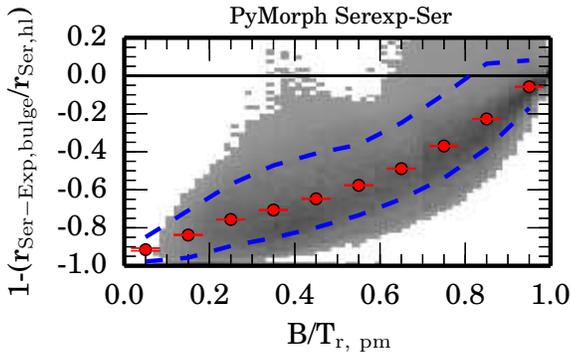}
\caption{The comparison of \Ser{} radius to the \SerExp{} bulge radius as a function of B/T for galaxies in this work. As B/T increases, the \SerExp{} bulge radius approaches the \Ser{} radius as expected. As B/T approaches 0, the ratio approaches 0
indicating that the bulge is shrinking in size with magnitude as would be expected if the bulges are properly fitting the central 
bulge of the galaxy. Median values for each bin are plotted in red with the errorbars representing a 95 per cent bootstrap CI on the median. The 68 per cent contours of the data are plotted as blue dashed lines. 
The density of points is plotted in grey-scale with the low end of the density representing 1 per cent of the maximum density. 
Bins with a density below the minimum density are plotted in white.} 
\label{fig:ser_serexp_bulgerad_meert}
\end{figure}

In this work, we favour the \SerExp{} model for both deciding on the structure of the galaxies (\ie one versus two components) 
and for estimating the total size and magnitude. Figure~\ref{fig:ser_serexp_meert} shows comparisons of fitted magnitude (top row) and half-light radius (bottom row) between the \Ser{} and \SerExp{} fits. The top row shows comparisons of the fitted magnitude as a function of apparent Petrosian magnitude (left), absolute Petrosian magnitude (centre), and \SerExp{} B/T (right). Magnitude 
differences are reported as \SerExp{} -- \Ser{} magnitude, therefore a positive magnitude indicates a brighter \Ser{} fit relative
to the corresponding \SerExp{} fit. Median values for each bin are plotted in red with the errorbars representing a 95 per cent bootstrap CI on the median. The 68 per cent contours of the data are plotted as blue dashed lines. 
The density of points is plotted in grey-scale with the low end of the density representing 1 per cent of the maximum density. 
Bins with a density below the minimum density are plotted in white.

Analogous plots are presented for the half-light radius in the second row.
In all the plots presented here, 
we bin galaxies by Petrosian magnitude in order to make consistent comparisons across the different models. 
The Petrosian magnitude provides a model-independent measure of the magnitude so that any comparison has the same 
distribution of galaxies among the bins. 

We observe larger differences in the magnitude at brighter apparent magnitude, brighter absolute magnitude, and intermediate B/T values. These differences are consistent with
the effects seen in the simulations of M13. The bias is caused 
by inappropriately fitting a one-component model to a more complicated light profile. This type of underfitting was shown to bias
measurements of the \Ser{} magnitude to be brighter than the combined \SerExp{} magnitude measured from the same galaxy. This effect is also 
reported by \cite{mosleh13} and \cite{Davari14}. 

Figure~1 of \cite{bernardi2013} also shows the difference in magnitudes for the \Ser{} and \SerExp{} fits of this work as a 
function of absolute magnitude. The figure also shows the comparison of the \Ser{} fit presented in this work to the \Ser{} fits
of S11 and the SDSS magnitudes. The \Ser{} magnitudes from this work are systematically brighter by up to 0.5 mag as you approach --24. This difference leads to large differences in the bright end of the LF. However, consistent difference of the
\Ser{} magnitudes is not an indication of bias in our fits because 
\cite{kravtsov2014} showed that the magnitudes reported here are more
consistent with their results than either SDSS or S11.

The bottom row of Figure~\ref{fig:ser_serexp_meert} shows comparisons of half-light radius between the \Ser{} and \SerExp{} fits.
We show the fractional difference in radii. 
For example, a value of 0 indicates no disagreement between the compared radii. A value of --1 would indicate that the \Ser{} model is 100 per cent larger compared to the \SerExp{} model. There is a difference of 5 per cent in radius and 0.05 mag at the intermediate B/T values. The lowest and highest B/T values show both agreement in the median difference between \Ser{} and \SerExp{} models as well as decreasing scatter. 

Figure~\ref{fig:ser_serexp_bulgerad_meert} also shows comparisons of bulge radius for the \Ser{} and \SerExp{} fits.
The \SerExp{} bulge radius gets smaller with decreasing B/T relative to the single-component \Ser{} radius, as one might expect if the \Ser{} component of the \SerExp{} model is fitting the central bulge of the galaxy. While this shows that the bulges are mostly compact, we discuss the correspondence between the \Ser{} component of the \SerExp{} fit and the central bulge of galaxies further in Sections~\ref{sec:ex:serexp} and \ref{sec:other_cmp}.

\section{Comparisons to literature} \label{sec:analysis}

In this section, we show several comparisons to the literature. We also direct the reader to the work presented in M13, 
in which simulations were used to test and verify the accuracy of the fitting code. Although the flagging was not applied to these simulations, the simulations demonstrate the accuracy of the fitting algorithm, particularly for total magnitude and radius. 

We describe the external catalogues used for comparison in Section~\ref{sec:ex_desc}. 
We compare the fits of SDSS, LG12, S11 and \citet[][hereafter Men14]{mendel2014} to our fits where
appropriate.
Section~\ref{sec:ex:dev} compares the \Dev{} fits to SDSS and LG12. 
Section~\ref{sec:ex:ser} compares the \Ser{} fits to S11 and LG12 \Ser{} fits. 
Section~\ref{sec:ex:devexp} compares the \DevExp{} fits to S11 and LG12 \DevExp{} fits.
Section~\ref{sec:ex:serexp} compares the \SerExp{}
fits to S11 \SerExp{} fits. Section~\ref{sec:ex:bestmod} presents a final comparison using the magnitudes measured
using the preferred model of this work, Men14, and S11 to commonly used SDSS magnitudes (\ie cModel and Petrosian).

\subsection{External catalogues used for comparison and analysis}\label{sec:ex_desc}

We use the fits of S11, SDSS DR7, and LG12 to make comparisons. 
S11 carried out fits of SDSS galaxies with \Ser{}, \DevExp{}, and \SerExp{} 
models using the \gimd{} program \citep{Groth2002}. \gimd{} uses the Metropolis search algorithm
\citep{Metropolis} to optimize the model parameters rather than a gradient
descent algorithm similar to \galfit{}. S11 also used \sextractor{} to mask out pixels dominated by neighbour galaxy light, which they refer to as `SEXTDEBL,' or \sextractor{} deblending to separate the light of neighbouring objects
from the target galaxy rather than the simultaneous fitting used in this work. They also 
chose a minimum number of pixels to use for
sky estimation (20\,000 pixels). The sky was fixed at this level during fitting rather than fitting the sky brightness as a free parameter. Fixing the sky was shown to provide the better fits than a fitted sky level for \gimd{} \citep{GEMS2007}.

Additionally, S11 provide a statistical probability of a given galaxy being \Ser{}, \DevExp{}, or \SerExp{} based on an {\em F}-test. 
The {\em F}-test statistic is used to analyse if increasing the number of free parameters in the fit is statistically justified. 
This probability is used to select the preferred `best-fitting' model from their data. S11 report two {\em F}-test probabilities, the $P_{pS}$ probability, which is the probability that a one-component \Ser{} model is preferred to a two-component \DevExp{} model, and the $P_{n4}$ probability, which is the probability that a two-component \SerExp{} model is preferred over a two-component \DevExp{} model when
attempting to explain the distribution of light in the observed galaxy. 

Men14 re-analysed the S11 fits and classified the galaxies by a different method. Men14 used the radial light profile of the 
\DevExp{} fit to separate galaxies into (1) bulge-dominated galaxies, (2) \Exp{} dominated galaxies, (3) two-component galaxies,
and (4) non-physical or unclear fits that do not fall into the previous categories.

LG12 provide an additional comparison to our data. The authors used a sample restricted to more nearby galaxies (0.003$<z<$0.05)
using SDSS DR8 \citep[][]{DR8} data. The catalogue contains \DevExp{} fits (referred to as `nb4' in LG12), pseudo-bulge (exponential bulge + exponential disc, referred to as `nb1' in LG12), \Dev{} (referred to as `dvc' in LG12), \Ser{} (referred to as `ser' in LG12), and exponential disc or \Exp{} (referred to as `exp' in LG12) fits for 71\,825 galaxies from the SDSS. We use the LG12 terminology for their fits throughout this paper to make it more clear which sample we are addressing. 

In addition to fitting the models listed above, LG12 also give a classification of the `best-fitting' model chosen from the five 
models they fit. LG12 assign a `best-fitting' model using a combination of statistical and other metrics. For \Exp{} models, 
statistical insignificance of the bulge and quality tests on the bulge magnitude, shape, and size is used to select \Exp{} models
rather than the \Ser{}, nb1, or nb4 models. \Dev{}, or dvc, models are selected in a similar manner. Additional galaxies are selected 
as \Dev{} galaxies based on the colours and shapes of the disc in the nb4 fits. nb1 and nb4 galaxies are chosen from the remaining 
galaxies using the statistical significance of the bulge and quality tests on the fitting parameters intended to identify bad fits
(\ie tests on bulge ellipticity and bulge size relative to disc size). Remaining galaxies that do not satisfy any of these criteria
are given the \Ser{} model as the `best-fitting' model.

LG12 report an absolute magnitude for each galaxy, which includes K-correction, extinction correction, and 
cosmological effects. While the assumed cosmology, the extinction correction, and the K-correction software used in this work 
are the same as those of LG12, the K-correction may be slightly different depending on the choice of input magnitudes (\ie the Petrosian, SDSS model magnitudes, or fitted magnitudes can be used to calculate a K-correction). Also the zeropoint of the magnitudes may vary from the values used here due to the small calibration differences between SDSS DR7 and DR8. Differences in the zeropoint calibration
are expected because the calibration procedure \citep[the `Ubercal' algorithm;][]{ubercal} is a global algorithm, using all
of the imaging data to determine the overall calibration rather than just a single frame. Since the volume of imaging data increased 
 between DR7 and DR8, this can cause slight differences in the calibration.

The overlap of LG12 and this work contain galaxies at the low-redshift end of the galaxy distribution in our catalogue. These galaxies generally have better resolution and are brighter than the full sample of our catalogue, so agreement between LG12 and this work only provide a lower bound on the bias and scatter of our full catalogue. However, this comparison provides a test 
of the most optimal fitting conditions where resolution effects are less of a concern.  

As in the last section, we bin all the plots presented here by Petrosian magnitude 
in order to make consistent comparisons across all the works. 
The Petrosian magnitude provides a model-independent measure of the magnitude so that any comparison has the same 
distribution of galaxies 

For the analysis presented in this section, we treat the `best-fitting' models given by S11 and LG12 as the most appropriate 
models to use in comparison. Therefore, we compare our fits to the best fits of the external works (\eg we compare our \DevExp{} fits to galaxies identified as \DevExp{} by S11 or to galaxies identified as `nb4' by LG12). This will reduce the bias introduced 
by fitting an incorrect model to the galaxy since we will be comparing the fits with the highest confidence of being correct. 

Sections~\ref{sec:ex:bestmod}~and~\ref{sec:other_cmp} compare the S11, Men14, and LG12 classifications to the choice of best model based on the flagging presented in Section~\ref{sec:validation}. In these sections, we examine the agreement between 
the various `best-fitting' models of these different works. 

\subsection{The \Dev{} fits}\label{sec:ex:dev}

\begin{figure*}
\centering
\includegraphics[width=0.329\linewidth]{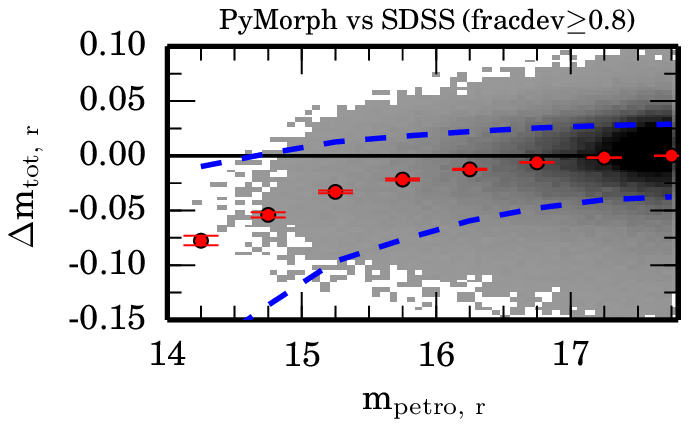}
\includegraphics[width=0.329\linewidth]{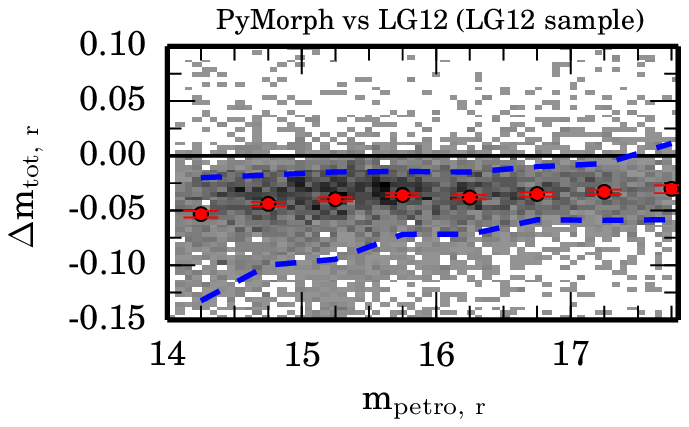}
\includegraphics[width=0.329\linewidth]{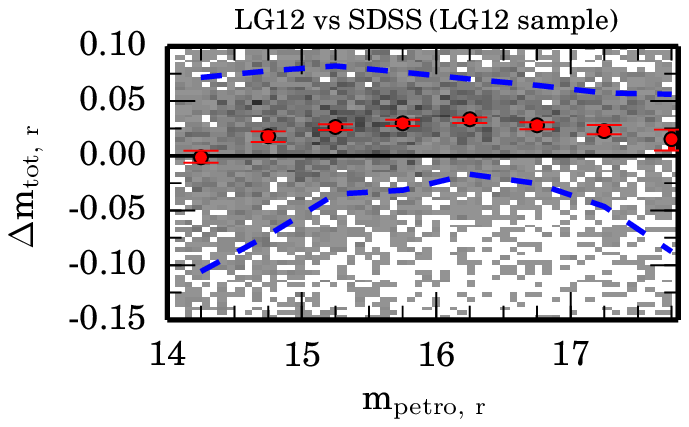}
\includegraphics[width=0.329\linewidth]{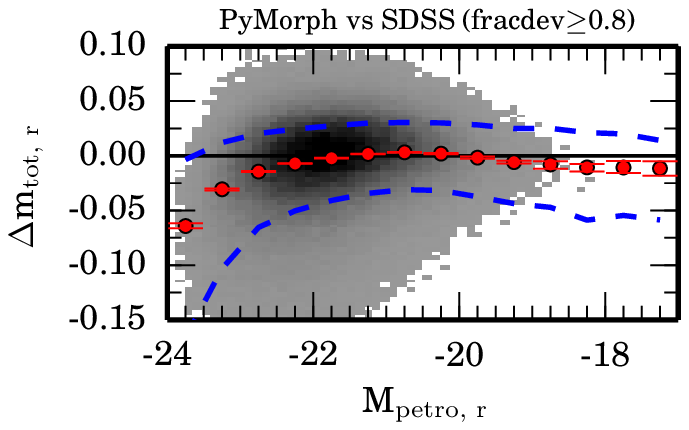}
\includegraphics[width=0.329\linewidth]{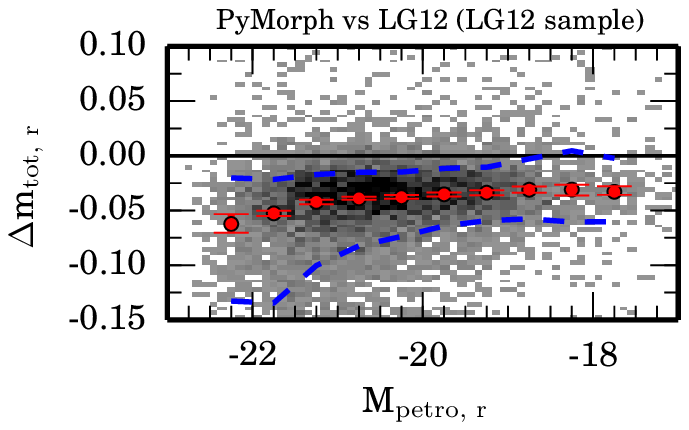}
\includegraphics[width=0.329\linewidth]{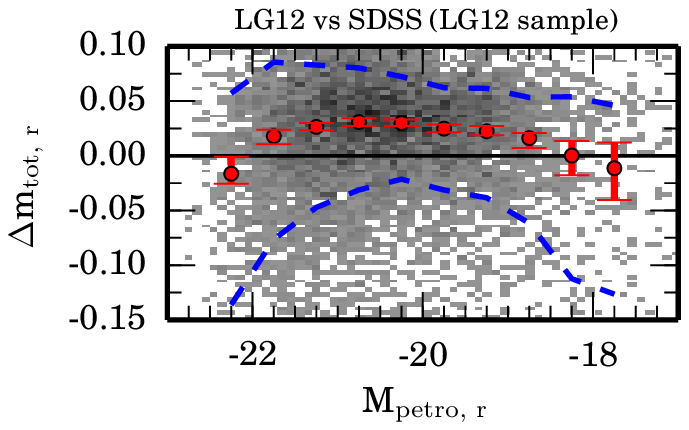}
\caption{A comparison of fitted \Dev{} magnitude from this work, SDSS, and LG12 as a function of apparent Petrosian magnitude (top row) and absolute Petrosian magnitude (bottom). SDSS \Dev{} fits and \pymorph{} \Dev{} fits are compared for the 217131 galaxies in SDSS that are best fitted by the \Dev{} profile as identified by \texttt{fracdev\_r}$\geq0.8$. LG12 comparisons use the $\sim9\,000$ galaxies    identified as `dvc' galaxies using the `best model' parameter of LG12. 
We also remove all poorly fitted galaxies with flag bit 20 set (\ie galaxies with fitting problems). 
This removes 2.82 per cent of the \Dev{} sample. 
Magnitude comparisons are shown as the first model -- the second model. For example, the top-left panel compares \pymorph{} with SDSS, so magnitude differences are  quoted as \pymorph{} -- SDSS and negative values indicate that \pymorph{} produces a brighter magnitude.  Median values for magnitude bins are plotted in red with the errorbars representing a 95 per cent bootstrap CI on the median. The 68 per cent contours of the data are plotted as blue dashed lines. 
The density of points is plotted in grey-scale with the low end of the density representing 1 per cent of the maximum density. 
Bins with a density below the minimum density are plotted in white. 
SDSS and \pymorph{} agree within 0.05 across the magnitude range, except at the bright end where we are consistently brighter than 
SDSS. LG12 exhibit an offset to both SDSS and our fits of up to 0.05. See Appendix~\ref{app:dev} (available online) for additional comparisons.}
\label{fig:dev_fits}
\end{figure*}

The SDSS pipeline computes PSF-convolved 2D \Dev{} fits in addition to other parametric and non-parametric measurements. The SDSS fits are truncated at $7\ r_{eff}$ to go smoothly to zero beyond $8\ r_{eff}$ and also employ some softening 
of the profile within $r = r_{eff}/50$ \citep{SDSS_EDR}. No truncation of the profile is imposed upon the fits presented 
in this paper. For a \Dev{} profile, 94 per cent of the light is contained within $8\ r_{eff}$ and 93 per cent of the light is contained 
within $7\ r_{eff}$. So, an offset between the \Dev{} model fit by SDSS and the \Dev{} fit presented here is expected. 
If different pipelines recover the same fitting parameters
for sky, radius, axis ratio, \etc the expected offset due to the profile truncation is 0.0716 mag, assuming that half the light between 7 and 8 $r_{eff}$, where the softening takes place, is also truncated.

For purposes of a fair comparison, we correct SDSS magnitudes 
by making the \Dev{} magnitudes of SDSS brighter by 0.07 mag.  LG12 also truncate the \Dev{} profile following the
same prescription as SDSS. However, LG12 do not soften the centre of the profile within $r = r_{eff}/50$.
A similar offset is expected for the LG12 `dvc' fits, so we apply the same correction of 0.07 mag.
No modification is made to the radii of the fits or any other fitting parameters. 

Figure~\ref{fig:dev_fits} shows a comparison of fitted \Dev{} magnitude from this work, SDSS, and LG12 as a function of apparent Petrosian magnitude (top row) and absolute Petrosian magnitude (bottom). SDSS \Dev{} fits and \pymorph{} \Dev{} fits are compared for the 217131 galaxies in SDSS that are best fit by the \Dev{} profile as identified by \texttt{fracdev\_r}$\geq$0.8. LG12 comparisons use the $\sim 9\,000$ galaxies identified as `dvc' galaxies using the `best model' parameter of LG12. 
We also remove all galaxies with flag bit 20 set (\ie galaxies with fitting problems). This removes 2.82 per cent of the \Dev{} sample. 

Magnitude comparisons are shown as the first model minus the second model. For example, the top-left panel compares \pymorph{} with SDSS, so magnitude differences are  quoted as \pymorph{} -- SDSS and negative values indicate that \pymorph{} produces a brighter magnitude.  Median values for magnitude bins are plotted in red with the errorbars representing a 95 per cent bootstrap CI on the median. The 68 per cent contours of the data are plotted as blue dashed lines. 
The density of points is plotted in grey-scale with the low end of the density representing 1 per cent of the maximum density. 
Bins with a density below the minimum density are plotted in white. 

In the left-hand column of Figure~\ref{fig:dev_fits}, \pymorph{} \Dev{} magnitudes agree with SDSS with a scatter of 0.05 
(using 68 per cent contours) across most of the magnitude range but show a systematic increase to about 0.07 mag brighter than SDSS values at the bright end of the apparent magnitude range (top row, left-hand column). While SDSS and this work differ at the bright end, the majority of the galaxies show no systematic bias. In contrast, LG12 have an offset of $-0.03$ to $-0.05$ across the entire magnitude range (centre column of Figure~\ref{fig:dev_fits}). LG12 comment on this systematic difference in their paper noting a $-0.025$ mag offset in their \Dev{} fits to SDSS galaxies with \texttt{fracdev\_r}$>$0.5 as well as a $\sim9$ per cent difference in the fitted radii. We investigated the source of this offset, but can find no reason for it. Although the zeropoint of DR8 and DR7 vary up to 0.2 mag, correcting for this effect does not reduce the offset.

Median differences in the measured radii are below 5 per cent [see Appendix~\ref{app:dev} (available online) for the radius comparisons] when comparing LG12 
and this work. The difference in radii agrees with the observed magnitude differences, suggesting that the \pymorph{} fits presented 
here are larger and brighter compared to the LG12 fits. 

The radii in SDSS are $5-10$ per cent smaller when compared to \pymorph{}. 
LG12 comment on this difference as well, attributing it to effects of the softening of the fitted profile 
in the centre (inside $r_{eff}/50$) which suppresses the half-light radius in SDSS. This cannot, however, be the source
of the disagreement between LG12 and this work since neither work implements such softening.

The level of the sky in our fitting is found to be, on average, 0.25 per cent dimmer when compared to SDSS
(see also Figure~\ref{fig:sky_ser} and related discussion). 
Sky level and the fitted magnitude have been shown to be correlated 
(see M13, and references therein for a full discussion). 
Bias in the sky level may explain the slight differences in magnitudes.  
In M13, overestimates of sky at the level of 0.5 per cent are shown to suppress 
fitted \Ser{} magnitude by $\approx$ 0.1 mag. 
We discuss the sky brightness further in Section~\ref{sec:ex:ser}.

\subsection{The \Ser{} fits}\label{sec:ex:ser}

\begin{figure*}
\centering
\includegraphics[width=0.32\linewidth]{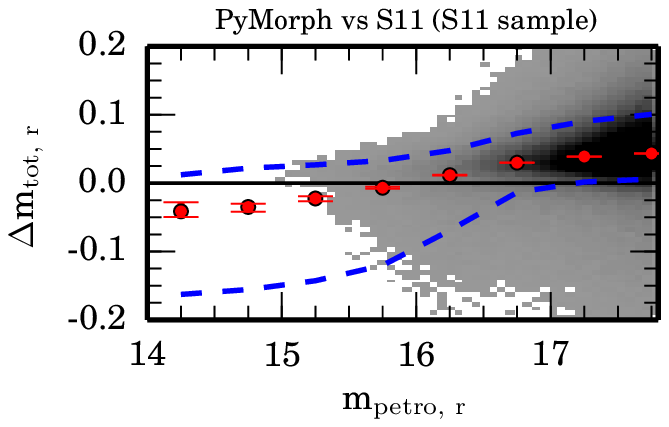}
\includegraphics[width=0.32\linewidth]{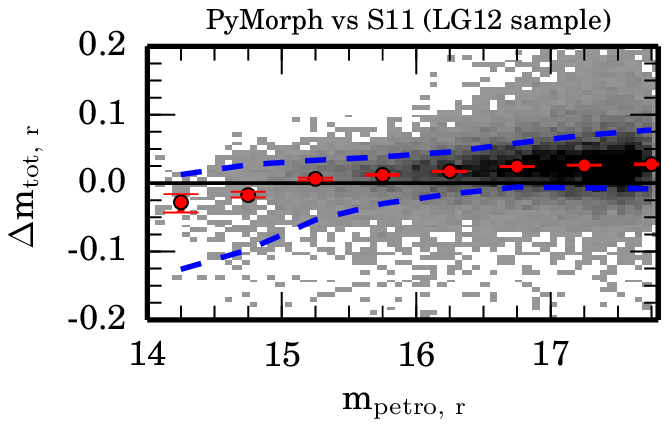}
\includegraphics[width=0.32\linewidth]{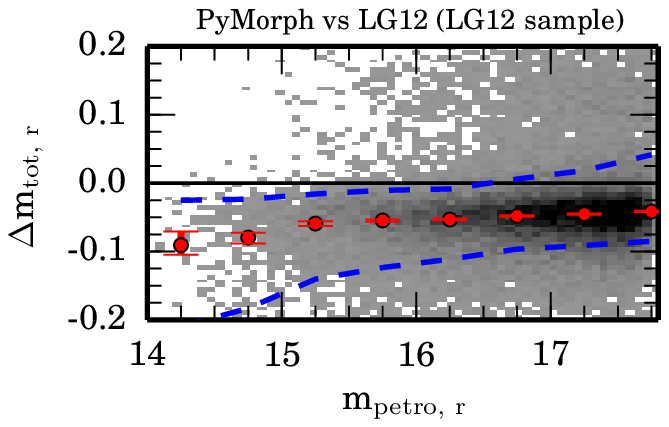}
\includegraphics[width=0.32\linewidth]{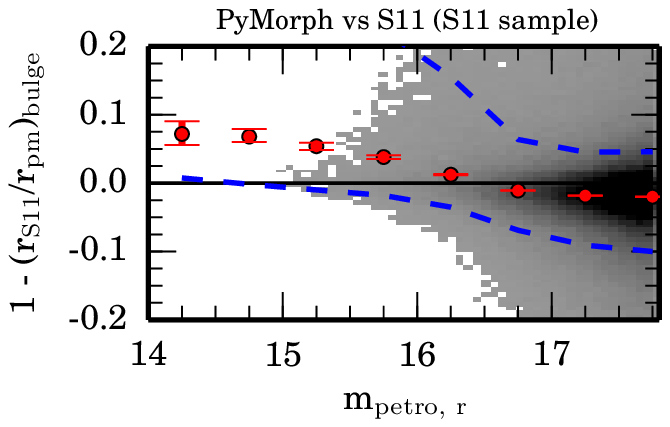}
\includegraphics[width=0.32\linewidth]{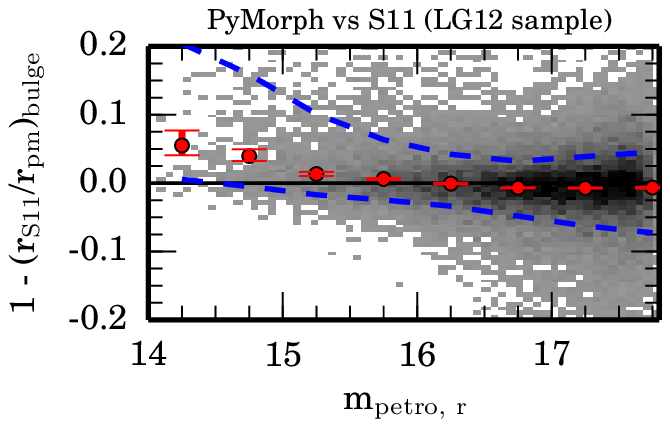}
\includegraphics[width=0.32\linewidth]{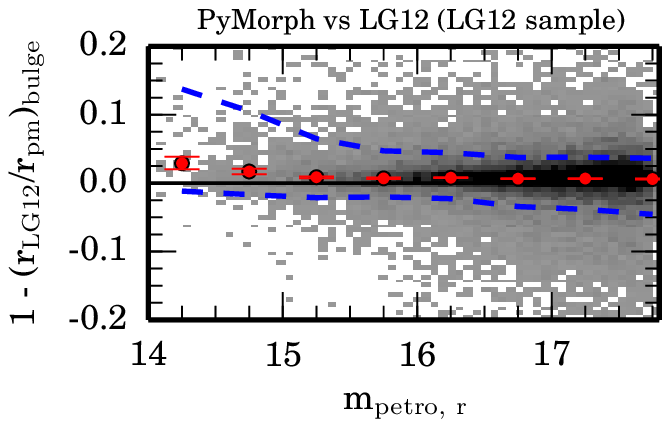}
\includegraphics[width=0.32\linewidth]{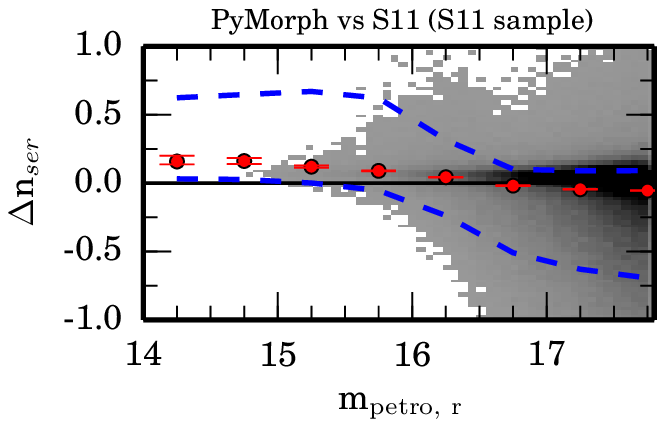}
\includegraphics[width=0.32\linewidth]{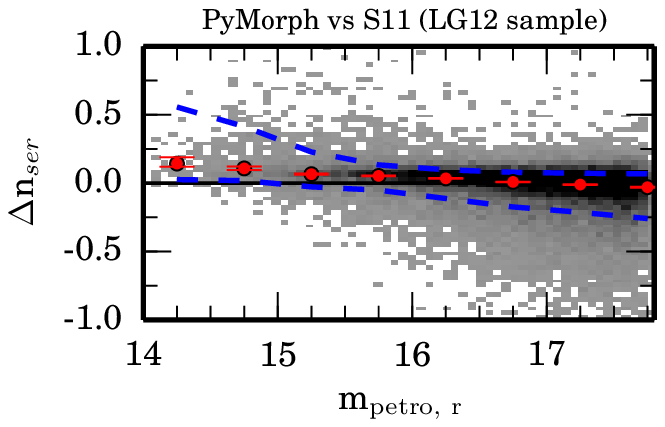}
\includegraphics[width=0.32\linewidth]{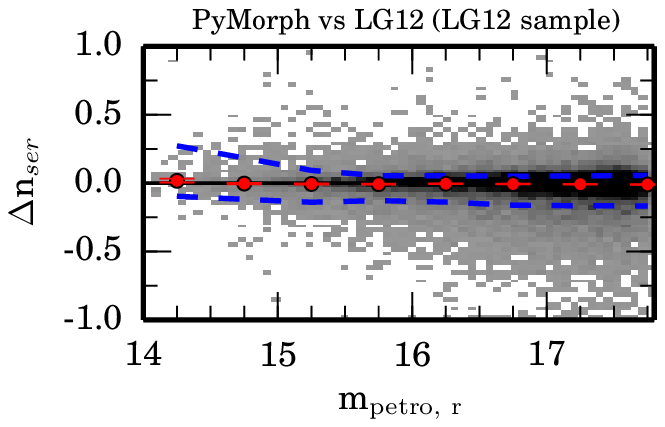}

\caption{The difference between the \Ser{} magnitude (first row), radius (second row), and S\'{e}rsic index (third row). The left-hand column shows the comparison of this work and S11 for the full sample. The centre column shows a comparison of this work and S11 for the galaxies of the LG12, low-redshift sample. The right column show comparisons for this work and LG12.  S11 and LG12 galaxies are identified as best fit `ser' models in S11. The format of the plot is the same as Figure~\ref{fig:dev_fits}. The LG12 fits exhibit an offset in magnitude similar to the offset seen in the \Dev{} fits. Differences between this work and LG12 for bright galaxies are reduced in comparison to the S11--\pymorph{} comparison. See Appendix~\ref{app:ser} (available online) for additional comparisons.}
\label{fig:ser_fits}
\end{figure*}

\begin{figure}
\begin{center}
\includegraphics[width=0.95\linewidth]{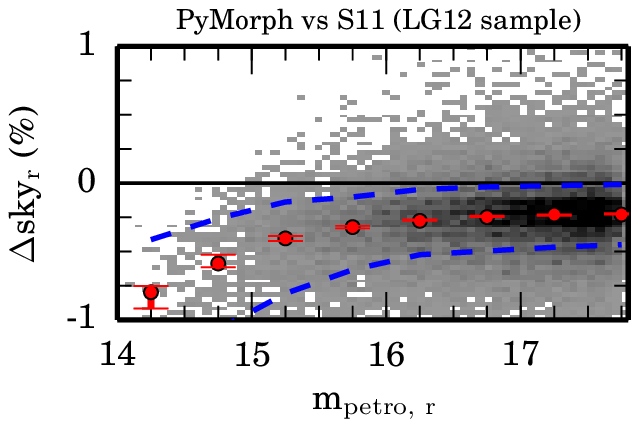}
\includegraphics[width=0.95\linewidth]{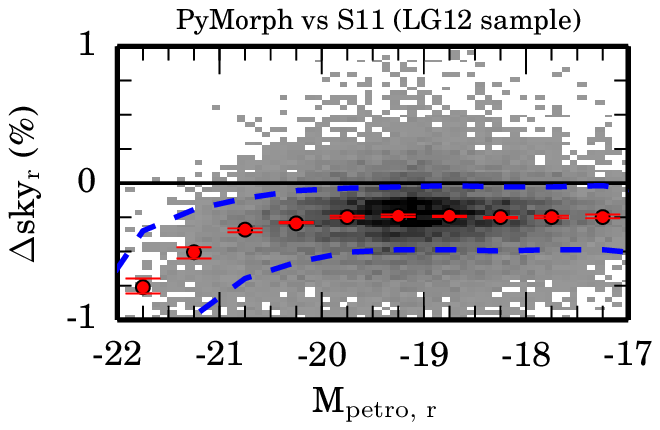}
\end{center}
\caption{The difference in the {\em r} band sky values of S11 \Ser{} fits and our \pymorph{} \Ser{} fits shown as
\pymorph{} -- S11 (negative values indicate that \pymorph{} produces a dimmer sky measurement compared to S11).
The top row shows the difference between the S11 and \pymorph{} \Ser{} sky brightness as a function of the apparent Petrosian magnitude. The second row shows the difference in sky versus absolute Petrosian magnitude. An offset in sky brightness is observed. This difference increases with increasing brightness, suggesting that the S11 sky values are systematically biased brighter for brighter and larger galaxies in both the observed and absolute frame, causing a 
systematic bias in the {\em r} band magnitudes of S11.}
\label{fig:sky_ser}
\end{figure}

We also compare our \Ser{} fits to those of LG12 and S11. Figure~\ref{fig:ser_fits} shows the difference between magnitude (first row), half-light radius (second row), and S\'{e}rsic index (third row). We compare \pymorph{} and S11 for the full sample (left-hand column), \pymorph{} and S11 for the LG12 sample (centre column), and \pymorph{} and LG12 (right-hand column)
Plots are in a format similar to that of Figure~\ref{fig:dev_fits}. For the \Ser{} fits we also examine the differences in  S\'{e}rsic index, $n_{ser}$. For the full comparison of the \Ser{} fits see Appendix~\ref{app:ser} (available online). 

For the plots presented in Figure~\ref{fig:ser_fits}, we select galaxies in both S11 and LG12 that are `best fit' by a \Ser{}
model according to S11. We use the S11 classification in this case because LG12 use the \Ser{} category as a default category 
for fits that are not well described by the other models they fit. The authors comment that the LG12 \Ser{} galaxies are primarily 
low S\'{e}rsic index. S11 sample a broader range of S\'{e}rsic indices because they do not have separate \Dev{} or \Exp{} categories.
LG12 do not mention any truncation of the \Ser{} profile, so we do not apply a correction here. The top-left panel of Figure~\ref{fig:ser_fits} shows that the LG12 magnitudes are offset again relative to \pymorph{}. Comparison with S11 (middle row, left) also show a zero-point offset, but in the opposite direction (\ie S11 is brighter than this work while LG12 is fainter). In addition, there is an overall trend in S11 across the magnitude range.

The radii of S11 are also smaller at the bright end of the apparent magnitude range when compared to this work (bottom row, centre column). This trend is reduced in the LG12 comparison (top row, centre). The  S\'{e}rsic index  also displays a trend in the S11 data (bottom row, right) that does not appear in the comparison of LG12 and this work (top row, right). Although this trend appears to be insignificant, there are larger biases in the S\'{e}rsic index when compared as a function of absolute magnitude (see Appendix~\ref{app:ser}, available online). This trend was also discussed in \cite{bernardi2014}, where the authors argue that
the behaviour of the S\'{e}rsic Index is consistent with systematic errors in the S\'{e}rsic index due to over subtracting sky.

We used the S11 sky values (Simard, private communication) to examine the effect of the sky on the \Ser{} fits. 
\pymorph{} prefers sky levels about 0.25 per cent lower than that of S11. S11 sky brightness is similar to SDSS. Figure~\ref{fig:sky_ser} shows a comparison of S11 and \pymorph{} sky levels as a function of apparent (top) and absolute (bottom) Petrosian magnitude. 
There is an offset in sky level of approximately 0.25 per cent, similar to the SDSS. However, this difference increases in the \Ser{} fits with \pymorph{} sky levels appearing up to  1 per cent dimmer in the brightest apparent and absolute magnitude bin.

The simulations in M13 showed that \pymorph{} estimates the sky with a bias $\sim0.1$ per cent 
which is a factor of $\approx2$ smaller than the observed difference seen for the SDSS measurements. 
Furthermore, the underestimate of sky brightness observed in M13 was not large enough to cause a measurable bias in recovered magnitude for the simulations. 
Since we expect an offset of 0.1 per cent in sky brightness if SDSS sky measurements are accurate,
the observed offset of 0.25 per cent indicates that SDSS sky levels are likely slight overestimates (about 0.15 per cent of sky brightness).
Since S11 sky levels have a similar offset, we expect that they are overestimates as well. Although \cite{GEMS2007} argue that such an overestimate of sky improves \gimd{} performance, offsetting the inefficient masking of neighbouring galaxies.

We propose a simple explanation for this effect in the S11 fits. S11 used at least 20\,000 pixels nearest to each galaxy to estimate the sky (the nearest 20\,000 pixels that are classified as neither source, nor neighbour pixels). This is, in general, a large number of pixels that sample the sky at many different radii. However, as the size of the target galaxy grows, the annuli that form the perimeter of the galaxy grow as well. This leads to a systematic sampling of the sky for the brighter and larger galaxies. For the extended objects studied here, this will lead to an overestimate of the sky and, as a consequence, a suppression of galaxy size and brightness. Using an image cutout that instead scales with the galaxy radius, as is used here, ensures that the same range of half-light radii are sampled for sky estimation and prevents this systematic effect. 

We note that two works have already shown that our measurements should be preferred to those of S11. 
First, \cite{kravtsov2014} performed detailed fits of approximately 10 BCGs and found that their measurements agree more with our measurements than with the measurements of S11. 
Secondly, we direct the reader to appendix~A of \cite{bernardi2014} which shows an unexpected redshift evolution of 
the S\'{e}rsic index in the S11 data. The trend in S\'{e}rsic index is not observed in the fits of this work. \cite{bernardi2014} argue that this is another reason to prefer the fits of this work to those of S11. Therefore, we recommend that caution should be exercised when using the \Ser{} fits from different works to properly account for such differences in fitting.

\subsection{The \DevExp{} fits}\label{sec:ex:devexp}

\begin{figure*}
\includegraphics[width=0.32\linewidth]{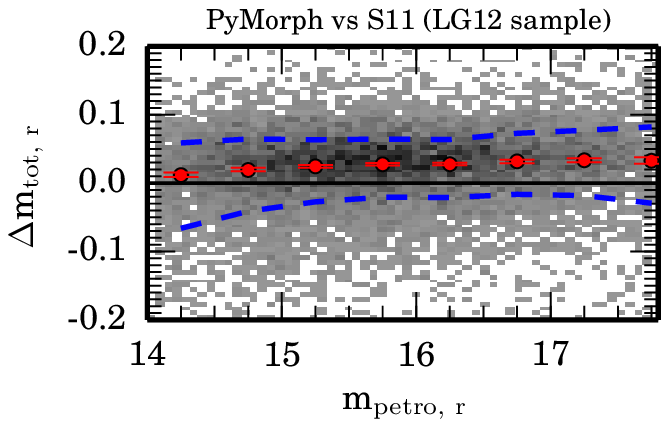}
\includegraphics[width=0.32\linewidth]{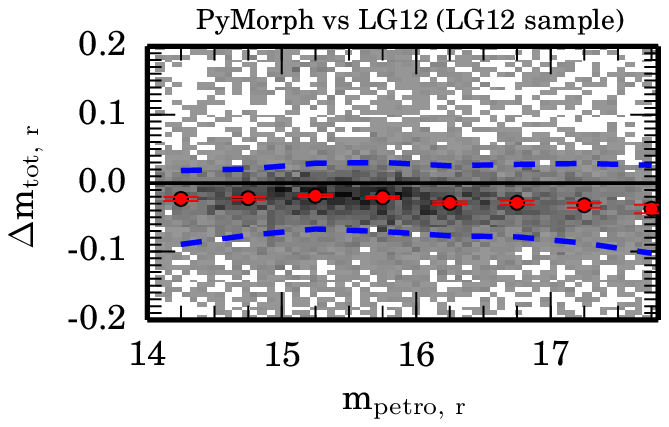}
\includegraphics[width=0.32\linewidth]{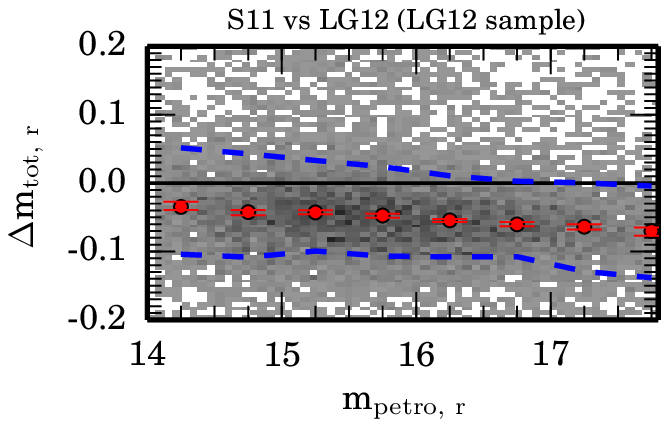}

\includegraphics[width=0.32\linewidth]{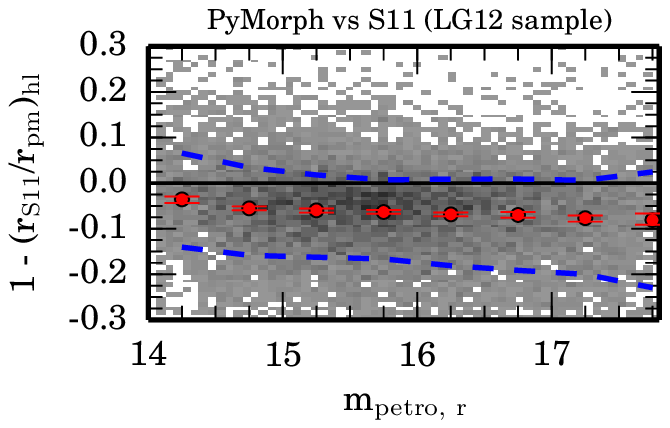}
\includegraphics[width=0.32\linewidth]{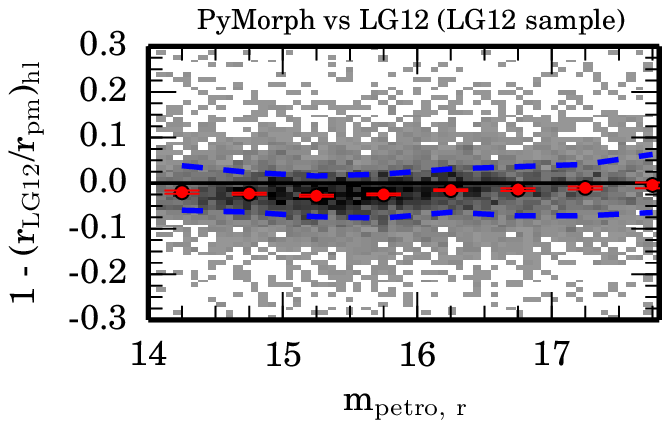}
\includegraphics[width=0.32\linewidth]{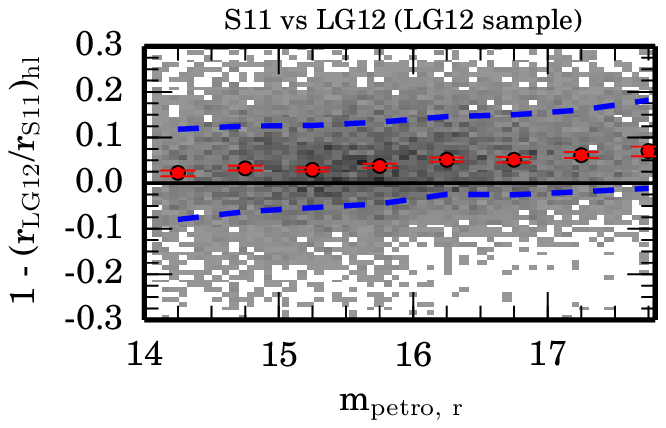}

\includegraphics[width=0.32\linewidth]{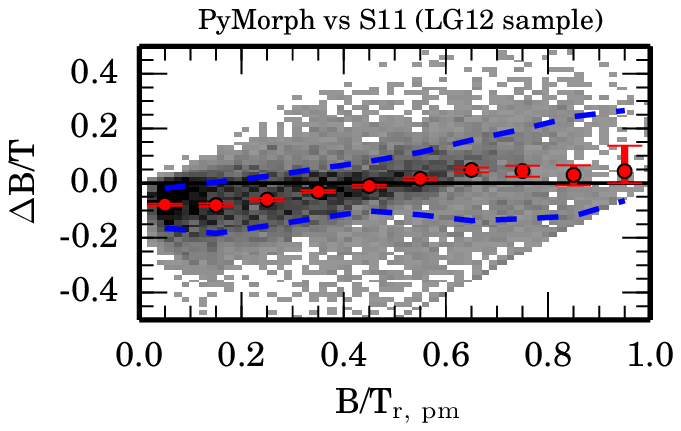}
\includegraphics[width=0.32\linewidth]{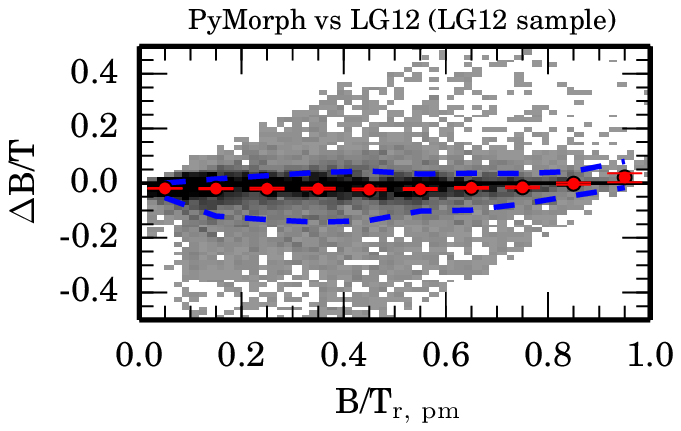}
\includegraphics[width=0.32\linewidth]{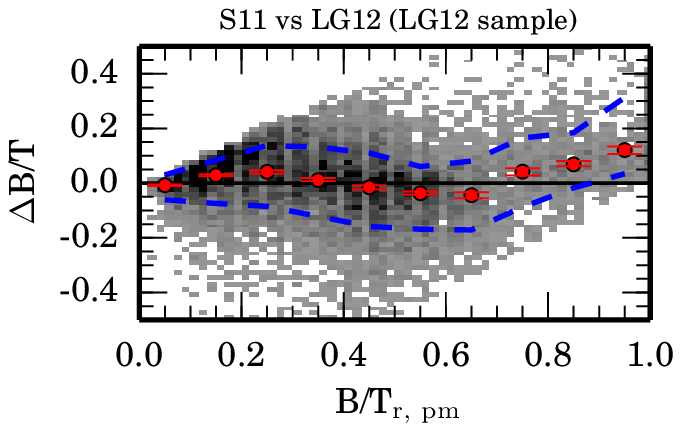}

\caption{The difference between the \pymorph{} \DevExp{} magnitude, half-light radius, and B/T and those of LG12 \DevExp{} and S11 \DevExp{} fits  for galaxies in the LG12 sample. The top row shows the comparison of magnitude between S11 and \pymorph{} (left), \pymorph{} and LG12 (centre), and S11 and LG12 (right). The second row shows a comparison of the total half-light radius as a function of apparent magnitude. The third row shows a comparison of B/T as a function of \pymorph{} B/T. S11 and LG12 galaxies are identified as best-fitting `nb4' models in LG12. The format of the plot is the same as Figure~\ref{fig:dev_fits}. S11 B/T has larger scatter with respect to either \pymorph{} or LG12.  See Appendix~\ref{app:devexp} (available online) for additional comparisons.}
\label{fig:devexp}
\end{figure*}

The \DevExp{} model is the final model that allows for a direct comparison of both the S11 and LG12 fits with our own. We see broad agreement between \pymorph{}, S11, and LG12 for the total half-light radius and total magnitude. Figure~\ref{fig:devexp} shows comparisons of total magnitude (top row), half-light radius (middle row), and B/T (bottom row) for this work compared to S11 (left-hand column), this work compared to  LG12 (centre column) and S11 compared to LG12 (right-hand column). 
The \pymorph{} and LG12 magnitude and half light radius agree with smaller scatter than either comparison to S11. B/T values of LG12 and this work are also in better agreement (bottom row).
S11 tend to overestimate B/T relative to this work by about 0.1 at the low B/T while the LG12 fits are in closer agreement to this work across the B/T range. 

The scatter in the sub components is quite broad making any conclusions difficult. When comparing this work to S11, bulge magnitudes have a scatter of $\sim0.3$ mag, bulges tend to be dimmer in this work compared to S11 by as much as 0.1 mag, and bulges are larger by almost 10 per cent.  There is better agreement in the disc magnitudes, but \pymorph{} discs tend to be 5 per cent larger when compared to S11. The scatter decreases when comparing this work with LG12, although a difference in bulge and disc parameters is still observed. We refer the reader to Appendix~\ref{app:devexp} (available online) which shows the full comparison of the sub components. 

\subsection{The \SerExp{} fits}\label{sec:ex:serexp}

\begin{figure*}
\includegraphics[width=0.32\linewidth]{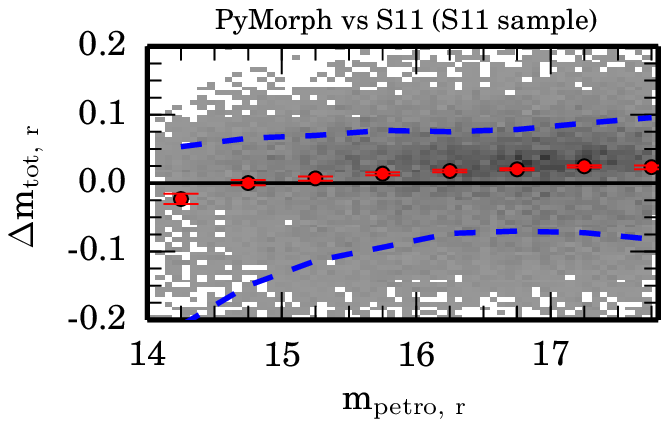}
\includegraphics[width=0.32\linewidth]{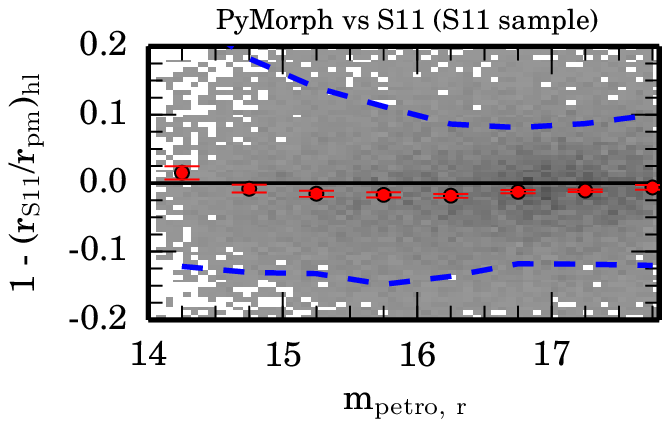}
\includegraphics[width=0.32\linewidth]{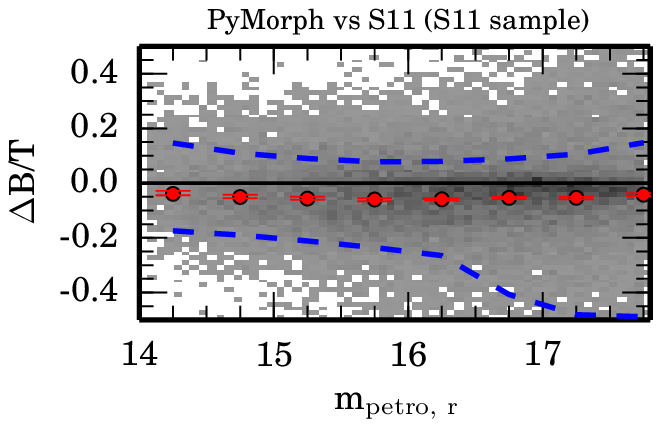}
\includegraphics[width=0.32\linewidth]{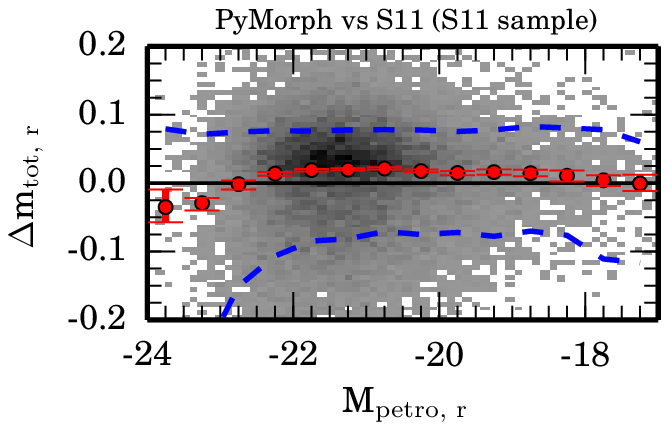}
\includegraphics[width=0.32\linewidth]{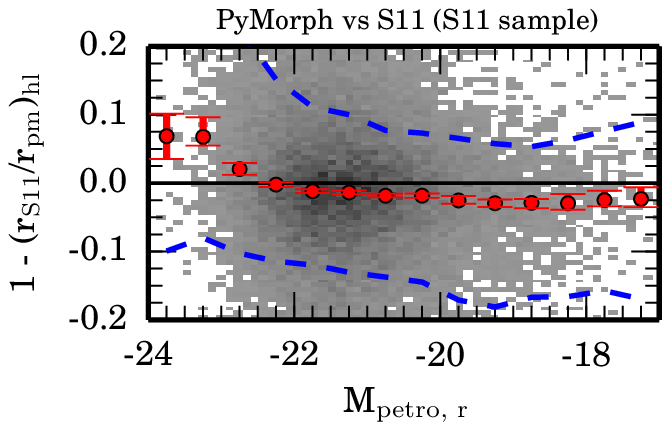}
\includegraphics[width=0.32\linewidth]{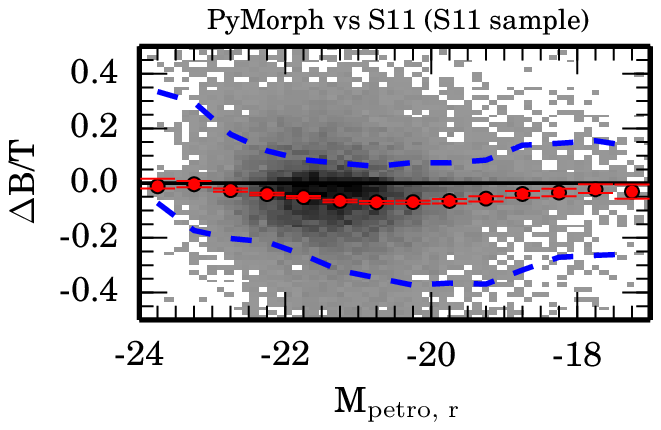}
\caption{The difference between the \pymorph{} \SerExp{} magnitude, half-light radius, and B/T for S11 \SerExp{} fits in the full catalogue. The top row shows the comparison of magnitude (left), half-light radius (centre), and B/T (right) between S11 and \pymorph{} as a function of apparent Petrosian magnitude. The second row shows the same comparisons as a function of absolute Petrosian magnitude. Galaxies  
 identified as best fit by \SerExp{} models in S11 are used in this plot. The format of the plot is the same as Figure~\ref{fig:dev_fits}. The \SerExp{} fits have wider scatter than the analogous parameters in the \DevExp{} fits.
  See Appendix~\ref{app:serexp} (available online) for additional comparisons.}
\label{fig:simard_serexp}
\end{figure*}

While S11 provide a \SerExp{} model fits, they expect less than 10 per cent of galaxies to support such a model according to the $\chi^2$ arguments in their analysis. Also, they apply a prior on the  S\'{e}rsic index  of the bulge during fitting that favours a traditional $n=4$ bulge. LG12 do not attempt to fit such a model to the low $z$ sample. As a result, our ability to compare the \SerExp{} fits is somewhat limited. 

Figure~\ref{fig:simard_serexp} shows a comparison of the total magnitude, radius and B/T values for galaxies selected as \SerExp{} according to S11. These galaxies have {\em F}-test probabilities below 0.32 for both the $P_{pS}$ and $P_{n4}$ probabilities, indicating that the \SerExp{} fit is significantly better than either the \DevExp{} or \Ser{} fits. The additional comparisons of the \SerExp{} model to S11 are presented in Section~\ref{app:serexp} for completeness. 

The total values are in agreement with wider scatter compared to that seen in Figures~\ref{fig:ser_fits} and \ref{fig:devexp}. However, the bright end of both the apparent magnitude and absolute magnitude distributions show trends similar to the \Ser{} fits. The S11 fits use the same fixed sky level for all fitted models. Since we see evidence of bias in the sky level for the \Ser{} galaxies of S11, it is likely that the same problems exist in the \SerExp{} fits. The components of the \SerExp{} fits have wider scatter than the analogous parameters in the \DevExp{} fits. 

More on the general comparisons with the S11 measurements is presented in Section~\ref{sec:other_cmp} which support 
the accuracy of our \SerExp{} fits. 

\subsection{Comparing the preferred models}\label{sec:ex:bestmod}

\begin{figure*}
\begin{center}
\includegraphics[width=0.32\linewidth]{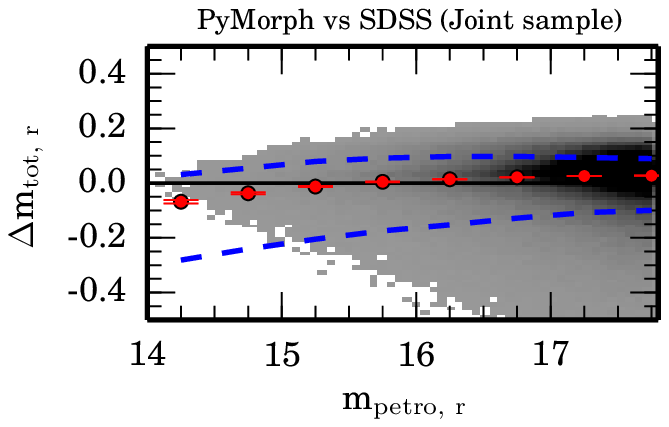}
\includegraphics[width=0.32\linewidth]{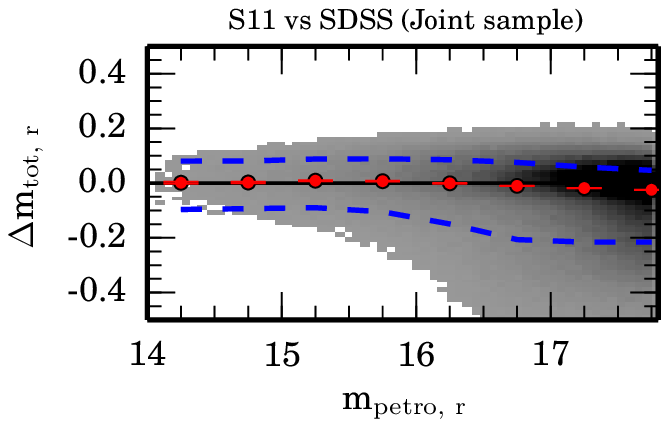}
\includegraphics[width=0.32\linewidth]{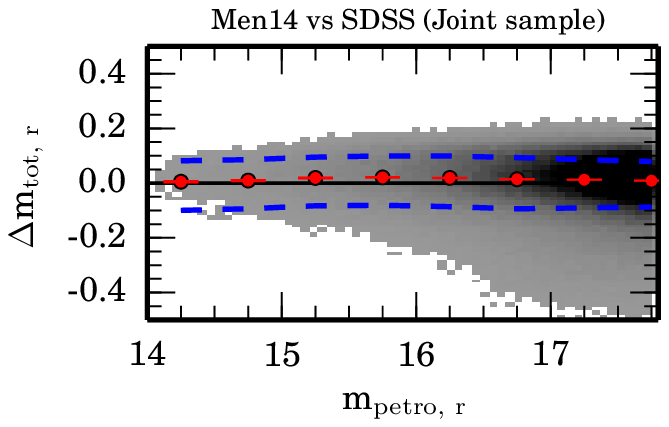}
\includegraphics[width=0.32\linewidth]{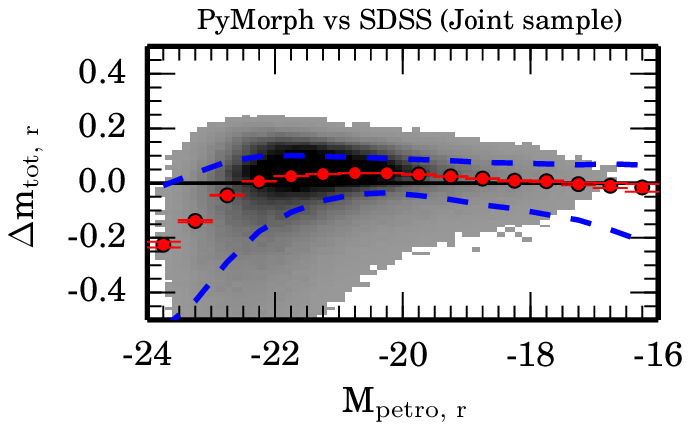}
\includegraphics[width=0.32\linewidth]{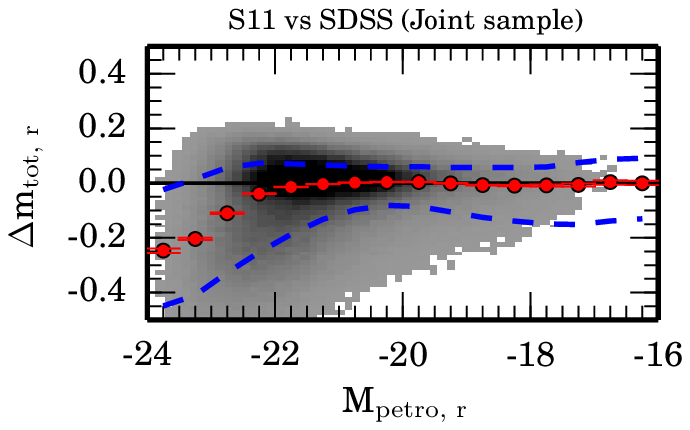}
\includegraphics[width=0.32\linewidth]{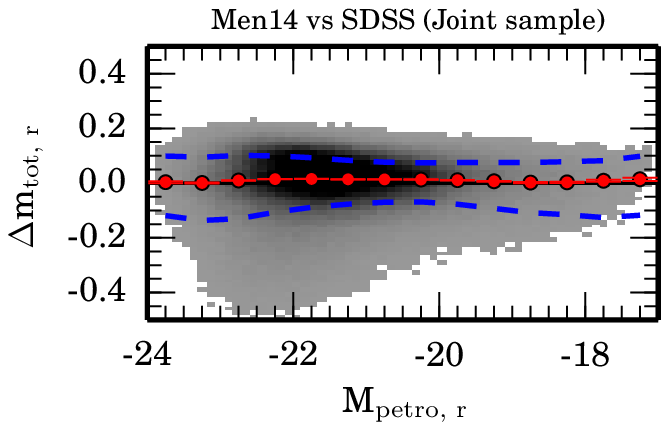}
\end{center}
\caption{The difference between SDSS cModel magnitude and the magnitude fit by \pymorph{}, S11, and Men14 for the galaxies appearing in all three catalogues. Galaxies classified
as an unknown profile in Men14 (\texttt{Proftype}$=4$) are excluded from the plots. This removes approximately 10 per cent of our original catalogue. We also exclude any failed fits from our catalogue.}
\label{fig:sdss_model_cmodel}
\end{figure*}

\begin{figure*}
\begin{center}
\includegraphics[width=0.32\linewidth]{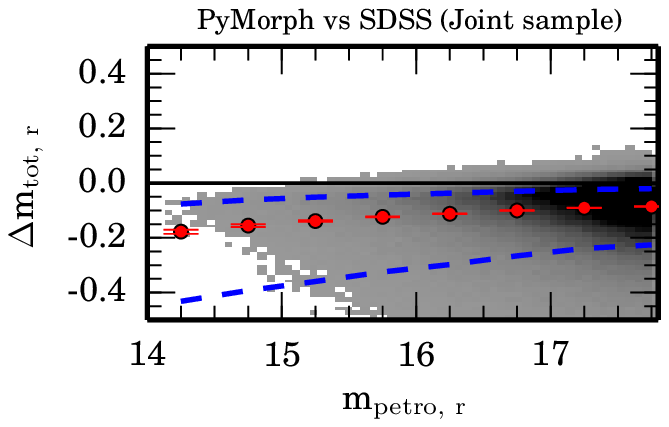}
\includegraphics[width=0.32\linewidth]{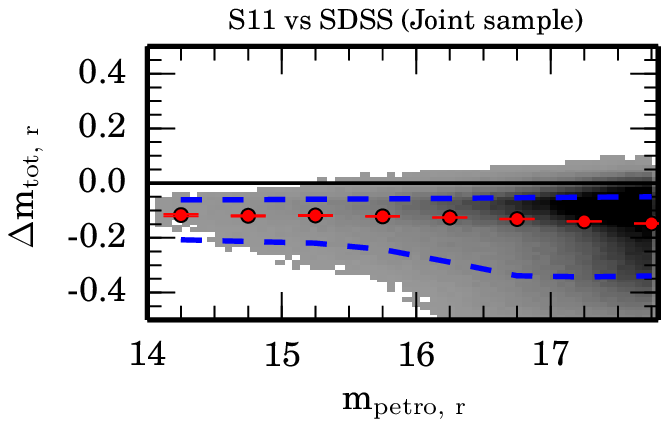}
\includegraphics[width=0.32\linewidth]{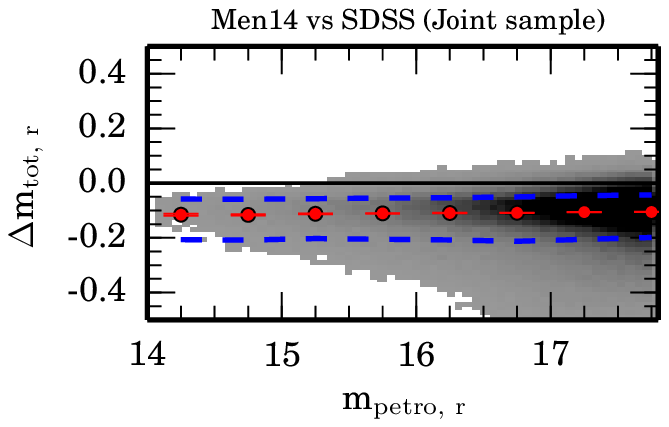}
\includegraphics[width=0.32\linewidth]{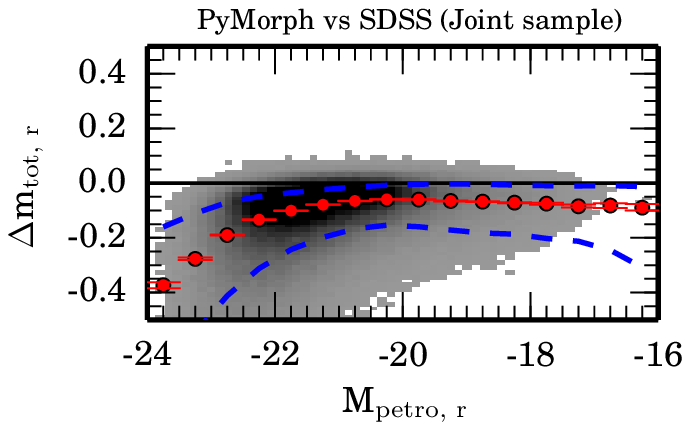}
\includegraphics[width=0.32\linewidth]{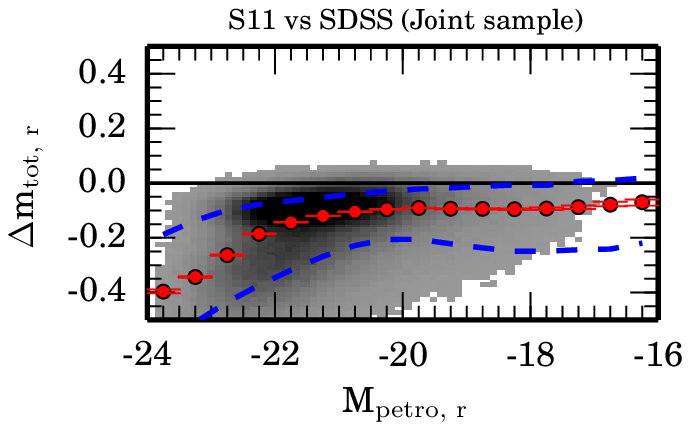}
\includegraphics[width=0.32\linewidth]{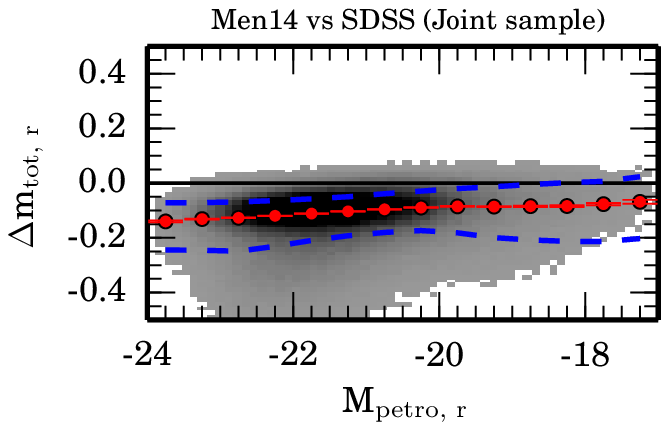}
\end{center}
\caption{The difference between SDSS Petrosian magnitude and the magnitude fit by \pymorph{}, S11, and Men14 for the galaxies appearing in all three catalogues. Galaxies classified
as an unknown profile in Men14 (\texttt{Proftype}$=4$) are excluded from the plots. This removes approximately 10 per cent of our original catalogue. We also exclude any failed fits from our catalogue.}
\label{fig:sdss_model_petro}
\end{figure*}

We conclude this section by comparing the `best model' magnitudes of this work, S11, and the Men14 classification of S11 fits to commonly used SDSS magnitudes. \cite{bernardi2013} showed that the choice of fitted model can substantially alter the bright end of
the LF. Here, we examine whether these differences can be eliminated by the use of the `best-fitting' magnitude 
rather than using magnitudes measured using only a single fitted model.

For the fits of this work, we construct a `best fit' from the combination of the \Ser{} and \SerExp{} catalogue as follows: first, 
galaxies with flag bit 10 or 14 set (galaxies shown as `2com' and `prob 2com' in Figure~\ref{fig:type_comp_pabsmag}) 
are given the total magnitude measured by fitting the \SerExp{} model. Galaxies with flag bits 1 or 4 set (\ie `bulge' or pure `disc' in Figure~\ref{fig:type_comp_pabsmag}) are given the total magnitude of the
\Ser{} fit. Finally, any galaxies with flag 20 set (\ie `bad' fits ) for the \SerExp{} fit are given the fitted \Ser{} magnitude 
if the \Ser{} fit does not have flag bit 20 set.

For S11 fits, we assign fitted \Ser{} magnitudes to galaxies with {\em F}-test probabilities $P_{pS}\geq$0.32. \DevExp{} fit magnitudes
are used for galaxies with {\em F}-test probabilities $P_{pS}<$0.32 and $P_{n4}\geq$0.32. Galaxies with {\em F}-test probabilities $P_{pS}<$0.32 and $P_{n4}<$0.32 are given the reported \SerExp{} magnitudes.

For the Men14 classification of S11 fits, galaxies are assigned \Ser{} magnitudes if they have \texttt{Proftype}$=$1 or 2 (\ie the bulge category or disc category, respectively) in the Men14 data. Galaxies are assigned the \DevExp{} fit magnitude if they have \texttt{Proftype}$=$3 (\ie the two-component category). \texttt{Proftype}$=$4 (\ie the problematic fits) are ignored.
After assigning the galaxies to these categories, we select only those galaxies present in all three data sets for comparison.

Figure~\ref{fig:sdss_model_cmodel} shows the comparison of \pymorph{}, S11, and Men14 to the SDSS cModel magnitudes for the full sample.  
The cModel magnitudes \citep[defined in][]{bernardi10} are calculated from the linear combination of the independently fit SDSS \Dev{} and \Exp{} models that best fits the galaxy. The parameter that sets the fraction of flux contributed
by the \Dev{} fit is defined as \texttt{fracdev}. A \texttt{fracdev}$=$1 galaxy is best fit by the \Dev{} model, while a \texttt{fracdev}$=$0 galaxy is best fit by the \Exp{} model. The \texttt{fracdev} parameter provides a very crude estimate similar 
to the B/T measured during simultaneous fitting. The cModel magnitudes are corrected for the offset due to profile truncation as was previously done in Figure~\ref{fig:dev_fits}. 

The difference between cModel magnitudes and S11 or this work increases at brighter magnitudes. This effect is not present using the Men14 selection of S11 (right-hand column). \cite{bernardi2013} showed that the systematic effects on the bright end of the luminosity and stellar mass functions can cause a substantial underestimate of the bright end of the LF. The lowest estimates of the bright end of the LF occur in the SDSS Petrosian and cModel measurements. Since Men14 agree more closely with SDSS at the bright end, this suggests that Men14 are selecting models that underestimate the brightest galaxies by a substantial amount. 

The large difference between Men14 and S11 in the brightest magnitude bins may be the result of Men14 strongly favouring the two-component \DevExp{} fit for all galaxies as opposed to the preference of this work and S11 for single-component \Ser{} fits at these magnitudes. The brightest, largest galaxies tend to have higher S\'{e}rsic index, which requires more light in the wings of the profile in order to account for the concentration observed at the centre of the galaxy. Similar concentrations can be reproduced 
with two lower concentration components (\Dev{} bulge and \Exp{} disc), but when integrated to infinity, these two-component galaxies have less total light compared to the broader \Ser{} profile. Fitting of a two-component \DevExp{} model appears to produce a dimmer estimate of the total magnitude relative to the \Ser{} model.

Figure~\ref{fig:sdss_model_petro} shows the similar comparison of \pymorph{}, S11, and Men14 to the SDSS Petrosian magnitudes. 
Differences approaching 0.5 mag in the brightest bins are observed in S11 and this work. These differences will increase the number
of bright galaxies in the brightest bins of the LF. While \cite{bernardi2013} only explored cases of pure fits (\ie fitting the entire sample with \Ser{} fits only or fitting the entire sample with \SerExp{} fits only), the large differences in the 
LF reported there likely persist based on the differences observed here where the `best-fitting' profile is used.

\section{Comparisons using morphological information}\label{sec:other_cmp}

Finally, we examine our selection of bulge, disc and two-component galaxies using morphological information from other catalogues. 
The work of \cite{nair} gives a detailed visual morphological classification of a set of about 10\,000 nearby galaxies. 
The GZ2 project \citep{galzoo01, galzoo02} provides another morphological classification of nearly half of our sample.
These catalogues give morphological classifications not dependent on fitting and provide another test of our profile-based flagging.
Section~\ref{sec:cmp_label} compares the preferred models of LG12, S11 and Men14 to our selection with respect to magnitude and {\em T}-types calculated using the work of \cite{nair}. 
Section~\ref{sec:intern_ttype} examines a few internal checks, including B/T and axis ratios separated by {\em T}-type.
Section~\ref{sec:mod_bar} discusses possible effects of bars on the fitting based on barred galaxies identified in GZ2.

\subsection{Preferred models as an indicator of morphological classification}\label{sec:cmp_label}

\subsubsection{Magnitude distribution of preferred models}\label{subsec:mag_cmp_models}

\begin{figure*}
\begin{center}
\includegraphics[width=0.95\linewidth]{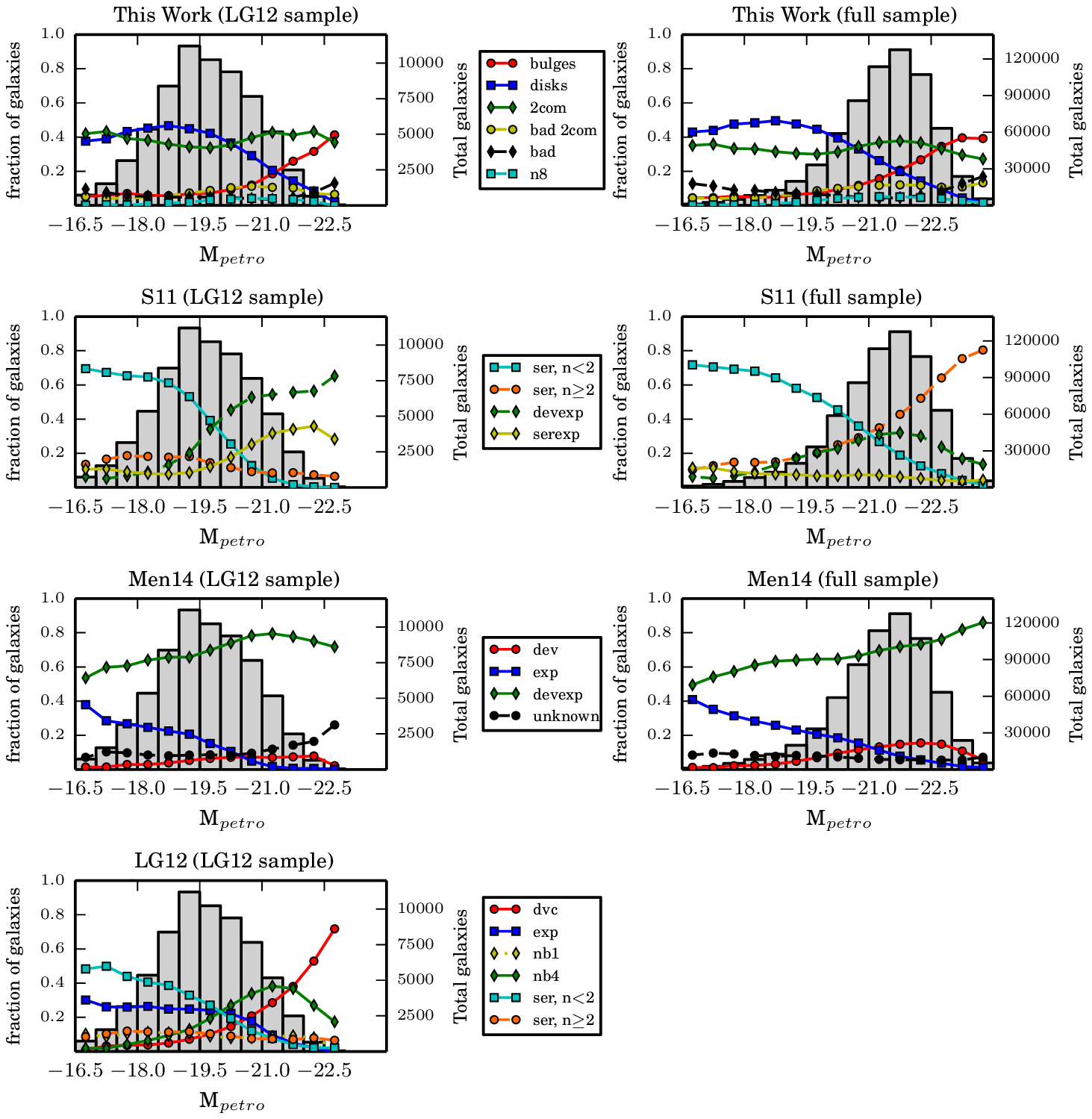}
\end{center}
\caption{Type comparisons as a function of absolute Petrosian magnitude for this work (top row), S11 (second row),
Men14 (third row), and LG12 (fourth row). The-hand left column shows the classifications for the smaller, low-$z$ LG12 sample. 
The right-hand column shows the distribution for the full sample presented in this work. Model types for each group are plotted. 
The distribution of galaxies as a function of magnitude is plotted in the background histogram and the histogram scale is on the right-hand axis of the plot. The fraction of galaxies reported is the fraction within the respective bin. For each panel, the sum of all models in any given bin is 1. Men14 are largely insensitive to the absolute magnitude of the galaxy. S11 (second row) have a large shift from fitting galaxies brighter than --21 with two-component models in the low-$z$ sample to predominantly fitting \Ser{} models to the same galaxies in the full sample. This indicates that S11 are more sensitive to resolution than to the actual morphology of the galaxy. LG12 find much higher proportions of one-component models at the low and high magnitudes compared to this work ($\sim80$ per cent for LG12 compared to $\sim40$ per cent for this work).}
\label{fig:type_comp_pabsmag}
\end{figure*}

In this section, we compare our selection of bulge, disc, and two-component galaxies presented in Section~\ref{sec:validation} with S11, LG12, and Men14, all of which include a `preferred' model in their catalogues. Figure~\ref{fig:type_comp_pabsmag} shows the percent of model types as a function of absolute Petrosian magnitude for two samples. 
The left-hand column shows the results for the low-$z$ sample of LG12. The right-hand column shows the results for the entire spectroscopic
catalogue used in this work. The panels show the models of this work (top row), S11 (second row),
Men14 (third row), and LG12 (fourth row). Model types for each group are plotted as lines with symbols. The fraction of galaxies reported is the fraction within the respective bin. For each panel, the sum of all models in any given bin is 1.
The total distribution of galaxies as a function of magnitude is plotted in the background histogram and the histogram scale is on the right-hand axis of the plot.  

The fits of this work show expected trends with almost no bulges (only a few percent) and a mixture of disc and two-component galaxies at low magnitudes (below about --19.5). Two-component fits are the dominant model between --20 and --22 and bulges dominate at magnitudes brighter than --22. This behaviour is visible in both samples and shows that the flag-based model selection is reasonably independent of apparent magnitude and resolution effects. Such effects would likely cause a different appearance in the model selection for the low-$z$ LG12 catalogue (left-hand column) when compared to the full sample (right-hand column).   

In contrast, S11 (second row) have a large shift from fitting galaxies brighter than --21 with two-component models in the low-$z$ sample (about 80 per cent of galaxies brighter than --21 in the LG12 sample are two-component) to predominantly fitting \Ser{} models to the galaxies with similar absolute magnitude in the full sample. This indicates that S11 are potentially more sensitive to observational effects (\eg resolution or S/N) than to the actual morphology of the galaxy. However, the general behaviour in the full sample is similar to the fits presented in this work. The single-component, disc-like fits dominate at faint magnitudes. Two-component fits are more prevalent at intermediate magnitudes (in the neighbourhood of -21) and single-component bulges make up a large fraction of galaxies at the brightest magnitudes.

Men14 (third row) are less sensitive to the absolute magnitude of the galaxy, with high percentages (between 60 and 80 per cent) of \DevExp{} galaxies across the entire magnitude range in the LG12 sample. Men14 do identify slightly more discs in the full sample, but the \DevExp{} model is still dominant across the magnitude range.  

LG12 (fourth row) cannot be compared to the full sample.  In the LG12 sample, the behaviour of both LG12 and this work is similar over the range of magnitudes examined here. Bulges dominate at the brightest magnitudes, discs are much more frequent at dimmer magnitudes, and two-component fits dominate in the region surrounding --20 mag. However,  LG12 find higher proportions of one-component models at the low and high magnitudes compared to this work ($\sim80$ per cent for LG12 compared to $\sim40$ per cent for this work). The authors discuss in their paper that an initial identification of bulges using fitting parameters and \chinu{} statistics does not produce a high enough percentage of bulges, so many galaxies are chosen to be `dvc' (\ie \Dev{} fits) based on colour information rather than the fitting parameters. Many of these galaxies have large diffuse components that are fitted by the \Exp{} component of the two-component models. 

As was briefly mentioned in Section~\ref{sec:fitting:basic}, we find similar behaviour in our fits (the incidence of two-component fits is higher than expected at the brightest magnitudes). 
We choose not to force these galaxies to be fit by a single component in order to avoid the magnitude bias reported in 
M13. Instead, we choose to relax our definition of the fitted models and caution the user that a significant \Exp{} component 
may be an indication of an extended halo component rather than a classical disc in early-type galaxies.

\subsubsection{Model selection as a function of {\em T}-type}\label{sec:mod_ttype}

\begin{figure*}
\begin{center}
\includegraphics[width=0.95\linewidth]{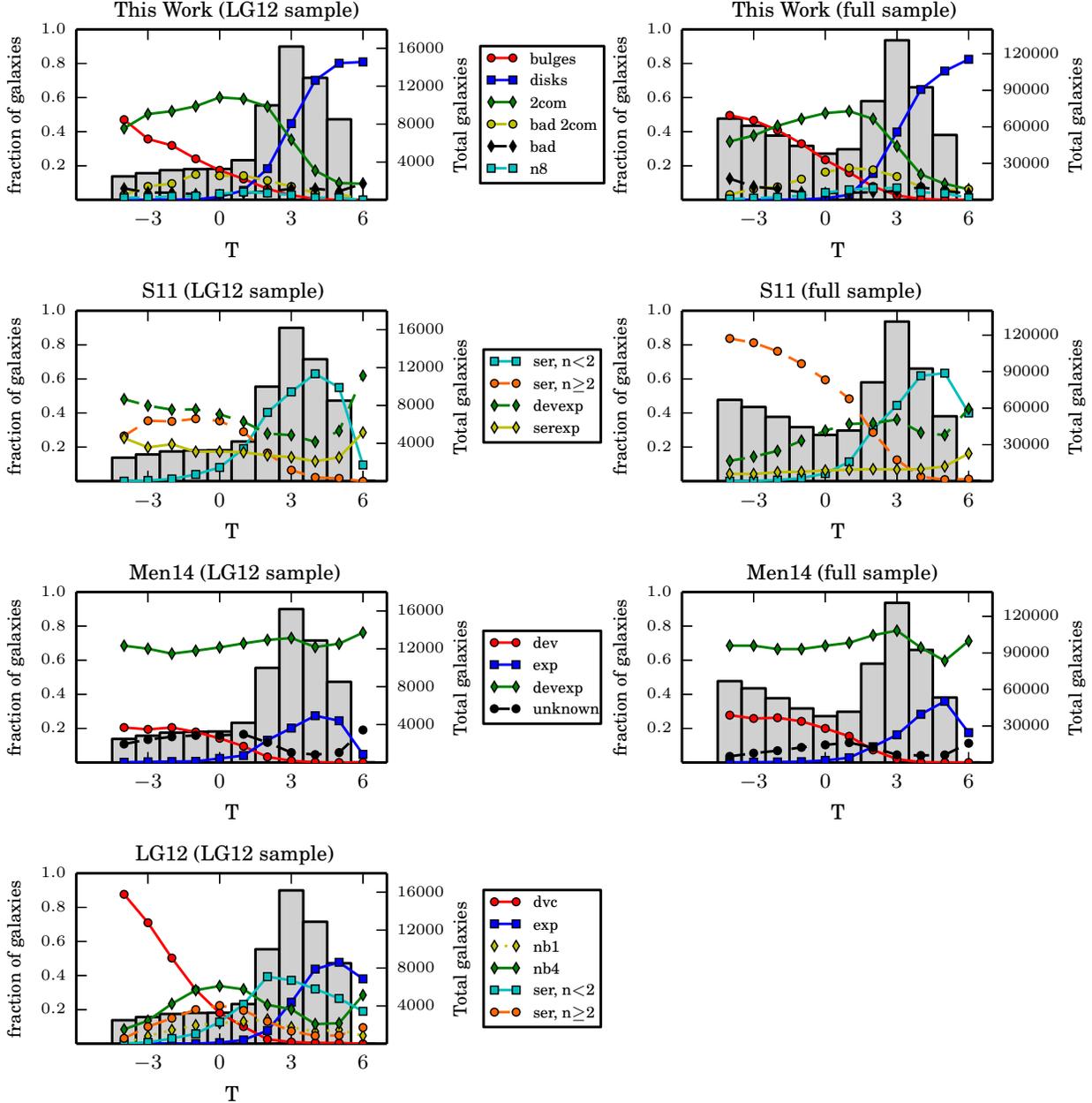}
\end{center}
\caption{Comparisons of the inferred {\em T}-type for this work (top row), S11 (second row),
Men14 (third row), and LG12 (fourth row). The left-hand column shows the classifications for the smaller, low-$z$ LG12 sample. 
The right-hand column shows the distribution for the full sample presented in this work.
Model types for each group are plotted. 
The distribution of galaxies as a function of {\em T}-type is plotted in the background histogram and the histogram scale is on the right-hand axis of the plot. The fraction of galaxies reported is the fraction within the respective bin. For each panel, the sum of all models in any given bin is 1. Men14 are again largely insensitive to the {\em T}-type of the galaxy. S11 (second row) have a large shift from fitting early type galaxies with two-component models in the low-$z$ sample to predominantly fitting \Ser{} models to the same galaxies in the full sample. This indicates that S11 are more sensitive to resolution than to the actual morphology of the galaxy.
 The bin normalization and background histogram are the same as in Figure~\protect\ref{fig:type_comp_pabsmag}}.
\label{fig:autotype_comparisons_dist}
\end{figure*}

We assign a {\em T}-type to each galaxy in the catalogue using the type probabilities (Ell, S0, Sab, and Scd) provided by the BAC 
(H2011). The BAC of H2011 used a Bayesian approach to assign probabilities of being one of four broad galaxy types (Ell, S0, Sab, or Scd) to each galaxy using colour, total axis ratio, and concentration as measured by the 
SDSS. An SVM algorithm was used to produce the probabilities, using the \cite{fukugita2007} sample as the training set. 

We calibrate the probabilities from H2011 to the {\em T}-types using a simple linear model:
\begin{equation}
\begin{aligned}
 T =& -4.6 \times P(Ell) -2.4 \times P(S0) \\
    &+2.5 \times P(Sab)+6.1 \times P(Scd)
\end{aligned}\label{eq:huertas_ttype}
\end{equation}
The coefficients of the equation are calibrated to the visually classified galaxies of \cite{nair} by an unweighted linear regression. A similar comparison was used in \cite{galzoo13}, although a symbolic regression was used rather than a linear regression. In \cite{galzoo13}, the result of the symbolic regression was a linear model in parameters relating only to the dominance of the bulge. The regression fit in this work estimates {\em T}-types between --5 and 4 with median bias of 0 in {\em T}-type and 68 per cent of the estimates within $\pm2$ in {\em T}-type. Although the scatter is relatively broad, we can  reliably separate early and late types (the difference in {\em T}-type for these galaxies is more than 4).  Since the Nair catalogue is quite small (only about 10\,000 galaxies from this work are present in Nair), we use this extension to estimate {\em T}-types for the entire sample common to S11, LG12, and this work. Since the parameters used in the H2011 model are not specific to any particular fitting method or fitting code, we expect them to be unbiased across the studies examined here. 

Figure~\ref{fig:autotype_comparisons_dist} shows the distributions of the preferred models of Men14, S11, LG12, and our \SerExp{} categories of bulge, disc, and two-component fits as a function of {\em T}-type for galaxies in the 
LG12 sample. Each plot shows the percentage of galaxies for each preferred model as a function of {\em T}-type. \Ser{} models are divided into two categories, n$<$2 and n$\geq$2, to  better understand whether the preferred models are more disc like or bulge like. Although this is the same S\'{e}rsic index cut used in this work, we also use other criteria (based on the B/T, bulge and disc axis ratio, \etc) in addition to S\'{e}rsic index to classify pure bulges and pure discs (see Section~\ref{sec:ex:bestmod}).

For this work, we plot the distribution of our three models (bulge, bulge+disc (2com), and disc) as well as the failed galaxies (called `bad'), the problematic two-component fits (called `bad 2com') and the two-component galaxies with n$=$8 bulges (called `n8'). For the S11 fits, the \DevExp{}, \SerExp{}, and \Ser{} models are plotted. For the LG12 galaxies, the five different models are shown with `nb1' referring to two-component galaxies with n$=$1 bulges and `nb4' referring to galaxies with n$=$4 bulges. The Men14 fits are separated into \Dev{}, \Exp{}, \DevExp{}, and unknown following the \texttt{Proftype} provided by the authors and used earlier in this paper (see Section~\ref{sec:ex:bestmod}). 

The classification used in this work (top row) performs largely as expected over the range of {\em T}-types. Late {\em T}-types (above T$=4$) are 70-80 per cent discs and the remainder is approximately equal parts two-component and failed fits. Pure bulges approach 0 percent of the sample at this end. The early end (below T$=-2$) has an increasing percentage of pure bulge systems, but has a large contribution from two-component models. This is discussed further in Section~\ref{sec:mod_ba}.  
These \Exp{} components are not indicative of the presence of a true disc
but rather a departure from a pure S\'{e}rsic profile. Finally, the failed fits (in black) and the n$=8$ bulges (in cyan) are uniformly distributed in {\em T}-type. This is expected if the failure rate is not correlated with {\em T}-type. A lack of correlation with {\em T}-type is preferable for evolutionary or environmental studies because it reduces the likelihood of introducing a bias by excluding these categories. 

The S11 classification (second row) again shows substantial differences in bulge preference between the LG12 and full samples. This difference occurs for the early types (T$<0$) similar to Figure~\ref{fig:type_comp_pabsmag} where the shift from two-component to one-component galaxies occurs at brighter absolute magnitude. When the \SerExp{} and \DevExp{} categories are combined into a class of two-component systems, 46 per cent of the entire LG12 sample are represented by this class. In contrast, the full S11 sample for has 35 per cent two-component galaxies (26 and 9 per cent are \DevExp{} and \SerExp{}, respectively). This shows a possible dependence on the image quality of the fitted galaxies (see second row, left). Nearby galaxies are more likely to be considered for two-component fits using the S11 criteria. This is expected since nearby galaxies are better resolved and more likely to have resolved structure. However, we would expect the well-resolved ellipticals to not be as strongly affected as the intermediate types where more structure is expected. These results suggest that while S11 fits are similar to this work (see Section~\ref{sec:analysis}) S11 model selection (\ie \Ser{}, \DevExp{}, or \SerExp{}) is possibly affected by observational effects and may not provide an accurate indication of galaxy type.

The Men14 classification (third row) is largely independent of {\em T}-type, similar to the behaviour with absolute magnitude 
discussed in Section~\ref{subsec:mag_cmp_models}. The dominant one-component model transitions from \Dev{} to \Exp{} across the {\em T}-types, but makes up nearly 20 per cent of the sample across the range.

The LG12 classification (bottom row) shows a large percentage of \Dev{} fits at the early end of the {\em T}-types. This is achieved by using a colour cut on the data. Forcing galaxies to be \Dev{} profiles based on the colour without considering the fitted models is a qualitatively different process compared to using the profiles to select preferred models, as is done in this work. There is dependence of S\'{e}rsic index across type, and we show in Section~\ref{sec:intern_bt_bulge} that Ell galaxies (\ie the most negative {\em T}-types) are dominant at higher values of S\'{e}rsic index. Choosing the \Dev{} model for these galaxies may then impose a bias on the fitted magnitudes for the brightest galaxies in the LG12 fits. LG12 have a similar percentage of pure disc systems at the late end compared to the fits of this work. Combining the \Ser{} galaxies with n$<$2 and the \Exp{} galaxies in LG12, LG12  pure discs at $T=$4-6  account for approximately 70 per cent of galaxies compared to 80 per cent in this work. Also, the \Ser{} models with $n\geq2$ (orange points) and the \texttt{nb4} (green points) models appear to be related by a simple constant fraction (\ie they are scaled versions of each other). The  similar shape of the green and orange curves suggests that the two samples may identify the same type of galaxy, with the distinction between one and two-component models being the result of observational limitations rather than intrinsic differences in the galaxies. 

\subsection{Internal consistency checks using {\em T}-type}\label{sec:intern_ttype}

\begin{figure*}
\begin{center}
\includegraphics[width=0.95\linewidth]{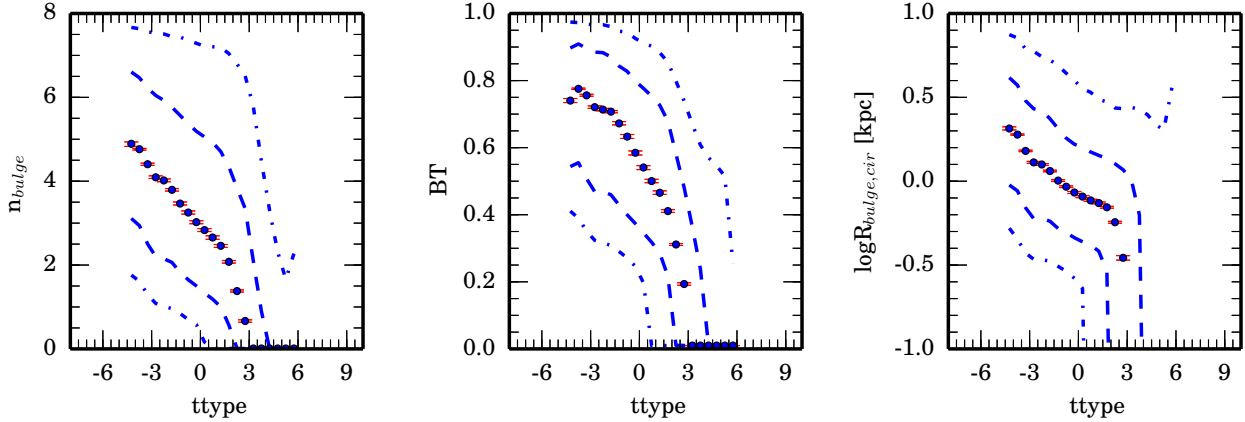}
\end{center}
\caption{The \SerExp{} bulge S\'{e}rsic index, B/T, and bulge radius for all galaxies as a function of {\em T}-type. Pure discs as denoted by our flags have been given B/T $=$ 0 and the radius and S\'{e}rsic index of the bulge are set to 0. The values trend towards zero at later {\em T}-types. Median values are plotted as the points. 68 and 95 per cent contours are plotted as dashed and dot--dashed lines. Errorbars on the median values represent 95 per cent CI obtained from bootstrap resampling.}
\label{fig:serexp_ttype_nBTr}
\end{figure*}

\subsubsection{Bulge behaviour with {\em T}-type}\label{sec:intern_bt_bulge}

We can examine a few internal checks on the component parameters of bulges and discs in our two-component models.  Figure~\ref{fig:serexp_ttype_nBTr} shows the median \SerExp{} bulge  S\'{e}rsic index, B/T, and bulge radius for all galaxies in our catalogue as a function of {\em T}-type calculated using Equation~\ref{eq:huertas_ttype}. 68 and 95 per cent contours are plotted as dashed and dot--dashed lines. Error bars on the median values represent 95 per cent CI obtained from bootstrap resampling. Pure discs, as denoted by our flags, have been given B/T $=$ 0 and the radius and  S\'{e}rsic index  of the bulge are set to 0, but not excluded from the sample. 

B/T, bulge radius, and bulge  S\'{e}rsic index  all decrease with increasing {\em T}-type. The median bulge S\'{e}rsic index (left-hand panel) for the earliest {\em T}-types is approximately $5\pm1.5$, using the 68 per cent contours. Median \SerExp{} bulge S\'{e}rsic index decreases to 2 by {\em T}-type of 2 before dropping rapidly to zero due to the increased presence of pure disc systems. Median B/T decreases from 0.8 to 0.2 over the same range while median bulge size also decreases. Since we do not consider the {\em T}-type at all during the flagging or fitting, and we also do not use the colour or any other source of morphological information during fitting, the behaviour of the bulges in our sample is good evidence for proper fitting and effective flagging. 

\subsubsection{Bulge and disk axis ratios}\label{sec:mod_ba}

\begin{figure}
\begin{center}
\includegraphics[width=0.95\linewidth]{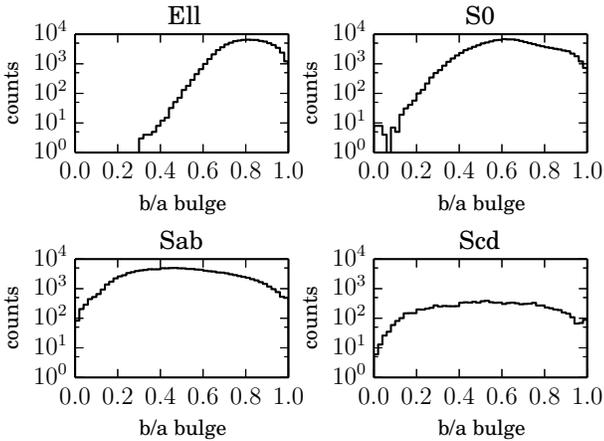}
\end{center}
\caption{The \SerExp{} bulge axis ratio for all good two-component or pure bulge galaxies separated by type.}
\label{fig:serexp_ba_bulge}
\end{figure}

 \begin{figure}
\begin{center}
\includegraphics[width=0.95\linewidth]{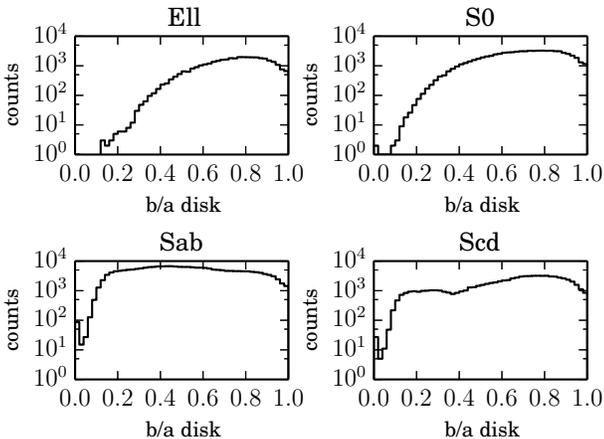}
\end{center}
\caption{The \SerExp{} disc axis ratio for all good two-component or pure disc galaxies separated by type.}
\label{fig:serexp_ba_disk}
\end{figure}

Disk and bulge axis ratios are expected to have different distributions if we are truly measuring bulges and discs as opposed to fitting unphysical components. For instance, bulges are expected to be rounder with lower ellipticities while discs should be more evenly distributed across the range of axis ratios. This distribution for discs is expected for thin discs with uniformly distributed disc inclination.  Figures~\ref{fig:serexp_ba_bulge} and \ref{fig:serexp_ba_disk} show the bulge and disc axial ratios for galaxies considered to be (1) two-component or pure bulge for the bulge axis ratio and (2) two-component or pure disc for the disc axis ratio as defined by our flags. We divide the galaxies using the {\em T}-type defined in Equation~\ref{eq:huertas_ttype}. We then bin galaxies as either Ell, S0, Sab or Scd using Equation~\ref{class_us}.
 \begin{equation}
 \textrm{Type} = \left\{
  \begin{array}{ll}
    \textrm{Ell} & : T \leq -3\\
    \textrm{S0} & : -3< T \leq 0.5\\
    \textrm{Sab} & : 0.5 < T \leq 4\\
    \textrm{Scd} & : 4 < T \\
  \end{array}
\right. \label{class_us}
\end{equation}

The axis ratio for early types is peaked near 0.8 for the Ell and 0.6 for the S0 galaxies. The distributions get progressively flatter as later types are considered. The increase of bulges with lower b/a seen in the S0, Sab, and Scd samples is, in part, explained by poor fitting due to contamination of a bar. We address this in Section~\ref{sec:mod_bar}. Overall, the distributions shown here are flatter than LG12 but similar to those reported in S11 for two-component galaxies. The distributions of disc axial ratios are flatter for the later types. We expect a flat distribution in disc axis ratio if galaxies have thin discs due to random orientation of galaxies with respect to the observer. 

Also, for the early-type galaxies, the distribution of disc axis ratios is flatter but tends to follow the bulge distribution. This is an additional indication that the second component fit by the disc is not a true disc (\ie a rotationally flattened disc) but an extended component similar to the bulge. A detailed study on this will be done in the future.  

\subsection{Effects of bars}\label{sec:mod_bar}

\begin{figure*}
\begin{center}
\includegraphics[width=0.95\linewidth]{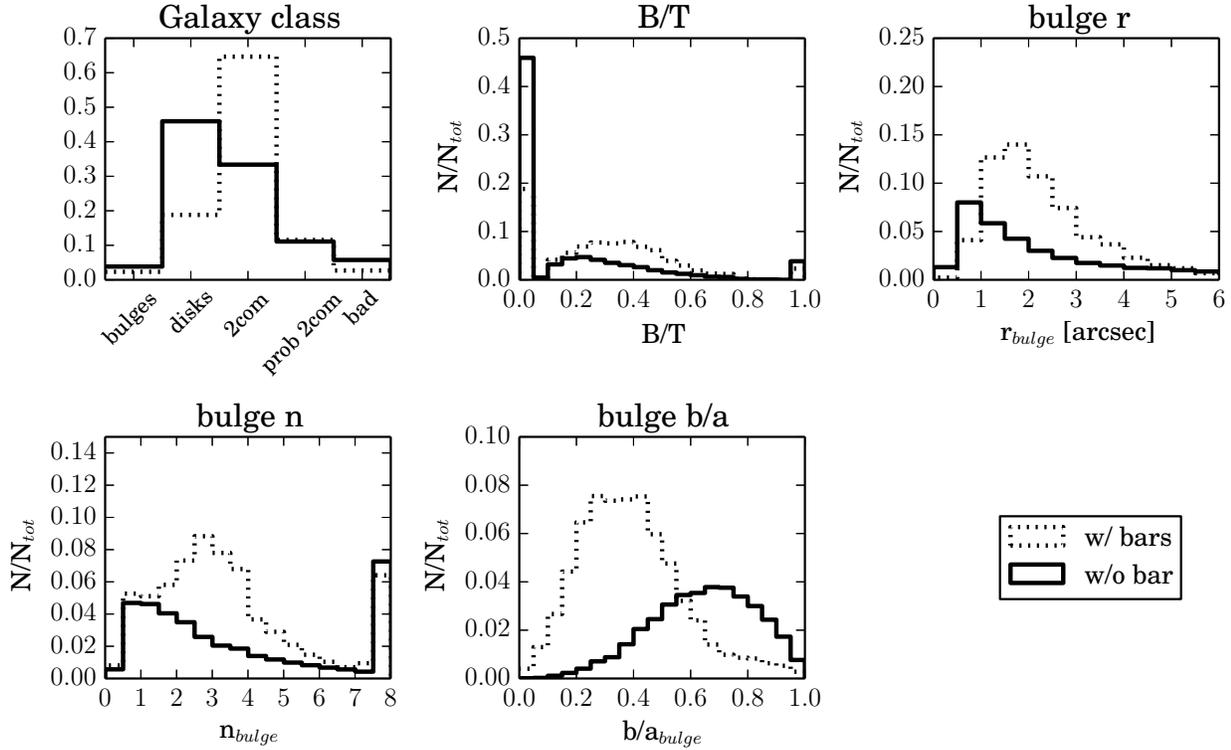}
\end{center}
\caption{The distribution of \SerExp{} fitting parameters for disc galaxies with bar components and those that do not have a bar, as identified by selecting on the bar versus no\_bar GZ2 t03 classification. Preferred models using our \SerExp{} flags (top left), fitted \SerExp{} B/T (top centre), \SerExp{} semi major bulge  halflight radius (top right), \SerExp{} bulge S\'{e}rsic index (bottom left), and \SerExp{} bulge axis ratio (bottom centre) are shown.}
\label{fig:bar_gals}
\end{figure*}

The simple models fit in this paper and the other works of LG12 and S11 neglect many often observed components of galaxies. The
effect of these components is a concern whenever the fitted values are used to test models of formation and evolution. \cite{gadotti08} showed that the effect of neglecting a bar or AGN component can still be substantial at lower resolution. A significant bar can increase B/T and the bulge effective radius by 20 per cent or more. \cite{gadotti08} also showed that AGN components had little effect on the bulge S\'{e}rsic index (this is most likely due to the loss of resolution which suppresses the S\'{e}rsic index, acting oppositely to the effects of the AGN).

Comparing galaxies with known bars to galaxies without bars can establish how strong the effect of bars is in our sample. 
GZ2 \citep{galzoo13} presents detailed morphologies for more than 300\,000 of the largest and brightest
SDSS galaxies ($m_r<17$). The galaxies were visually classified by thousands of citizen-scientists and corrected for any classification bias introduced by the citizen-scientists using spectroscopic information. GZ2 presents a debiased classification as well as a simple binary flag that is intended to select a pure, but possibly incomplete sample of many morphological classifications. 

Figure~\ref{fig:bar_gals} presents the distribution of fitting parameters for disc galaxies with bar components and those that do not have a bar, as identified by GZ (by selecting on the bar versus no\_bar GZ2 t03 classification). Preferred models using our \SerExp{} flags (top left), fitted \SerExp{} B/T (top centre), \SerExp{} semi major bulge half-light radius (top right), \SerExp{} bulge S\'{e}rsic index (bottom left), and \SerExp{} bulge axis ratio (bottom centre) for these galaxies are shown.  

Figure~\ref{fig:bar_gals} shows that we find approximately 50 per cent of the galaxies without bars to be classified as pure discs 
(\ie no measurable bulge component). However, when a bar is present, about 65 per cent of barred galaxies are fitted with two components and the percentage of pure disc galaxies drops to $\sim20$ per cent. In addition, the B/T and  S\'{e}rsic index of the \SerExp{} bulge in barred galaxies is increased. 

The distribution of semi major bulge halflight radii is skewed towards smaller values for bar-free galaxies compared to galaxies that have bar component. This is especially true when considering that almost 50 per cent of galaxies without bars are classified as pure disc versus only 20 per cent of galaxies with bars. These galaxies are included in the normalization of the plots, but do not add to the visible part of this distribution. Cases with the presence of a bar also complicate the interpretation of the bulge component, since the relative brightness of the bar and bulge must be considered when interpreting the radius of this component.

The axis ratio of galaxies with bars is much more bar like (close to a median of 0.4) than bulge like. This is clear evidence that \SerExp{} bulges of late type galaxies can be strongly affected 
by bar contributions and caution should be taken to separate bulges and bars in the \SerExp{} fits. The distribution of \SerExp{} bulge axis ratio does offer a potential method of separating galaxies with bars from the non-barred galaxies. The distribution
of \SerExp{} bulge axis ratios of the two samples are distinct in their median b/a, although the overlap is quite large. Galaxies without a visually confirmed bar have round bulges while barred galaxies have \SerExp{} bulges of much higher ellipticity. This may also explain the relative flatness of the bulge axis ratio distribution in Figure~\ref{fig:serexp_ba_bulge}. Proper modelling of the distribution may provide a constraint on the bulge fraction in our sample. However, this is beyond the scope of this work.

\section{Discussion and use of the catalogue} \label{sec:discussion}

\begin{figure}
\begin{center}
\includegraphics[width=0.95\linewidth]{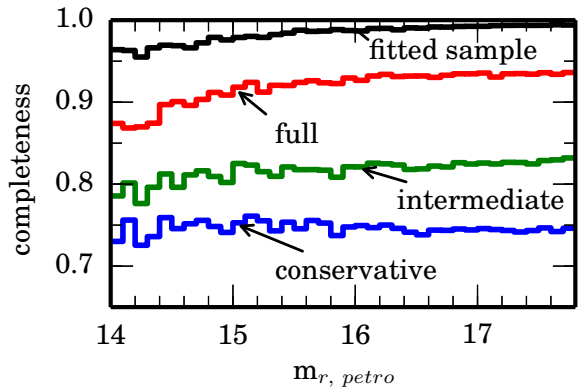}
\end{center}
\caption{The completeness of the samples described in Section~\protect\ref{sec:discussion}. The `fitted sample' represents our
selection after the cuts used in Section~\protect\ref{sec:data}. 
All completeness calculations are relative to the original
magnitude-limited galaxy sample downloaded from SDSS DR7.}
\label{fig:completeness}
\end{figure}

When using the catalogue, we recommend removing galaxies flagged as bad (flag 20) as these galaxies have catastrophically bad estimates of total magnitude and radius. Additional galaxies may be removed depending on how conservative the user seeks to be. The problematic two-component fits (flag 14) or the two-component fits with bulge  S\'{e}rsic index n$=$8 may be used for total magnitude and radius measurements, but the subcomponents are not reliable. 

The user should also be aware that we have swapped the bulge and disc components of galaxies with bit 6, 7, or 13 set (which were flagged as inverted profiles in the \SerExp{} fit). These galaxies have B/T inverted and the components reversed relative to the \SerExp{} fit. Therefore, no additional alterations must be made to 
account for the inverted nature of the profile. However, using the `raw' fit produced prior to flagging requires swapping bulge and disc parameters and inverting the B/T. This alteration has been done in all catalogues.

We provide our fits and flags for each of the four models (\Dev{}, \Ser{}, \DevExp{}, and \SerExp). In addition, we provide the preferred catalogue used in Sections~\ref{sec:ex:bestmod}~and~\ref{sec:other_cmp} for public use. This preferred catalogue 
has \Ser{} fits for galaxies flagged as pure bulge or pure disc. The remaining galaxies have \SerExp{} fits. The derived {\em T}-types used in Section~\ref{sec:other_cmp} are not included in these tables, but can be calculated directly using Equation~\ref{eq:huertas_ttype}.

We suggest one of these composite samples drawn from the preferred fit catalogue described in the previous paragraph:
\begin{description}
 \item[{\bf The conservative catalogue}] Select all galaxies with final flag bits 11, 12, or 13 set and bulge  S\'{e}rsic index  $<$8. In addition, the user should select galaxies with \SerExp{} final flag bits 1 or 4 set. These galaxies will have B/T of 1 (for bulges; final flag bit 1 set) or a B/T of 0 (for discs; final flag bit 4 set) and the relevant \Ser{} parameters are reported in the catalogue.
 \item[{\bf The intermediate catalogue}] Use the catalogue above plus all galaxies with final flag bit 10 set and bulge  S\'{e}rsic index $=$8.
 \item[{\bf The full catalogue}] Use the catalogue above plus all galaxies with final flag bit 14 set. This is the least restrictive 
 version of the catalogue but may include galaxies with strange, difficult-to-interpret fit parameters.
\end{description}

Figure~\ref{fig:completeness} shows the completeness of the three samples described above. The `fitted sample' represents our
selection after the cuts made in Section~\protect\ref{sec:data}. All completeness calculations are relative to the original
magnitude-limited galaxy sample downloaded from SDSS DR7. Not surprisingly, the completeness drops with more conservative catalogue choices. However, the completeness is largely flat across the magnitude range with a slight decrease of the order of 0.05 at magnitudes brighter than 14.5 extinction-corrected Petrosian {\em r} band magnitude. 

While the \Dev{} and \DevExp{} models are not a focus of this work, we recognize that there may be cases in which these 
models are preferred for analysis, especially when comparing to prior works. The breakdown of flags for these models is shown in 
the online Appendix~\ref{app:autoflag}. If the \DevExp{} model is desired for two-component galaxies rather than the \SerExp{} used 
here, the same criteria used to draw the composite samples described in this section may be applied to the \DevExp{} fits. This may
be done at the user's discretion. We point out here that all flagging is carried out in the same manner for the \SerExp{} and 
\DevExp{} models, except where the fits do not allow such treatment (\eg there can be no cases of bulges appearing disc-like in the \DevExp{} sample since the S\'{e}rsic index is fixed at 4 in all \DevExp{} fits). The single-component \Dev{} and \Ser{} fits also 
receive the same treatment during flagging.

The data files for this catalogue are available online at \url{http://www.physics.upenn.edu/~ameert/SDSS_PhotDec/}. 
We also provide an interface for generating panels similar to Figure~\ref{fig:vis_panel}, \ie postage stamp images of the 
2D model and residual as well as the 1D profile. These panels can be generated for a user-uploaded list of galaxies on demand.

Tables~\ref{tab:modtab1} and \ref{tab:modtab2} describe the format of the data tables released as part of this work. We distribute
the data as a binary table using the \fits{} standard. The first binary extension contains the model-independent measurements
for each galaxy (\eg \sextractor{} measurements, the number of fitted neighbours, \etc{}). The following extensions contain the
`best model' (the combination of \Ser{} and \SerExp{} fits described in Sections~\ref{sec:ex:bestmod}~and~\ref{sec:other_cmp}), the \Dev{} model, the \Ser{} model, the \DevExp{} model, and the \SerExp{} model in that order.
These extensions include the fitted values for magnitude, radius, and other parameters as well as the flags described in Section~\ref{sec:flag_auto} (labelled as \texttt{finalflag}, in column 34). A separate table containing RA/Dec/$z$ information 
and other identifying information is also available to allow matching between this catalogue and external works.

\begin{table*}
\centering
 \begin{tabular}{c c p{9 cm} c }
 \textbf{Column Number} & \textbf{Column Name} & \textbf{Explanation} & \textbf{Data Type} \\ \hline
 0  & SExMag    & The  \sextractor{} magnitude (mag)& Float \\
 1  & SExMagErr & The  \sextractor{} magnitude error (mag)& Float \\
 2  & SExHrad & The  \sextractor{} half-light radii (arcsec)& Float \\
 3  & SExSky    & The  \sextractor{} sky brightness (mag arcsec$^{-2}$)& Float\\
 4  & num\_targets & the number of targets & Int\\ 
 5  & num\_neighborfit & the number neighbour sources fitted with \Ser{} profiles& Int\\
 6  &C & The  concentration& Float \\
 7  &C\_err & The  concentration error& Float \\
 8  &A & The  asymmetry & Float \\
 9 &A\_err & The  asymmetry error& Float \\
 10 &S & The  smoothness & Float \\
 11 &S\_err & The  smoothness error& Float \\
 12 &G & The  gini coefficient & Float \\
 13 &M20 & The  M$_{20}$ value & Float \\
 14 & extinction & The SDSS-provided galactic extinction (magnitude) & Float \\
 15 & dismod   & The calculated distance modulus & Float \\
 16 & kpc\_per\_arcsec & The angular scale (kpc/arcsec) & Float \\
 17 & Vmax & The volume used for Vmax corrections (Mpc$^{3}$) & Float \\
 18 & SN & The average S/N per pixel inside the half-light radius & Float\\ 
 19 & kcorr & k-correction calculated using the SDSS model magnitudes & Float 
 \end{tabular}
\caption{Description of columns in the electronic table UPenn\_PhotDec\_nonParam\_rband. 
The data are model-independent measurements fitted by \pymorph{}. 
Problematic data or parameters are replaced with -999.}
\label{tab:modtab1}
\end{table*}
 
\begin{table*}
\centering
 \begin{tabular}{c c p{9 cm} c }
 \textbf{Column Number} & \textbf{Column Name} & \textbf{Explanation} & \textbf{Data Type} \\ \hline
0& m\_tot & Total fitted apparent magnitude & Float \\
1&BT & The B/T (bulge-to-total light ratio) of the fit & Float \\
2&r\_tot & The half-light radius (arcsec) of the total fit & Float\\
3&ba\_tot & The axis ratio (semi-minor/semi-major of the total fit & Float \\
4&xctr\_bulge & The bulge x centre (pixels) & Float \\
5&xctr\_bulge\_err & The bulge x centre error (pixels) & Float \\
6& yctr\_bulge & The  bulge y centre (pixels) & Float \\
7& yctr\_bulge\_err & The  bulge y centre error (pixels) & Float \\
8&m\_bulge & The  bulge magnitude & Float \\
9& m\_bulge\_err & The  bulge magnitude error& Float \\
10& r\_bulge & The  bulge halflight radius (arcsec) & Float \\
11& r\_bulge\_err & The  bulge radius error (arcsec)& Float \\
12& n\_bulge & The  bulge S\'{e}rsic index & Float \\
13& n\_bulge\_err & The  bulge S\'{e}rsic index error & Float \\
14& ba\_bulge & The  bulge b/a & Float \\
15& ba\_bulge\_err & The  bulge b/a error & Float \\
16& pa\_bulge & The  bulge position angle (degrees) & Float \\
17& pa\_bulge\_err & The  bulge position angle error (degrees) & Float\\
18& xctr\_disc & The  disc x centre (pixels) & Float \\
19& xctr\_disc\_err & The  disc x centre error (pixels) & Float \\
20& yctr\_disc & The  disc y centre (pixels) & Float \\
21& yctr\_disc\_err & The  disc y centre error (pixels) & Float \\
22& m\_disc & The  disc magnitude & Float \\
23& m\_disc\_err & The  disc magnitude error& Float \\
24& r\_disc & The  disc  halflight radius (arcsec) & Float \\
25& r\_disc\_err & The  disc radius error (arcsec)& Float \\
26& n\_disc & The  disc S\'{e}rsic index & Float \\
27& n\_disc\_err & The  disc S\'{e}rsic index error & Float \\
28& ba\_disc & The  disc b/a & Float \\
29& ba\_disc\_err & The  disc b/a error & Float \\
30& pa\_disc & The  disc position angle (degrees) & Float \\
31& pa\_disc\_err & The  disc position angle error (degrees) & Float \\
32&GalSky    & The  \pymorph{} sky brightness (mag/arcsec$^2$)& Float \\
33&GalSky\_err    & The  \pymorph{} sky brightness (mag/arcsec$^2$)& Float \\
34&chi2nu & The  $\chi^2/$DOF & Float \\
35&finalflag & The primary quality flag described in this work & Float \\
36&autoflag & The intermediate, visually calibrated, automated flag described in this work & Float \\
37&pyflag & The \pymorph{} run flag  & Float \\
38&pyfitflag & The \pymorph{} fit flag & Float 
\end{tabular}
\caption{Description of columns in the electronic table 
UPenn\_PhotDec\_Models\_rband. 
The data are the `best model', \Dev{}, \Ser{},
\DevExp{}, and \SerExp{} model fit parameters fitted by \pymorph{}. 
Unfit parameters or missing data are replaced with values of -999.}
\label{tab:modtab2}
\end{table*}

\section{Conclusions} \label{sec:conclusion}

A catalogue of \Dev{}, \Ser{}, \DevExp{}, and \SerExp{} galaxies was constructed for the SDSS DR7 spectroscopic sample. We used 
the \pymorph{} pipeline, including the \sextractor{} and \galfit{} programs, to perform 2D decompositions (see Sections~\ref{sec:data}~and~\ref{sec:fitting}). 
We developed a physically motivated flagging system  that removes poor fits and accurately identifies pure bulge, pure disc, and two-component systems (Sections~\ref{sec:flag_auto} and \ref{sec:cmp_label}). After applying 
the flagging system to the \SerExp{} fit, we identified about 94 per cent of our fitted sample as having reliable total 
magnitude and half-light measurements. About 39 per cent of the sample
are two-component galaxies with well-behaved components. An additional $\approx11$ per cent may be two-component fits, but with 
difficult-to-interpret components. The remaining 44 per cent are pure bulge and disc galaxies. 

We compared the fits to the low-redshift sample of LG12 and the larger samples from S11, Men14, and SDSS. 
We showed that some measurable systematic differences exist in sky brightness as well as 
percent-level systematic differences in size and magnitude. Sections~\ref{sec:ex:dev} and \ref{sec:ex:ser} discuss some of 
these differences on a catalogue-by-catalogue basis. This catalogue consistently fits brighter magnitudes to the brightest galaxies (in absolute magnitude) when compared to previous works (\eg SDSS, S11, LG12, and Men14) (Section~\ref{sec:ex_desc}). Recent work suggests that the magnitudes reported in this work are more accurate than previously measured magnitudes \citep[][]{bernardi2014,kravtsov2014}. 

Throughout this work, we have focused on the \SerExp{} model. Simulations \citep[\eg M13, ][]{mosleh13,Davari14} have shown 
that there are cases where the \SerExp{} fit produces more reliable measurements of the total magnitude and half-light radius compared to the \Ser{} fit.
Examination of the \SerExp{} fits and the flagged categories show that the morphological classes assigned in this catalogue using the flags (\ie bulge, disc, or two component) also correlate well with estimated {\em T}-types (Section~\ref{sec:mod_ttype}). Our fits also behave appropriately in B/T, bulge size, and bulge S\'{e}rsic index with respect to {\em T}-type (Section~\ref{sec:intern_bt_bulge}). 
 The different methods of model selection used by S11, LG12, Men14, and this work show some differences in the breakdown of preferred models, particularly at brighter magnitudes and for nearby galaxies. However, the judgement of which model selection method is `best' is 
dependent on individual science goals, and each selection method may be optimal for different studies.

The two-component fits are difficult to interpret in many circumstances. We examined several potential fitting systematics. 
We make several observations of potential bias in our catalogue.
\begin{itemize}[leftmargin=0.25in]
 \item Bias due to resolution effects for the \Ser{} bulge components of the \SerExp{} fits is likely present in the catalogue. 
 The majority of galaxies affected by this are the small, low B/T galaxies. Galaxies with B/T$>0.5$, including pure bulge galaxies
 have a size distribution similar to \cite{gadotti09}, and we expect the resolution to have little effect on the determination of bulge size for the pure bulge and B/T$>0.5$ as a result(Section~\ref{sec:bulge_rad_psf}).
\item Bars can strongly alter the measured parameters for the bulges of galaxies with lower B/T values, 
although using the b/a distribution can help to separate barred and non-barred galaxies (Section~\ref{sec:mod_bar}).
\item At the bright end, two-component models should not be interpreted as traditional bulge+disc systems, even though the \SerExp{} fit provides a more accurate measurement of galaxy half-light radius and magnitude. The axis-ratio of discs in the early type (see Section~\ref{sec:mod_ba} and Figures~\ref{fig:serexp_ba_bulge}~and~\ref{fig:serexp_ba_disk}) suggest that these components are 
similar to the bulge component and may represent departures from a single \Ser{} profile or an extended halo around the galaxy. 
\end{itemize}

This catalogue will be extended to {\em g} and {\em i} bands in the near future to allow for multi wavelength analysis.

\section*{Acknowledgements}
We would like to thank our referee for many helpful comments and suggestions. These have greatly improved the paper.

We would also like to thank Luc Simard for providing sky brightness values for the fits 
published in their paper \citep[][]{simard11}.
We would also like to thank Mike Jarvis and Joseph Clampitt
for many helpful discussions.

This work was supported in part by NASA grant ADP/NNX09AD02G
and NSF/0908242.

Funding for the SDSS and SDSS-II has been provided by the Alfred P. Sloan
Foundation, the Participating Institutions, the National Science Foundation, the U.S.
Department of Energy, the National Aeronautics and Space Administration, the Japanese
Monbukagakusho, the Max Planck Society, and the Higher Education Funding Council for
England. The SDSS website is http://www.sdss.org/.

The SDSS is managed by the Astrophysical Research Consortium for the
Participating Institutions. The Participating Institutions are the American Museum of
Natural History, Astrophysical Institute Potsdam, University of Basel, University of
Cambridge, Case Western Reserve University, University of Chicago, Drexel University,
Fermilab, the Institute for Advanced Study, the Japan Participation Group, Johns
Hopkins University, the Joint Institute for Nuclear Astrophysics, the Kavli Institute
for Particle Astrophysics and Cosmology, the Korean Scientist Group, the Chinese
Academy of Sciences (LAMOST), Los Alamos National Laboratory, the
Max-Planck-Institute for Astronomy (MPIA), the Max-Planck-Institute for Astrophysics
(MPA), New Mexico State University, Ohio State University, University of Pittsburgh,
University of Portsmouth, Princeton University, the US Naval Observatory,
and the University of Washington.

\bibliography{bibliography}

\begin{thebibliography}{66}
\expandafter\ifx\csname natexlab\endcsname\relax\def\natexlab#1{#1}\fi

\bibitem[{{Abazajian} {et~al}\mbox{.}(2009){Abazajian}, {Adelman-McCarthy},
  {Ag{\"u}eros}, {Allam}, {Allende Prieto}, {An}, {Anderson}, {Anderson},
  {Annis}, {Bahcall}, {Bailer-Jones}, {Barentine}, {Bassett}, {Becker},
  {Beers}, {Bell}, {Belokurov}, {Berlind}, {Berman}, {Bernardi}, {Bickerton},
  {Bizyaev}, {Blakeslee}, {Blanton}, {Bochanski}, {Boroski}, {Brewington},
  {Brinchmann}, {Brinkmann}, {Brunner}, {Budav{\'a}ri}, {Carey}, {Carliles},
  {Carr}, {Castander}, {Cinabro}, {Connolly}, {Csabai}, {Cunha}, {Czarapata},
  {Davenport}, {de Haas}, {Dilday}, {Doi}, {Eisenstein}, {Evans}, {Evans},
  {Fan}, {Friedman}, {Frieman}, {Fukugita}, {G{\"a}nsicke}, {Gates},
  {Gillespie}, {Gilmore}, {Gonzalez}, {Gonzalez}, {Grebel}, {Gunn},
  {Gy{\"o}ry}, {Hall}, {Harding}, {Harris}, {Harvanek}, {Hawley}, {Hayes},
  {Heckman}, {Hendry}, {Hennessy}, {Hindsley}, {Hoblitt}, {Hogan}, {Hogg},
  {Holtzman}, {Hyde}, {Ichikawa}, {Ichikawa}, {Im}, {Ivezi{\'c}}, {Jester},
  {Jiang}, {Johnson}, {Jorgensen}, {Juri{\'c}}, {Kent}, {Kessler}, {Kleinman},
  {Knapp}, {Konishi}, {Kron}, {Krzesinski}, {Kuropatkin}, {Lampeitl},
  {Lebedeva}, {Lee}, {Lee}, {French Leger}, {L{\'e}pine}, {Li}, {Lima}, {Lin},
  {Long}, {Loomis}, {Loveday}, {Lupton}, {Magnier}, {Malanushenko},
  {Malanushenko}, {Mandelbaum}, {Margon}, {Marriner},
  {Mart{\'{\i}}nez-Delgado}, {Matsubara}, {McGehee}, {McKay}, {Meiksin},
  {Morrison}, {Mullally}, {Munn}, {Murphy}, {Nash}, {Nebot}, {Neilsen},
  {Newberg}, {Newman}, {Nichol}, {Nicinski}, {Nieto-Santisteban}, {Nitta},
  {Okamura}, {Oravetz}, {Ostriker}, {Owen}, {Padmanabhan}, {Pan}, {Park},
  {Pauls}, {Peoples}, {Percival}, {Pier}, {Pope}, {Pourbaix}, {Price},
  {Purger}, {Quinn}, {Raddick}, {Re Fiorentin}, {Richards}, {Richmond},
  {Riess}, {Rix}, {Rockosi}, {Sako}, {Schlegel}, {Schneider}, {Scholz},
  {Schreiber}, {Schwope}, {Seljak}, {Sesar}, {Sheldon}, {Shimasaku}, {Sibley},
  {Simmons}, {Sivarani}, {Allyn Smith}, {Smith}, {Smol{\v c}i{\'c}}, {Snedden},
  {Stebbins}, {Steinmetz}, {Stoughton}, {Strauss}, {SubbaRao}, {Suto},
  {Szalay}, {Szapudi}, {Szkody}, {Tanaka}, {Tegmark}, {Teodoro}, {Thakar},
  {Tremonti}, {Tucker}, {Uomoto}, {Vanden Berk}, {Vandenberg}, {Vidrih},
  {Vogeley}, {Voges}, {Vogt}, {Wadadekar}, {Watters}, {Weinberg}, {West},
  {White}, {Wilhite}, {Wonders}, {Yanny}, {Yocum}, {York}, {Zehavi}, {Zibetti},
  \& {Zucker}}]{DR7}
{Abazajian} K.~N. {et~al.}, 2009, \apjs, 182, 543

\bibitem[{{Aihara} {et~al}\mbox{.}(2011){Aihara}, {Allende Prieto}, {An},
  {Anderson}, {Aubourg}, {Balbinot}, {Beers}, {Berlind}, {Bickerton},
  {Bizyaev}, {Blanton}, {Bochanski}, {Bolton}, {Bovy}, {Brandt}, {Brinkmann},
  {Brown}, {Brownstein}, {Busca}, {Campbell}, {Carr}, {Chen}, {Chiappini},
  {Comparat}, {Connolly}, {Cortes}, {Croft}, {Cuesta}, {da Costa}, {Davenport},
  {Dawson}, {Dhital}, {Ealet}, {Ebelke}, {Edmondson}, {Eisenstein},
  {Escoffier}, {Esposito}, {Evans}, {Fan}, {Femen{\'{\i}}a Castell{\'a}},
  {Font-Ribera}, {Frinchaboy}, {Ge}, {Gillespie}, {Gilmore}, {Gonz{\'a}lez
  Hern{\'a}ndez}, {Gott}, {Gould}, {Grebel}, {Gunn}, {Hamilton}, {Harding},
  {Harris}, {Hawley}, {Hearty}, {Ho}, {Hogg}, {Holtzman}, {Honscheid}, {Inada},
  {Ivans}, {Jiang}, {Johnson}, {Jordan}, {Jordan}, {Kazin}, {Kirkby}, {Klaene},
  {Knapp}, {Kneib}, {Kochanek}, {Koesterke}, {Kollmeier}, {Kron}, {Lampeitl},
  {Lang}, {Le Goff}, {Lee}, {Lin}, {Long}, {Loomis}, {Lucatello}, {Lundgren},
  {Lupton}, {Ma}, {MacDonald}, {Mahadevan}, {Maia}, {Makler}, {Malanushenko},
  {Malanushenko}, {Mandelbaum}, {Maraston}, {Margala}, {Masters}, {McBride},
  {McGehee}, {McGreer}, {M{\'e}nard}, {Miralda-Escud{\'e}}, {Morrison},
  {Mullally}, {Muna}, {Munn}, {Murayama}, {Myers}, {Naugle}, {Neto}, {Nguyen},
  {Nichol}, {O'Connell}, {Ogando}, {Olmstead}, {Oravetz}, {Padmanabhan},
  {Palanque-Delabrouille}, {Pan}, {Pandey}, {P{\^a}ris}, {Percival},
  {Petitjean}, {Pfaffenberger}, {Pforr}, {Phleps}, {Pichon}, {Pieri}, {Prada},
  {Price-Whelan}, {Raddick}, {Ramos}, {Reyl{\'e}}, {Rich}, {Richards}, {Rix},
  {Robin}, {Rocha-Pinto}, {Rockosi}, {Roe}, {Rollinde}, {Ross}, {Ross},
  {Rossetto}, {S{\'a}nchez}, {Sayres}, {Schlegel}, {Schlesinger}, {Schmidt},
  {Schneider}, {Sheldon}, {Shu}, {Simmerer}, {Simmons}, {Sivarani}, {Snedden},
  {Sobeck}, {Steinmetz}, {Strauss}, {Szalay}, {Tanaka}, {Thakar}, {Thomas},
  {Tinker}, {Tofflemire}, {Tojeiro}, {Tremonti}, {Vandenberg}, {Vargas
  Maga{\~n}a}, {Verde}, {Vogt}, {Wake}, {Wang}, {Weaver}, {Weinberg}, {White},
  {White}, {Yanny}, {Yasuda}, {Yeche}, \& {Zehavi}}]{DR8}
{Aihara} H. {et~al.}, 2011, \apjs, 193, 29

\bibitem[{{Allen} {et~al}\mbox{.}(2006){Allen}, {Driver}, {Graham}, {Cameron},
  {Liske}, \& {de Propris}}]{Allen2006}
{Allen} P.~D., {Driver} S.~P., {Graham} A.~W., {Cameron} E., {Liske} J., {de
  Propris} R., 2006, \mnras, 371, 2

\bibitem[{{Andrae}, {Schulze-Hartung} \& {Melchior}(2010){Andrae},
  {Schulze-Hartung}, \& {Melchior}}]{Andrae2010}
{Andrae} R., {Schulze-Hartung} T., {Melchior} P., 2010,
  arXiv:astro-ph/1012.3754

\bibitem[{{Barden} {et~al}\mbox{.}(2012){Barden}, {H{\"a}u{\ss}ler}, {Peng},
  {McIntosh}, \& {Guo}}]{barden2012}
{Barden} M., {H{\"a}u{\ss}ler} B., {Peng} C.~Y., {McIntosh} D.~H., {Guo} Y.,
  2012, \mnras, 422, 449

\bibitem[{{Bernardi} {et~al}\mbox{.}(2003{\natexlab{a}}){Bernardi}, {Sheth},
  {Annis}, {Burles}, {Eisenstein}, {Finkbeiner}, {Hogg}, {Lupton}, {Schlegel},
  {SubbaRao}, {Bahcall}, {Blakeslee}, {Brinkmann}, {Castander}, {Connolly},
  {Csabai}, {Doi}, {Fukugita}, {Frieman}, {Heckman}, {Hennessy}, {Ivezi{\'c}},
  {Knapp}, {Lamb}, {McKay}, {Munn}, {Nichol}, {Okamura}, {Schneider}, {Thakar},
  \& {York}}]{bernardi03b}
{Bernardi} M. {et~al.}, 2003{\natexlab{a}}, \aj, 125, 1849

\bibitem[{{Bernardi} {et~al}\mbox{.}(2003{\natexlab{b}}){Bernardi}, {Sheth},
  {Annis}, {Burles}, {Eisenstein}, {Finkbeiner}, {Hogg}, {Lupton}, {Schlegel},
  {SubbaRao}, {Bahcall}, {Blakeslee}, {Brinkmann}, {Castander}, {Connolly},
  {Csabai}, {Doi}, {Fukugita}, {Frieman}, {Heckman}, {Hennessy}, {Ivezi{\'c}},
  {Knapp}, {Lamb}, {McKay}, {Munn}, {Nichol}, {Okamura}, {Schneider}, {Thakar},
  \& {York}}]{bernardi03c}
{Bernardi} M. {et~al.}, 2003{\natexlab{b}}, \aj, 125, 1866

\bibitem[{{Bernardi} {et~al}\mbox{.}(2003{\natexlab{c}}){Bernardi}, {Sheth},
  {Annis}, {Burles}, {Finkbeiner}, {Lupton}, {Schlegel}, {SubbaRao}, {Bahcall},
  {Blakeslee}, {Brinkmann}, {Castander}, {Connolly}, {Csabai}, {Doi},
  {Fukugita}, {Frieman}, {Heckman}, {Hennessy}, {Ivezi{\'c}}, {Knapp}, {Lamb},
  {McKay}, {Munn}, {Nichol}, {Okamura}, {Schneider}, {Thakar}, \&
  {York}}]{bernardi03d}
{Bernardi} M. {et~al.}, 2003{\natexlab{c}}, \aj, 125, 1882

\bibitem[{{Bernardi}(2009)}]{Bernardi2009}
{Bernardi} M., 2009, \mnras, 395, 1491

\bibitem[{{Bernardi} {et~al}\mbox{.}(2010){Bernardi}, {Shankar}, {Hyde}, {Mei},
  {Marulli}, \& {Sheth}}]{bernardi10}
{Bernardi} M., {Shankar} F., {Hyde} J.~B., {Mei} S., {Marulli} F., {Sheth}
  R.~K., 2010, \mnras, 404, 2087

\bibitem[{{Bernardi} {et~al}\mbox{.}(2013){Bernardi}, {Meert}, {Sheth},
  {Vikram}, {Huertas-Company}, {Mei}, \& {Shankar}}]{bernardi2013}
{Bernardi} M., {Meert} A., {Sheth} R.~K., {Vikram} V., {Huertas-Company} M.,
  {Mei} S., {Shankar} F., 2013, \mnras, 436, 697

\bibitem[{{Bernardi} {et~al}\mbox{.}(2014){Bernardi}, {Meert}, {Vikram},
  {Huertas-Company}, {Mei}, {Shankar}, \& {Sheth}}]{bernardi2014}
{Bernardi} M., {Meert} A., {Vikram} V., {Huertas-Company} M., {Mei} S.,
  {Shankar} F., {Sheth} R.~K., 2014, \mnras, 443, 874

\bibitem[{{Bertin} \& {Arnouts}(1996)}]{sex}
{Bertin} E., {Arnouts} S., 1996, \aaps, 117, 393

\bibitem[{{Blanton} {et~al}\mbox{.}(2001){Blanton}, {Dalcanton}, {Eisenstein},
  {Loveday}, {Strauss}, {SubbaRao}, {Weinberg}, {Anderson}, {Annis}, {Bahcall},
  {Bernardi}, {Brinkmann}, {Brunner}, {Burles}, {Carey}, {Castander},
  {Connolly}, {Csabai}, {Doi}, {Finkbeiner}, {Friedman}, {Frieman}, {Fukugita},
  {Gunn}, {Hennessy}, {Hindsley}, {Hogg}, {Ichikawa}, {Ivezi{\'c}}, {Kent},
  {Knapp}, {Lamb}, {Leger}, {Long}, {Lupton}, {McKay}, {Meiksin}, {Merelli},
  {Munn}, {Narayanan}, {Newcomb}, {Nichol}, {Okamura}, {Owen}, {Pier}, {Pope},
  {Postman}, {Quinn}, {Rockosi}, {Schlegel}, {Schneider}, {Shimasaku},
  {Siegmund}, {Smee}, {Snir}, {Stoughton}, {Stubbs}, {Szalay}, {Szokoly},
  {Thakar}, {Tremonti}, {Tucker}, {Uomoto}, {Vanden Berk}, {Vogeley},
  {Waddell}, {Yanny}, {Yasuda}, \& {York}}]{blanton2001}
{Blanton} M.~R. {et~al.}, 2001, \aj, 121, 2358

\bibitem[{{Blanton} {et~al}\mbox{.}(2005){Blanton}, {Eisenstein}, {Hogg},
  {Schlegel}, \& {Brinkmann}}]{blanton2005}
{Blanton} M.~R., {Eisenstein} D., {Hogg} D.~W., {Schlegel} D.~J., {Brinkmann}
  J., 2005, \apj, 629, 149

\bibitem[{{Blanton} \& {Roweis}(2007)}]{blantonKcorr}
{Blanton} M.~R., {Roweis} S., 2007, \aj, 133, 734

\bibitem[{{Blanton} {et~al}\mbox{.}(2011){Blanton}, {Kazin}, {Muna}, {Weaver},
  \& {Price-Whelan}}]{blanton2011}
{Blanton} M.~R., {Kazin} E., {Muna} D., {Weaver} B.~A., {Price-Whelan} A.,
  2011, \aj, 142, 31

\bibitem[{{Capaccioli}(1989)}]{capaccioli1989}
{Capaccioli} M., 1989, in World of Galaxies (Le Monde des Galaxies), {Corwin}
  Jr. H.~G., {Bottinelli} L., eds., Springer-Verlag, Berlin, p. 208

\bibitem[{{Davari} {et~al}\mbox{.}(2014){Davari}, {Ho}, {Peng}, \&
  {Huang}}]{Davari14}
{Davari} R., {Ho} L.~C., {Peng} C.~Y., {Huang} S., 2014, \apj, 787, 69

\bibitem[{{de Vaucouleurs}(1948)}]{DeVacouleurs1948}
{de Vaucouleurs} G., 1948, Ann. Astrophys, 11, 247

\bibitem[{{Djorgovski} \& {Davis}(1987)}]{FundamentalPlane}
{Djorgovski} S., {Davis} M., 1987, \apj, 313, 59

\bibitem[{{Donzelli}, {Muriel} \& {Madrid}(2011){Donzelli}, {Muriel}, \&
  {Madrid}}]{Donzelli2011}
{Donzelli} C.~J., {Muriel} H., {Madrid} J.~P., 2011, \apjs, 195, 15

\bibitem[{{Dressler}(1980)}]{DresslerMDR}
{Dressler} A., 1980, \apj, 236, 351

\bibitem[{{Faber} \& {Jackson}(1976)}]{FaberJackson}
{Faber} S.~M., {Jackson} R.~E., 1976, \apj, 204, 668

\bibitem[{{Fukugita} {et~al}\mbox{.}(2007){Fukugita}, {Nakamura}, {Okamura},
  {Yasuda}, {Barentine}, {Brinkmann}, {Gunn}, {Harvanek}, {Ichikawa}, {Lupton},
  {Schneider}, {Strauss}, \& {York}}]{fukugita2007}
{Fukugita} M. {et~al.}, 2007, \aj, 134, 579

\bibitem[{{Gadotti}(2008)}]{gadotti08}
{Gadotti} D.~A., 2008, \mnras, 384, 420

\bibitem[{{Gadotti}(2009)}]{gadotti09}
{Gadotti} D.~A., 2009, \mnras, 393, 1531

\bibitem[{{Gonzalez}, {Zabludoff} \& {Zaritsky}(2005){Gonzalez}, {Zabludoff},
  \& {Zaritsky}}]{gonzalez2005}
{Gonzalez} A.~H., {Zabludoff} A.~I., {Zaritsky} D., 2005, \apj, 618, 195

\bibitem[{{Graham} \& {Driver}(2005)}]{Graham2005}
{Graham} A.~W., {Driver} S.~P., 2005, \pasa, 22, 118

\bibitem[{{Graham} \& {Driver}(2007)}]{graham2007}
{Graham} A.~W., {Driver} S.~P., 2007, \apj, 655, 77

\bibitem[{{H{\"a}ussler} {et~al}\mbox{.}(2007){H{\"a}ussler}, {McIntosh},
  {Barden}, {Bell}, {Rix}, {Borch}, {Beckwith}, {Caldwell}, {Heymans},
  {Jahnke}, {Jogee}, {Koposov}, {Meisenheimer}, {S{\'a}nchez}, {Somerville},
  {Wisotzki}, \& {Wolf}}]{GEMS2007}
{H{\"a}ussler} B. {et~al.}, 2007, \apjs, 172, 615

\bibitem[{{H{\"a}u{\ss}ler} {et~al}\mbox{.}(2013){H{\"a}u{\ss}ler}, {Bamford},
  {Vika}, {Rojas}, {Barden}, {Kelvin}, {Alpaslan}, {Robotham}, {Driver},
  {Baldry}, {Brough}, {Hopkins}, {Liske}, {Nichol}, {Popescu}, \&
  {Tuffs}}]{Haussler2012}
{H{\"a}u{\ss}ler} B. {et~al.}, 2013, \mnras, 430, 330

\bibitem[{{Huang} {et~al}\mbox{.}(2013){Huang}, {Ho}, {Peng}, {Li}, \&
  {Barth}}]{huang13}
{Huang} S., {Ho} L.~C., {Peng} C.~Y., {Li} Z.-Y., {Barth} A.~J., 2013, \apj,
  766, 47

\bibitem[{{Huertas-Company} {et~al}\mbox{.}(2011){Huertas-Company}, {Aguerri},
  {Bernardi}, {Mei}, \& {S{\'a}nchez Almeida}}]{huertas10}
{Huertas-Company} M., {Aguerri} J.~A.~L., {Bernardi} M., {Mei} S., {S{\'a}nchez
  Almeida} J., 2011, \aap, 525, A157+

\bibitem[{{Huertas-Company} {et~al}\mbox{.}(2013){Huertas-Company}, {Shankar},
  {Mei}, {Bernardi}, {Aguerri}, {Meert}, \& {Vikram}}]{huertas12}
{Huertas-Company} M., {Shankar} F., {Mei} S., {Bernardi} M., {Aguerri}
  J.~A.~L., {Meert} A., {Vikram} V., 2013, \apj, 779, 29

\bibitem[{{Kauffmann} {et~al}\mbox{.}(2003){Kauffmann}, {Heckman}, {White},
  {Charlot}, {Tremonti}, {Brinchmann}, {Bruzual}, {Peng}, {Seibert},
  {Bernardi}, {Blanton}, {Brinkmann}, {Castander}, {Cs{\'a}bai}, {Fukugita},
  {Ivezic}, {Munn}, {Nichol}, {Padmanabhan}, {Thakar}, {Weinberg}, \&
  {York}}]{Kauffman_SFR}
{Kauffmann} G. {et~al.}, 2003, \mnras, 341, 33

\bibitem[{{Kelvin} {et~al}\mbox{.}(2012){Kelvin}, {Driver}, {Robotham}, {Hill},
  {Alpaslan}, {Baldry}, {Bamford}, {Bland-Hawthorn}, {Brough}, {Graham},
  {H{\"a}ussler}, {Hopkins}, {Liske}, {Loveday}, {Norberg}, {Phillipps},
  {Popescu}, {Prescott}, {Taylor}, \& {Tuffs}}]{Kelvin2011}
{Kelvin} L.~S. {et~al.}, 2012, \mnras, 421, 1007

\bibitem[{{Kormendy}(1977)}]{Kormendy}
{Kormendy} J., 1977, \apj, 218, 333

\bibitem[{{Kormendy} {et~al}\mbox{.}(2009){Kormendy}, {Fisher}, {Cornell}, \&
  {Bender}}]{kormendy2009}
{Kormendy} J., {Fisher} D.~B., {Cornell} M.~E., {Bender} R., 2009, \apjs, 182,
  216

\bibitem[{{Kravtsov}, {Vikhlinin} \& {Meshscheryakov}(2014){Kravtsov},
  {Vikhlinin}, \& {Meshscheryakov}}]{kravtsov2014}
{Kravtsov} A., {Vikhlinin} A., {Meshscheryakov} A., 2014,
  arXiv:astro-ph/1401.7329

\bibitem[{{Lackner} \& {Gunn}(2012)}]{Lackner2012}
{Lackner} C.~N., {Gunn} J.~E., 2012, \mnras, 421, 2277

\bibitem[{{Lintott} {et~al}\mbox{.}(2008){Lintott}, {Schawinski}, {Slosar},
  {Land}, {Bamford}, {Thomas}, {Raddick}, {Nichol}, {Szalay}, {Andreescu},
  {Murray}, \& {Vandenberg}}]{galzoo01}
{Lintott} C.~J. {et~al.}, 2008, \mnras, 389, 1179

\bibitem[{{Lintott} {et~al}\mbox{.}(2011){Lintott}, {Schawinski}, {Bamford},
  {Slosar}, {Land}, {Thomas}, {Edmondson}, {Masters}, {Nichol}, {Raddick},
  {Szalay}, {Andreescu}, {Murray}, \& {Vandenberg}}]{galzoo02}
{Lintott} C. {et~al.}, 2011, \mnras, 410, 166

\bibitem[{{LSST Science Collaboration} {et~al}\mbox{.}(2009){LSST Science
  Collaboration}, {Abell}, {Allison}, {Anderson}, {Andrew}, {Angel}, {Armus},
  {Arnett}, {Asztalos}, {Axelrod}, {Bailey}, {Ballantyne}, {Bankert},
  {Barkhouse}, {Barr}, {Barrientos}, {Barth}, {Bartlett}, {Becker}, {Becla},
  {Beers}, {Bernstein}, {Biswas}, {Blanton}, {Bloom}, {Bochanski}, {Boeshaar},
  {Borne}, {Bradac}, {Brandt}, {Bridge}, {Brown}, {Brunner}, {Bullock},
  {Burgasser}, {Burge}, {Burke}, {Cargile}, {Chandrasekharan}, {Chartas},
  {Chesley}, {Chu}, {Cinabro}, {Claire}, {Claver}, {Clowe}, {Connolly}, {Cook},
  {Cooke}, {Cooray}, {Covey}, {Culliton}, {de Jong}, {de Vries}, {Debattista},
  {Delgado}, {Dell'Antonio}, {Dhital}, {Di Stefano}, {Dickinson}, {Dilday},
  {Djorgovski}, {Dobler}, {Donalek}, {Dubois-Felsmann}, {Durech},
  {Eliasdottir}, {Eracleous}, {Eyer}, {Falco}, {Fan}, {Fassnacht}, {Ferguson},
  {Fernandez}, {Fields}, {Finkbeiner}, {Figueroa}, {Fox}, {Francke}, {Frank},
  {Frieman}, {Fromenteau}, {Furqan}, {Galaz}, {Gal-Yam}, {Garnavich},
  {Gawiser}, {Geary}, {Gee}, {Gibson}, {Gilmore}, {Grace}, {Green}, {Gressler},
  {Grillmair}, {Habib}, {Haggerty}, {Hamuy}, {Harris}, {Hawley}, {Heavens},
  {Hebb}, {Henry}, {Hileman}, {Hilton}, {Hoadley}, {Holberg}, {Holman},
  {Howell}, {Infante}, {Ivezic}, {Jacoby}, {Jain}, {R}, {Jedicke}, {Jee},
  {Garrett Jernigan}, {Jha}, {Johnston}, {Jones}, {Juric}, {Kaasalainen},
  {Styliani}, {Kafka}, {Kahn}, {Kaib}, {Kalirai}, {Kantor}, {Kasliwal},
  {Keeton}, {Kessler}, {Knezevic}, {Kowalski}, {Krabbendam}, {Krughoff},
  {Kulkarni}, {Kuhlman}, {Lacy}, {Lepine}, {Liang}, {Lien}, {Lira}, {Long},
  {Lorenz}, {Lotz}, {Lupton}, {Lutz}, {Macri}, {Mahabal}, {Mandelbaum},
  {Marshall}, {May}, {McGehee}, {Meadows}, {Meert}, {Milani}, {Miller},
  {Miller}, {Mills}, {Minniti}, {Monet}, {Mukadam}, {Nakar}, {Neill}, {Newman},
  {Nikolaev}, {Nordby}, {O'Connor}, {Oguri}, {Oliver}, {Olivier}, {Olsen},
  {Olsen}, {Olszewski}, {Oluseyi}, {Padilla}, {Parker}, {Pepper}, {Peterson},
  {Petry}, {Pinto}, {Pizagno}, {Popescu}, {Prsa}, {Radcka}, {Raddick},
  {Rasmussen}, {Rau}, {Rho}, {Rhoads}, {Richards}, {Ridgway}, {Robertson},
  {Roskar}, {Saha}, {Sarajedini}, {Scannapieco}, {Schalk}, {Schindler},
  {Schmidt}, {Schmidt}, {Schneider}, {Schumacher}, {Scranton}, {Sebag},
  {Seppala}, {Shemmer}, {Simon}, {Sivertz}, {Smith}, {Allyn Smith}, {Smith},
  {Spitz}, {Stanford}, {Stassun}, {Strader}, {Strauss}, {Stubbs}, {Sweeney},
  {Szalay}, {Szkody}, {Takada}, {Thorman}, {Trilling}, {Trimble}, {Tyson}, {Van
  Berg}, {Vanden Berk}, {VanderPlas}, {Verde}, {Vrsnak}, {Walkowicz},
  {Wandelt}, {Wang}, {Wang}, {Warner}, {Wechsler}, {West}, {Wiecha},
  {Williams}, {Willman}, {Wittman}, {Wolff}, {Wood-Vasey}, {Wozniak}, {Young},
  {Zentner}, \& {Zhan}}]{lsst2009}
{LSST Science Collaboration} {et~al.}, 2009, arXiv:astro-ph/0912.0201

\bibitem[{{Meert}, {Vikram} \& {Bernardi}(2013){Meert}, {Vikram}, \&
  {Bernardi}}]{meert2013}
{Meert} A., {Vikram} V., {Bernardi} M., 2013, \mnras, 433, 1344

\bibitem[{{Mendel} {et~al}\mbox{.}(2014){Mendel}, {Simard}, {Palmer},
  {Ellison}, \& {Patton}}]{mendel2014}
{Mendel} J.~T., {Simard} L., {Palmer} M., {Ellison} S.~L., {Patton} D.~R.,
  2014, \apjs, 210, 3

\bibitem[{{Metropolis} {et~al}\mbox{.}(1953){Metropolis}, {Rosenbluth},
  {Rosenbluth}, {Teller}, \& {Teller}}]{Metropolis}
{Metropolis} N., {Rosenbluth} A.~W., {Rosenbluth} M.~N., {Teller} A.~H.,
  {Teller} E., 1953, \jcp, 21, 1087

\bibitem[{Mosleh, Williams \& Franx(2013)Mosleh, Williams, \& Franx}]{mosleh13}
Mosleh M., Williams R.~J., Franx M., 2013, \apj, 777, 117

\bibitem[{{Nair} \& {Abraham}(2010)}]{nair}
{Nair} P.~B., {Abraham} R.~G., 2010, \apjs, 186, 427

\bibitem[{{Padmanabhan} {et~al}\mbox{.}(2008){Padmanabhan}, {Schlegel},
  {Finkbeiner}, {Barentine}, {Blanton}, {Brewington}, {Gunn}, {Harvanek},
  {Hogg}, {Ivezi{\'c}}, {Johnston}, {Kent}, {Kleinman}, {Knapp}, {Krzesinski},
  {Long}, {Neilsen}, {Nitta}, {Loomis}, {Lupton}, {Roweis}, {Snedden},
  {Strauss}, \& {Tucker}}]{ubercal}
{Padmanabhan} N. {et~al.}, 2008, \apj, 674, 1217

\bibitem[{{Peng} {et~al}\mbox{.}(2002){Peng}, {Ho}, {Impey}, \& {Rix}}]{galfit}
{Peng} C.~Y., {Ho} L.~C., {Impey} C.~D., {Rix} H.-W., 2002, \aj, 124, 266

\bibitem[{{S{\'e}rsic}(1963)}]{Sersic1963}
{S{\'e}rsic} J.~L., 1963, Bol. Asociacion Argentina Astron. La Plata Argentina,
  6, 41

\bibitem[{{Shankar} {et~al}\mbox{.}(2010{\natexlab{a}}){Shankar}, {Marulli},
  {Bernardi}, {Dai}, {Hyde}, \& {Sheth}}]{shankar10a}
{Shankar} F., {Marulli} F., {Bernardi} M., {Dai} X., {Hyde} J.~B., {Sheth}
  R.~K., 2010{\natexlab{a}}, \mnras, 403, 117

\bibitem[{{Shankar} {et~al}\mbox{.}(2010{\natexlab{b}}){Shankar}, {Marulli},
  {Bernardi}, {Boylan-Kolchin}, {Dai}, \& {Khochfar}}]{shankar10b}
{Shankar} F., {Marulli} F., {Bernardi} M., {Boylan-Kolchin} M., {Dai} X.,
  {Khochfar} S., 2010{\natexlab{b}}, \mnras, 405, 948

\bibitem[{{Shankar} {et~al}\mbox{.}(2013){Shankar}, {Marulli}, {Bernardi},
  {Mei}, {Meert}, \& {Vikram}}]{Shankar}
{Shankar} F., {Marulli} F., {Bernardi} M., {Mei} S., {Meert} A., {Vikram} V.,
  2013, \mnras, 428, 109

\bibitem[{{Shen} {et~al}\mbox{.}(2003){Shen}, {Mo}, {White}, {Blanton},
  {Kauffmann}, {Voges}, {Brinkmann}, \& {Csabai}}]{Shen2003}
{Shen} S., {Mo} H.~J., {White} S.~D.~M., {Blanton} M.~R., {Kauffmann} G.,
  {Voges} W., {Brinkmann} J., {Csabai} I., 2003, \mnras, 343, 978

\bibitem[{{Simard} {et~al}\mbox{.}(2002){Simard}, {Willmer}, {Vogt},
  {Sarajedini}, {Phillips}, {Weiner}, {Koo}, {Im}, {Illingworth}, \&
  {Faber}}]{Groth2002}
{Simard} L. {et~al.}, 2002, \apjs, 142, 1

\bibitem[{{Simard} {et~al}\mbox{.}(2011){Simard}, {Mendel}, {Patton},
  {Ellison}, \& {McConnachie}}]{simard11}
{Simard} L., {Mendel} J.~T., {Patton} D.~R., {Ellison} S.~L., {McConnachie}
  A.~W., 2011, \apjs, 196, 11

\bibitem[{{Stoughton} {et~al}\mbox{.}(2002){Stoughton}, {Lupton}, {Bernardi},
  {Blanton}, {Burles}, {Castander}, {Connolly}, {Eisenstein}, {Frieman},
  {Hennessy}, {Hindsley}, {Ivezi{\'c}}, {Kent}, {Kunszt}, {Lee}, {Meiksin},
  {Munn}, {Newberg}, {Nichol}, {Nicinski}, {Pier}, {Richards}, {Richmond},
  {Schlegel}, {Smith}, {Strauss}, {SubbaRao}, {Szalay}, {Thakar}, {Tucker},
  {Vanden Berk}, {Yanny}, {Adelman}, {Anderson}, {Anderson}, {Annis},
  {Bahcall}, {Bakken}, {Bartelmann}, {Bastian}, {Bauer}, {Berman},
  {B{\"o}hringer}, {Boroski}, {Bracker}, {Briegel}, {Briggs}, {Brinkmann},
  {Brunner}, {Carey}, {Carr}, {Chen}, {Christian}, {Colestock}, {Crocker},
  {Csabai}, {Czarapata}, {Dalcanton}, {Davidsen}, {Davis}, {Dehnen},
  {Dodelson}, {Doi}, {Dombeck}, {Donahue}, {Ellman}, {Elms}, {Evans}, {Eyer},
  {Fan}, {Federwitz}, {Friedman}, {Fukugita}, {Gal}, {Gillespie}, {Glazebrook},
  {Gray}, {Grebel}, {Greenawalt}, {Greene}, {Gunn}, {de Haas}, {Haiman},
  {Haldeman}, {Hall}, {Hamabe}, {Hansen}, {Harris}, {Harris}, {Harvanek},
  {Hawley}, {Hayes}, {Heckman}, {Helmi}, {Henden}, {Hogan}, {Hogg}, {Holmgren},
  {Holtzman}, {Huang}, {Hull}, {Ichikawa}, {Ichikawa}, {Johnston}, {Kauffmann},
  {Kim}, {Kimball}, {Kinney}, {Klaene}, {Kleinman}, {Klypin}, {Knapp},
  {Korienek}, {Krolik}, {Kron}, {Krzesi{\'n}ski}, {Lamb}, {Leger},
  {Limmongkol}, {Lindenmeyer}, {Long}, {Loomis}, {Loveday}, {MacKinnon},
  {Mannery}, {Mantsch}, {Margon}, {McGehee}, {McKay}, {McLean}, {Menou},
  {Merelli}, {Mo}, {Monet}, {Nakamura}, {Narayanan}, {Nash}, {Neilsen},
  {Newman}, {Nitta}, {Odenkirchen}, {Okada}, {Okamura}, {Ostriker}, {Owen},
  {Pauls}, {Peoples}, {Peterson}, {Petravick}, {Pope}, {Pordes}, {Postman},
  {Prosapio}, {Quinn}, {Rechenmacher}, {Rivetta}, {Rix}, {Rockosi}, {Rosner},
  {Ruthmansdorfer}, {Sandford}, {Schneider}, {Scranton}, {Sekiguchi}, {Sergey},
  {Sheth}, {Shimasaku}, {Smee}, {Snedden}, {Stebbins}, {Stubbs}, {Szapudi},
  {Szkody}, {Szokoly}, {Tabachnik}, {Tsvetanov}, {Uomoto}, {Vogeley}, {Voges},
  {Waddell}, {Walterbos}, {Wang}, {Watanabe}, {Weinberg}, {White}, {White},
  {Wilhite}, {Wolfe}, {Yasuda}, {York}, {Zehavi}, \& {Zheng}}]{SDSS_EDR}
{Stoughton} C. {et~al.}, 2002, \aj, 123, 485

\bibitem[{{Strauss} {et~al}\mbox{.}(2002){Strauss}, {Weinberg}, {Lupton},
  {Narayanan}, {Annis}, {Bernardi}, {Blanton}, {Burles}, {Connolly},
  {Dalcanton}, {Doi}, {Eisenstein}, {Frieman}, {Fukugita}, {Gunn},
  {Ivezi{\'c}}, {Kent}, {Kim}, {Knapp}, {Kron}, {Munn}, {Newberg}, {Nichol},
  {Okamura}, {Quinn}, {Richmond}, {Schlegel}, {Shimasaku}, {SubbaRao},
  {Szalay}, {Vanden Berk}, {Vogeley}, {Yanny}, {Yasuda}, {York}, \&
  {Zehavi}}]{Strauss2002}
{Strauss} M.~A. {et~al.}, 2002, \aj, 124, 1810

\bibitem[{{The Dark Energy Survey Collaboration}(2005)}]{des_whitepaper}
{The Dark Energy Survey Collaboration}, 2005, arXiv:astro-ph/0510346

\bibitem[{{Tremonti} {et~al}\mbox{.}(2004){Tremonti}, {Heckman}, {Kauffmann},
  {Brinchmann}, {Charlot}, {White}, {Seibert}, {Peng}, {Schlegel}, {Uomoto},
  {Fukugita}, \& {Brinkmann}}]{tremonti04}
{Tremonti} C.~A. {et~al.}, 2004, \apj, 613, 898

\bibitem[{{Tully} \& {Fisher}(1977)}]{TullyFisher}
{Tully} R.~B., {Fisher} J.~R., 1977, \aap, 54, 661

\bibitem[{{Vikram} {et~al}\mbox{.}(2010){Vikram}, {Wadadekar}, {Kembhavi}, \&
  {Vijayagovindan}}]{pymorph}
{Vikram} V., {Wadadekar} Y., {Kembhavi} A.~K., {Vijayagovindan} G.~V., 2010,
  \mnras, 409, 1379

\bibitem[{{Willett} {et~al}\mbox{.}(2013){Willett}, {Lintott}, {Bamford},
  {Masters}, {Simmons}, {Casteels}, {Edmondson}, {Fortson}, {Kaviraj}, {Keel},
  {Melvin}, {Nichol}, {Raddick}, {Schawinski}, {Simpson}, {Skibba}, {Smith}, \&
  {Thomas}}]{galzoo13}
{Willett} K.~W. {et~al.}, 2013, \mnras, 435, 2835

\bibitem[{{York} {et~al}\mbox{.}(2000){York}, {Adelman}, {Anderson},
  {Anderson}, {Annis}, {Bahcall}, {Bakken}, {Barkhouser}, {Bastian}, {Berman},
  {Boroski}, {Bracker}, {Briegel}, {Briggs}, {Brinkmann}, {Brunner}, {Burles},
  {Carey}, {Carr}, {Castander}, {Chen}, {Colestock}, {Connolly}, {Crocker},
  {Csabai}, {Czarapata}, {Davis}, {Doi}, {Dombeck}, {Eisenstein}, {Ellman},
  {Elms}, {Evans}, {Fan}, {Federwitz}, {Fiscelli}, {Friedman}, {Frieman},
  {Fukugita}, {Gillespie}, {Gunn}, {Gurbani}, {de Haas}, {Haldeman}, {Harris},
  {Hayes}, {Heckman}, {Hennessy}, {Hindsley}, {Holm}, {Holmgren}, {Huang},
  {Hull}, {Husby}, {Ichikawa}, {Ichikawa}, {Ivezi{\'c}}, {Kent}, {Kim},
  {Kinney}, {Klaene}, {Kleinman}, {Kleinman}, {Knapp}, {Korienek}, {Kron},
  {Kunszt}, {Lamb}, {Lee}, {Leger}, {Limmongkol}, {Lindenmeyer}, {Long},
  {Loomis}, {Loveday}, {Lucinio}, {Lupton}, {MacKinnon}, {Mannery}, {Mantsch},
  {Margon}, {McGehee}, {McKay}, {Meiksin}, {Merelli}, {Monet}, {Munn},
  {Narayanan}, {Nash}, {Neilsen}, {Neswold}, {Newberg}, {Nichol}, {Nicinski},
  {Nonino}, {Okada}, {Okamura}, {Ostriker}, {Owen}, {Pauls}, {Peoples},
  {Peterson}, {Petravick}, {Pier}, {Pope}, {Pordes}, {Prosapio},
  {Rechenmacher}, {Quinn}, {Richards}, {Richmond}, {Rivetta}, {Rockosi},
  {Ruthmansdorfer}, {Sandford}, {Schlegel}, {Schneider}, {Sekiguchi}, {Sergey},
  {Shimasaku}, {Siegmund}, {Smee}, {Smith}, {Snedden}, {Stone}, {Stoughton},
  {Strauss}, {Stubbs}, {SubbaRao}, {Szalay}, {Szapudi}, {Szokoly}, {Thakar},
  {Tremonti}, {Tucker}, {Uomoto}, {Vanden Berk}, {Vogeley}, {Waddell}, {Wang},
  {Watanabe}, {Weinberg}, {Yanny}, {Yasuda}, \& {SDSS
  Collaboration}}]{sdss_tech}
{York} D.~G. {et~al.}, 2000, \aj, 120, 1579

\end{thebibliography}

\onecolumn

\appendix

\appendixpage

\section{Automated Flagging Tables}\label{app:autoflag}

In this appendix we provide more detailed descriptions of the flags used in this catalogue. Table~\ref{tab:useful_flag_definitions} gives a short description of the fitting parameters used to set the value of each flag. The majority of the flags use the B/T, S\'{e}rsic index, and radial light profiles of
the galaxy to set the flags. We also consider the axis ratio and the difference in position angle of the bulge and disc because small axis ratios combined 
with large differences in position angle of the two components is an unphysical fit for most cases. B/T({\em r}) in flags 3, 7, 8, \etc{} refers to the radial B/T profile. For the radial profile, we plot the local B/T ratio (\ie the ratio of light contributed by the bulge at the annulus of width 1 arcsec centred on radius $r$ to the light contributed by the disc). $r_{sextractor}$ is the half-light radius reported by \sextractor{} during fitting. We use this radius to set the scale of the galaxy during fitting and to test whether components have become unphysically large after fitting. 
\clearpage
\begin{table*}
\begin{tabular}{l l l l p{7.5cm}}
\textbf{Flag bit} & \multicolumn{3}{l}{\textbf{Analysis flag}} &  \textbf{Flag criteria} \\ \hline \hline
-- & \multicolumn{3}{l}{\textbf{Good total and component magnitudes and sizes}} & \\ \hline
10 & & \multicolumn{2}{l}{\textbf{Two-component galaxies}} &  \\
11 & & & No flags & No flags are present\\
12 & & & Good \Ser{}, Good \Exp\ (some flags) & Some minor automated flags\\
13 & & & Flip components & \Exp{} component fitting the inner and \Ser{} component fitting the outer part of the profile \\ \hline
-- & \multicolumn{3}{l}{\textbf{Good total magnitudes and sizes only}} & \\ \hline
1 & & \multicolumn{2}{l}{\textbf{Bulge galaxies}} & \\
2 & & & No \Exp{} component, n$_{\Ser}>$2 & 1000(B/T-0.8)$^3$+(m$_{disc}$-19)$>$0.5 AND n$_{\Ser}\geq$2\\
3 & & & \Ser{} dominates always & B/T(r) $\geq$ 0.5 for all r AND  n$_{\Ser}\geq$2\\
4 & & \multicolumn{2}{l}{\textbf{Disc galaxies}} & \\
5 & & & No \Ser{} component & 1000(0.2-B/T)$^3$+(m$_{bulge}$-19)$>$0.5\\
6 & & & No \Exp{}, n$_{\Ser}<$2, Flip Components & 1000(B/T-0.8)$^3$+(m$_{disc}$-19)$>$0.5 AND n$_{\Ser}<$2\\
7 & & & \Ser{} dominates always, n$_{\Ser}<$2 &  B/T(r) $\geq$ 0.5 for all r AND  n$_{\Ser}<$2\\
8 & & & \Exp{} dominates always &  B/T(r) $<$ 0.5 for all r\\
9 & & & Parallel components &  $RMS\left[(B/T - \mu(B/T)): 0<r<r(0.9L_{tot}\right] <0.1$ AND n$_{bulge}<2.0$\\
14 & & \multicolumn{2}{l}{\textbf{Problematic two-component galaxies}} & \\
15 & & & \Ser{} outer only & B/T$<0.7$ AND B/T(r(0.9$L_{tot}$)$>$0.5 AND n$_{bulge}\geq2.0$ AND B/T(0)$>$0.5\\
16 & & & \Exp\ inner only & B/T(0)$<$0.5 AND B/T(r: r$<2.7$r$_{hl}$) $>$ 0.5 AND n$_{bulge}\geq2.0$ AND NOT Ser\ Outer Only \\
17 & & & Good \Ser{}, bad \Exp{}, B/T$>=$0.5 & B/T$>$0.75 AND $\Delta\phi>45$ AND $b/a_{bulge}<0.75$ AND $b/a_{disc}<0.4$\\
18 & & & Bad \Ser{}, good \Exp{}, B/T$<$0.5 & B/T$<$0.25 AND $\Delta\phi>45$ AND $b/a_{bulge}<0.4$ AND $b/a_{disc}<0.75$\\ 
19 & & & Bulge is point& Circularized bulge radius $<$0.188 arcsec\\ \hline
20 & \multicolumn{3}{l}{\textbf{Bad total magnitudes and sizes}} &  \\ \hline
21 & & \multicolumn{2}{l}{Centering problems} &  (galaxy centroid-SDSS galaxy centroid)$>0.7\times$r$_{sextractor}$ \\
22 & & \multicolumn{2}{l}{\Ser{} component contamination by neighbours or sky} & $r_{bulge,cir}/r_{sextractor}>4.0$ \\
23 & & \multicolumn{2}{l}{\Exp{} component contamination by neighbours or sky} & $r_{bulge,cir}/r_{sextractor}>4.0$ \\
24 & & \multicolumn{2}{l}{Bad \Ser{} and bad \Exp{} components} & Failure of measurements  \\
25 & & \multicolumn{2}{l}{\galfit{} failure} & \galfit{} fails to converge or other failure of the pipeline\\
26 & & \multicolumn{2}{l}{Polluted or fractured} & CasJobs neighbours not properly masked/masked or target is separated into 2 or more objects \\
\end{tabular}
\caption{The description of our categories as described in the main text. The major flag bits used to select different catalogues are flag bits 10 (good two-component fits), 1 (good bulge galaxy), 4 (good disc galaxy), 14 (problematic two-component fit), and 20 (bad fit). We describe how to use these flags in 
the main text in Section~\ref{sec:flag_auto}.}
\label{tab:useful_flag_definitions}
\end{table*}

\begin{table*}
\centering
\tiny
\begin{tabular}{l l l l c  c  c c c}
\textbf{Flag bit} & \multicolumn{3}{l}{\textbf{Descriptive Category}} &  \textbf{ Per cent Dev}&  \textbf{ Per cent Ser}& \textbf{ Per cent DevExp}  & \textbf{ Per cent SerExp} &\textbf{ Per cent Test} \\ \hline \hline
-- & \multicolumn{3}{p{4.0cm}}{\textbf{Trust total and component magnitudes and sizes}}&      -- & -- & 42.223 & 39.055 & 39.167 \\ \hline
10& & \multicolumn{2}{p{4.0cm}}{\textbf{Two-component galaxies}} &  -- & -- & 42.223 & 39.055 & 39.167     \\
11& & & No flags &  --&-- & 30.152 & 18.095 & 17.917  \\
12& & & Good \Ser{}, good \Exp{} (some flags) &      -- & -- & 10.701 & 19.417 & 18.333  \\
13& & & Flip components &    -- & -- & 1.369 & 1.543 & 2.917  \\ \hline
--& \multicolumn{3}{p{4.0cm}}{\textbf{Trust total magnitudes and sizes only}} &    97.653 & 97.380 & 52.444 & 54.945 & 54.375 \\ \hline
1& & \multicolumn{2}{p{4.0cm}}{\textbf{Bulge galaxies}} &   97.653 & 58.378 & 14.636 & 18.964 & 18.958 \\
2& & &No \Exp{} component, n$_{\Ser}>$2& 97.653 & 58.378 & 4.646 & 7.074 & 7.917 \\
3& & &\Ser{} dominates always &   -- & -- & 9.989 & 11.889 & 11.042 \\
4& & \multicolumn{2}{p{4.0cm}}{\textbf{Disc galaxies}} &  -- & 39.001 & 28.462 & 25.146 & 23.958 \\
5& & & No \Ser{} component &    -- & -- & 25.041 & 16.876 & 15.625\\
6& & & No \Exp{}, n$_{Ser}<$2, flip components &    -- & 39.001 & -- & 0.551 & 0.208 \\
7& & & \Ser{} dominates always, n$_{\Ser}<$2 & -- & -- & -- & 0.103 & 0.625\\
8& & & \Exp{} dominates always &   -- & -- & 3.421 & 2.872 & 2.917\\
9& & & Parallel components &   -- & -- & -- & 4.745 & 4.583 \\
14& & \multicolumn{2}{p{4.0cm}}{\textbf{Problematic two-component galaxies}} & -- & -- & 9.346 & 10.835 & 11.458\\
15& & & \Ser{} outer only &   -- & -- & 5.289 & 7.504 & 8.125 \\
16& & & \Exp\ inner only &   -- & -- & 0.514 & 0.425 & 0.625\\
17& & & Good \Ser{}, bad \Exp, B/T$>=$0.5 &   -- & -- & 0.011 & 0.017 & 0.000 \\
18& & & Bad \Ser{}, good \Exp, B/T$<$0.5 & -- & -- & 0.884 & 0.652 & 0.625 \\ 
26& & & Bulge is point & -- & -- & 2.648 & 2.237 & 2.083 \\ \hline \hline
19& \multicolumn{3}{p{4.0cm}}{\textbf{Bad total magnitudes and sizes}} &  2.347 & 2.620 & 5.334 & 6.000 & 6.458\\
20& & \multicolumn{2}{p{4.0cm}}{Centering problems} & 0.338 & 0.357 & 0.599 & 0.557 & 0.625 \\
21& & \multicolumn{2}{p{4.0cm}}{\Ser{} component contamination by neighbours or sky} &  1.302 & 1.618 & 1.251 & 2.129 & 3.333\\
22& & \multicolumn{2}{p{4.0cm}}{\Exp\ component contamination by neighbours or sky} & -- & -- & 2.788 & 2.392 & 2.083  \\ 
23& & \multicolumn{2}{p{4.0cm}}{Bad \Ser{} and bad \Exp{} components} & -- & -- & 0.177 & 0.239 & 0.417\\
24& & \multicolumn{2}{p{4.0cm}}{\galfit{} failure} &  0.187 & 0.124 & 0.249 & 0.355 & 0.208 \\
25& &  \multicolumn{2}{p{4.0cm}}{Polluted or fractured} & 0.679 & 0.681 & 0.677 & 0.676 & 0.833 \\
\end{tabular}
\caption{A breakdown of the descriptive categories useful for analysis. We show percentages of the total catalogue for each of the fitted models and our visually classified test set. The one component models (\Ser{} and \Dev{}) can not be classified as two-component models, by definition. For the \Ser{} and
\Dev{} models, many of the categories are empty. The major distinction for the one-component models is whether the fits have major problems (\ie flag bit 20 is set). }
\label{tab:finalflag_percents}
\end{table*}

Flag bit 9 is used to determine whether the bulge and disc are fitting the same component of the galaxy. This is most common in very late-type galaxies (\ie {\em T}-types greater than 4) that have no visible bulge. The flag criteria reference $r$(0.9$L_{tot}$), which is the radius enclosing 90 per cent of the total light of the galaxy. The flag also measures the quantity $RMS(B/T - \mu(B/T)): 0<r<r(0.9L_{tot})$. The quantity $RMS(B/T - \mu(B/T))$ is the standard deviation of B/T about the mean B/T value. The average is taken over the radii interior to the radius enclosing 90 per cent of the light. If the standard deviation is less than 0.1, then the B/T ratio is 
approximately the same  at all radii and the components are parallel. Two parallel components, each with S\'{e}rsic index $<$2, are indistinguishable from one brighter exponential component, so we flag these as discs and declare the components unreliable.
The \Ser{} model is the most appropriate model in this case, similar to all the bulge and disc flag categories.

Flag bit 24 (bad \Ser{} and bad \Exp{}) also includes failures that indicate strange fitting behaviour. This category includes galaxies where the B/T $>$ 0.5 and the measured B/T is lower than 0.5, or vice versa. These cases occur when the bulge or disc parameters get very unusual (radii approaching 0 or b/a approaching 0).

The remaining flag bits are self-explanatory, so we do not describe them here. Table~\ref{tab:finalflag_percents} shows the percentage of galaxies in each flag type. The one-component models (\Ser{} and \Dev{}) cannot be classified as two-component models, by definition, so many of the categories are empty. The major distinction for the one-component models is whether the fits have major problems (\ie flag bit 20 is set). The failure rate (flag bit 20 set) increases with the complexity of the fits, but stays below 7 per cent for the \SerExp{} model and 8 per cent for the test sample. An additional $\sim10$ per cent of the two-component fits have problematic interpretations. The remaining $\sim80--85$ per cent of two-component galaxies appear to
have good fits that can be used in analysis. As discussed in the main text, this does not guarantee that the fits can be interpreted as a physical bulge 
and a physical disc model. Many caveats to this interpretation still exist.

 \clearpage
\section{Examples of flagged galaxies}\label{app:example_gals}

In this appendix, we provide sample \SerExp{} fits to galaxies for each of the flagging categories described in  
Table~\ref{tab:useful_flag_breakdown}. The details of each category are described in 
Appendix~\ref{app:autoflag} and Tables~\ref{tab:useful_flag_definitions}~and~\ref{tab:finalflag_percents}.
Although we show examples only for the \SerExp{} fits, the same flags are applied to the \DevExp{} fit. For the \Dev{} and \Ser{}
fits, a subset of these flags is used that include only the flags applicable to one-component fits (\ie the `bad' flags and pure 
bulge or pure disc flags).

Each figure in this section shows two example galaxies that have the flag described in the figure caption. 
For each example galaxy, we show the 2D fitted model and 2D data (top left and right, respectively). We also
show the 1D fitted profile and 2D masked residual image (bottom left and right, respectively).
We show the total magnitude, B/T, and fitted, PSF-corrected half-light radius in arcseconds in the 2D model image.
The 2D data image has the \galcount{} (our unique galaxy identification number), apparent Petrosian magnitude, 
and Petrosian half-light radius in arcseconds shown for comparison. 

The 1D fitted profiles show the measured 1D profile using elliptical annuli as black dots. The total profile is plotted as the 
solid blue line. The \SerExp{} bulge is plotted as the dashed blue line and the \SerExp{} disc is plotted as the dotted blue line.
The radial profiles are plotted in mag arcsec$^{-2}$ and the x-axis shows the radius in units of the fitted half-light
radius.

\begin{figure*}
\begin{minipage}{\linewidth}
\begin{center}
 \subfloat[Galaxy 35521]{\includegraphics[width=0.45\linewidth]{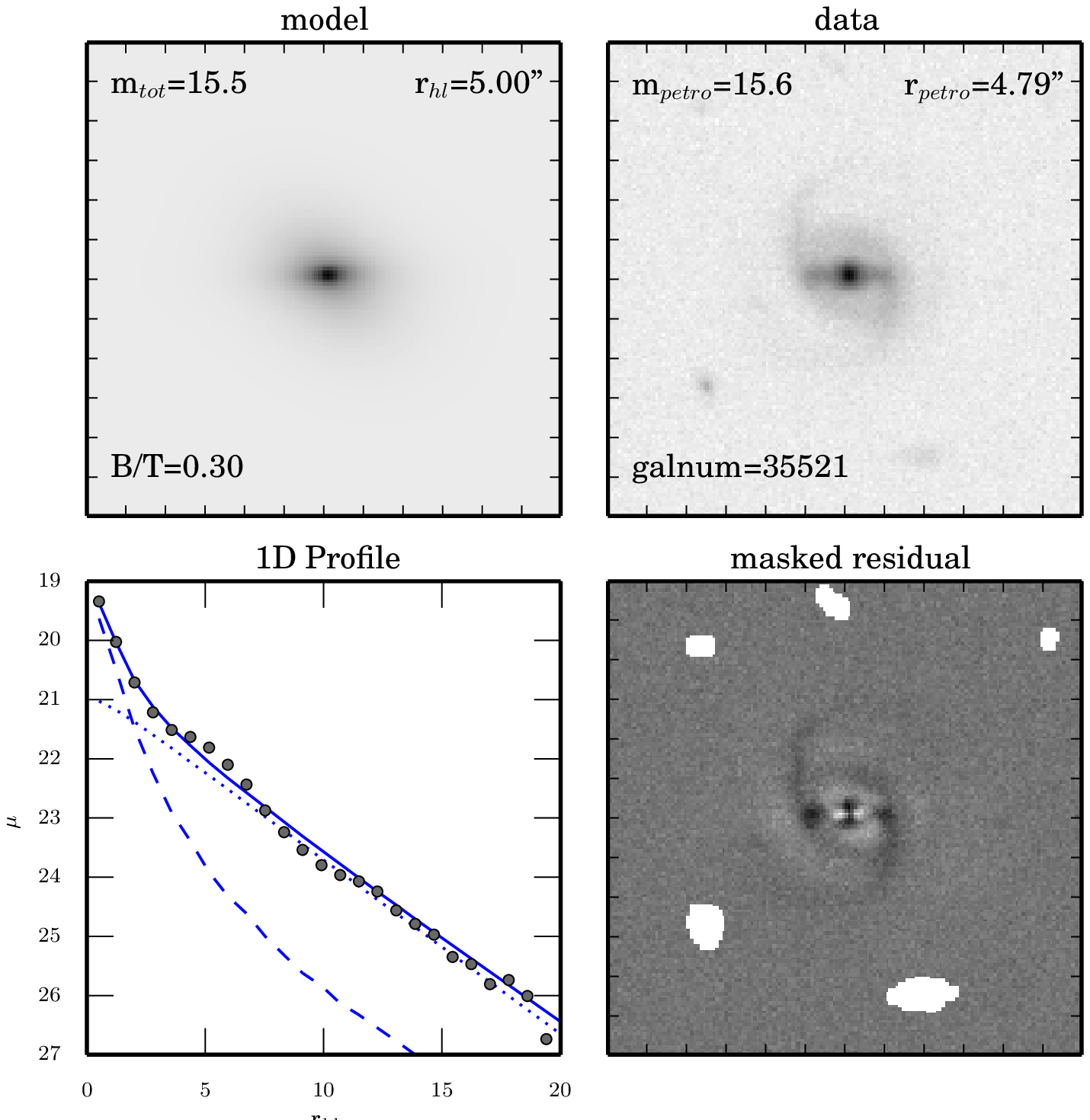}\label{subfig:good_fits:no_flag1}}  \,
 \subfloat[Galaxy 408774]{\includegraphics[width=0.45\linewidth]{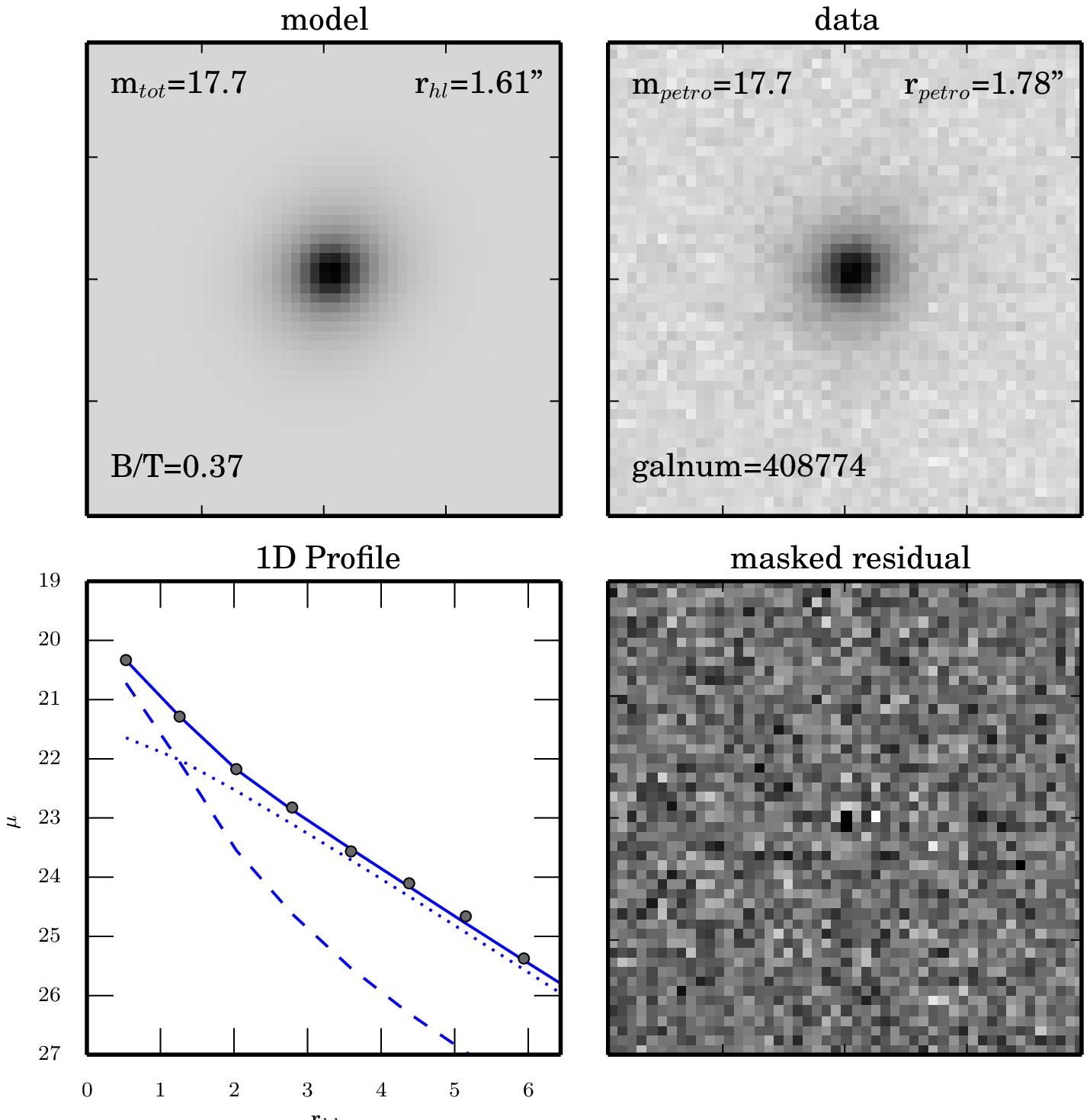}\label{subfig:good_fits:no_flag2}}  \,
\end{center}
\end{minipage}
 \caption{Example \SerExp{} fits considered good using our flagging criteria. For these galaxies, we claim that the two components of the fit are real and without fitting problems. These galaxies are in category `no flags' in Table~\ref{tab:useful_flag_breakdown} and have flag bits 10 and 11 set.}\label{fig:finalflag_good}
\end{figure*}

\begin{figure*}
\begin{minipage}{\linewidth}
\begin{center}
\subfloat[Galaxy 86055]{\includegraphics[width=0.45\linewidth]{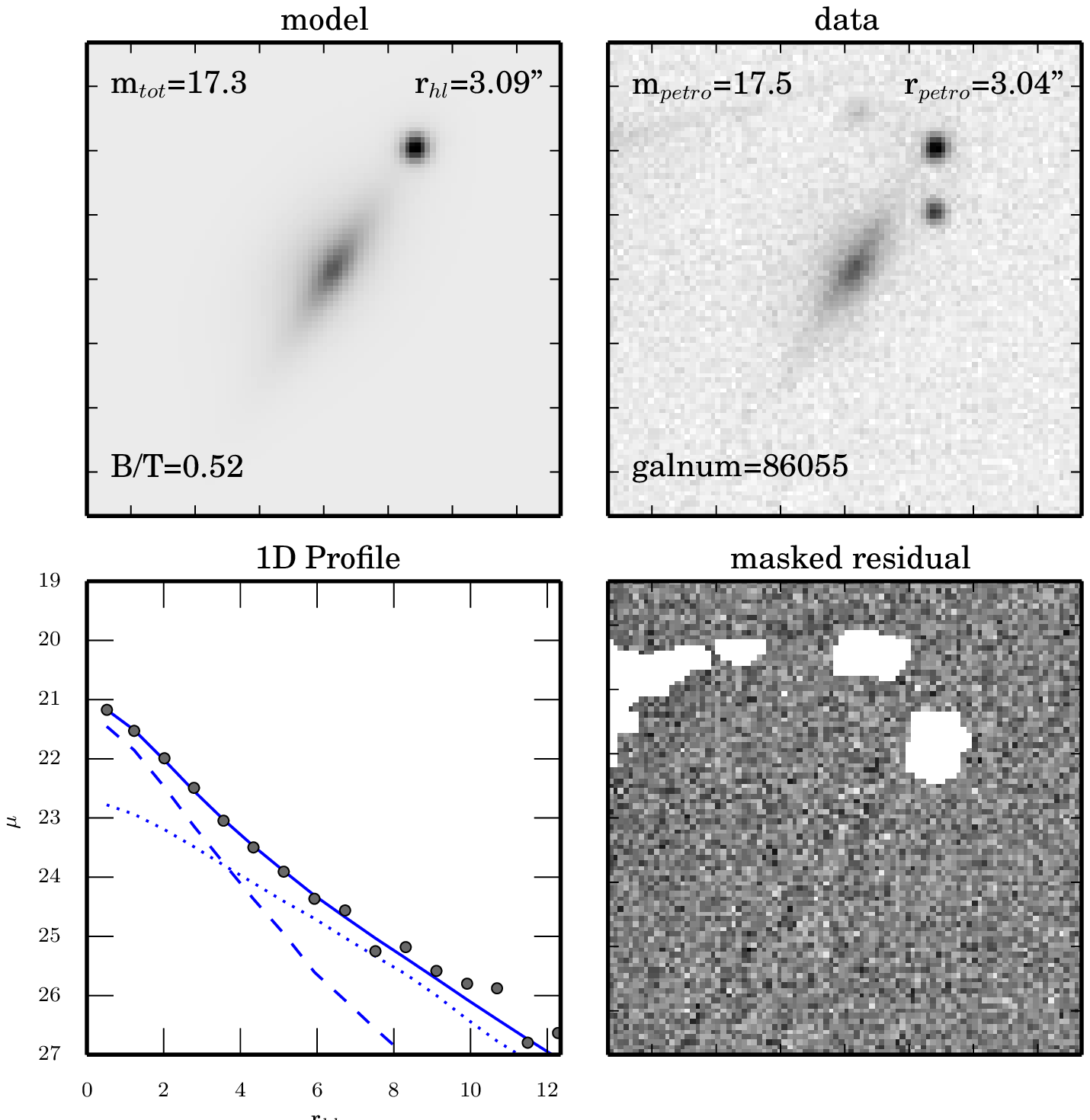}\label{subfig:good_fits:some_flag1}}  \,
 \subfloat[Galaxy 163409]{\includegraphics[width=0.45\linewidth]{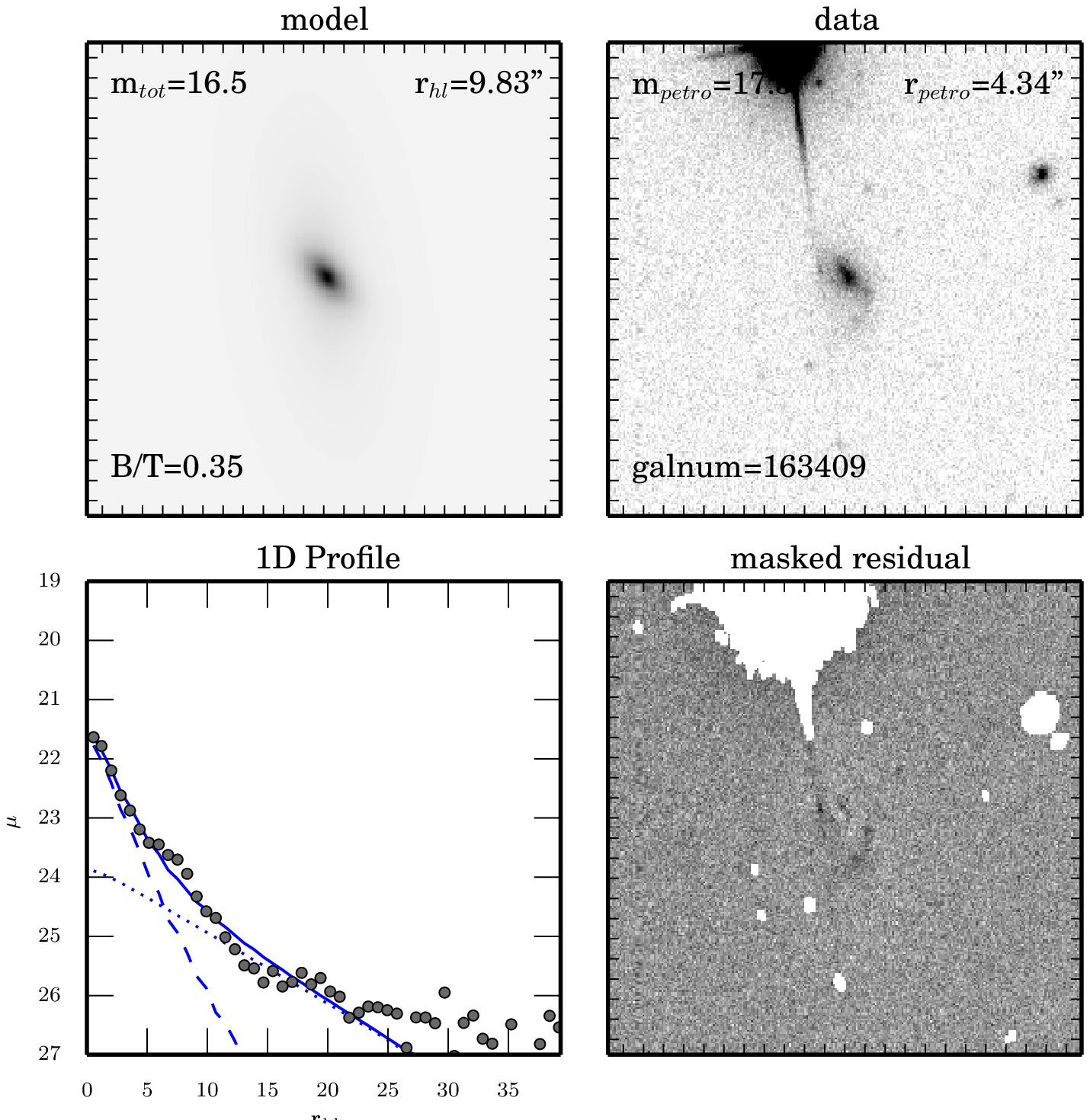}\label{subfig:good_fits:some_flag2}}  \,
\end{center}
\end{minipage}
 \caption{Example \SerExp{} fits considered good using our flagging criteria. For these galaxies, we claim that the two components of the fit are real and largely free of fitting problems. However, there can be some isolated problems with the fits. For example, galaxy 86055 (left) has a high ellipticity bulge (b/a$\approx$0.2) and galaxy 163409 (right) has a bulge with low S\'{e}rsic index (n$_{\mathrm{bulge}}\approx1.1$) and the bulge and disc position angles differ by a large amount ($\approx 45^{\circ}$). However, we do not find enough reason to remove the fit as unphysical. These galaxies are in category `good \Ser{}, good \Exp{} (some flags)' in Table~\ref{tab:useful_flag_breakdown}  and have flag bits 10 and 12 set.}\label{fig:some_flag}
\end{figure*}

\begin{figure*}
\begin{minipage}{\linewidth}
\begin{center}
\subfloat[Galaxy 48927]{\includegraphics[width=0.45\linewidth]{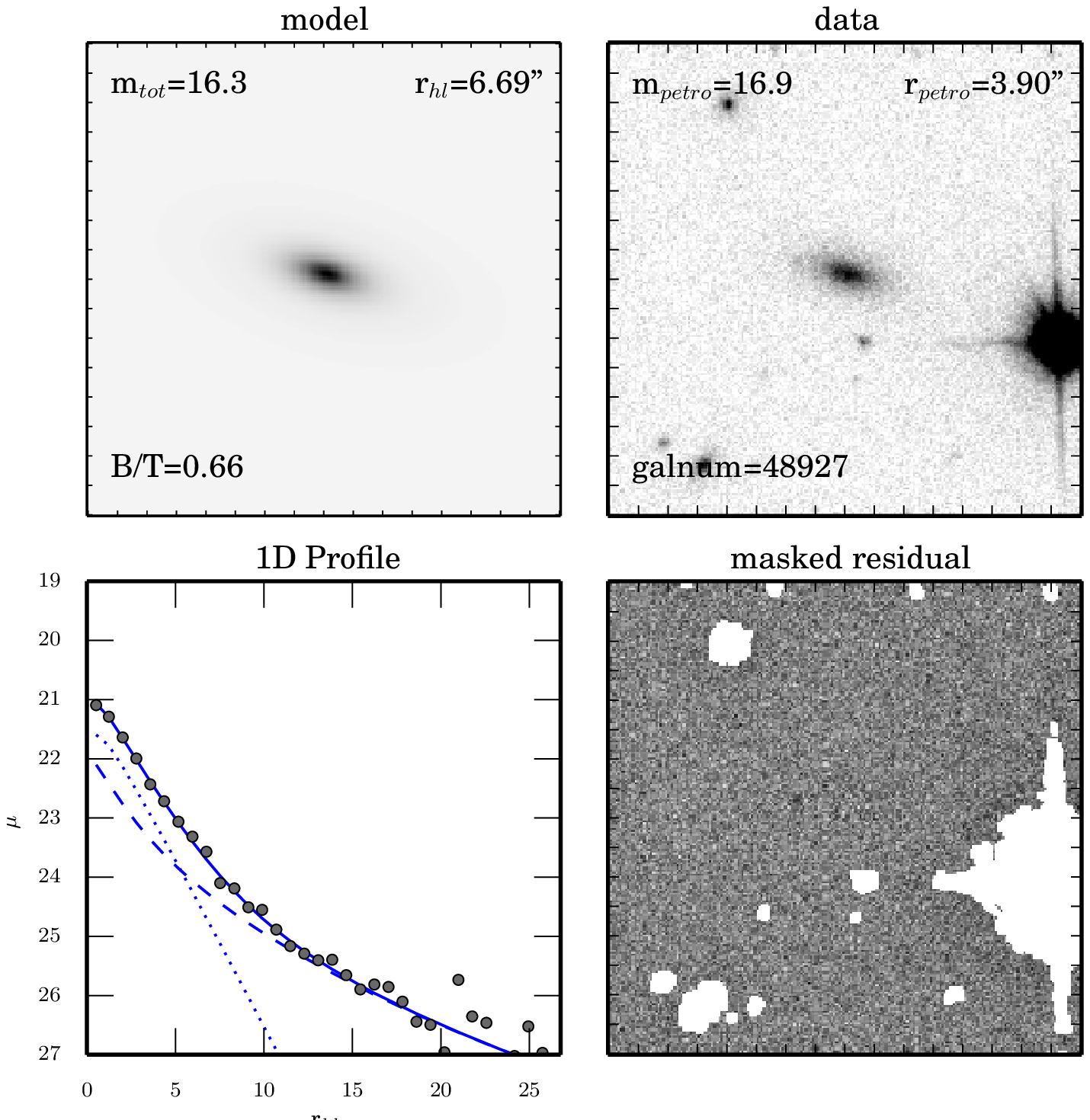} 
\label{subfig:good_fits:flip_com1} } \,
\subfloat[Galaxy 645918]{\includegraphics[width=0.45\linewidth]{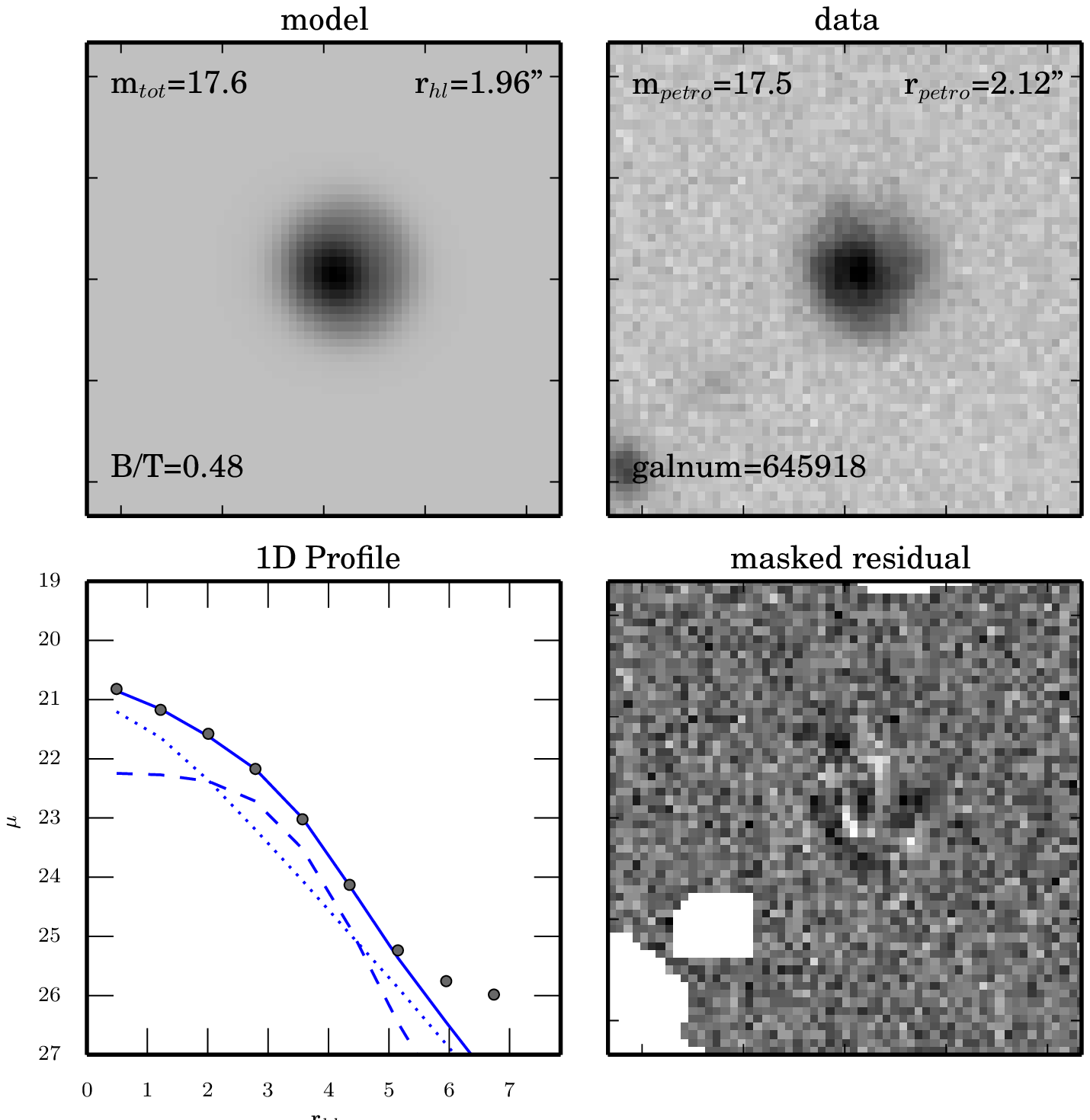} 
\label{subfig:good_fits:flip_com2} } \,
\end{center}
\end{minipage}
 \caption{Example \SerExp{} fits considered good using our flagging criteria. For these galaxies, we claim that the two components of the fit are real. However, the components need to be interchanged to fit with the standard interpretation of the fits that we use throughout the paper.These galaxies are in category `flip components, n$_{\Ser}<$2' in Table~\ref{tab:useful_flag_breakdown}  and have flag bits 10 and 13 set.}\label{fig:flip_com}
\end{figure*}

\begin{figure*}
\begin{minipage}{\linewidth}
\begin{center}
 \subfloat[Galaxy 246717]{\includegraphics[width=0.45\linewidth]{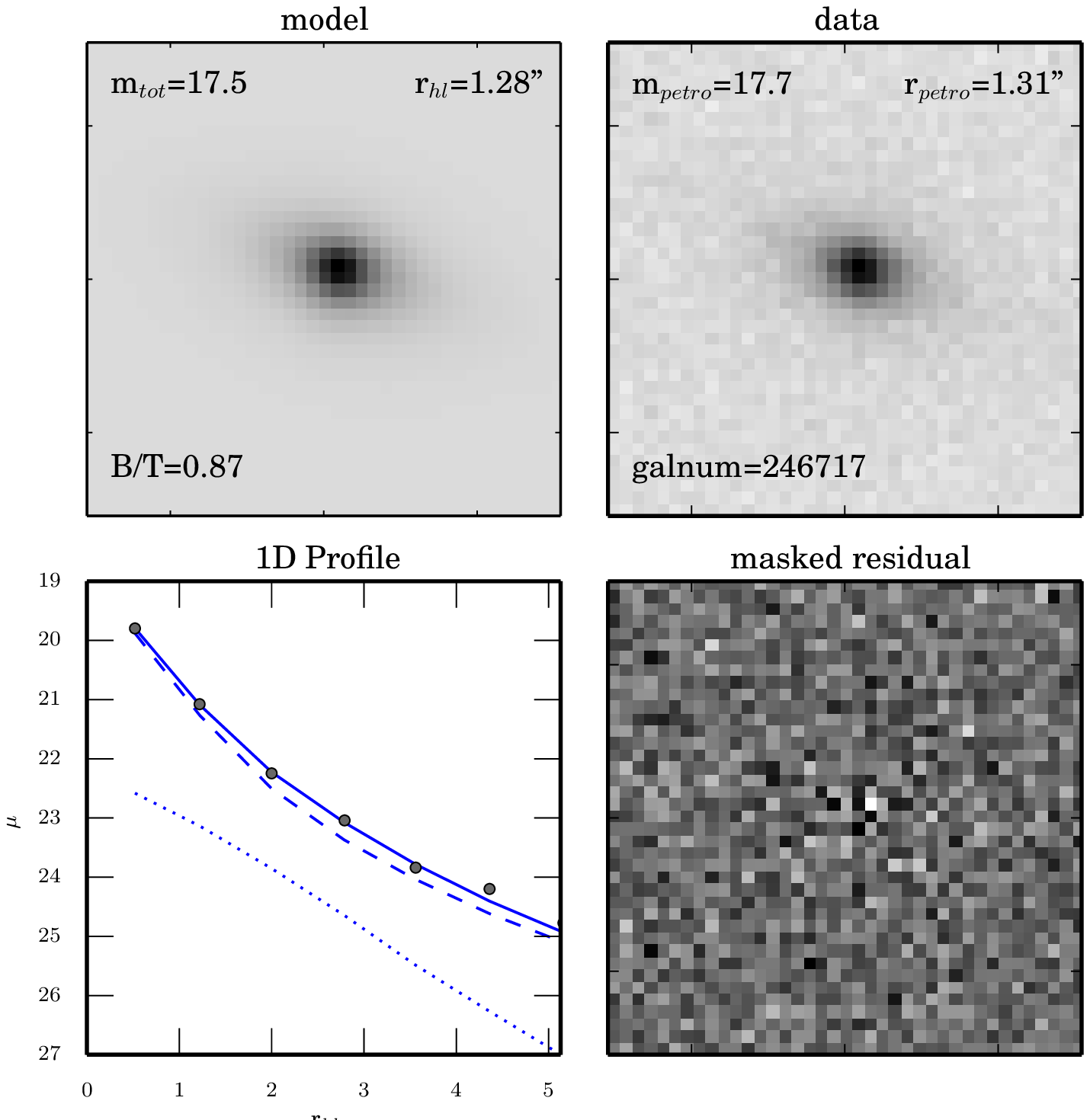}\label{subfig:finalflag_flags_bulge:no_exp1}}  \,
 \subfloat[Galaxy 167097]{\includegraphics[width=0.45\linewidth]{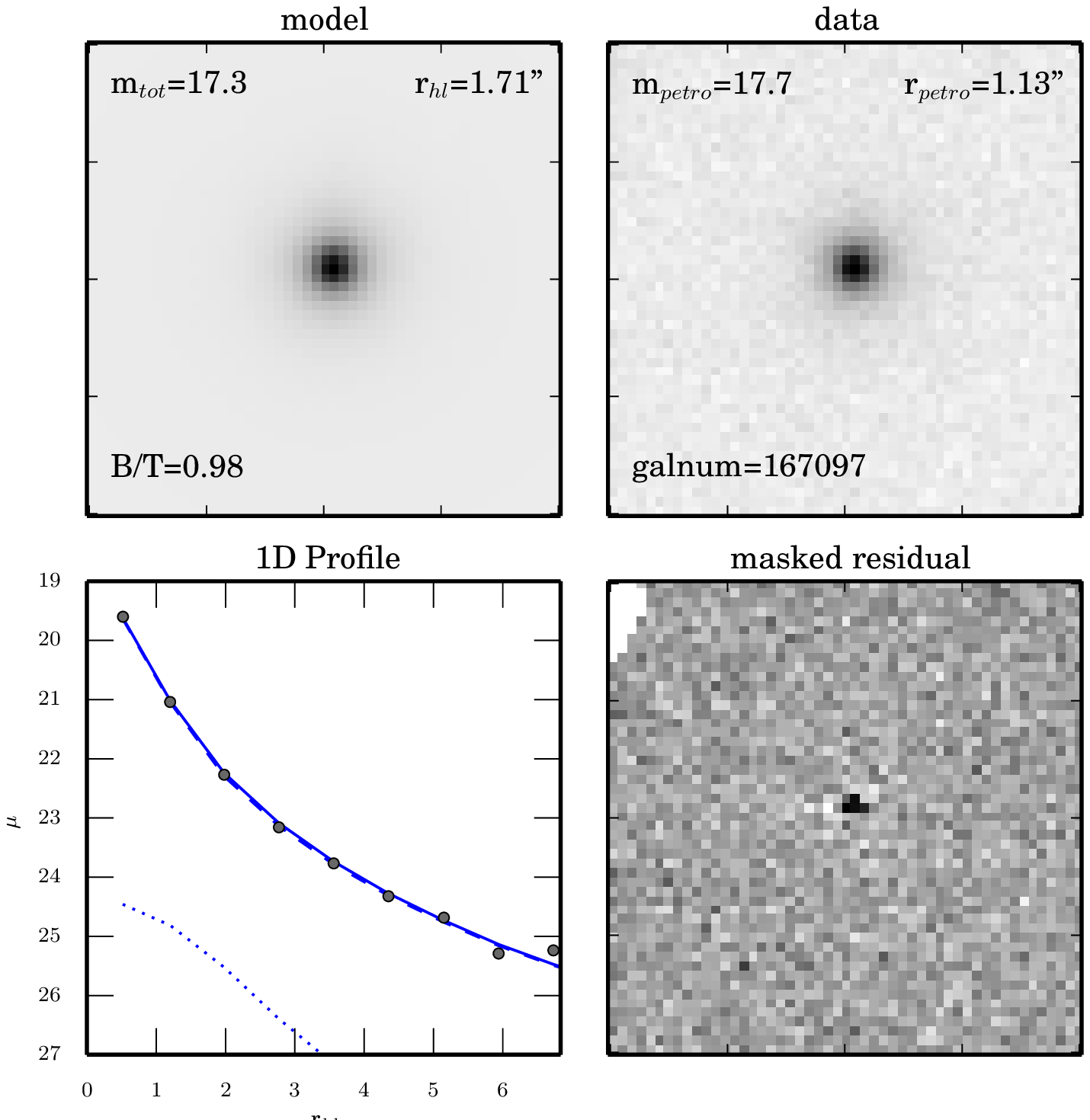}\label{subfig:finalflag_flags_bulge:no_exp2}}  \,
 \end{center}
\end{minipage}
 \caption{Example \SerExp{} galaxies considered single-component galaxies with bulge-like profiles. The profiles have little or no \Exp\ component contributing light to the galaxy. These galaxies are in category `no \Exp{} component, n$_{\Ser}>$2' in Table~\ref{tab:useful_flag_breakdown}  and have flag bits 1 and 2 set.}\label{fig:finalflag_flags_bulge:no_exp}
\end{figure*}

\begin{figure*}
\begin{minipage}{\linewidth}
\begin{center}
 \subfloat[Galaxy 255297]{\includegraphics[width=0.45\linewidth]{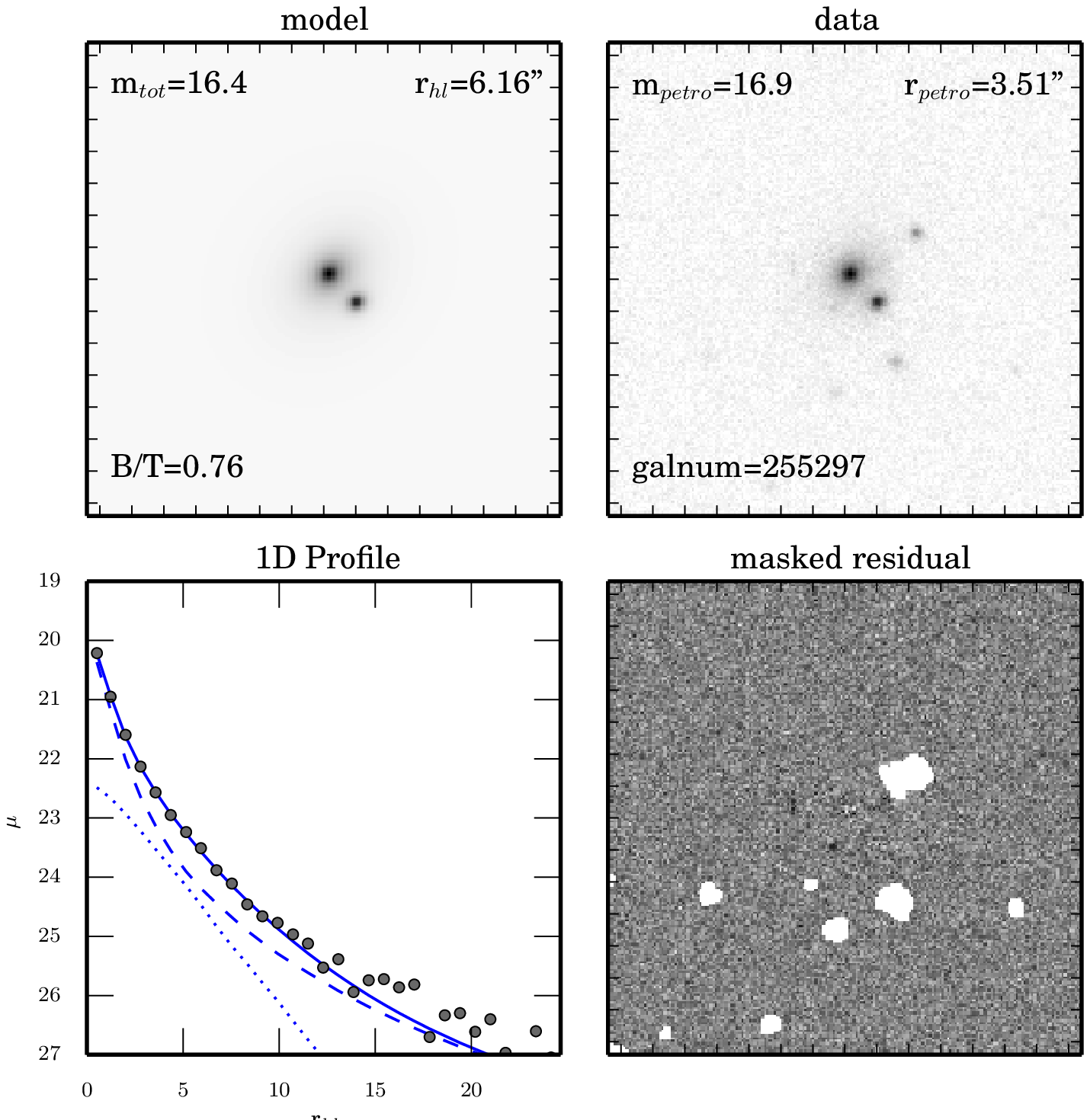}\label{subfig:finalflag_flags_bulge:ser_dom1}} \,
 \subfloat[Galaxy 557050]{\includegraphics[width=0.45\linewidth]{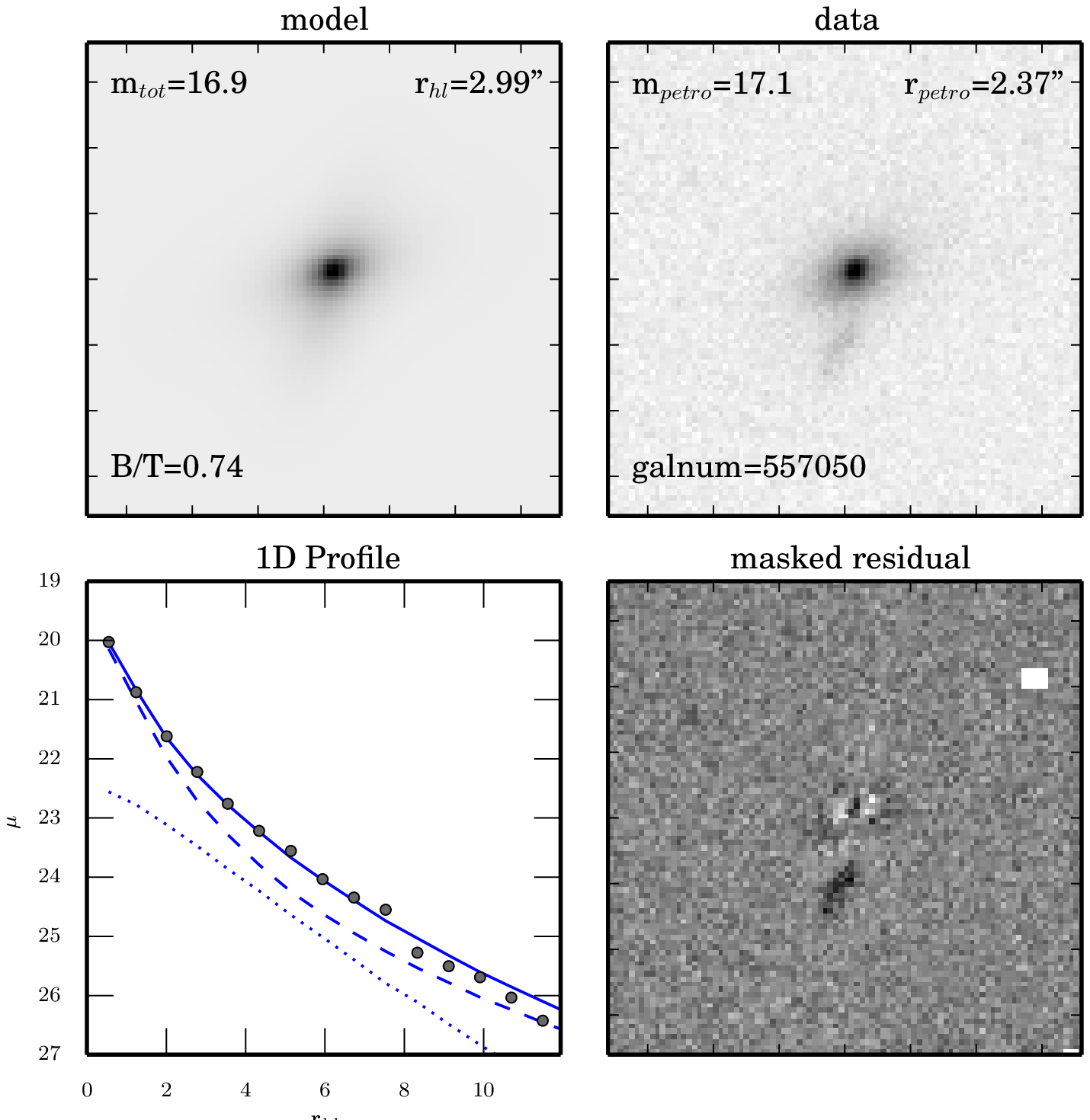}\label{subfig:finalflag_flags_bulge:ser_dom2}} \,
 \end{center}
\end{minipage}
 \caption{Example \SerExp{} galaxies considered single-component galaxies with bulge-like profiles. The profiles tend to have some small \Exp\ component, but the \Ser{} component dominates at all radii. These galaxies are in category `\Ser{} dominates always' in Table~\ref{tab:useful_flag_breakdown}  and have flag bits 1 and 3 set.}\label{fig:finalflag_flags_bulge:ser_dom}
\end{figure*}

\begin{figure*}
\begin{minipage}{\linewidth}
\begin{center}
\subfloat[Galaxy 12474]{\includegraphics[width=0.45\linewidth]{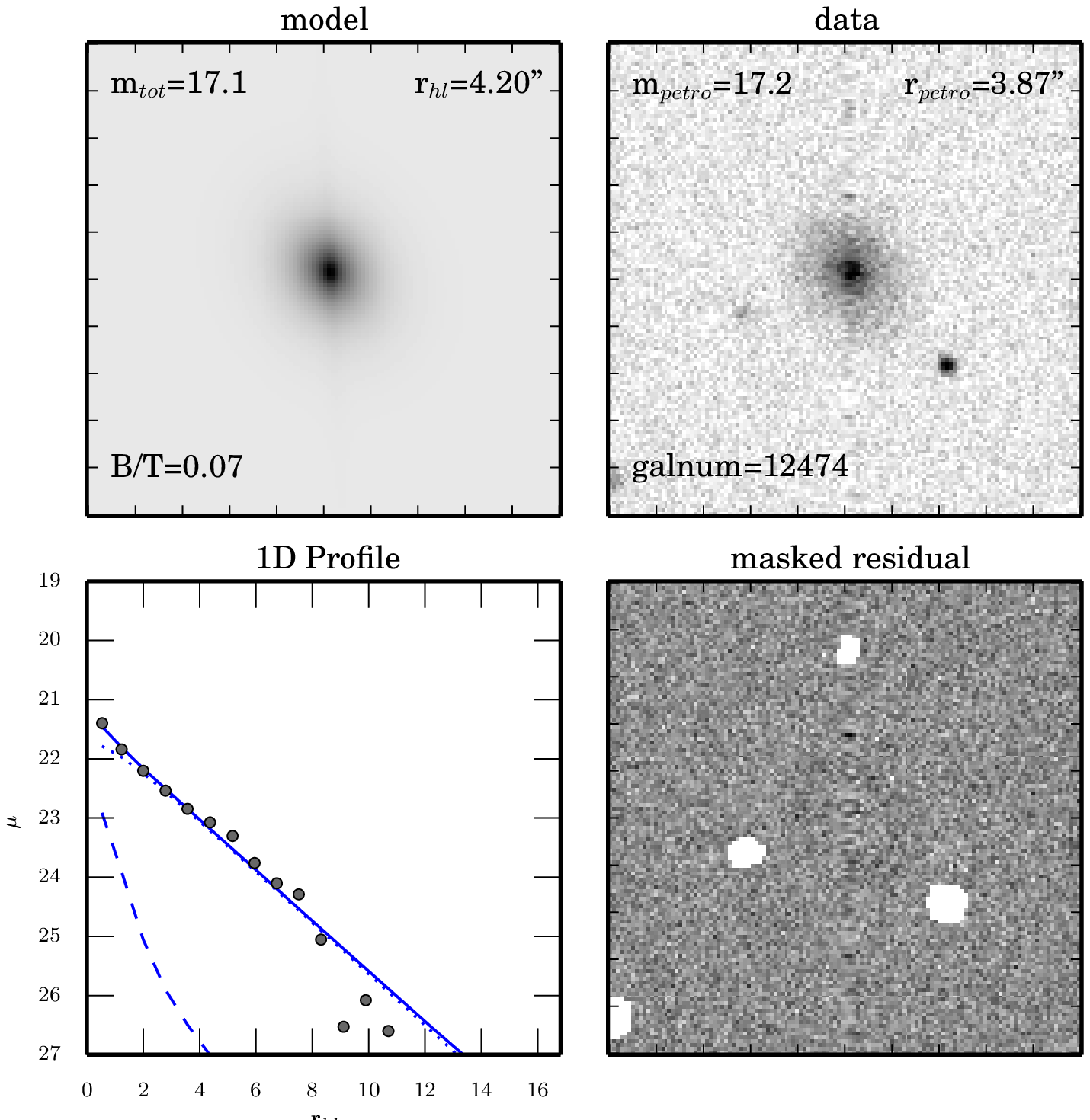}\label{fig:finalflag_flags_disk:no_ser1}}  \,
\subfloat[Galaxy 223566]{\includegraphics[width=0.45\linewidth]{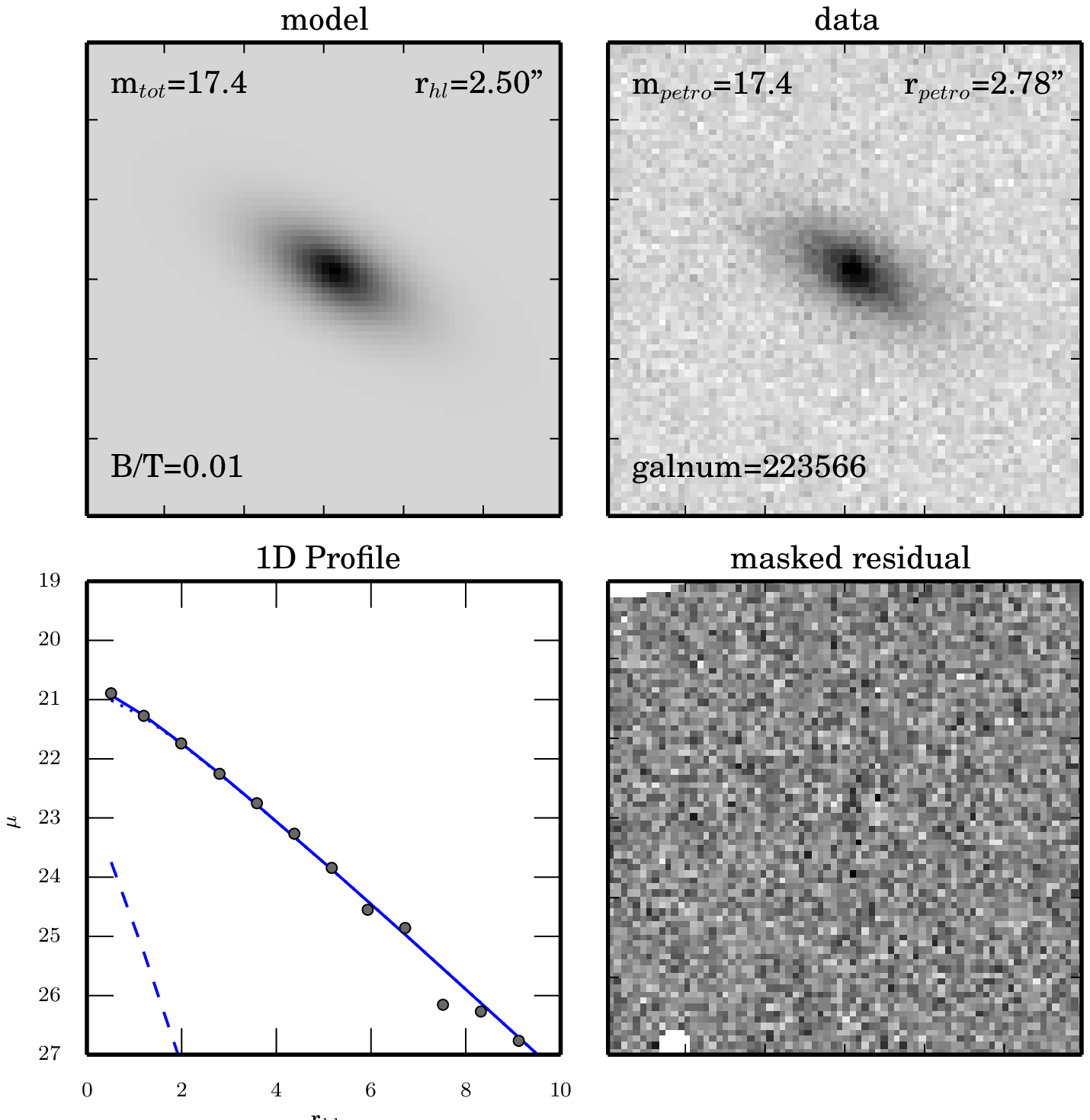}\label{fig:finalflag_flags_disk:no_ser2}} \,
 \end{center}
\end{minipage}
 \caption{Example \SerExp{} galaxies considered single-component galaxies with disc-like profiles. The profiles tend to have little or no contribution to the total light by the \Ser{} component. These galaxies are in category `no \Ser{} component' in Table~\ref{tab:useful_flag_breakdown}  and have flag bits 4 and 5 set.}\label{fig:finalflag_flags_disk:no_ser}
\end{figure*}

\begin{figure*}
\begin{minipage}{\linewidth}
\begin{center}
\subfloat[Galaxy 355]{\includegraphics[width=0.45\linewidth]{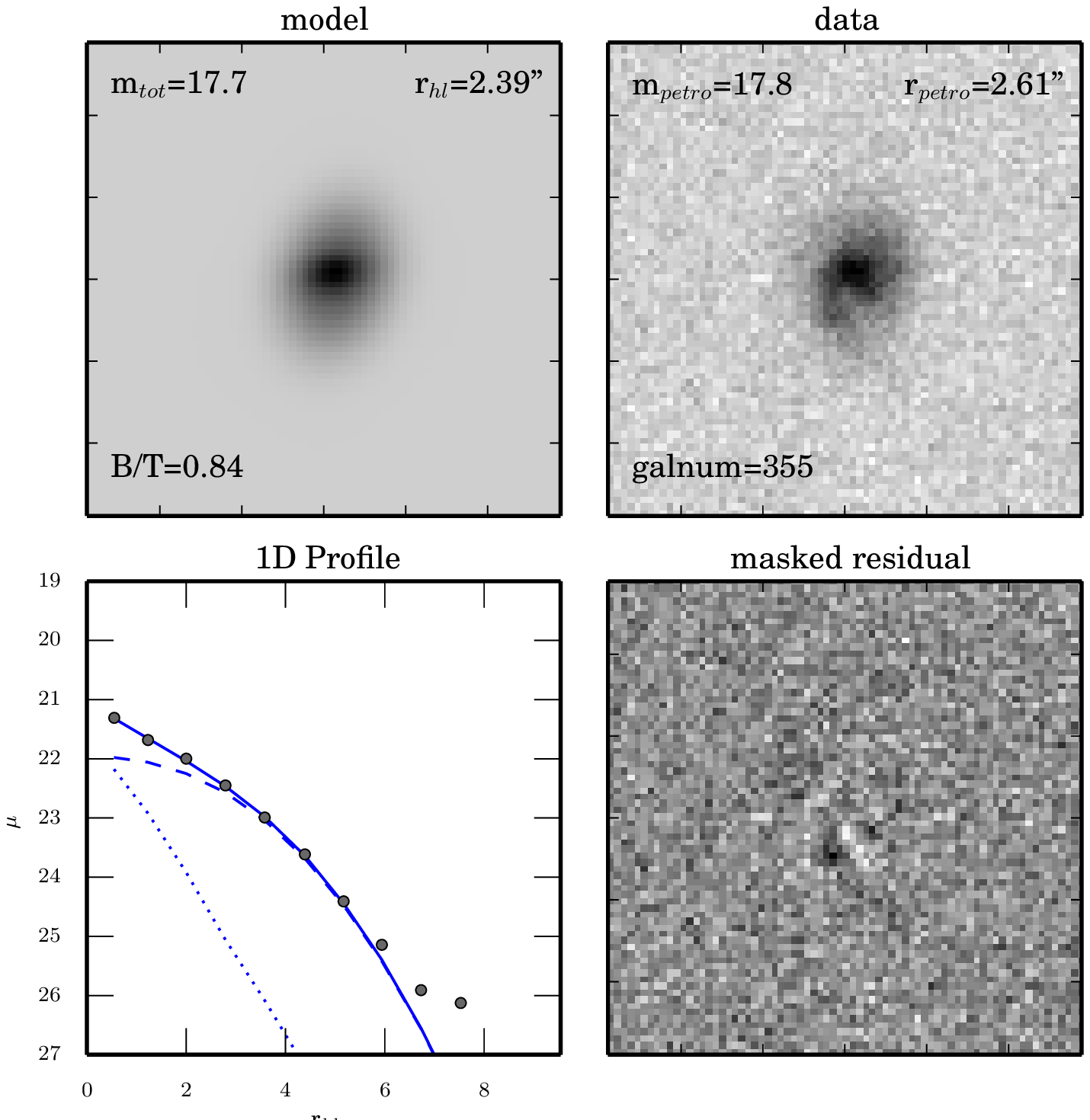}\label{fig:finalflag_flags_disk:no_exp_flip1}} \,
\subfloat[Galaxy 84055]{\includegraphics[width=0.45\linewidth]{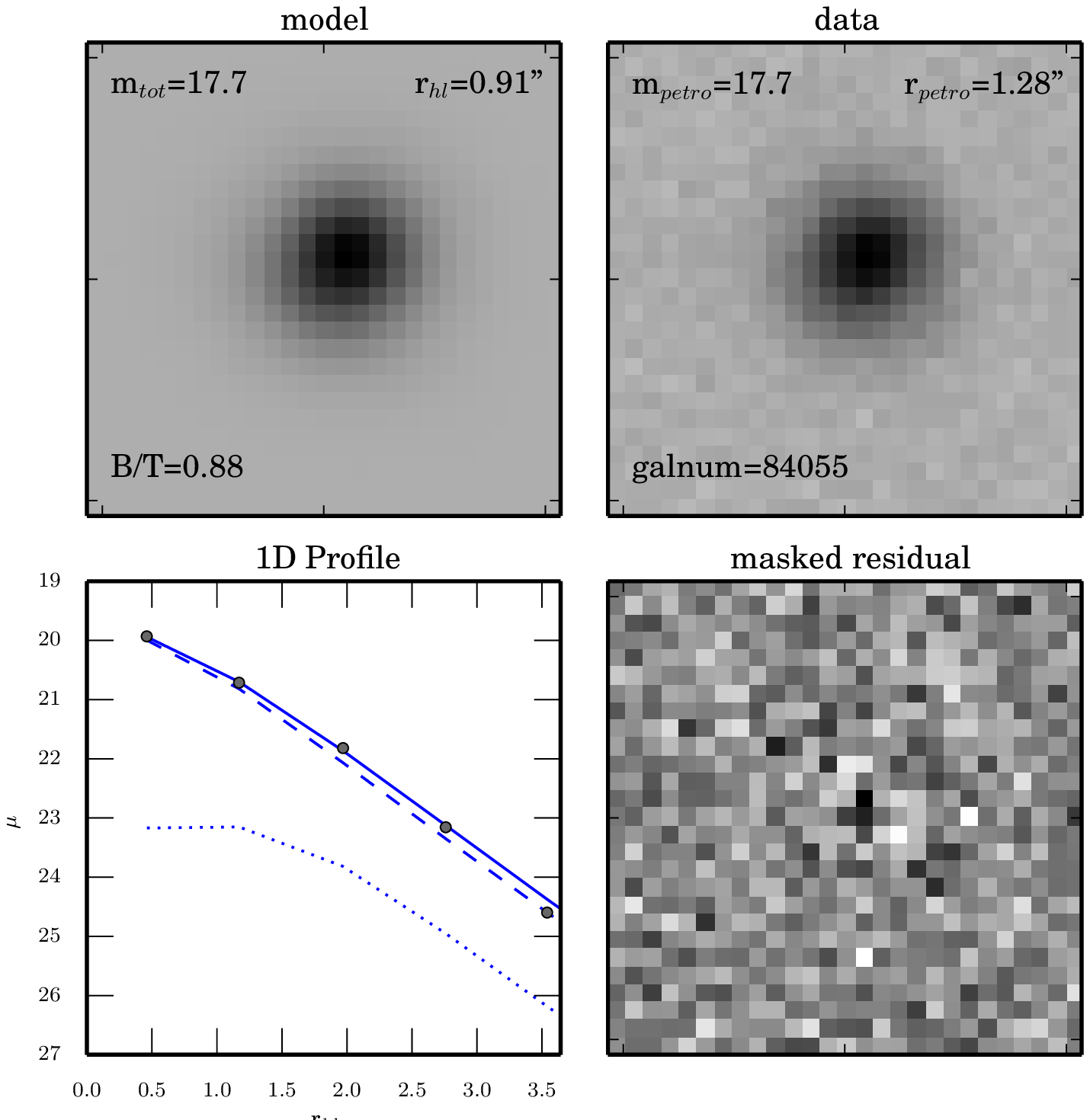}\label{fig:finalflag_flags_disk:no_exp_flip2}} \,
 \end{center}
\end{minipage}
 \caption{Example \SerExp{} galaxies considered single-component galaxies with disc-like profiles. The profiles tend to have little or no \Exp\ component, but the \Ser{} component is disc-like with a S\'{e}rsic index less than 2. These galaxies are in category `no \Exp, n$_{Ser}<$2, flip components' in Table~\ref{tab:useful_flag_breakdown} and have flag bits 4 and 6 set.}\label{fig:finalflag_flags_disk:no_exp_flip}
\end{figure*}

\begin{figure*}
\begin{minipage}{\linewidth}
\begin{center}
\subfloat[Galaxy 135981]{\includegraphics[width=0.45\linewidth]{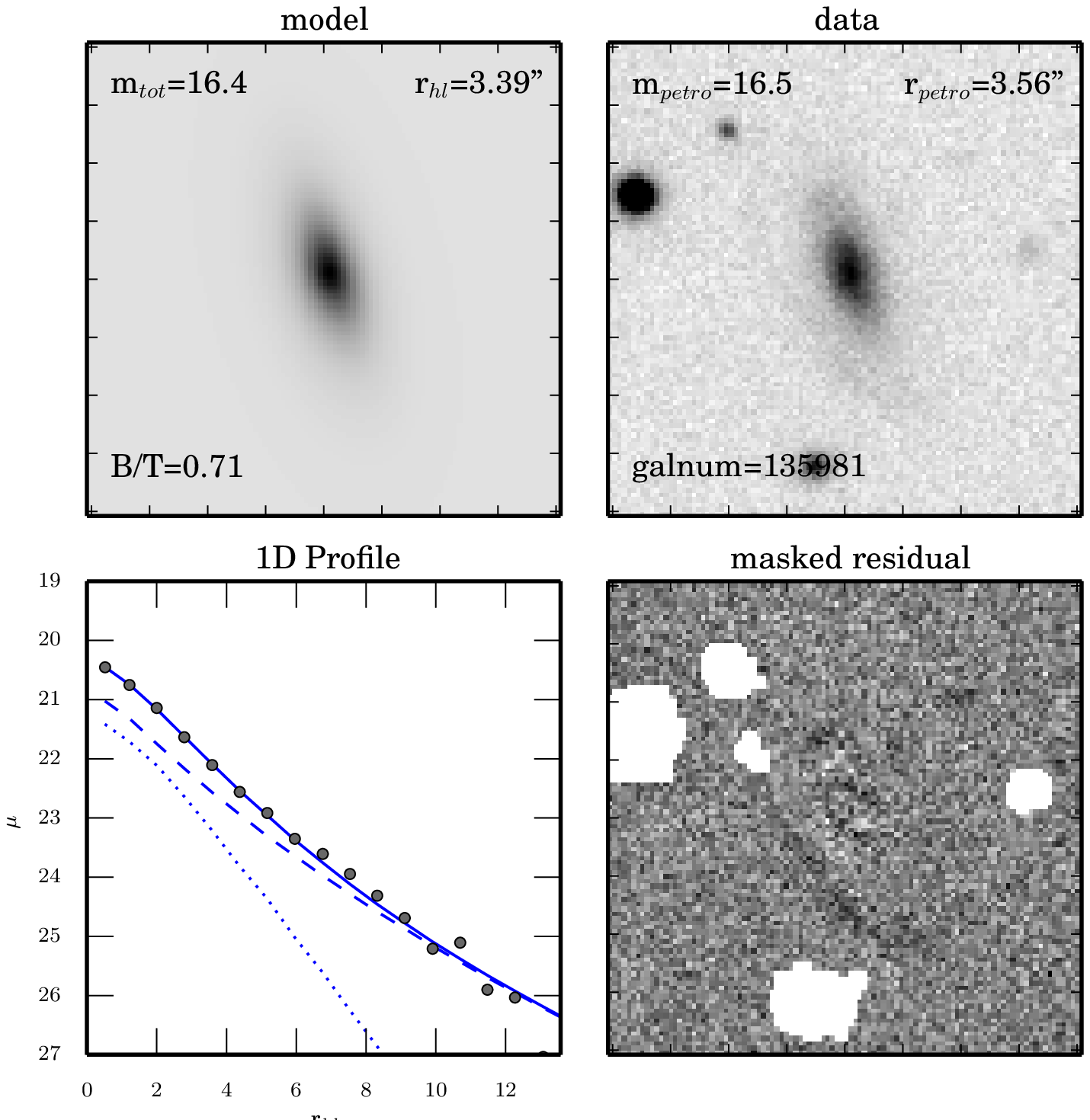}\label{fig:finalflag_flags_disk:ser_dom_flip1}} \, 
\subfloat[Galaxy 413052]{\includegraphics[width=0.45\linewidth]{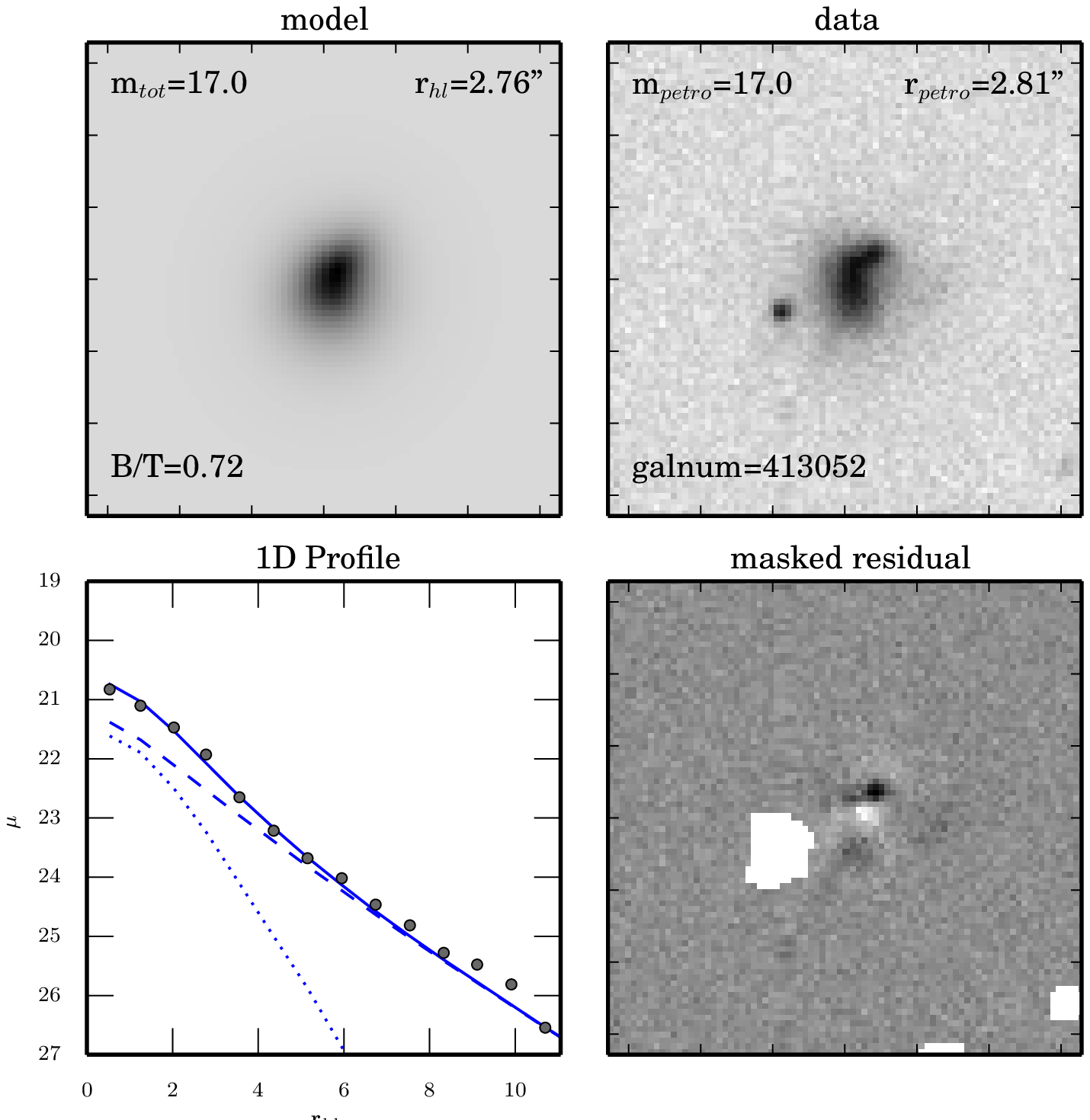}\label{fig:finalflag_flags_disk:ser_dom_flip2}} \, 
 \end{center}
\end{minipage}
 \caption{Example \SerExp{} galaxies considered single-component galaxies with disc-like profiles. The profiles tend to have some small \Exp{} component, but the \Ser{} component dominates at all radii and is disc like with a S\'{e}rsic index less than 2. These galaxies are in category `\Ser{} dominates always, n$_{\Ser}<$2' in Table~\ref{tab:useful_flag_breakdown} and have flag bits 4 and 7 set.}\label{fig:finalflag_flags_disk:ser_dom_flip}
\end{figure*}

\begin{figure*}
\begin{minipage}{\linewidth}
\begin{center}
\subfloat[Galaxy 157120]{\includegraphics[width=0.45\linewidth]{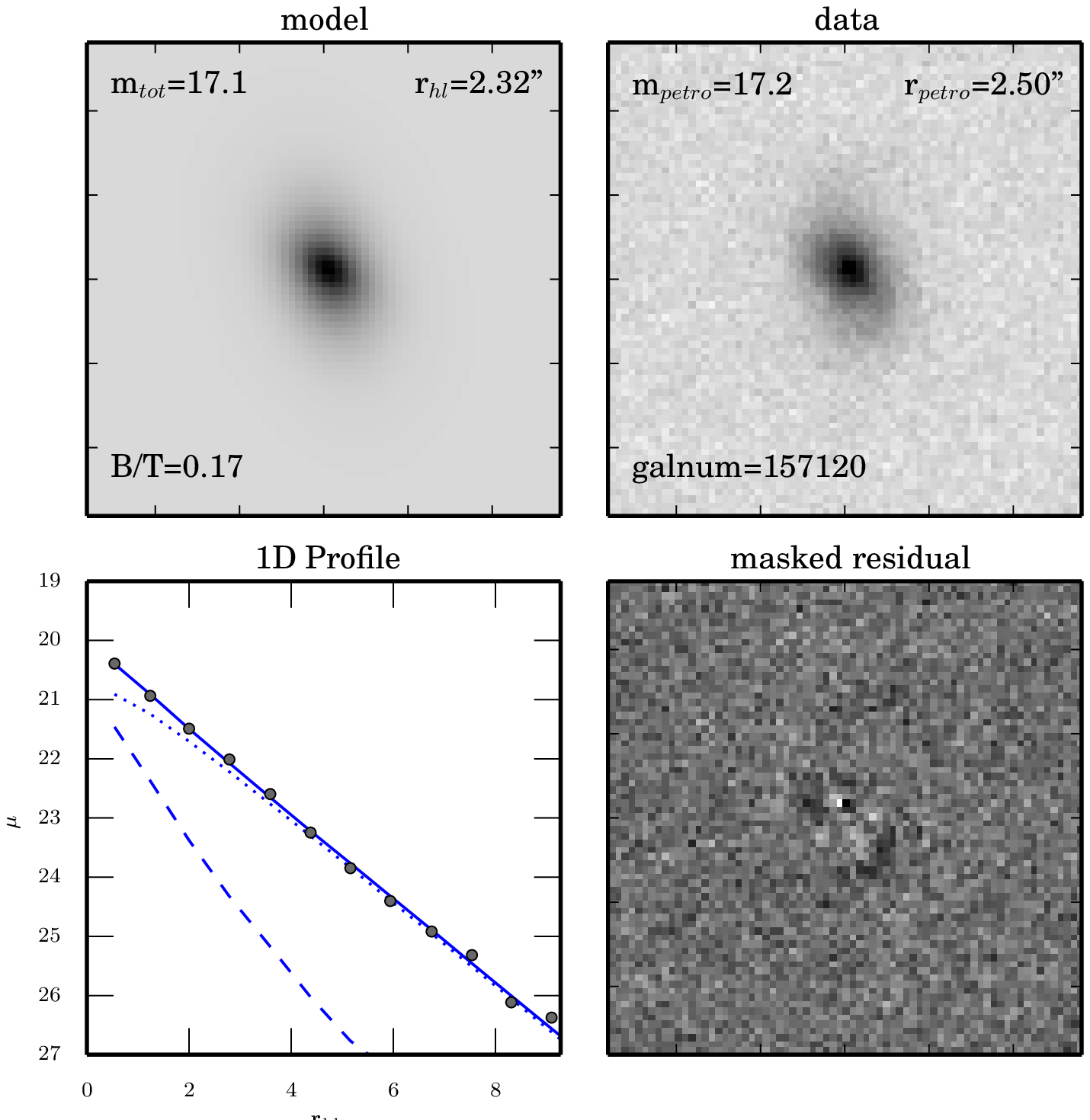}\label{fig:finalflag_flags_disk:exp_dom1}}  \,
\subfloat[Galaxy 656426]{\includegraphics[width=0.45\linewidth]{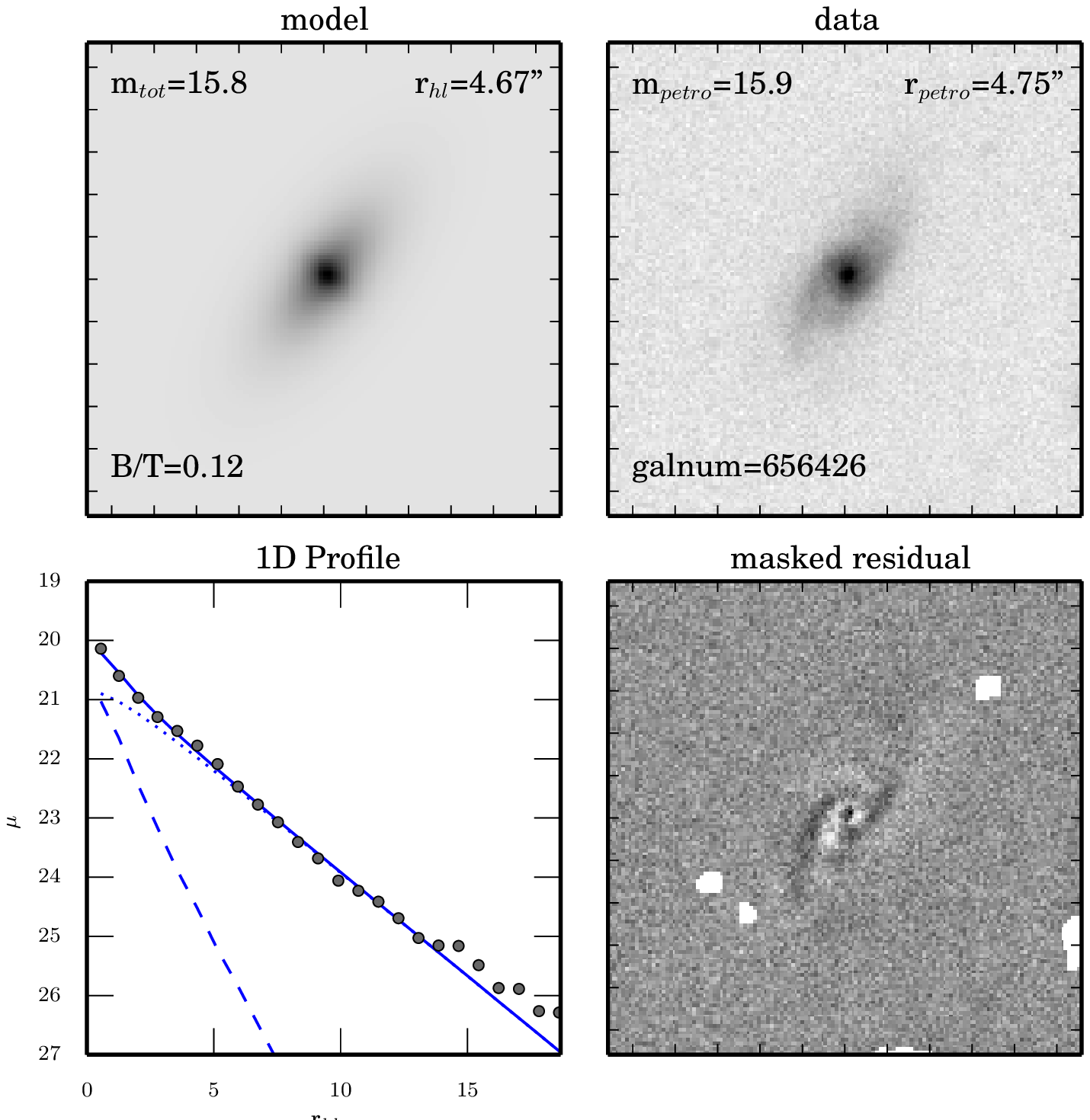}\label{fig:finalflag_flags_disk:exp_dom2}}  \,
 \end{center}
\end{minipage}
 \caption{Example \SerExp{} galaxies considered single-component galaxies with disc-like profiles. The profiles tend to have some small \Ser{} component, but the \Exp{} component dominates at all radii. These galaxies are in category `\Exp{} dominates always' in Table~\ref{tab:useful_flag_breakdown} and have flag bits 4 and 8 set.}\label{fig:finalflag_flags_disk:exp_dom}
\end{figure*}

\begin{figure*}
\begin{minipage}{\linewidth}
\begin{center}
\subfloat[Galaxy 543174]{\includegraphics[width=0.45\linewidth]{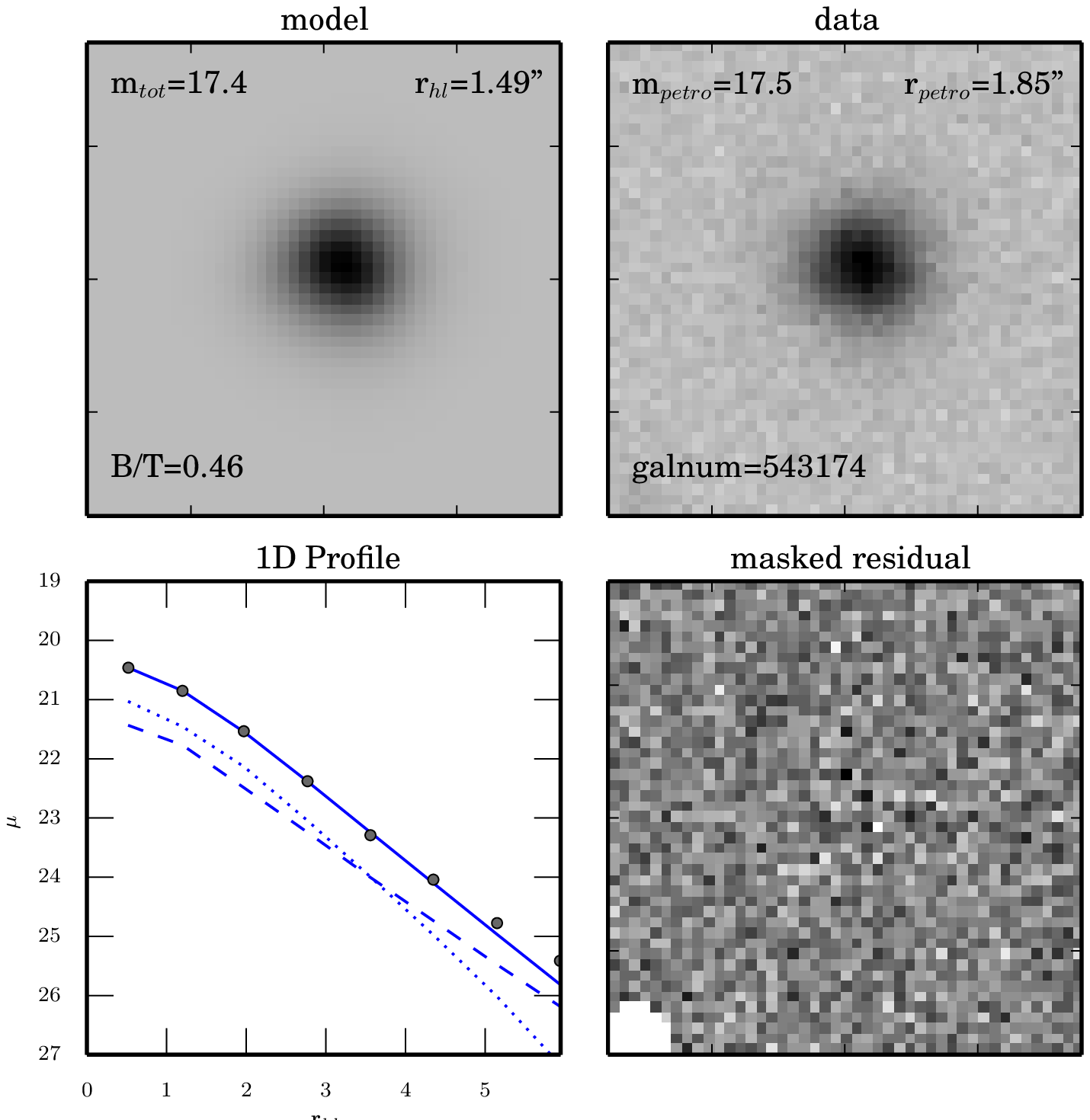}\label{fig:finalflag_flags_disk:parallel_com1}}  \,
\subfloat[Galaxy 642141]{\includegraphics[width=0.45\linewidth]{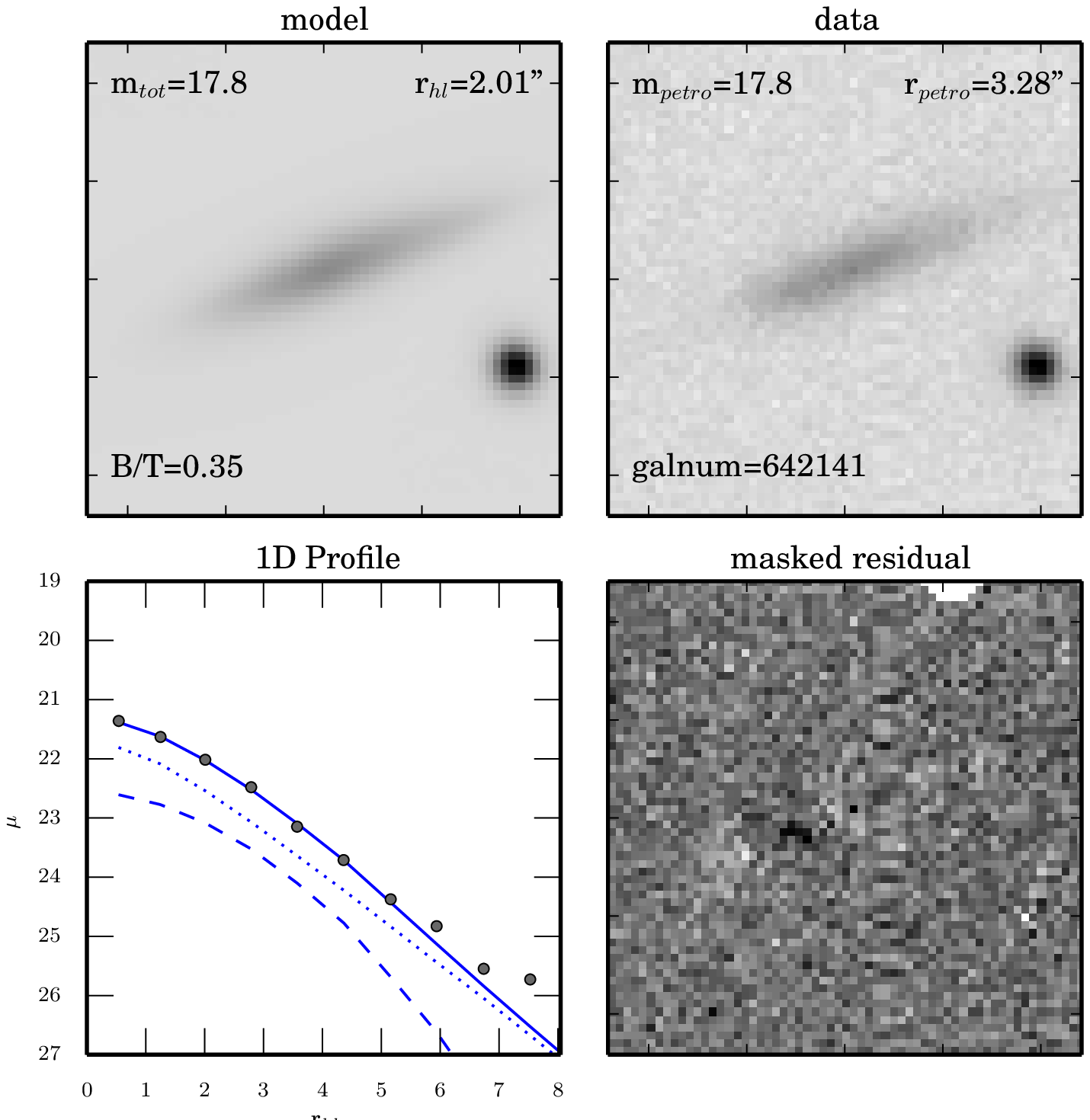}\label{fig:finalflag_flags_disk:parallel_com2}}  \,
 \end{center}
\end{minipage}
 \caption{Example \SerExp{} galaxies considered single-component galaxies with disc-like profiles. The profiles tend to have both \Ser{} and \Exp\ components, but the \Ser{} component is disc-like with S\'{e}rsic index less than 2 and components contribute similar amounts of light to the profile across a range of radii (they appear to parallel each other in the 1D fit). These galaxies are in category `parallel components' in Table~\ref{tab:useful_flag_breakdown} and have flag bits 4 and 9 set.}\label{fig:finalflag_flags_disk:parallel_com}
\end{figure*}

\begin{figure*}
\begin{minipage}{\linewidth}
\begin{center}
\subfloat[Galaxy 182763]{\includegraphics[width=0.45\linewidth]{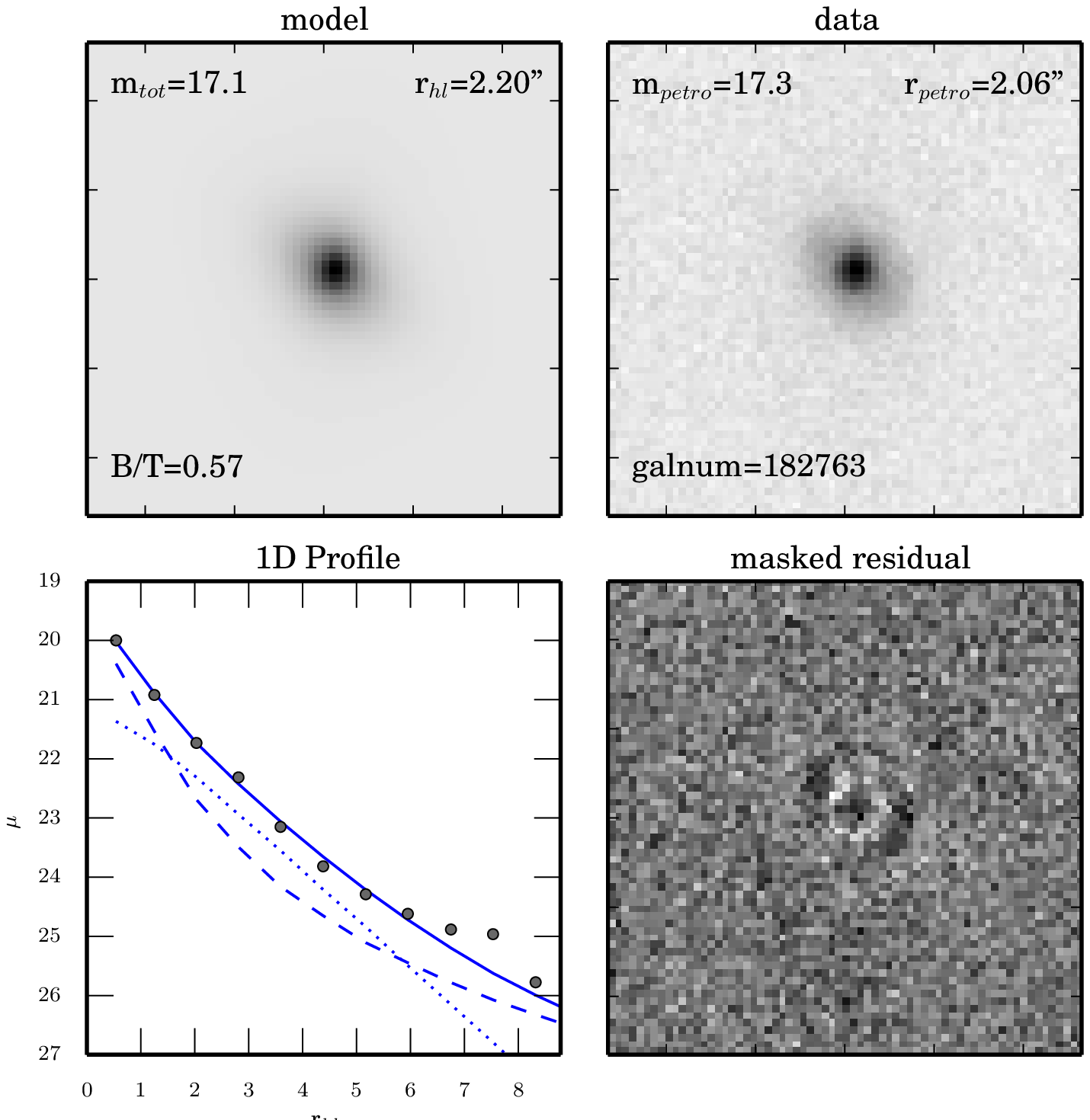}\label{fig:finalflag_flags_2comprob:ser_outer1}}  \,
\subfloat[Galaxy 532728]{\includegraphics[width=0.45\linewidth]{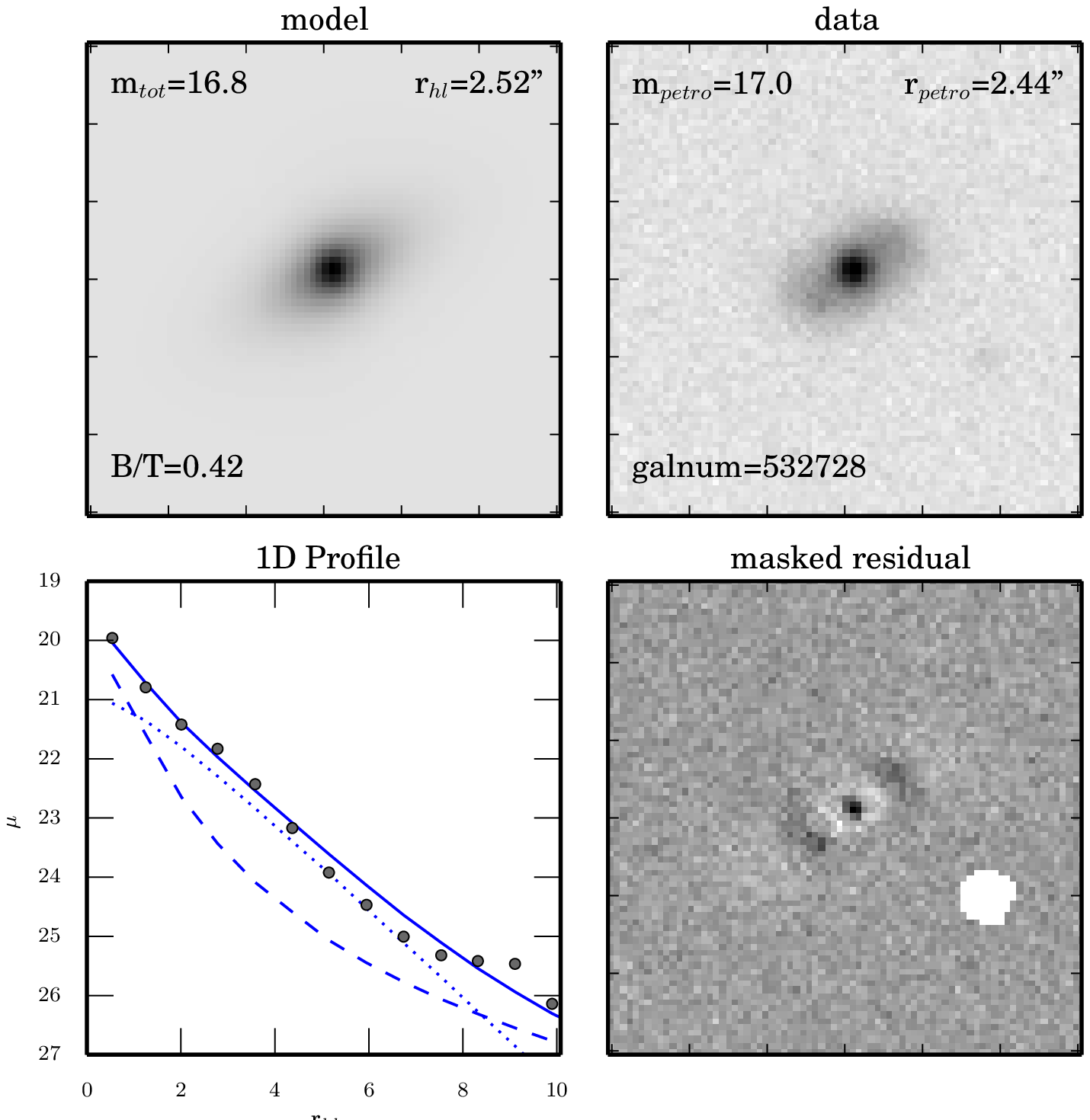}\label{fig:finalflag_flags_2comprob:ser_outer2}}  \,
 \end{center} 
\end{minipage}
 \caption{Example \SerExp{} galaxies considered two-component galaxies although the components may have problematic interpretations. The profiles tend to have both \Ser{} and \Exp\ components, but the \Ser{} component dominates at large radii. This can be related to sky problems and artificially increase the bulge radius and magnitude. These galaxies are in category `\Ser{} outer only' in Table~\ref{tab:useful_flag_breakdown} and have flag bits 14 and 15 set.}\label{fig:finalflag_flags_2comprob:ser_outer}
\end{figure*}

\begin{figure*}
\begin{minipage}{\linewidth}
\begin{center}
\subfloat[Galaxy 520456]{\includegraphics[width=0.45\linewidth]{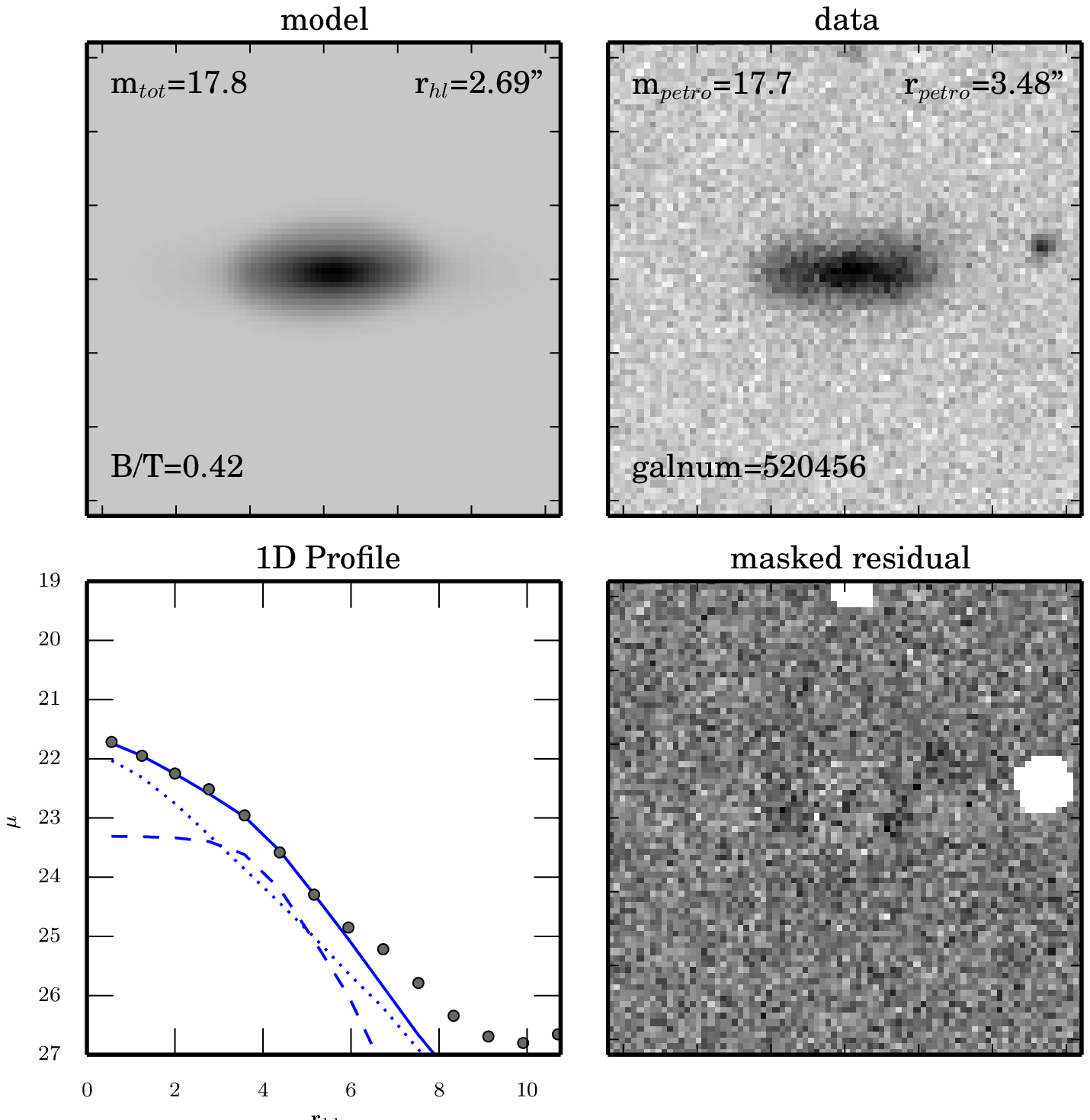}\label{fig:finalflag_flags_2comprob:exp_inner1}} \,
\subfloat[Galaxy 541105]{\includegraphics[width=0.45\linewidth]{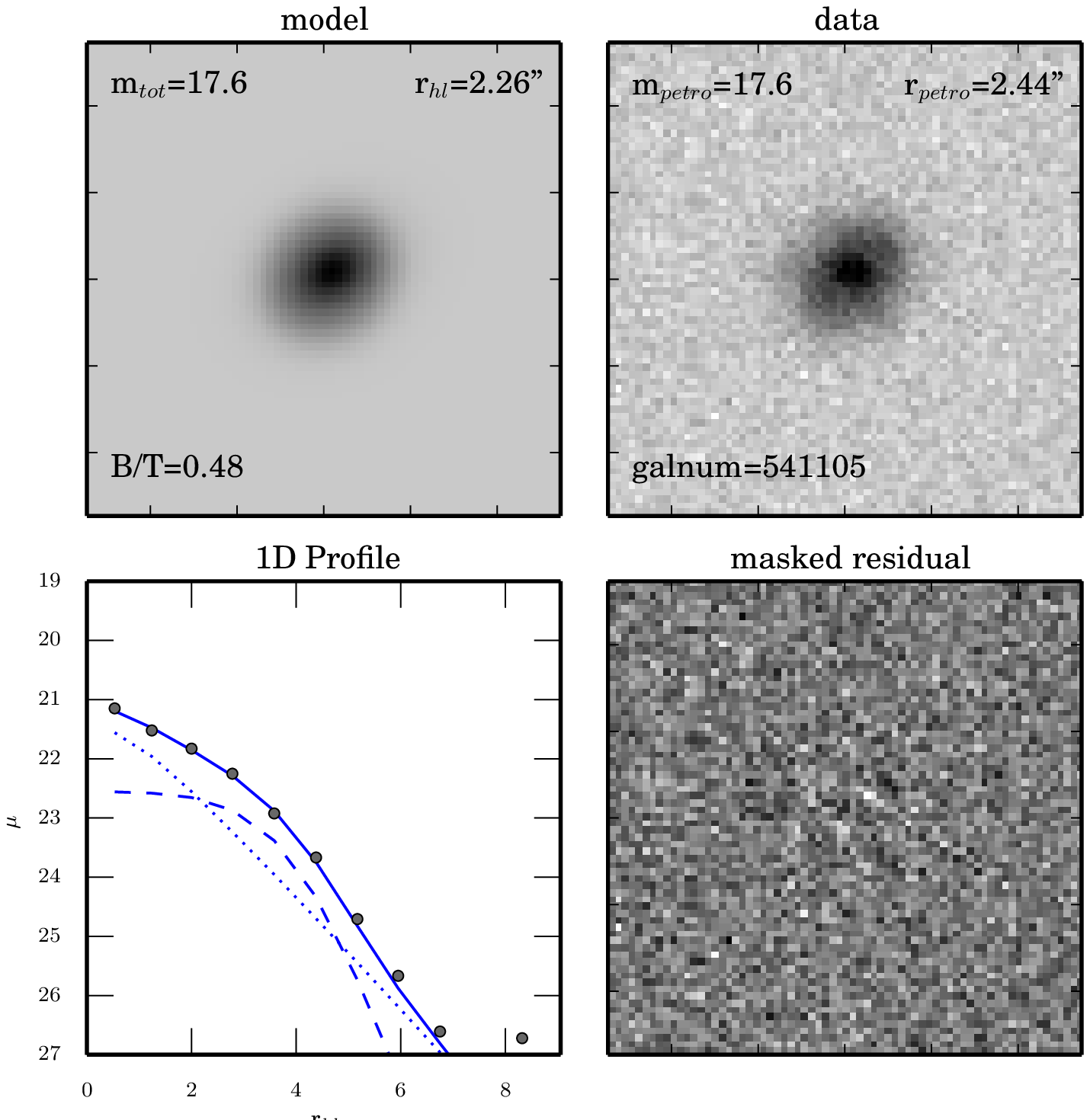}\label{fig:finalflag_flags_2comprob:exp_inner2}} \,
 \end{center} 
\end{minipage}
 \caption{Example \SerExp{} galaxies considered two-component galaxies although the components may have problematic interpretations. The profiles tend to have both \Ser{} and \Exp\ components, but the \Exp\ component dominates at small radii and the \Ser{} component is not necessarily disc like. These galaxies are in category `\Exp{} inner only' in Table~\ref{tab:useful_flag_breakdown} and have flag bits 14 and 16 set.}\label{fig:finalflag_flags_2comprob:exp_inner}
\end{figure*}

\begin{figure*}
\begin{minipage}{\linewidth}
\begin{center}
\subfloat[Galaxy 117380]{\includegraphics[width=0.45\linewidth]{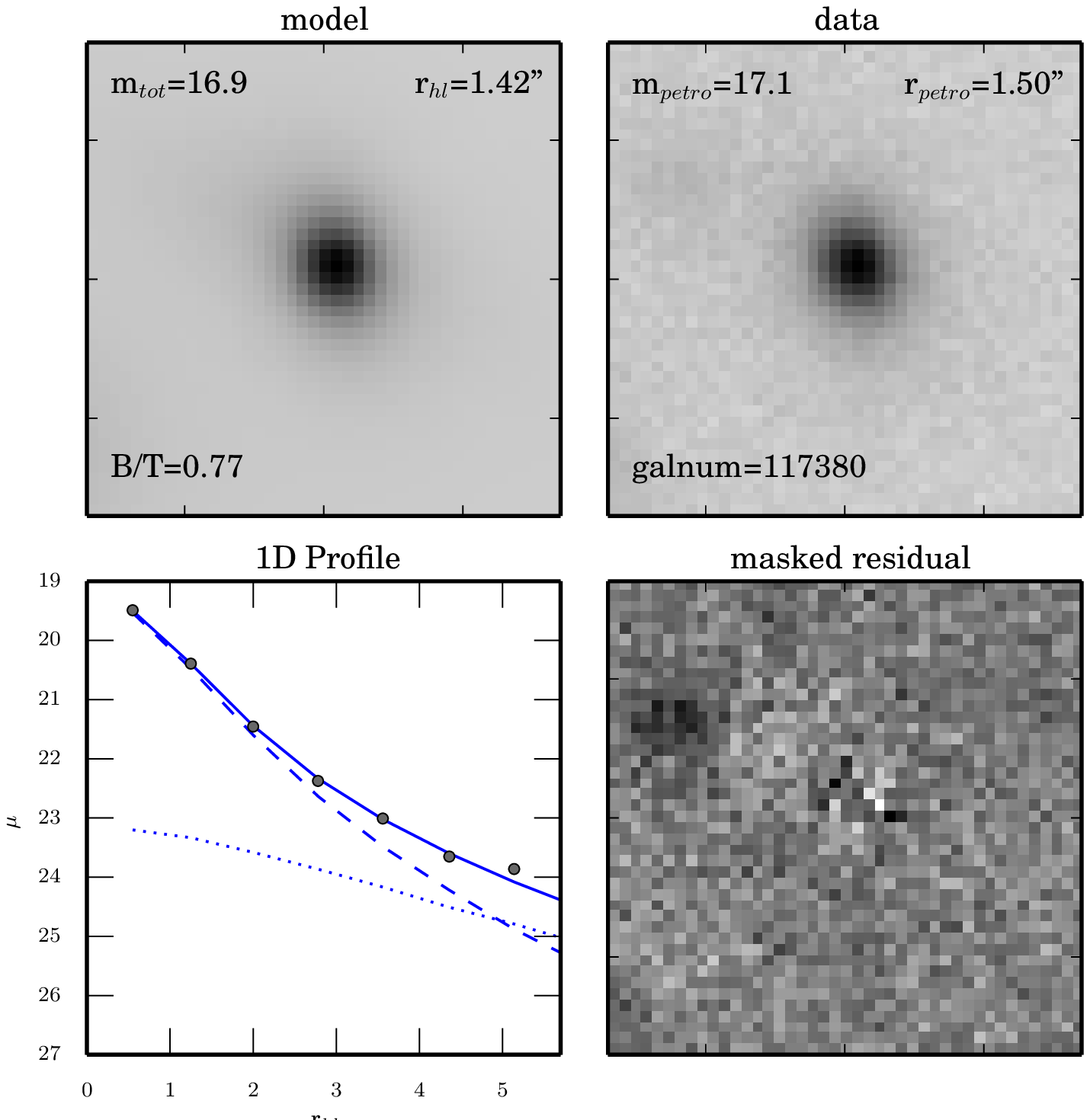}\label{fig:finalflag_flags_2comprob:good_ser_bad_exp1}}  \,
\subfloat[Galaxy 242460]{\includegraphics[width=0.45\linewidth]{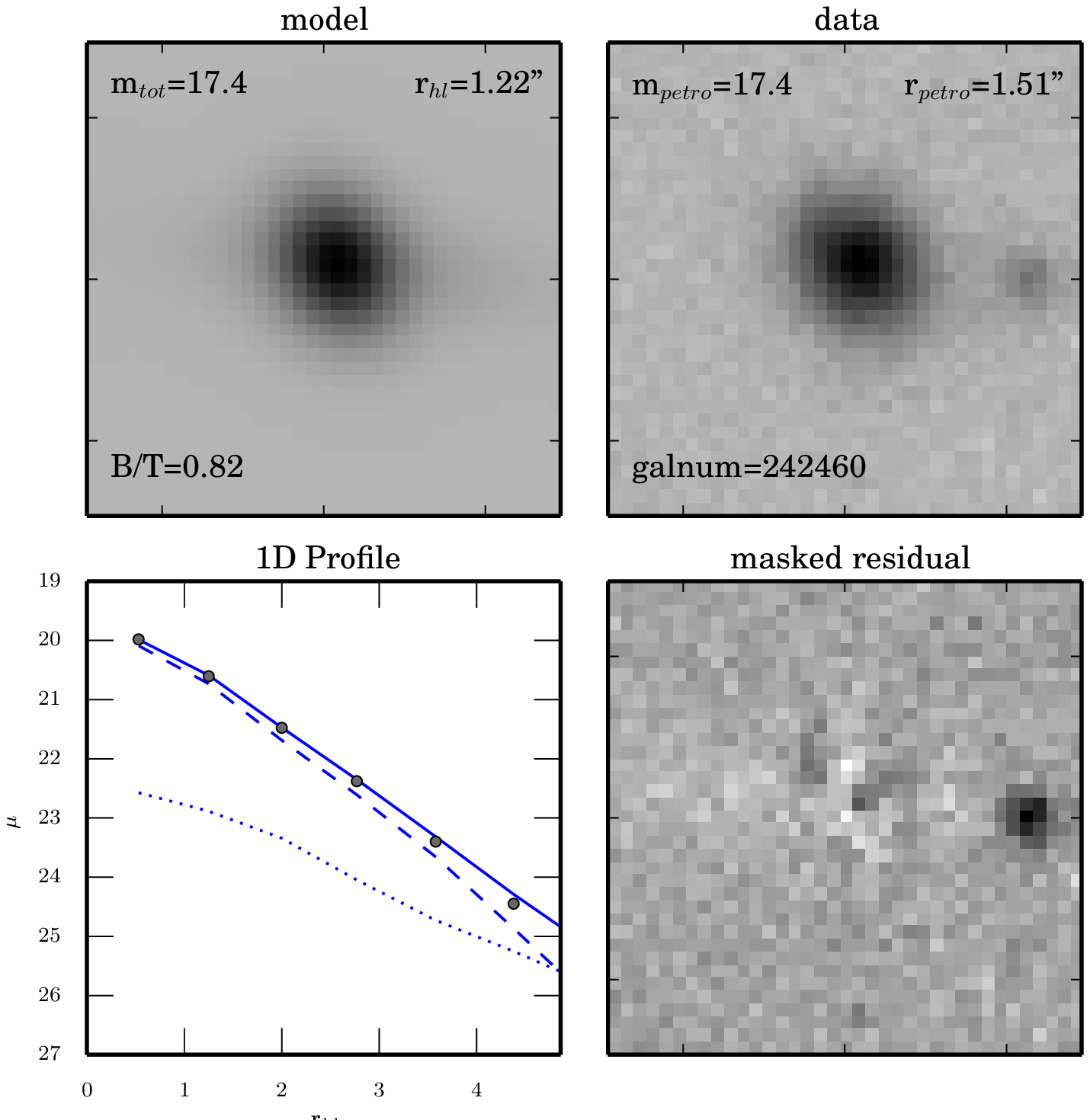}\label{fig:finalflag_flags_2comprob:good_ser_bad_exp2}}  \,
 \end{center} 
\end{minipage}
 \caption{Example \SerExp{} galaxies considered two-component galaxies although the components may have problematic interpretations. The profiles tend to have both \Ser{} and \Exp\ components, but the \Exp\ component is highly elliptical and misaligned with the galaxy. These galaxies are in category `good \Ser{}, bad \Exp{}, B/T$>=$0.5'' in Table~\ref{tab:useful_flag_breakdown} and have flag bits 14 and 17 set.}\label{fig:finalflag_flags_2comprob:good_ser_bad_exp}
\end{figure*}

\begin{figure*}
\begin{minipage}{\linewidth}
\begin{center}
\subfloat[Galaxy 22342]{\includegraphics[width=0.45\linewidth]{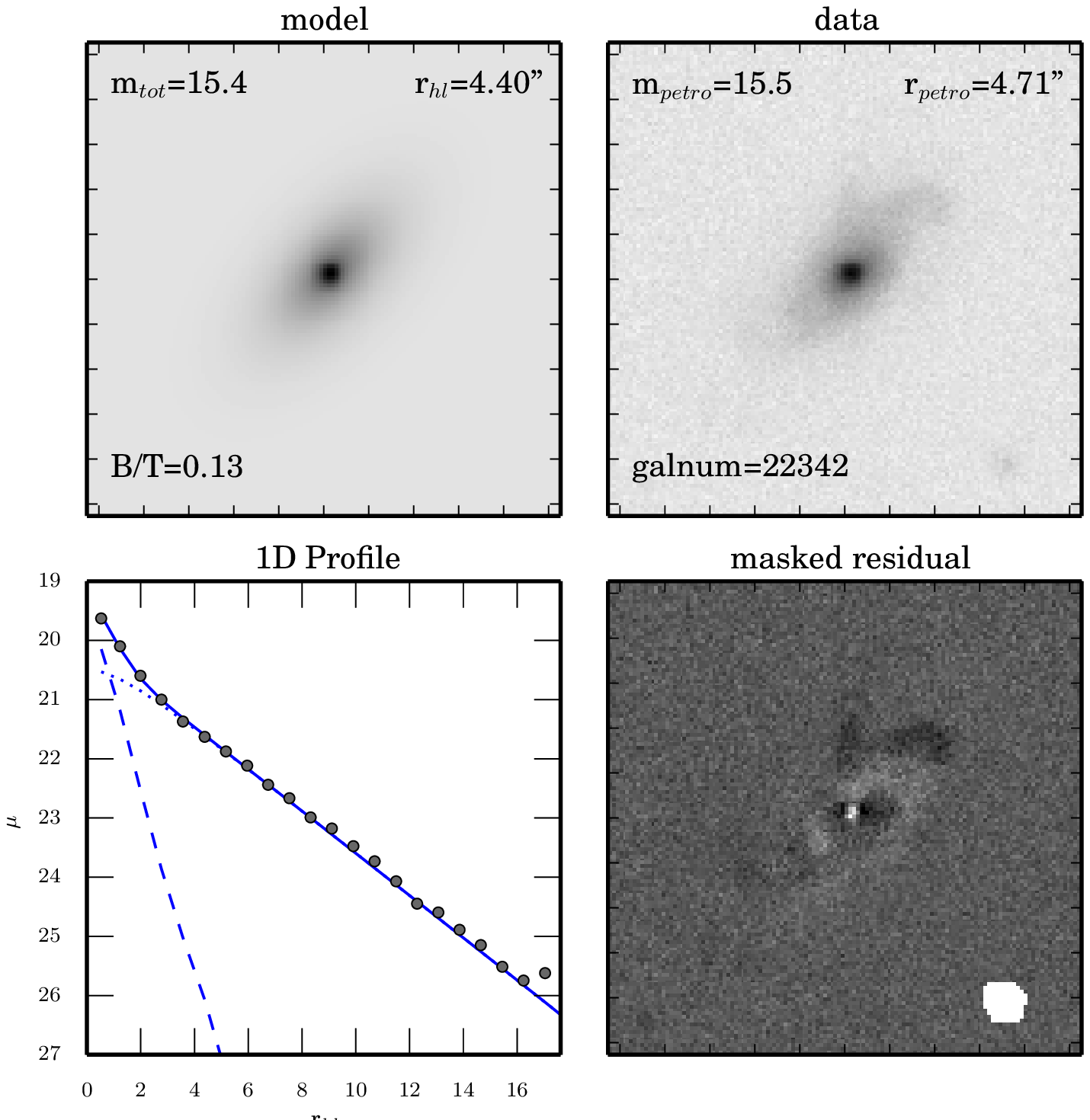}\label{fig:finalflag_flags_2comprob:bad_ser_good_exp1}} \,
\subfloat[Galaxy 554147]{\includegraphics[width=0.45\linewidth]{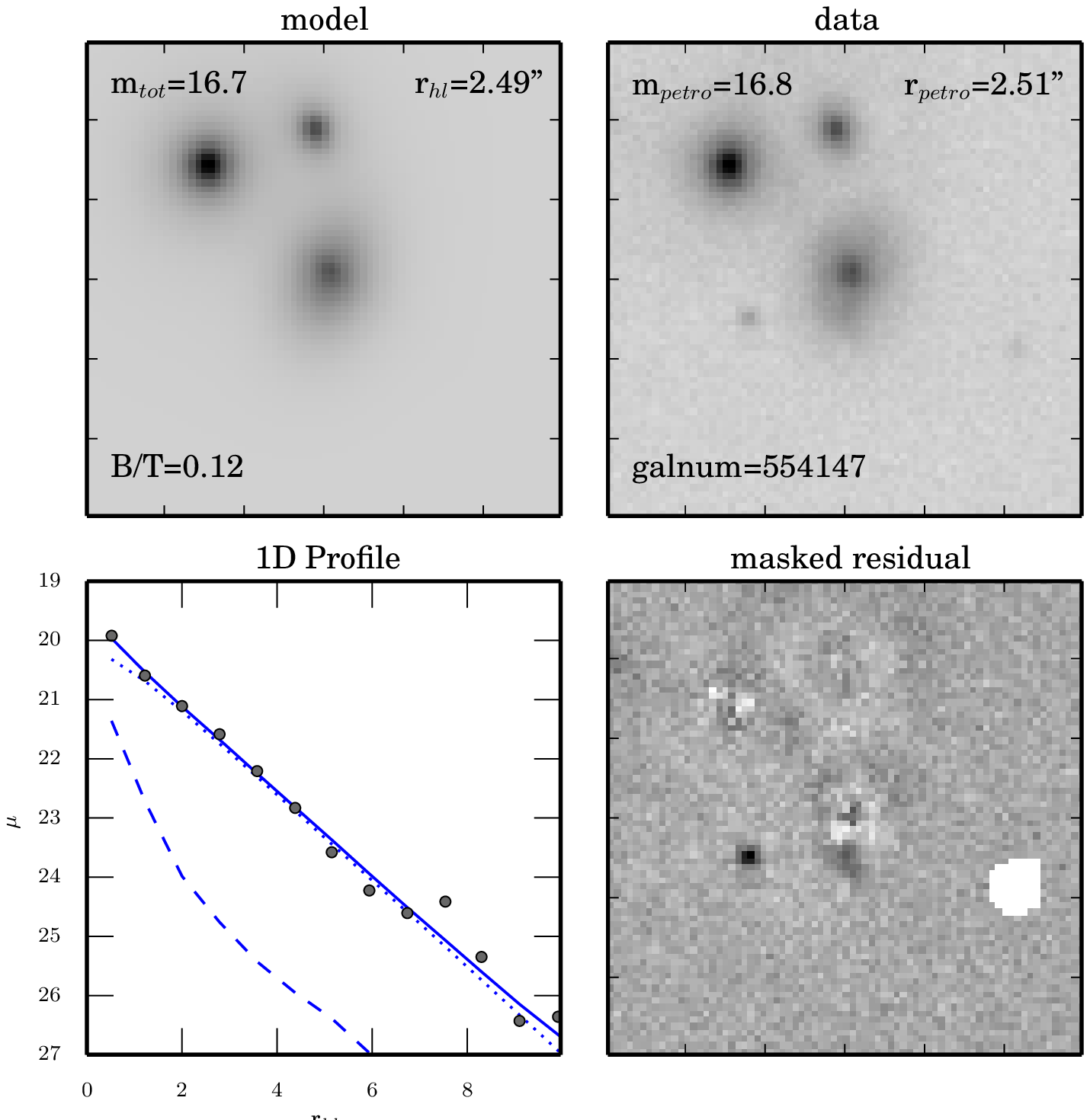}\label{fig:finalflag_flags_2comprob:bad_ser_good_exp2}} \,
 \end{center} 
\end{minipage}
 \caption{Example \SerExp{} galaxies considered two-component galaxies although the components may have problematic interpretations. The profiles tend to have both \Ser{} and \Exp\ components, but the \Ser{} component is highly elliptical and misaligned with the galaxy. These galaxies are in category `bad \Ser{}, good \Exp{}, B/T$<$0.5' in Table~\ref{tab:useful_flag_breakdown}  and have flag bits 14 and 18 set.}\label{fig:finalflag_flags_2comprob:bad_ser_good_exp}
\end{figure*}

\begin{figure*}
\begin{minipage}{\linewidth}
\begin{center}
\subfloat[Galaxy 642128]{\includegraphics[width=0.45\linewidth]{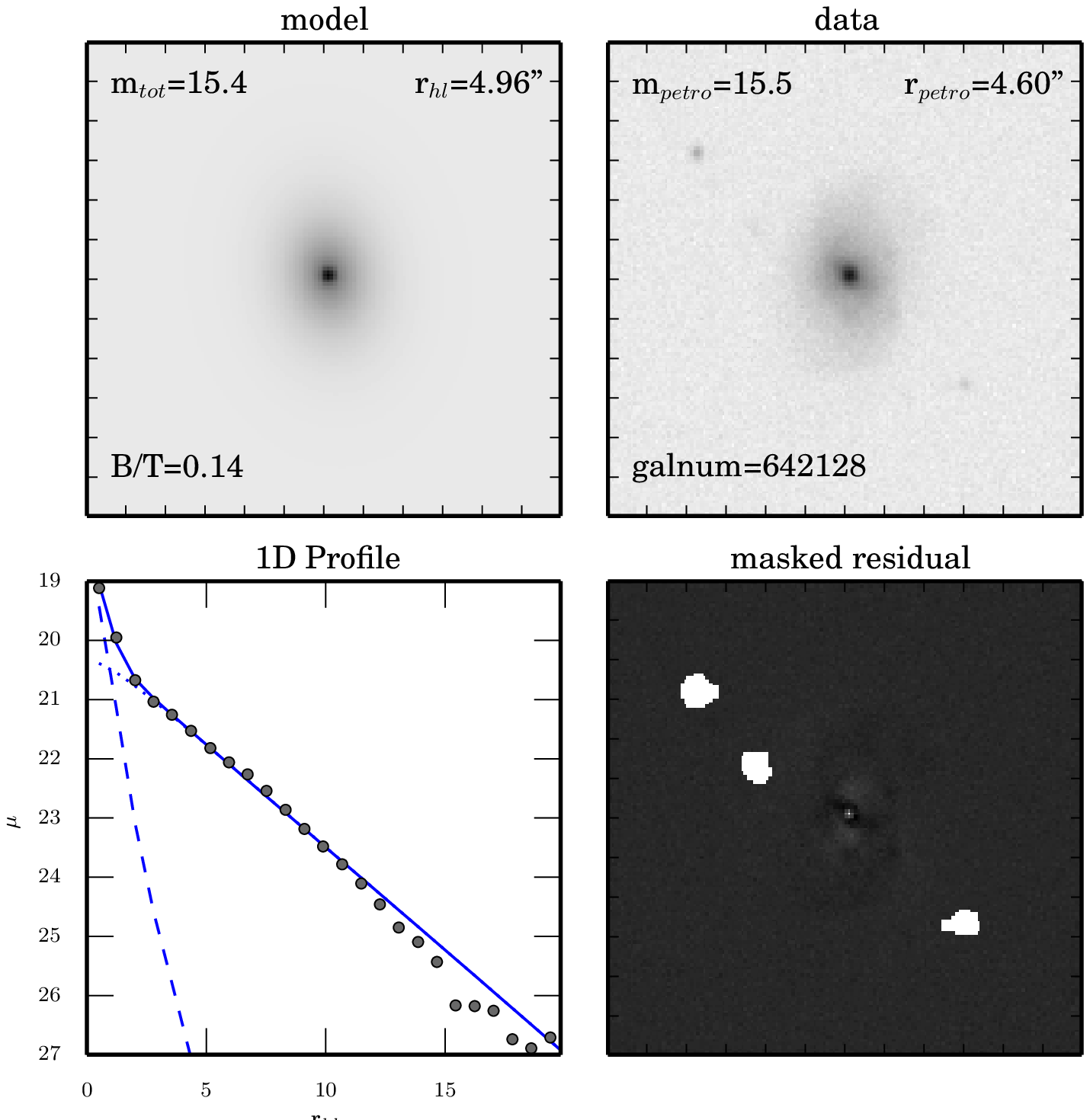}\label{fig:finalflag_flags_2comprob:tinybulge1}} \,
\subfloat[Galaxy 670464]{\includegraphics[width=0.45\linewidth]{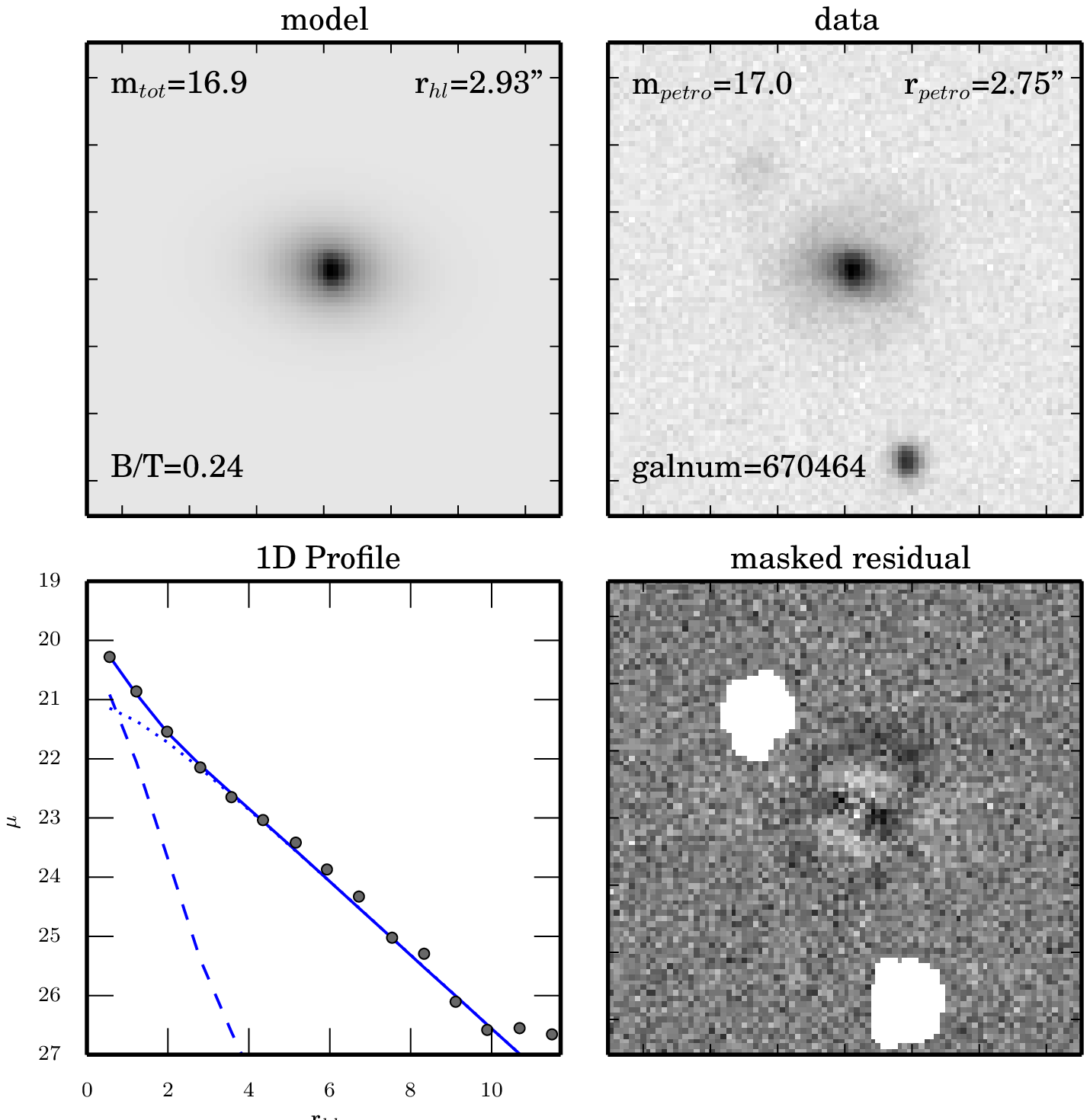}\label{fig:finalflag_flags_2comprob:tinybulge2}} \,
 \end{center} 
\end{minipage}
 \caption{Example \SerExp{} galaxies considered two-component galaxies although the components may have problematic interpretations. The profiles tend to have both \Ser{} and \Exp\ components, but the \Ser{} component has shrunk to a point source. These galaxies are in category `bulge is point' in Table~\ref{tab:useful_flag_breakdown}  and have flag bits 14 and 19 set.}\label{fig:finalflag_flags_2comprob:tinybulge}
\end{figure*}

Figures \ref{fig:finalflag_good}, \ref{fig:some_flag}, and \ref{fig:flip_com} show various types of good two-component fits. Figures \ref{fig:finalflag_flags_bulge:no_exp} and \ref{fig:finalflag_flags_bulge:ser_dom} show bulge fits. Figures \ref{fig:finalflag_flags_disk:no_ser}, \ref{fig:finalflag_flags_disk:no_exp_flip}, \ref{fig:finalflag_flags_disk:ser_dom_flip}, \ref{fig:finalflag_flags_disk:exp_dom}, and \ref{fig:finalflag_flags_disk:parallel_com}
show disc galaxies. Finally, Figures \ref{fig:finalflag_flags_2comprob:ser_outer}, \ref{fig:finalflag_flags_2comprob:exp_inner}, \ref{fig:finalflag_flags_2comprob:good_ser_bad_exp}, \ref{fig:finalflag_flags_2comprob:bad_ser_good_exp},  and \ref{fig:finalflag_flags_2comprob:tinybulge}  show two-component fits where the components are difficult to interpret.

\clearpage

\section{Additional fit comparisons}

\subsection{The \Dev{} fits}\label{app:dev}

\begin{figure*}
\centering
\includegraphics[width=0.24\linewidth]{./figures/cmp_plots/sdss/r_band_sdss_dev_petromag_mtot.eps}
\includegraphics[width=0.24\linewidth]{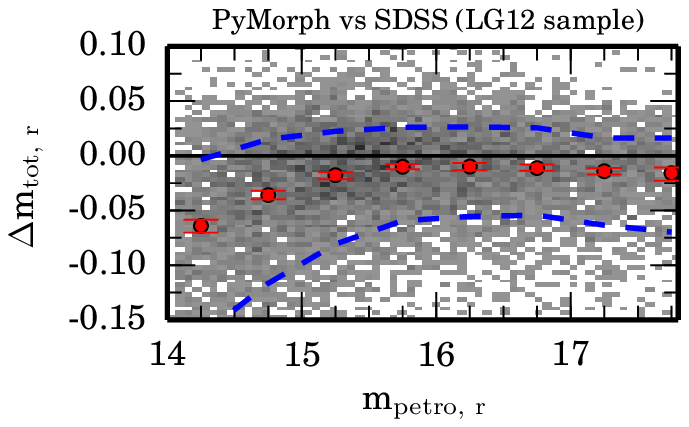}
\includegraphics[width=0.24\linewidth]{./figures/cmp_plots/lackner/r_band_lackner_dev_petromag_mtot.eps}
\includegraphics[width=0.24\linewidth]{./figures/cmp_plots/sdss/r_lackner_sdss_dev_petromag_mtot.eps}
\includegraphics[width=0.24\linewidth]{./figures/cmp_plots/sdss/r_band_sdss_dev_petromag_abs_mtot.eps}
\includegraphics[width=0.24\linewidth]{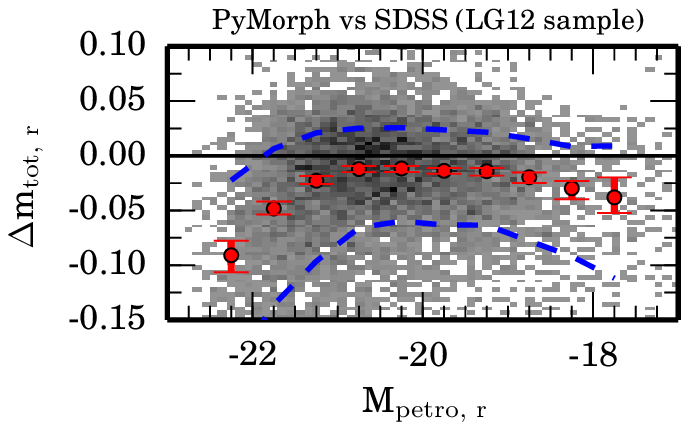}
\includegraphics[width=0.24\linewidth]{./figures/cmp_plots/lackner/r_band_lackner_dev_petromag_abs_mtot.eps}
\includegraphics[width=0.24\linewidth]{./figures/cmp_plots/sdss/r_lackner_sdss_dev_petromag_abs_mtot.eps}
\includegraphics[width=0.24\linewidth]{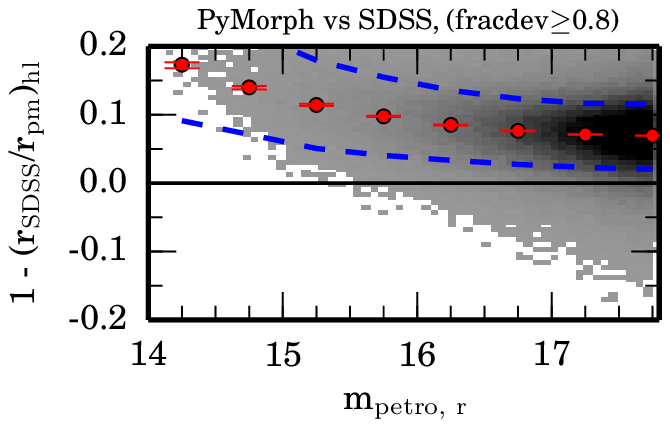}
\includegraphics[width=0.24\linewidth]{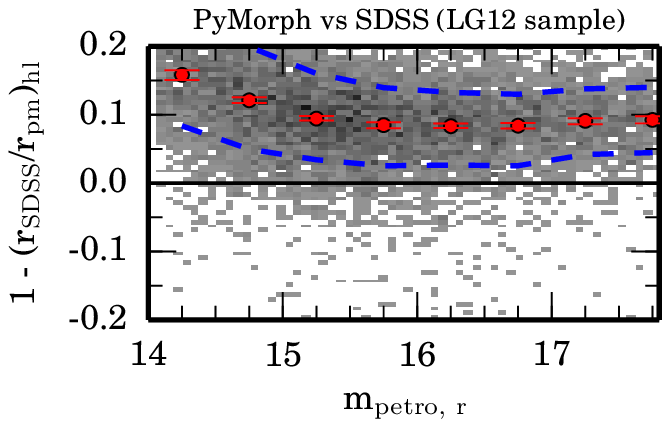}
\includegraphics[width=0.24\linewidth]{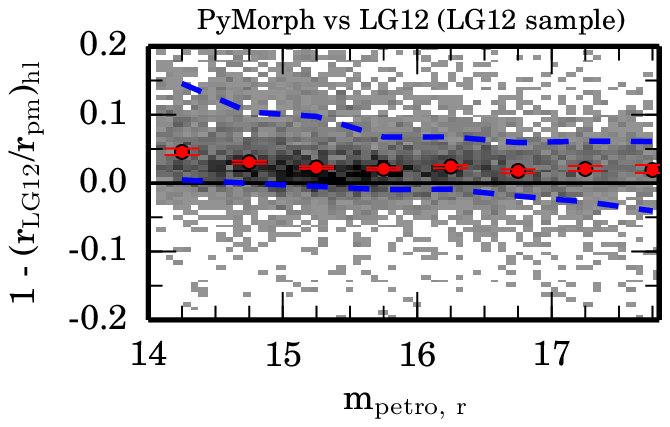}
\includegraphics[width=0.24\linewidth]{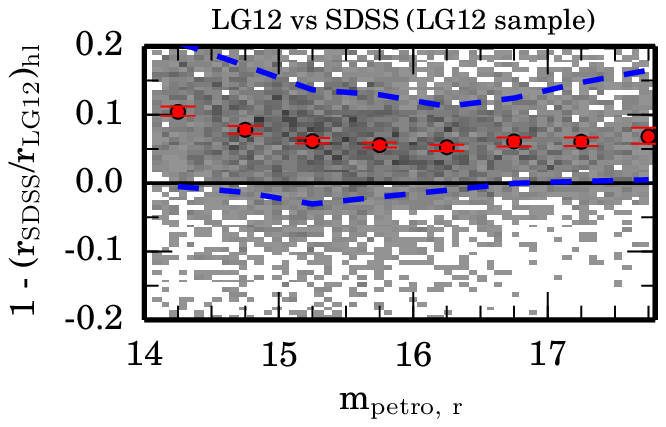}
\includegraphics[width=0.24\linewidth]{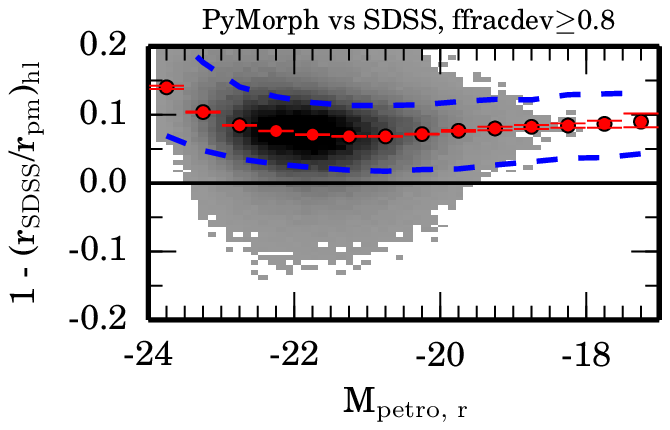}
\includegraphics[width=0.24\linewidth]{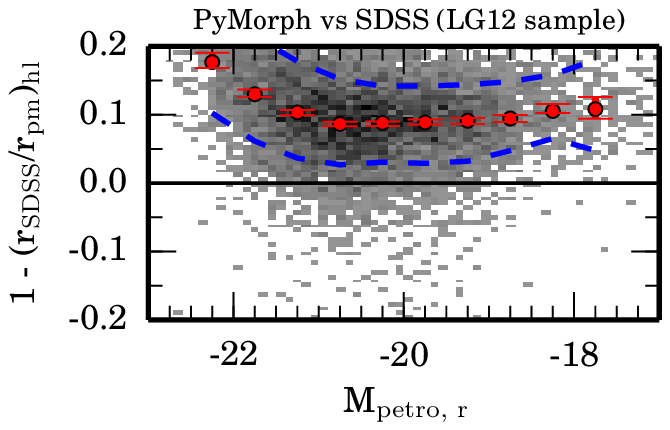}
\includegraphics[width=0.24\linewidth]{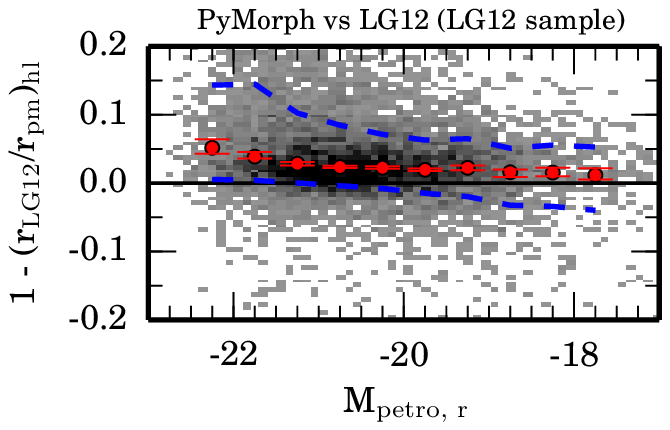}
\includegraphics[width=0.24\linewidth]{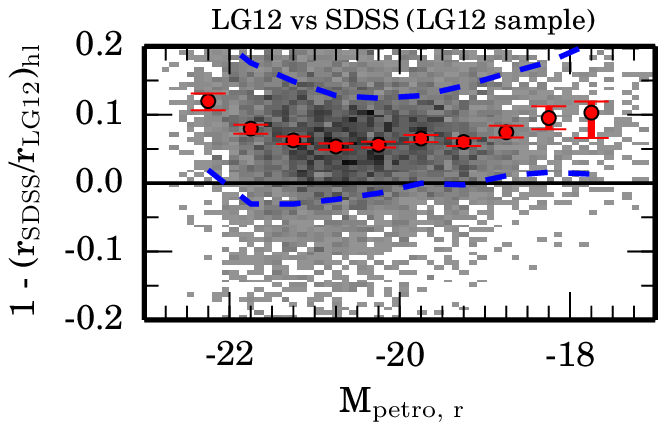}

\caption{Similar to Figure~\ref{fig:dev_fits}, but now also including the comparison of SDSS with LG12 galaxies 
(far-right column).
The panels show the difference in magnitude and radius 
as a function of apparent or absolute Petrosian magnitude. The offsets are discussed
in the main text.}
\label{fig:dev_fits_all}
\end{figure*}

This appendix includes an extended comparison of the \Dev{} fits. In Figure~\ref{fig:dev_fits_all},
we show comparisons between this work and SDSS for all galaxies with SDSS \texttt{fracdev\_r}$\geq$0.8 (left-hand column),
between this work and SDSS for LG12 galaxies best fit by a \Dev{} profile according to LG12 (centre left column),
between this work and LG12 for LG12 galaxies best fit by a \Dev{} profile according to LG12 (centre right column),
and between SDSS and LG12 for LG12 galaxies best fit by a \Dev{} profile according to LG12 (right-hand column).
The first row compares the difference in total magnitude as a function of apparent Petrosian magnitude. 
The second row compares the difference in total magnitude as a function of absolute Petrosian magnitude.
The third row compares the difference in half-light radius as a function of apparent Petrosian magnitude. 
The fourth row compares the difference in half-light radius as a function of absolute Petrosian magnitude.

The offsets between this work and SDSS in magnitude discussed in the main text are present. 
The magnitude offset with LG12 is also present as discussed in the main text.
The half-light radius also shows an offset which is consistent with a systematic difference in the fitted profiles
rather than a zero-point offset. This difference between LG12 and SDSS is explained in the text as the result of the softening 
in the central part of the profile. There may also be differences in the sky brightness. The offset between LG12 and this work 
can not be the result of the softening as neither work uses the softening at radii smaller than $r=r_{eff}/50$. This offset is 
consistent with a systematic difference in the fitting, perhaps due to differences in sky subtraction.

\clearpage

\subsection{The \Ser{} fits}\label{app:ser}

\begin{figure*}
\centering
\includegraphics[width=0.24\linewidth]{./figures/cmp_plots/simard/r_band_simard_ser_petromag_mtot.eps}
\includegraphics[width=0.24\linewidth]{./figures/cmp_plots/simard/r_band_simard_ser_petromag_mtot_lackner_only.eps}
\includegraphics[width=0.24\linewidth]{./figures/cmp_plots/lackner/r_band_lackner_ser_petromag_mtot.eps}
\includegraphics[width=0.24\linewidth]{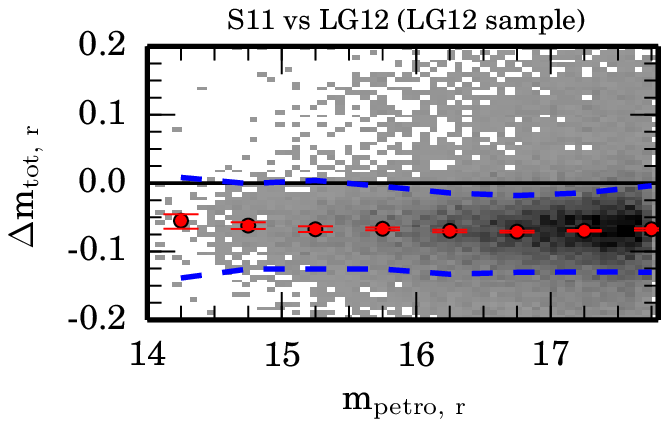}
\includegraphics[width=0.24\linewidth]{./figures/cmp_plots/simard/r_band_simard_ser_petromag_rbulge.eps}
\includegraphics[width=0.24\linewidth]{./figures/cmp_plots/simard/r_band_simard_ser_petromag_rbulge_lackner_only.eps}
\includegraphics[width=0.24\linewidth]{./figures/cmp_plots/lackner/r_band_lackner_ser_petromag_rbulge.eps}
\includegraphics[width=0.24\linewidth]{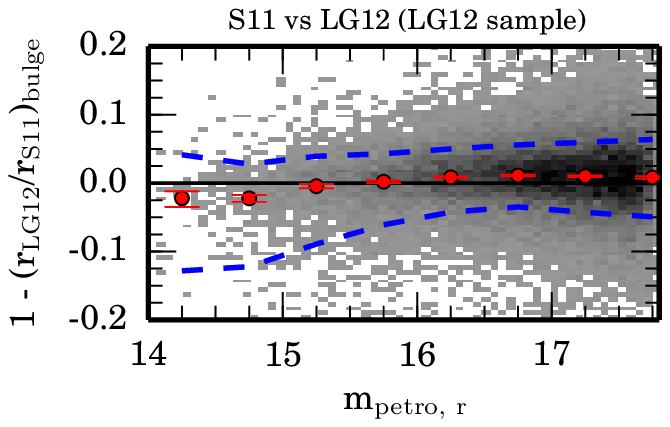}
\includegraphics[width=0.24\linewidth]{./figures/cmp_plots/simard/r_band_simard_ser_petromag_nbulge.eps}
\includegraphics[width=0.24\linewidth]{./figures/cmp_plots/simard/r_band_simard_ser_petromag_nbulge_lackner_only.eps}
\includegraphics[width=0.24\linewidth]{./figures/cmp_plots/lackner/r_band_lackner_ser_petromag_nbulge.eps}
\includegraphics[width=0.24\linewidth]{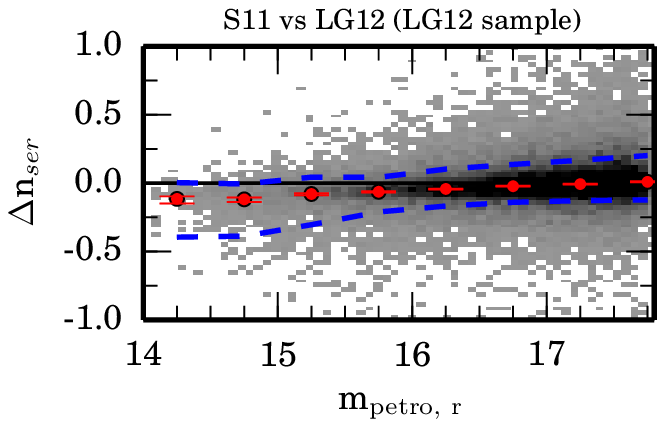}
\includegraphics[width=0.24\linewidth]{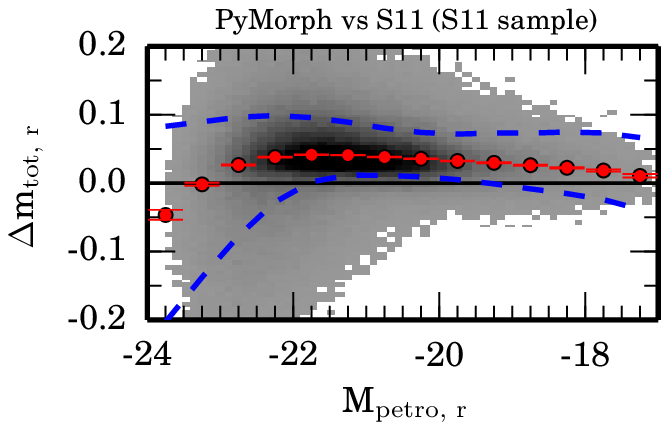}
\includegraphics[width=0.24\linewidth]{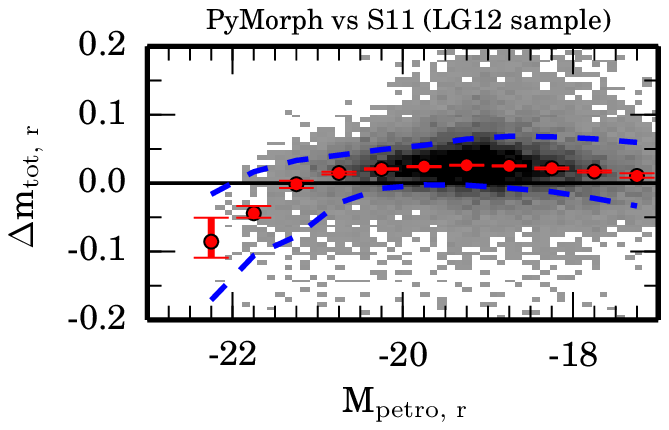}
\includegraphics[width=0.24\linewidth]{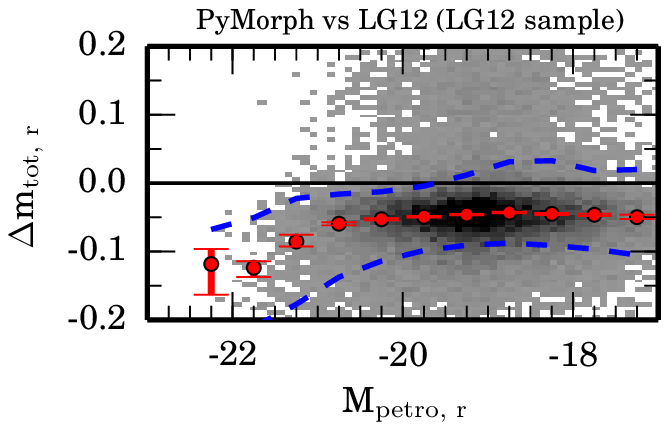}
\includegraphics[width=0.24\linewidth]{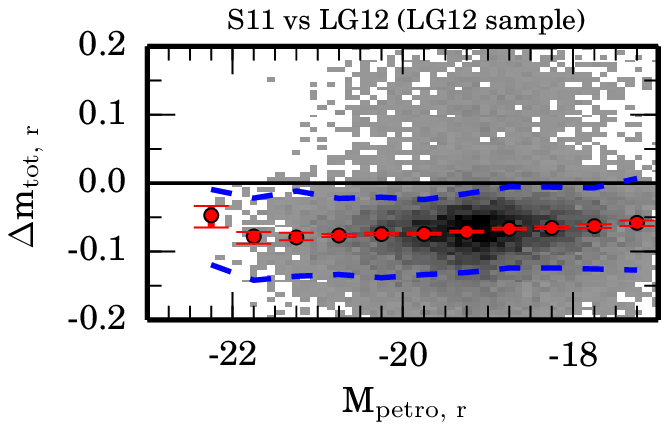}
\includegraphics[width=0.24\linewidth]{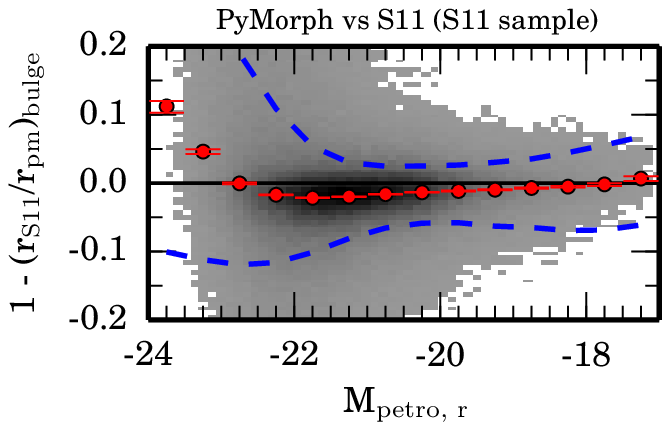}
\includegraphics[width=0.24\linewidth]{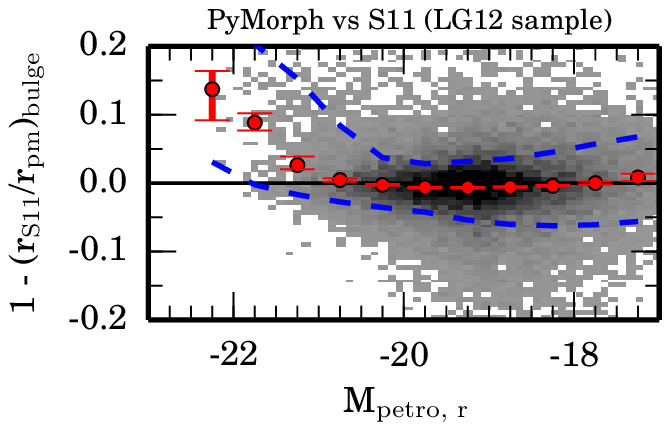}
\includegraphics[width=0.24\linewidth]{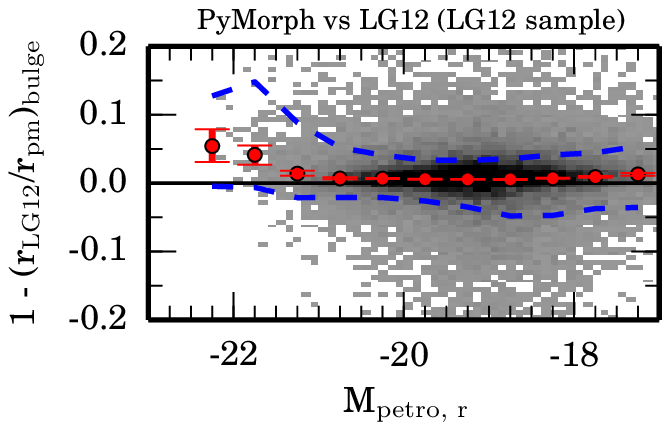}
\includegraphics[width=0.24\linewidth]{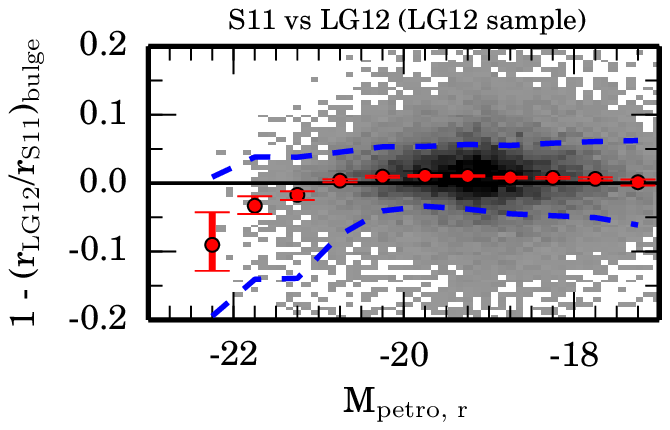}
\includegraphics[width=0.24\linewidth]{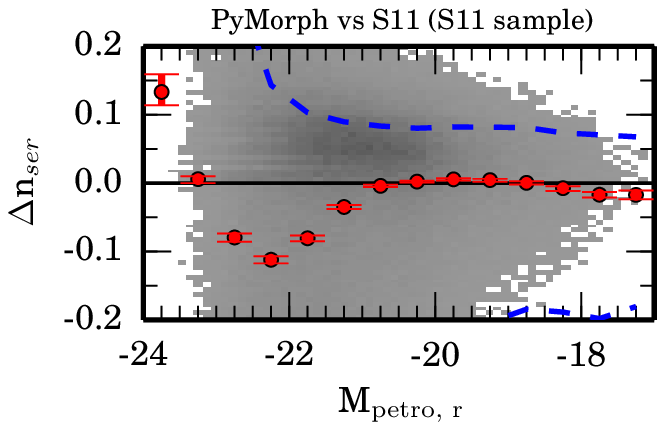}
\includegraphics[width=0.24\linewidth]{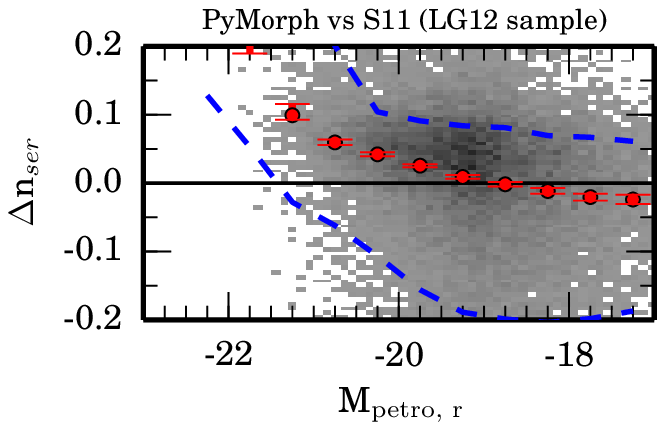}
\includegraphics[width=0.24\linewidth]{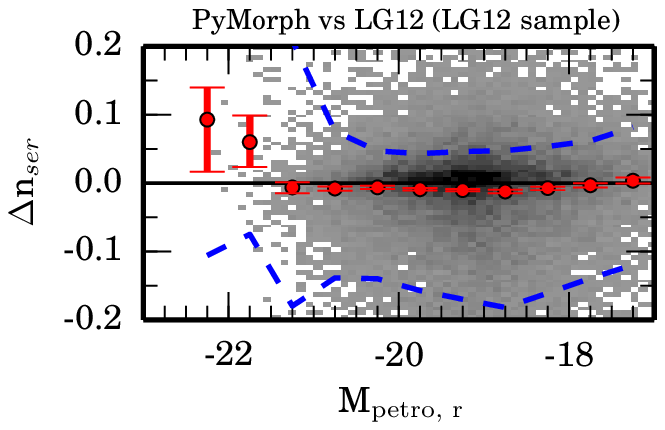}
\includegraphics[width=0.24\linewidth]{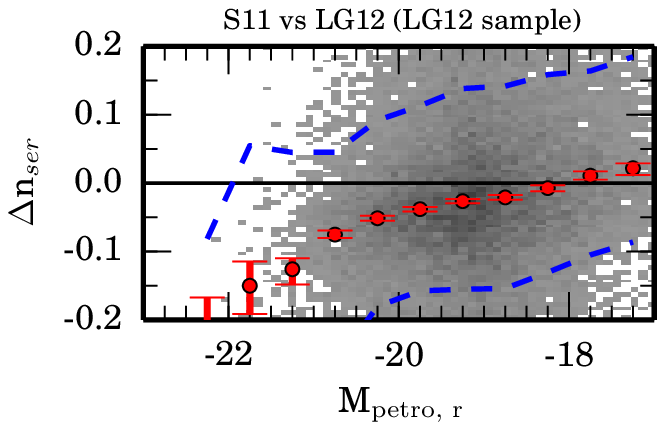}

\caption{Similar to Figure~\ref{fig:ser_fits}, but now also including the comparison of S11 with LG12 galaxies 
(far-right column).
The panels show the difference in magnitude, half-light radius, and S\'{e}rsic index 
as a function of apparent or absolute Petrosian magnitude. Note the trends in S11 when compared to LG12 or this work. 
Similar trends in S\'{e}rsic index, magnitude, and radius are not present in the comparison of this work with LG12.}
\label{fig:ser_fits_all}
\end{figure*}

This appendix includes an extended comparison of the \Ser{} fits. In Figure~\ref{fig:ser_fits_all},
we show comparisons between this work and S11 for all galaxies best fit by a \Ser{} profile according to S11 (left-hand column),
between this work and S11 for LG12 galaxies best fit by a \Ser{} profile according to S11 (centre left column),
between this work and LG12 for LG12 galaxies best fit by a \Ser{} profile according to S11 (centre right column),
and between S11 and LG12 for LG12 galaxies best fit by a \Ser{} profile according to S11 (right-hand column).
The rows one, two, and three compare the difference in total magnitude, half-light radius, and S\'{e}rsic index 
as a function of apparent Petrosian magnitude. 
The rows four, five, and six compare the difference in total magnitude, half-light radius, and S\'{e}rsic index 
as a function of absolute Petrosian magnitude. 

The comparison of this work and LG12 shows a small offset in magnitude that is independent of apparent or absolute magnitude.
There is also little or no offset in the half-light radius and S\'{e}rsic index. Comparisons of S11 to LG12 and this work show significant trends in S\'{e}rsic index and magnitude. These trends and the other issues discussed in the main text lead us to prefer 
the fits of this work over those of S11.

\clearpage

\subsection{The \DevExp{} Fits}\label{app:devexp}

\begin{figure*}
\includegraphics[width=0.24\linewidth]{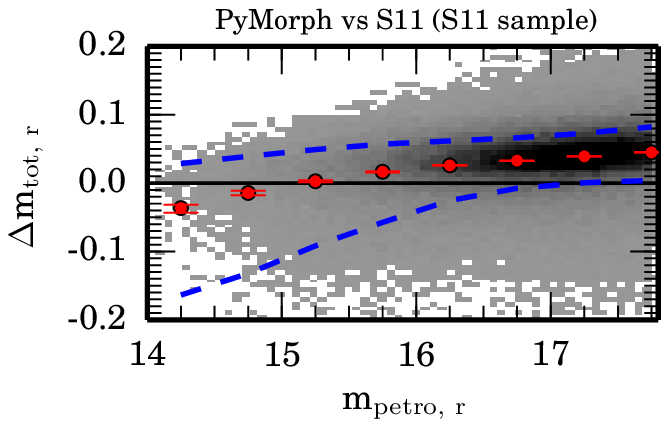}
\includegraphics[width=0.24\linewidth]{./figures/cmp_plots/simard/r_band_simard_devexp_petromag_mtot_lackner_only.eps}
\includegraphics[width=0.24\linewidth]{./figures/cmp_plots/lackner/r_band_lackner_devexp_petromag_mtot.eps}
\includegraphics[width=0.24\linewidth]{./figures/cmp_plots/lackner/r_simard_lackner_devexp_petromag_mtot.eps}
\includegraphics[width=0.24\linewidth]{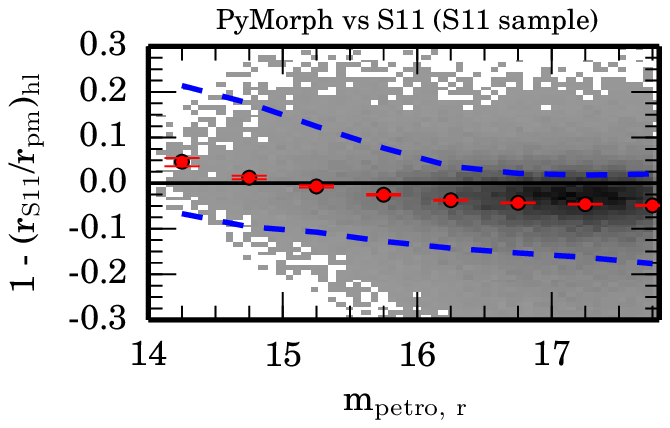}
\includegraphics[width=0.24\linewidth]{./figures/cmp_plots/simard/r_band_simard_devexp_petromag_hrad_lackner_only.eps}
\includegraphics[width=0.24\linewidth]{./figures/cmp_plots/lackner/r_band_lackner_devexp_petromag_hrad.eps}
\includegraphics[width=0.24\linewidth]{./figures/cmp_plots/lackner/r_simard_lackner_devexp_petromag_hrad.eps}
\includegraphics[width=0.24\linewidth]{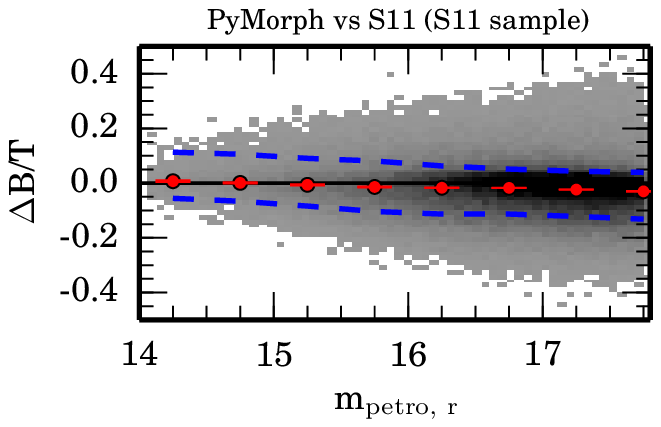}
\includegraphics[width=0.24\linewidth]{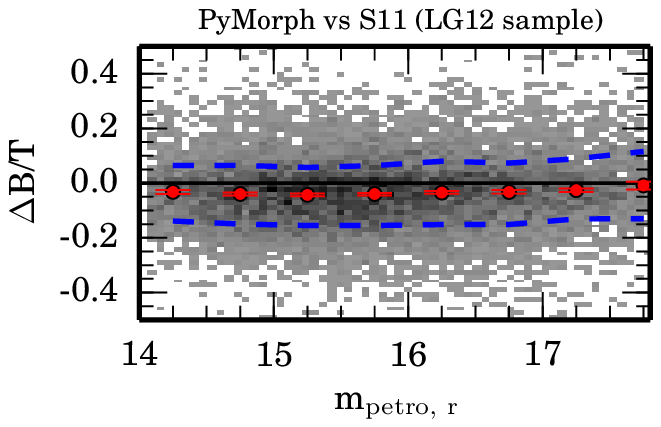}
\includegraphics[width=0.24\linewidth]{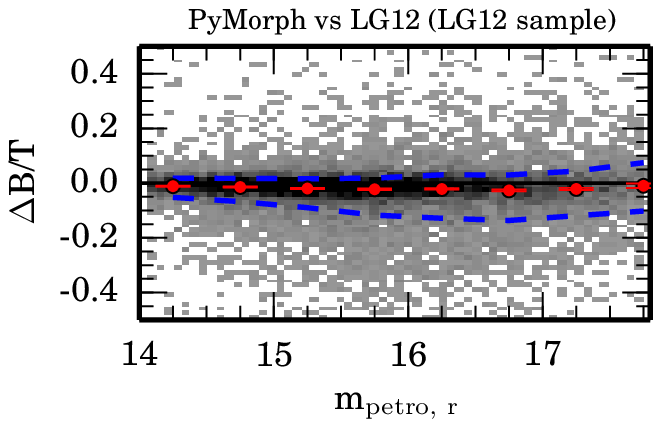}
\includegraphics[width=0.24\linewidth]{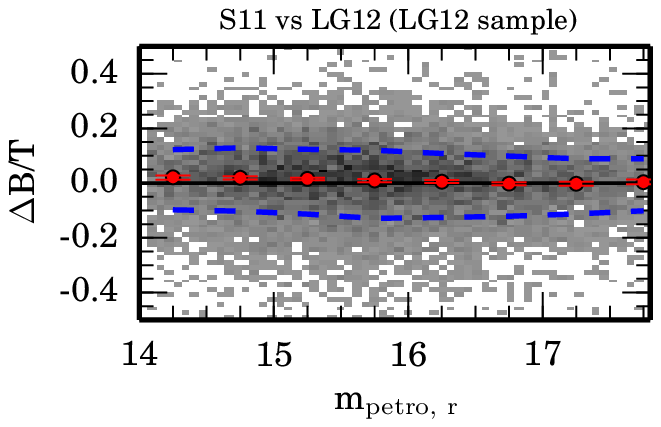}

\includegraphics[width=0.24\linewidth]{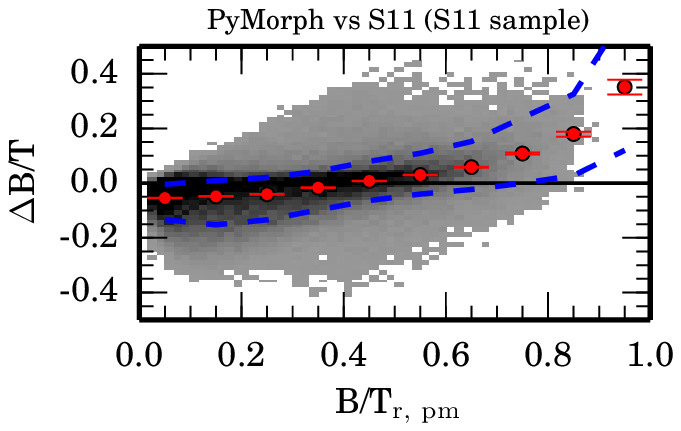}
\includegraphics[width=0.24\linewidth]{./figures/cmp_plots/simard/r_band_simard_devexp_BT_BT_lackner_only.eps}
\includegraphics[width=0.24\linewidth]{./figures/cmp_plots/lackner/r_band_lackner_devexp_BT_BT.eps}
\includegraphics[width=0.24\linewidth]{./figures/cmp_plots/lackner/r_simard_lackner_devexp_BT_BT.eps}

\includegraphics[width=0.24\linewidth]{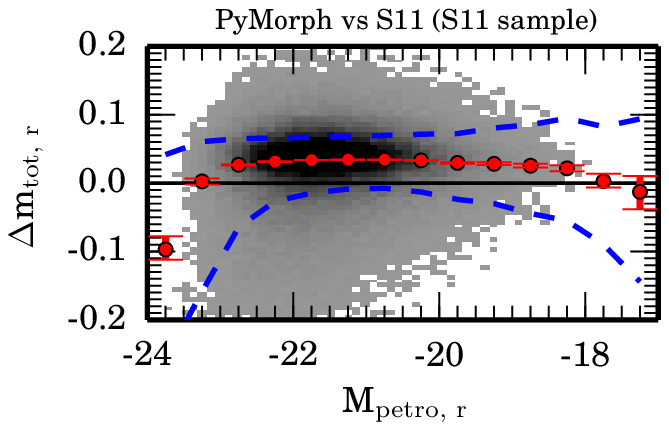}
\includegraphics[width=0.24\linewidth]{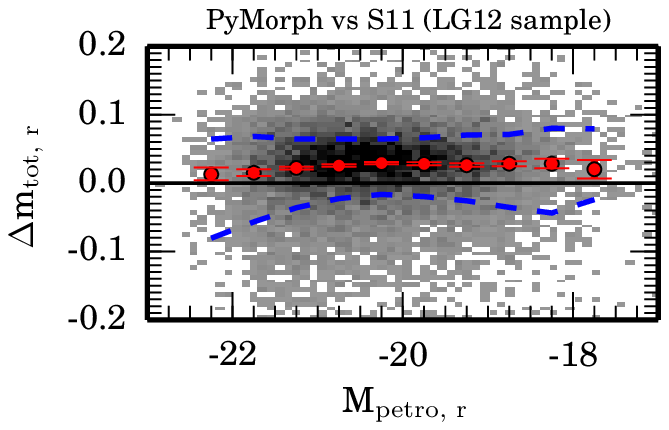}
\includegraphics[width=0.24\linewidth]{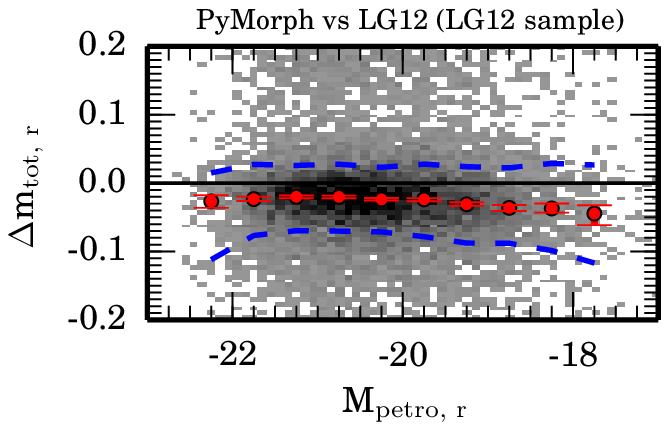}
\includegraphics[width=0.24\linewidth]{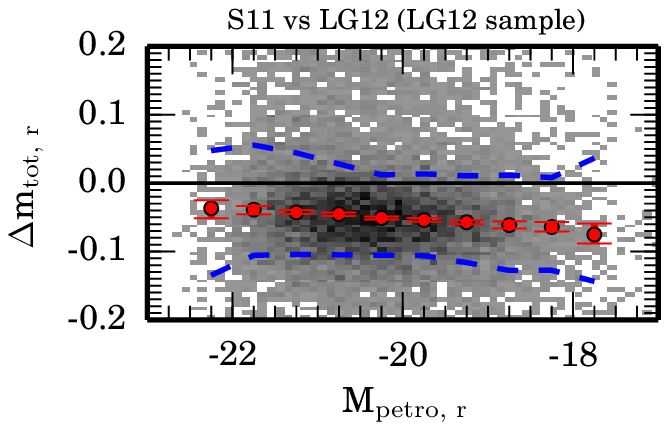}
\includegraphics[width=0.24\linewidth]{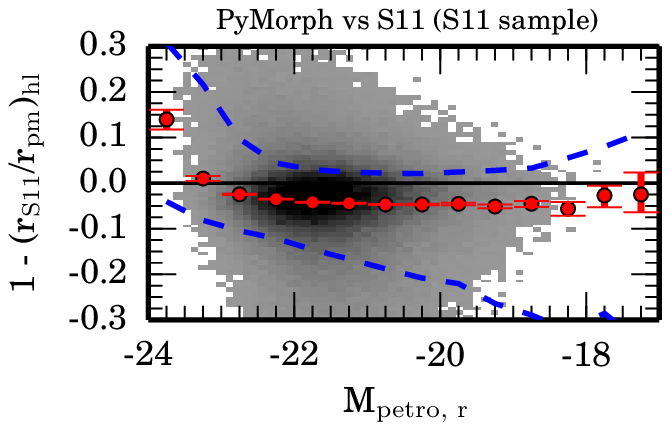}
\includegraphics[width=0.24\linewidth]{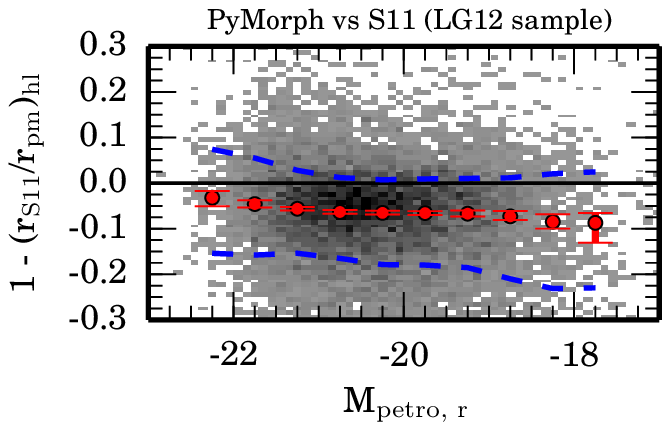}
\includegraphics[width=0.24\linewidth]{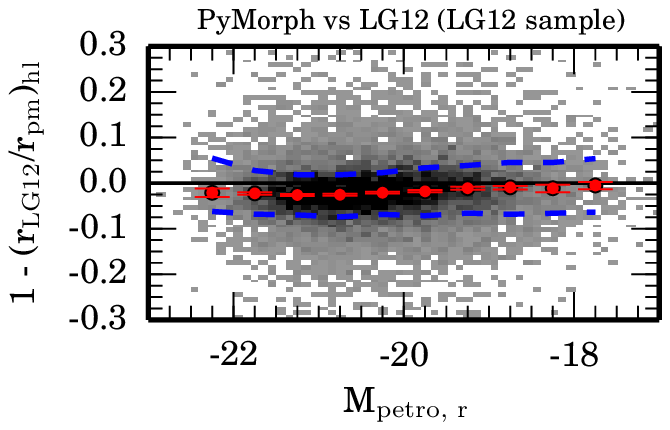}
\includegraphics[width=0.24\linewidth]{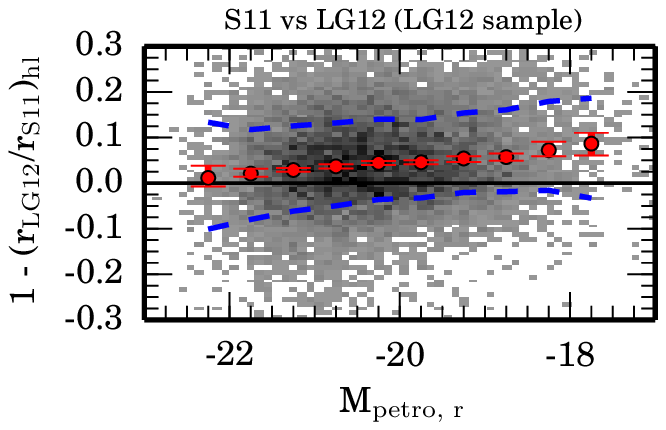}
\includegraphics[width=0.24\linewidth]{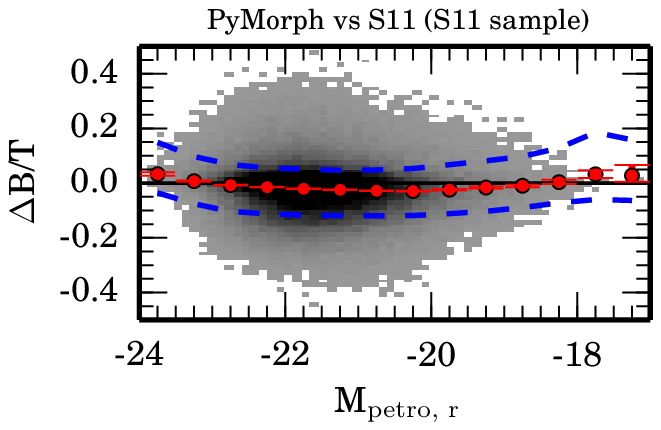}
\includegraphics[width=0.24\linewidth]{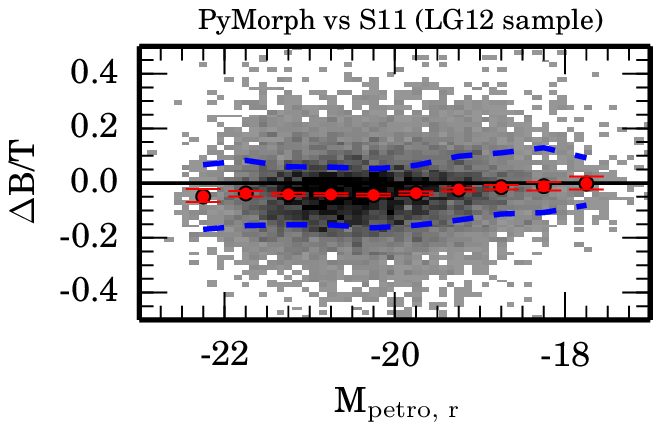}
\includegraphics[width=0.24\linewidth]{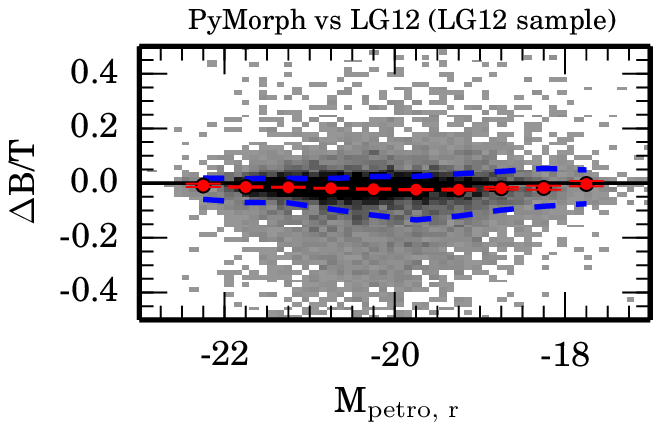}
\includegraphics[width=0.24\linewidth]{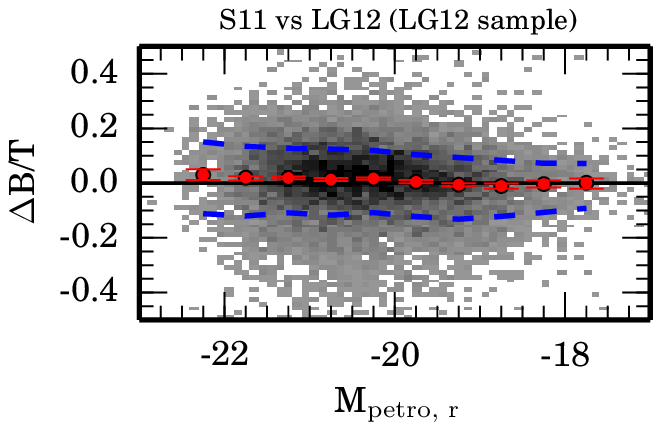}

\caption{Additional plots comparing the \DevExp{} fits of this work, S11, and LG12. See Section~\ref{sec:ex:devexp} of the main text
for discussion of the fits.
The panels show the difference in total magnitude, half-light radius, and B/T 
as a function of apparent Petrosian magnitude (rows 1, 2, and 3). The panels also show the difference in B/T 
as a function of B/T (row 4) and the difference in total magnitude, half-light radius, and B/T 
as a function of absolute Petrosian magnitude (rows 5, 6, and 7). Note the trends in S11 when compared to LG12 or this work. 
Similar trends in B/T, magnitude, and radius are not present in the comparison of this work with LG12.}

\label{fig:devexp_all}
\end{figure*}

\begin{figure*}
\includegraphics[width=0.24\linewidth]{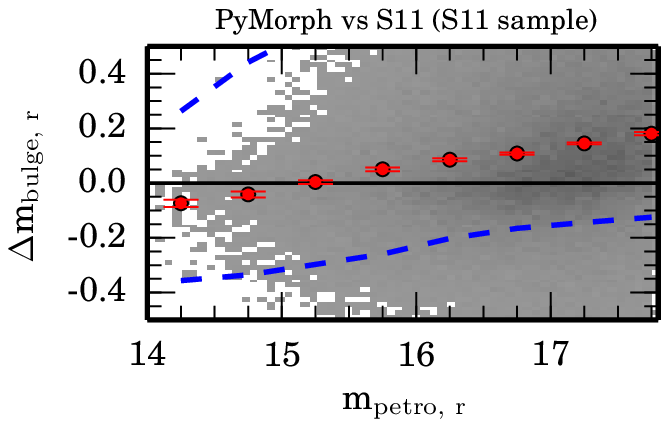}
\includegraphics[width=0.24\linewidth]{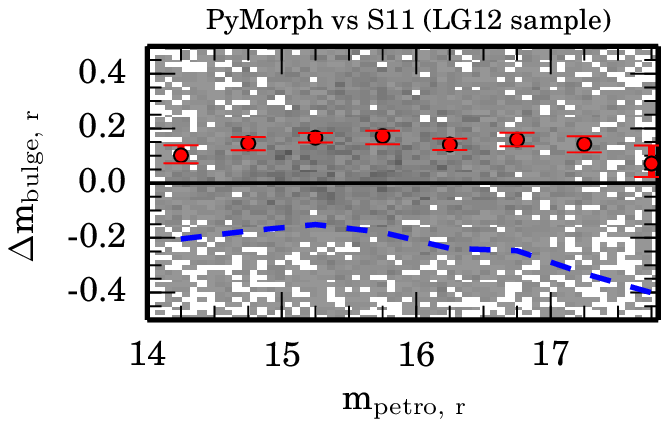}
\includegraphics[width=0.24\linewidth]{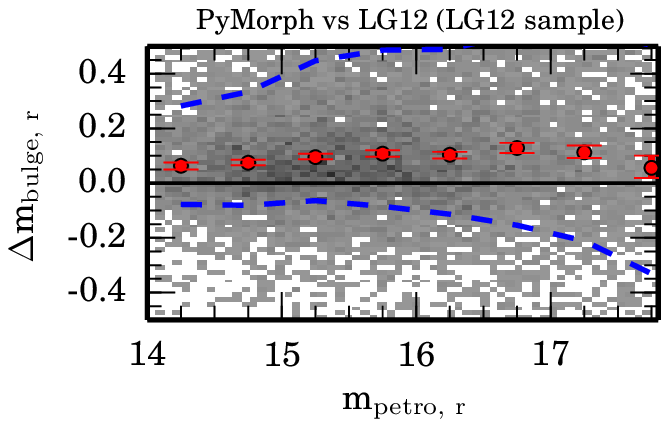}
\includegraphics[width=0.24\linewidth]{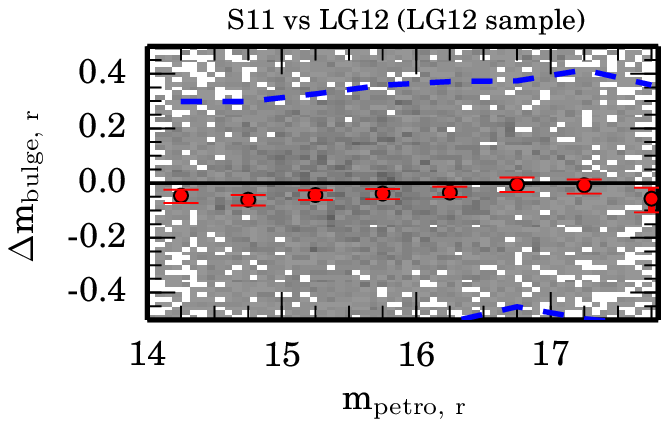}
\includegraphics[width=0.24\linewidth]{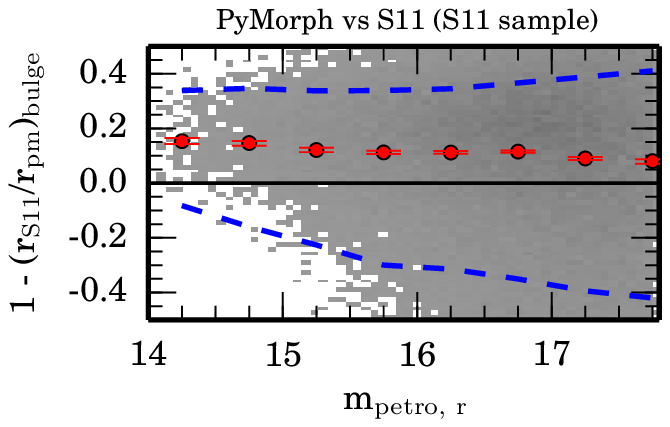}
\includegraphics[width=0.24\linewidth]{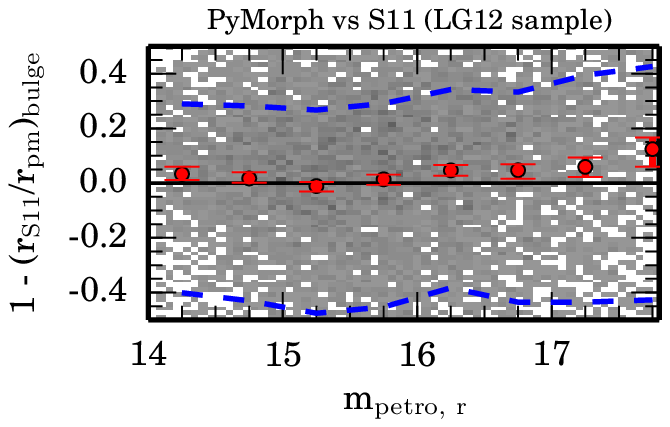}
\includegraphics[width=0.24\linewidth]{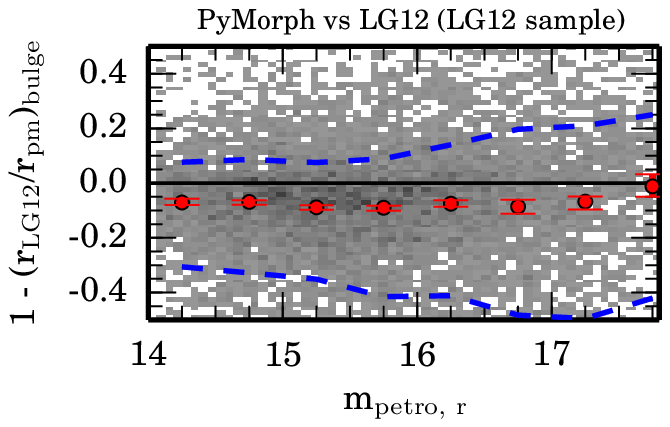}
\includegraphics[width=0.24\linewidth]{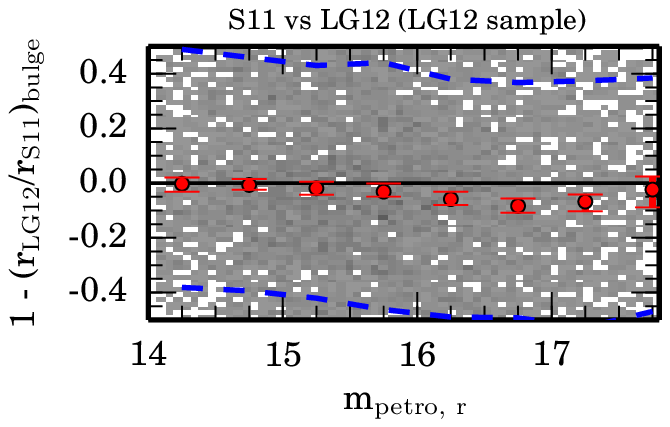}
\includegraphics[width=0.24\linewidth]{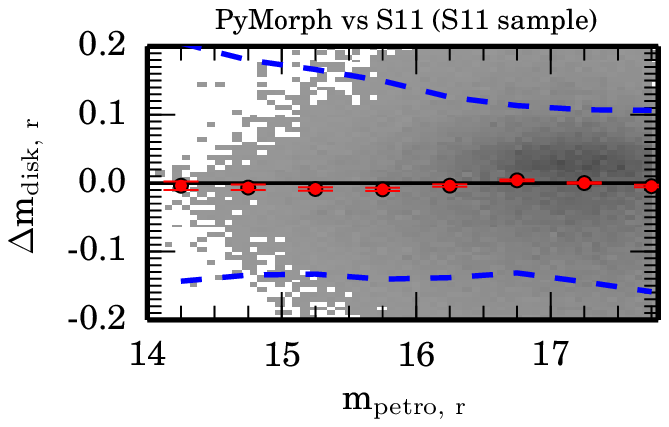}
\includegraphics[width=0.24\linewidth]{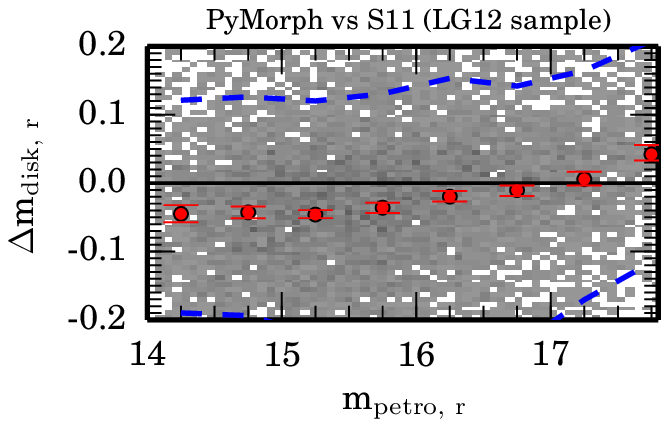}
\includegraphics[width=0.24\linewidth]{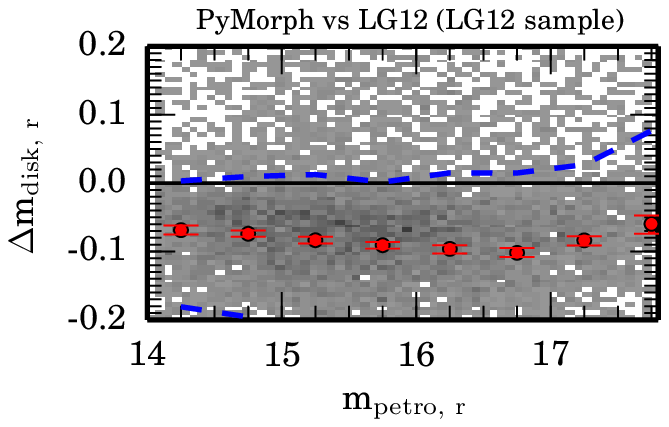}
\includegraphics[width=0.24\linewidth]{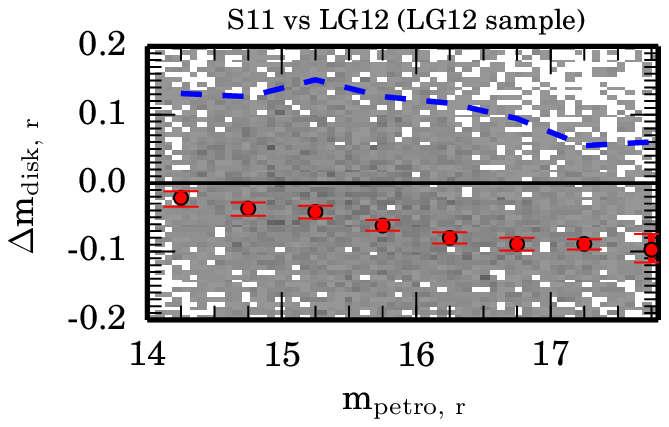}
\includegraphics[width=0.24\linewidth]{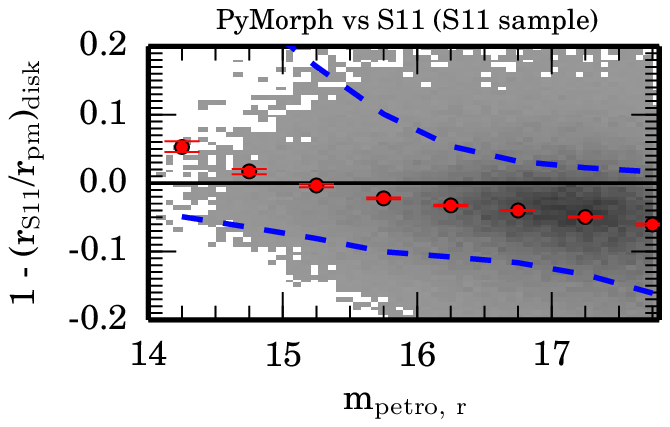}
\includegraphics[width=0.24\linewidth]{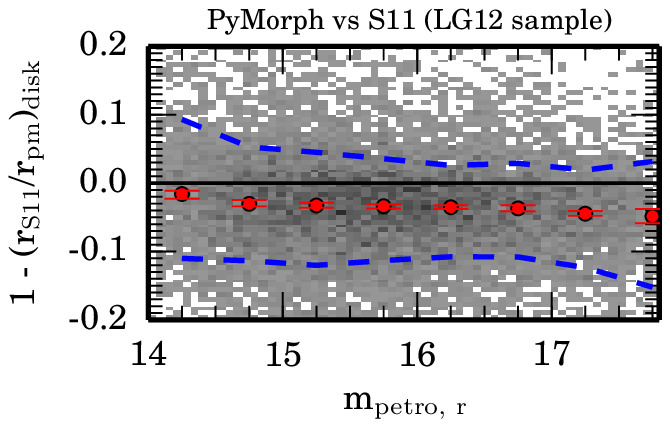}
\includegraphics[width=0.24\linewidth]{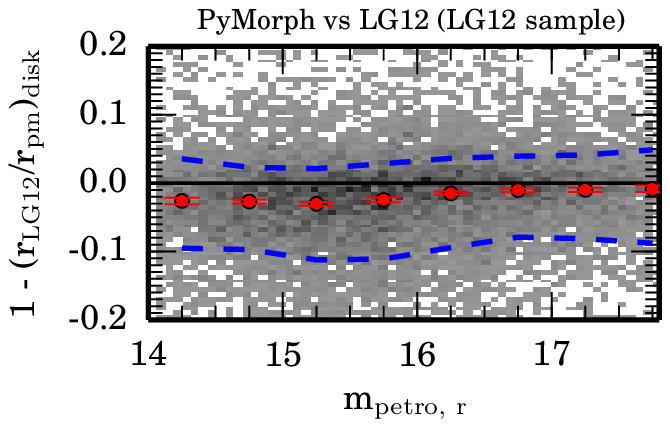}
\includegraphics[width=0.24\linewidth]{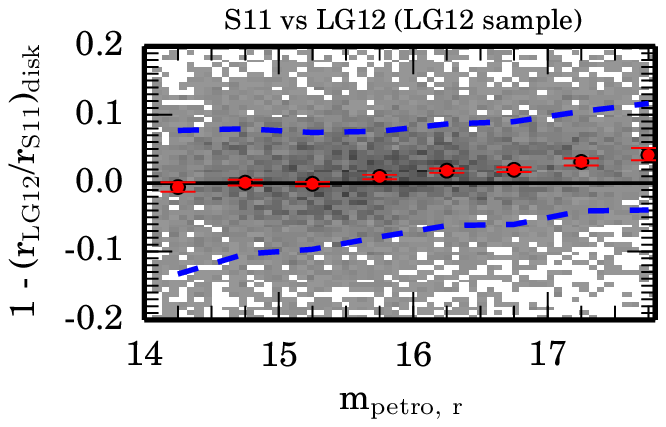}

\includegraphics[width=0.24\linewidth]{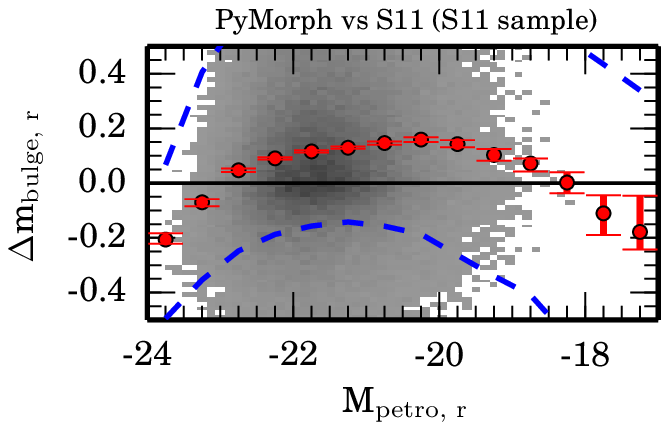}
\includegraphics[width=0.24\linewidth]{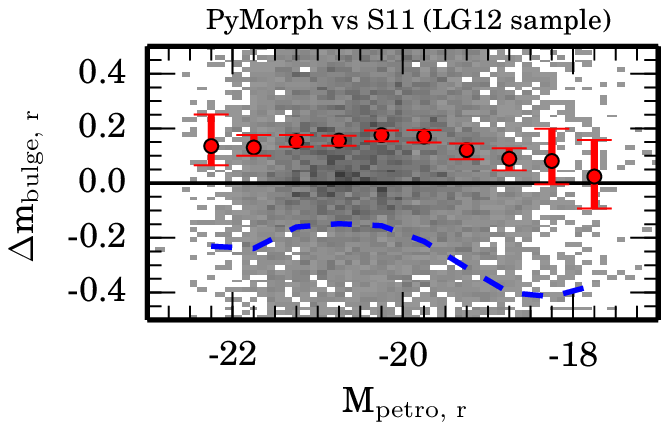}
\includegraphics[width=0.24\linewidth]{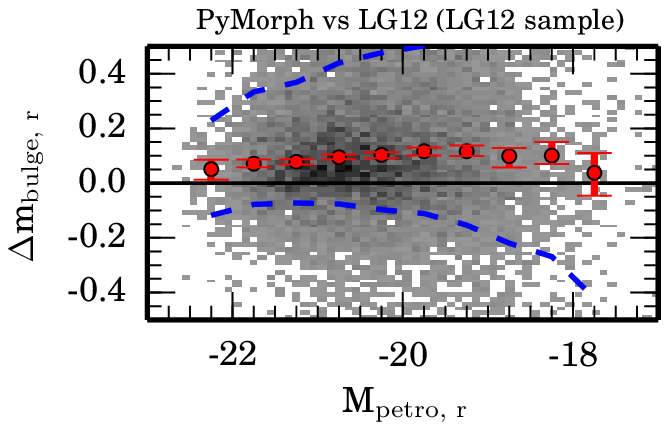}
\includegraphics[width=0.24\linewidth]{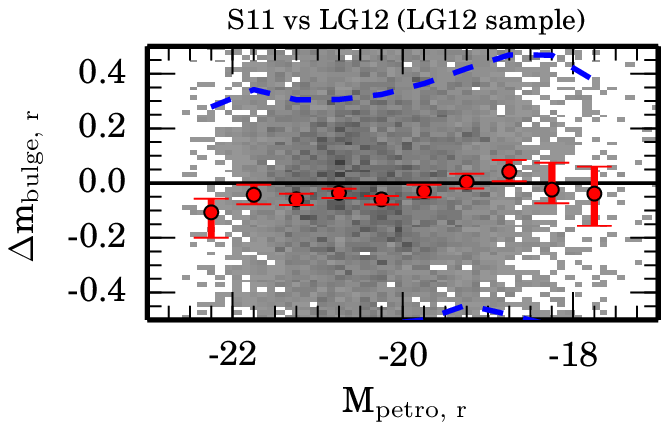}
\includegraphics[width=0.24\linewidth]{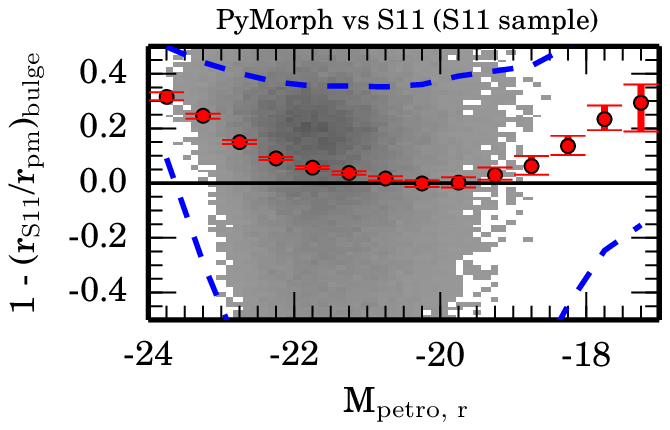}
\includegraphics[width=0.24\linewidth]{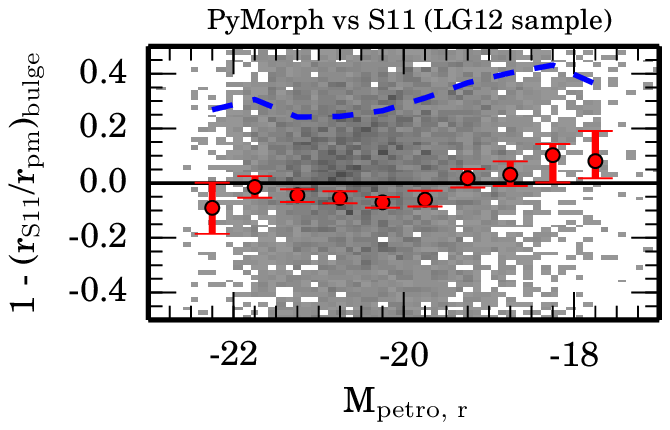}
\includegraphics[width=0.24\linewidth]{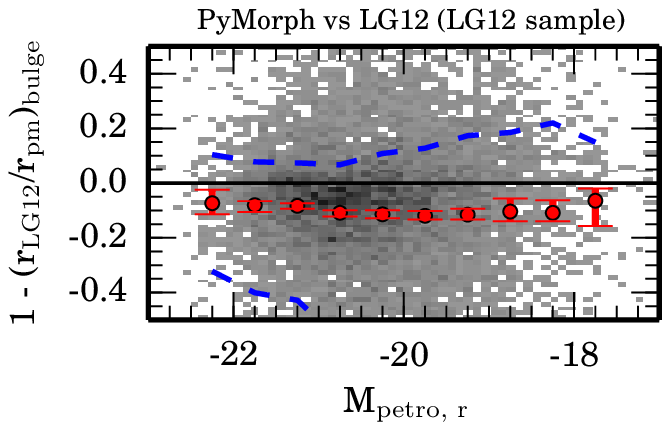}
\includegraphics[width=0.24\linewidth]{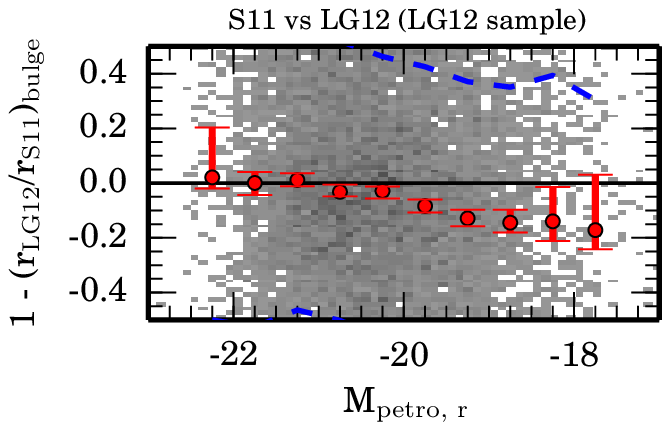}
\includegraphics[width=0.24\linewidth]{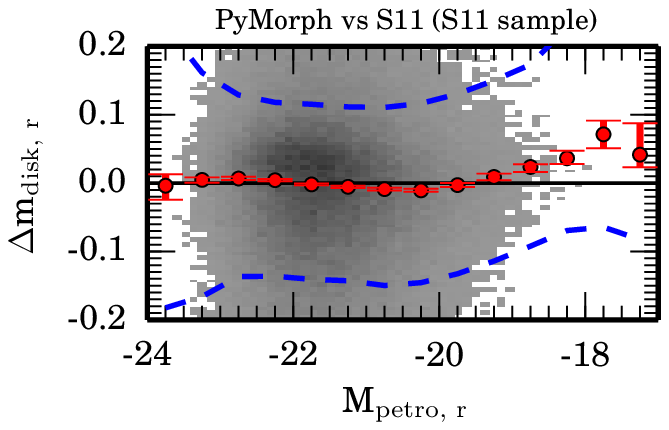}
\includegraphics[width=0.24\linewidth]{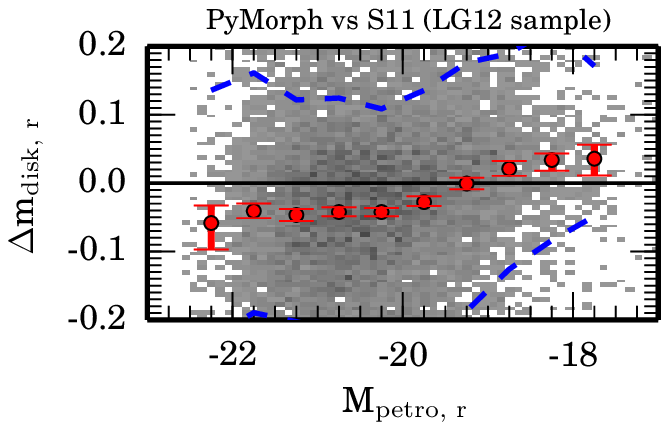}
\includegraphics[width=0.24\linewidth]{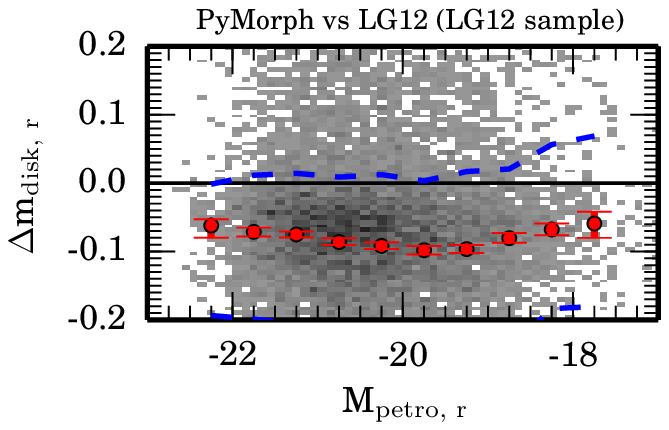}
\includegraphics[width=0.24\linewidth]{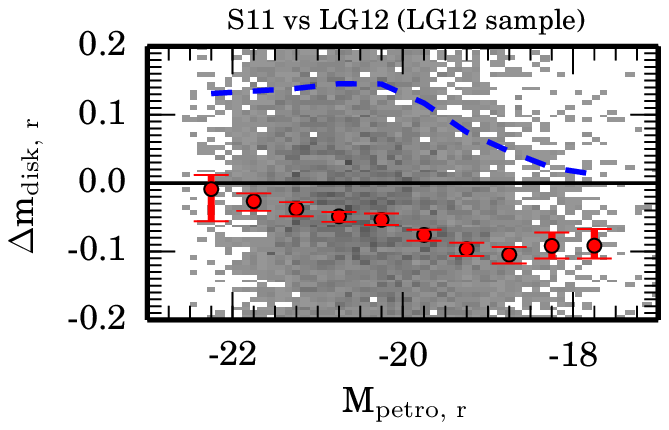}
\includegraphics[width=0.24\linewidth]{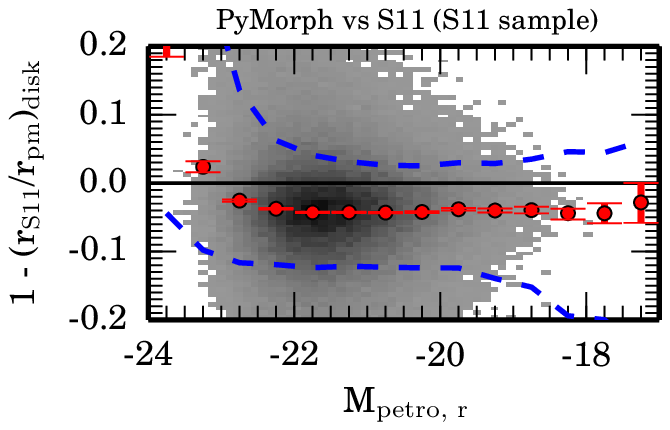}
\includegraphics[width=0.24\linewidth]{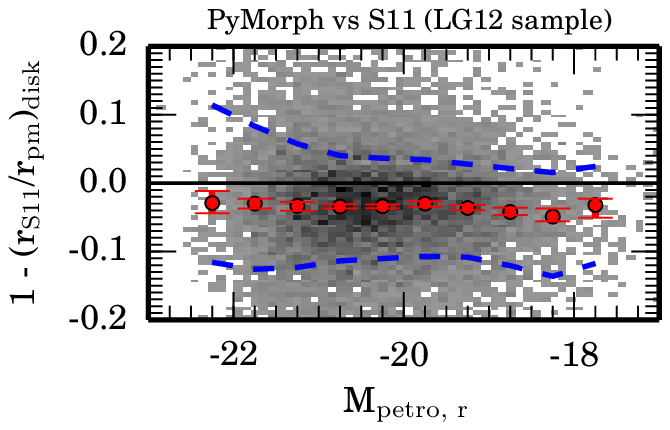}
\includegraphics[width=0.24\linewidth]{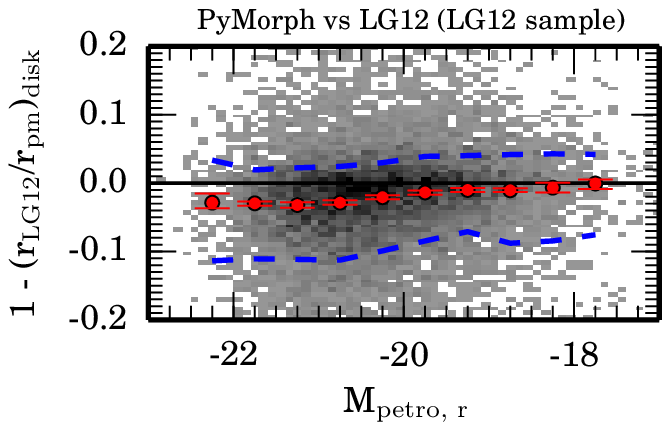}
\includegraphics[width=0.24\linewidth]{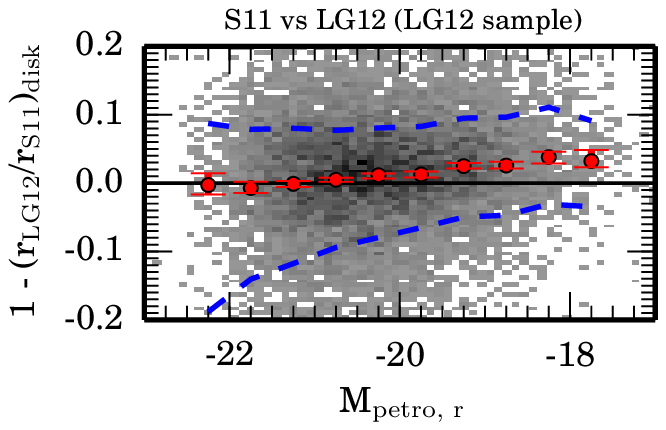}

\caption{Additional plots comparing the \DevExp{} bulge and disc magnitude and radius of this work, S11, and LG12. 
See Section~\ref{sec:ex:devexp} of the main text for discussion of the fits.
The panels show the difference in magnitude and half-light radius as a function of apparent Petrosian magnitude  for the 
bulge (rows 1 and 2) and for the 
disc (rows 3 and 4). The panels also show the difference in magnitude and half-light 
radius as a function of absolute Petrosian magnitude  for the 
bulge (rows 5 and 6) and for the 
disc (rows 7 and 8).
The scatter is quite broad, making it difficult to draw any conclusions.}

\label{fig:devexp:2com_all}
\end{figure*}

This appendix includes an extended comparison of the \DevExp{} fits. In Figure~\ref{fig:devexp_all},
we show comparisons between this work and S11 for all galaxies best fit by a \DevExp{} profile according to S11 (left-hand column),
between this work and S11 for LG12 galaxies best fit by a \DevExp{} profile according to LG12 (centre left column),
between this work and LG12 for LG12 galaxies best fit by a \DevExp{} profile according to LG12 (centre right column),
and between S11 and LG12 for LG12 galaxies best fit by a \DevExp{} profile according to LG12 (right-hand column).
The panels show the difference in total magnitude, half-light radius, and B/T 
as a function of apparent Petrosian magnitude (rows 1, 2, and 3). The panels also show the difference in B/T 
as a function of B/T (row 4) and the difference in total magnitude, half-light radius, and B/T 
as a function of absolute Petrosian magnitude (rows 5, 6, and 7). The scatter in B/T is wide and there are systematic shifts in the difference between S11 and LG12. The S11 fits strongly favour values of B/T between 0.2 and 0.4. We do not see agreement with these values here. This suggests that the uncertainty in the B/T parameter is quite large between fitting codes.

Figure~\ref{fig:devexp:2com_all} also
shows comparisons between this work and S11 for all galaxies best fit by a \DevExp{} profile according to S11 (left-hand column),
between this work and S11 for LG12 galaxies best fit by a \DevExp{} profile according to LG12 (centre left column),
between this work and LG12 for LG12 galaxies best fit by a \DevExp{} profile according to LG12 (centre right column),
and between S11 and LG12 for LG12 galaxies best fit by a \DevExp{} profile according to LG12 (right-hand column).
The panels show the difference in magnitude and half-light radius as a function of apparent Petrosian magnitude  for the 
bulge (rows 1 and 2) and for the 
disc (rows 3 and 4). The panels also show the difference in magnitude and half-light 
radius as a function of absolute Petrosian magnitude  for the 
bulge (rows 5 and 6) and for the 
disc (rows 7 and 8).

The scatter in the subcomponents is broad (0.2-0.4 mag in the bulge magnitude and 10--20 per cent in the radius) 
making any conclusions difficult. However, we do note that the scatter is smallest in the comparison of this work with LG12. 
The disc radii are better constrained relative to the bulge radii. 
This reflects the decreased sensitivity to incorrect estimation of the sky brightness in the \Exp{} disc profile. 
Reduced sensitivity to sky brightness for the \SerExp{} fits is visible in Figure~15 of M13.  M13 showed that the radius of 
the bulge of the \SerExp{} fit was about five times more sensitive to the sky level than the disc radius. 
We expect similar behaviour for the \DevExp{} galaxies because the higher S\'{e}rsic index of the bulge 
component makes the bulge more sensitive to adding flux to the wings of the profile. 

\clearpage

\subsection{The \SerExp{} fit}\label{app:serexp}

\begin{figure*}
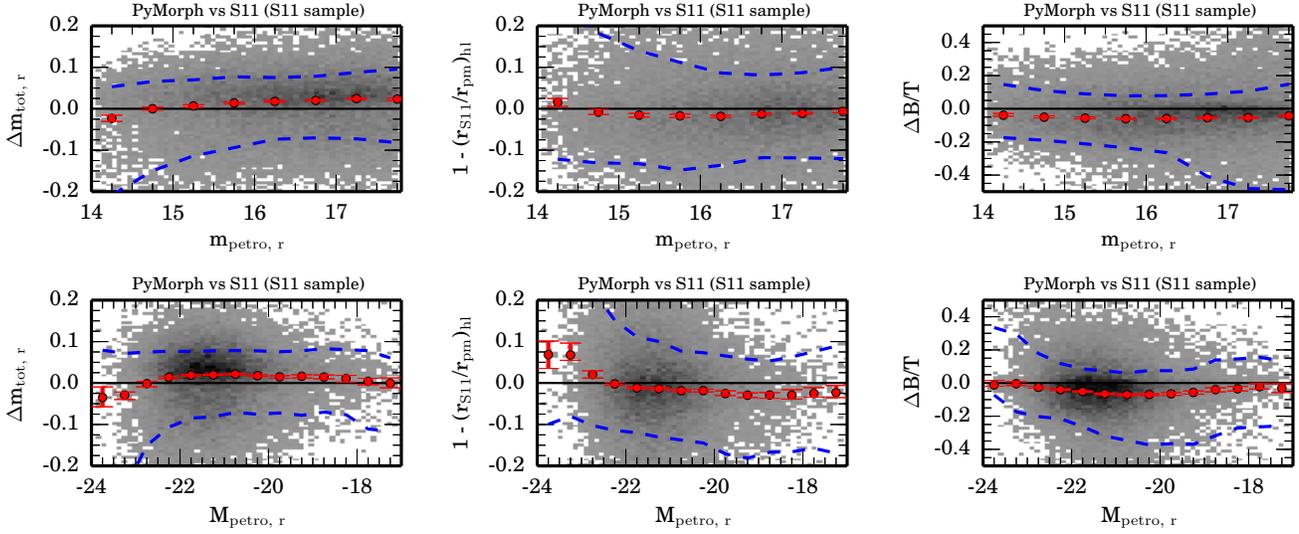

\includegraphics[width=0.329\linewidth]{./figures/cmp_plots/simard/r_band_simard_serexp_petromag_mtot.eps}
\includegraphics[width=0.329\linewidth]{./figures/cmp_plots/simard/r_band_simard_serexp_petromag_hrad.eps}
\includegraphics[width=0.329\linewidth]{./figures/cmp_plots/simard/r_band_simard_serexp_petromag_BT.eps}
\includegraphics[width=0.329\linewidth]{./figures/cmp_plots/simard/r_band_simard_serexp_petromag_abs_mtot.eps}
\includegraphics[width=0.329\linewidth]{./figures/cmp_plots/simard/r_band_simard_serexp_petromag_abs_hrad.eps}
\includegraphics[width=0.329\linewidth]{./figures/cmp_plots/simard/r_band_simard_serexp_petromag_abs_BT.eps}
\caption{Additional plots comparing the \SerExp{} fits of this work and S11. See Section~\ref{sec:ex:serexp} 
of the main text for discussion of the fits.
The panels show the difference in total magnitude, half-light radius, and B/T (left-hand, centre, and right-hand columns, respectively)
as a function of apparent Petrosian magnitude (row 1) and as a function of absolute Petrosian magnitude (row 2). See Section~\ref{sec:ex:serexp} of the main text for discussion of the fits.
}
\label{fig:simard_serexp_all}
\end{figure*}

\begin{figure*}
\includegraphics[width=0.329\linewidth]{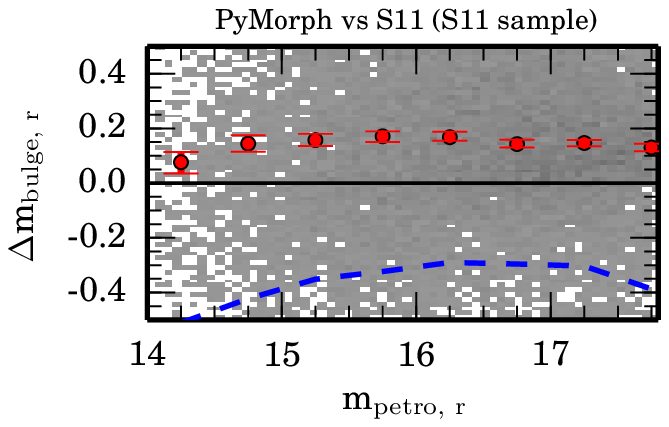}
\includegraphics[width=0.329\linewidth]{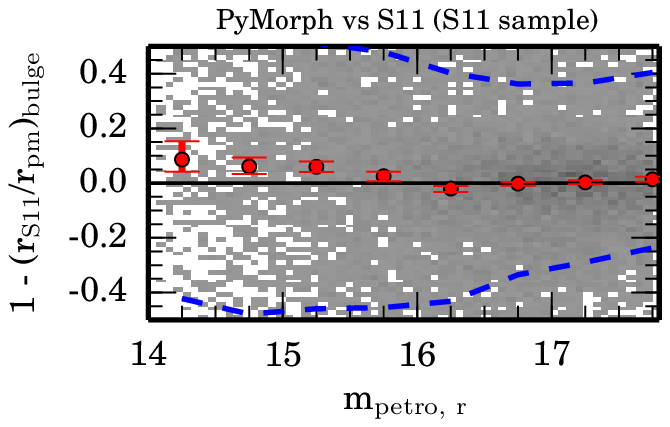}
\includegraphics[width=0.329\linewidth]{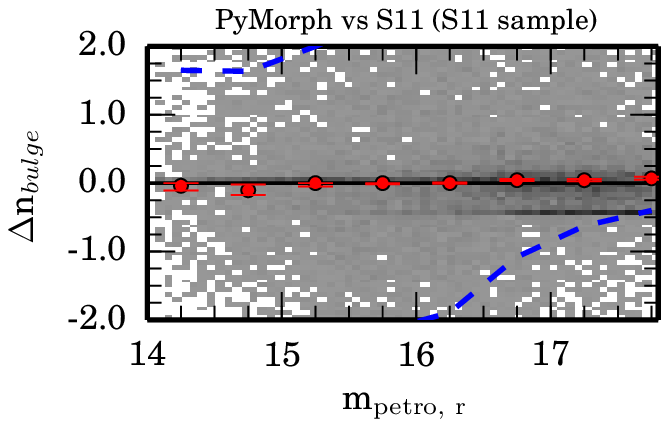}
\includegraphics[width=0.329\linewidth]{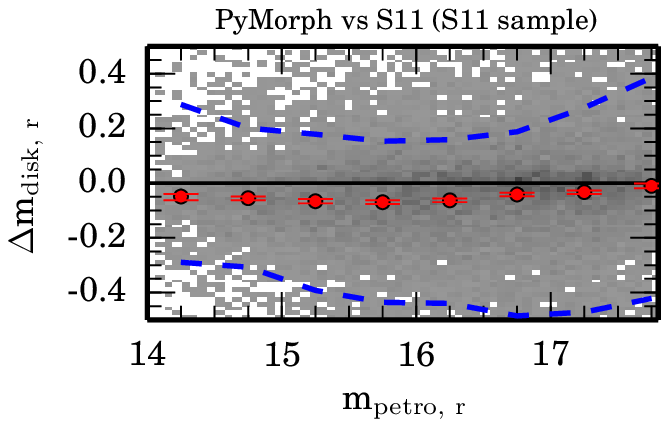}
\includegraphics[width=0.329\linewidth]{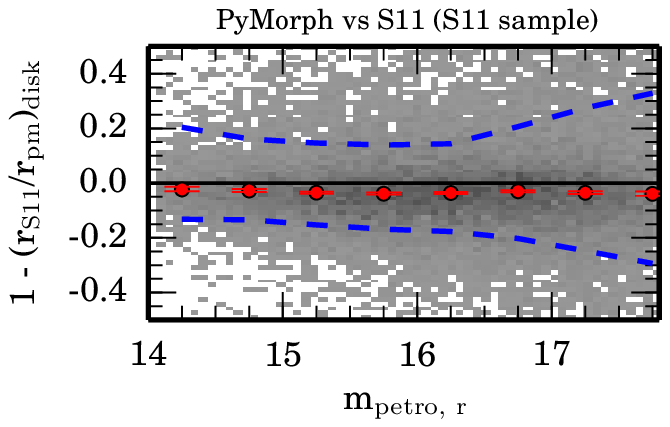}\\
\includegraphics[width=0.329\linewidth]{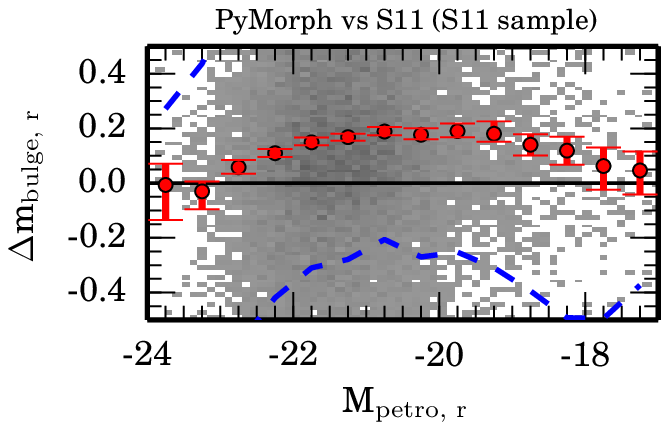}
\includegraphics[width=0.329\linewidth]{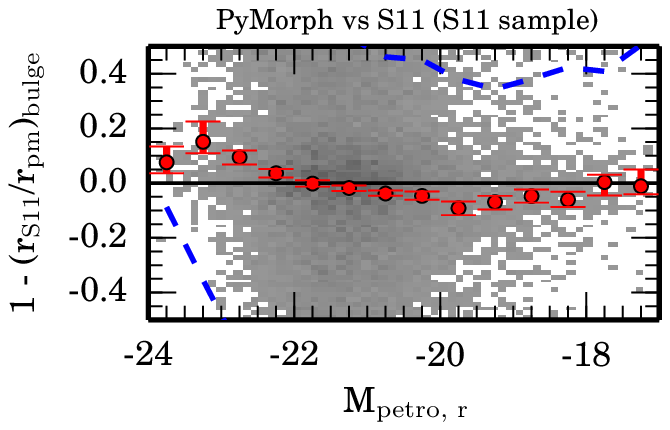}
\includegraphics[width=0.329\linewidth]{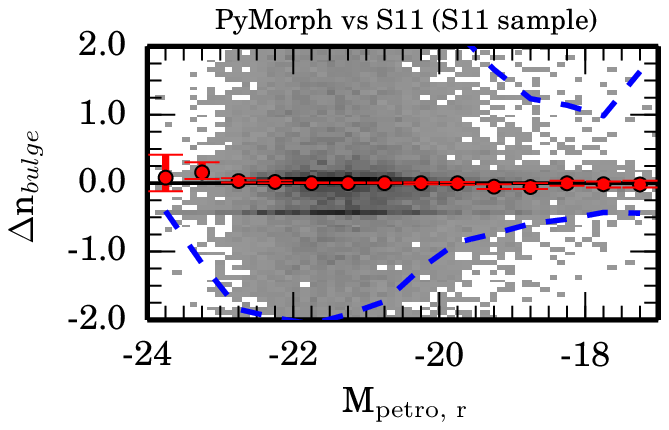}
\includegraphics[width=0.329\linewidth]{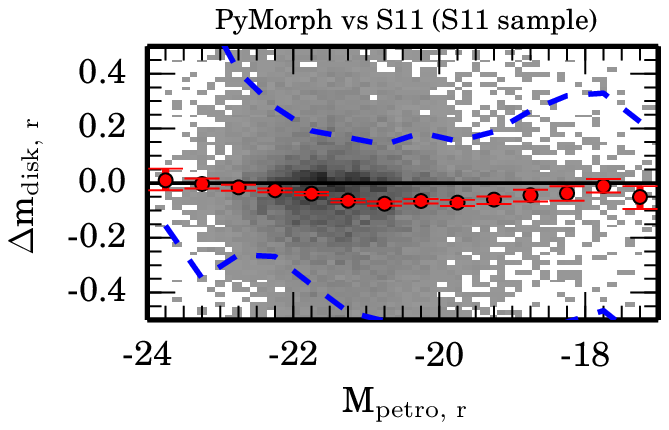}
\includegraphics[width=0.329\linewidth]{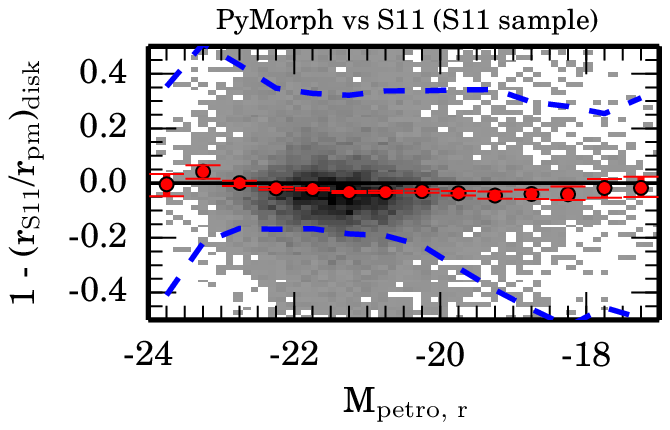}\\

\caption{Additional plots comparing the \SerExp{} bulge Petrosian magnitude, radius, and S\'{e}rsic index as a function 
of apparent magnitude (row 1); disc magnitude and radius as a function of apparent Petrosian magnitude (row 2); bulge magnitude, radius, and S\'{e}rsic index as a function of absolute Petrosian magnitude
(row 3); disc magnitude and radius as a function of absolute Petrosian magnitude (row 4) of this work and S11. 
See Section~\ref{sec:ex:serexp} of the main text for discussion of the fits.
The scatter is quite broad again, making it difficult to draw any conclusions.}

\label{fig:simard_serexp:2com_all}
\end{figure*}

This appendix includes an extended comparison of the \SerExp{} fits. In Figure~\ref{fig:simard_serexp_all},
we show comparisons between this work and S11 for all galaxies best fit by a \SerExp{} profile according to S11.
The panels show the difference in total magnitude (left), half-light radius (centre), and B/T (right) 
as a function of apparent Petrosian magnitude (row 1) and 
as a function of absolute Petrosian magnitude (row 2).

Figure~\ref{fig:simard_serexp:2com_all} 
shows additional comparisons of the \SerExp{} bulge Petrosian magnitude, radius, and S\'{e}rsic index as a function 
of apparent magnitude (row 1); disc magnitude and radius as a function of apparent Petrosian magnitude (row 2); bulge magnitude, radius, and S\'{e}rsic index as a function of absolute Petrosian magnitude
(row 3); disc magnitude and radius as a function of absolute Petrosian magnitude (row 4) of this work and S11.
The scatter in the sub-components is broad again (0.4 mag in both the bulge magnitude and 40 per cent in the radius) 
making any conclusions difficult. 
The disc radii are still better constrained relative to the bulge radii. As discussed in Appendix~\ref{app:devexp},
this reflects the decreased sensitivity to incorrect estimation of the sky brightness in the \Exp{} disc profile. 
Reduced sensitivity to sky brightness for the \SerExp{} fits is visible in Figure~15 of M13.  M13 showed that the radius of 
the bulge of the \SerExp{} fit was about five times more sensitive to the sky level than the disc radius. 

\clearpage
\section{Example Catalogue Data Table}
This appendix includes sample tables of the data released as part of this work. The meaning of the table columns are described in Table~\ref{tab:modtab1} and Table~\ref{tab:modtab2}. The full tables are available on-line in machine-readable format.

\begin{table}
\centering
\begin{tabular}{c c c c c c c c c}
\textbf{SExMag}& \textbf{SExMagErr}& \textbf{SExHrad}& \textbf{SExSky}& 
\ldots& 
\textbf{kpc\_per\_arcsec}& \textbf{Vmax}& \textbf{SN}& \textbf{kcorr}\\ \hline  
17.67 &     0.03 & 2.15 & 20.79 & \ldots &
0.61& 2800960 & 120.20 & 0.05 \\
17.43 &      0.03 & 2.58 & 20.79 &\ldots&
1.47& 58645700& 128.13 & 0.08 \\
17.11 &     0.02 & 3.96 & 20.79 & \ldots&
2.76& 699989000& 123.77 & 0.19 \\
16.93 &     0.02 & 2.75 & 20.79 & \ldots&
1.36& 94692800&181.81 & 0.09 \\
16.70 &      0.02 & 3.29 & 20.79 &\ldots&
1.37& 123920000 &193.85 & 0.13 \\
\ldots &\ldots &\ldots &\ldots &\ldots &\ldots &\ldots 
& \ldots & \ldots \\
\end{tabular}
\caption{Sample of the model-independent measurements described in Table~\ref{tab:modtab1}. The full table has 20 columns and is available in machine-readable format on-line.}
\end{table}

\begin{table}
\centering
\begin{tabular}{c c c c c c c c c}
\textbf{m\_tot}& \textbf{BT}& \textbf{r\_tot}& \textbf{ba\_tot}& 
\ldots & \textbf{finalflag}& \textbf{autoflag}&  \textbf{pyflag}&
\textbf{pyfitflag}\\ \hline  
17.65 &  0.21 &   1.76 &     0.32 & \ldots &  5121 & 2048&62 & 0 \\
17.37 & 0.05 &   2.28 &    0.42 &  \ldots & 49 &4&62 &     128 \\
17.01 &  0.14 &   4.11 &    0.66 & \ldots &  273&131072& 62 &     128 \\
16.81 & 0.09 &   2.79 &    0.66 &  \ldots & 49&67108868&62 &     384 \\
16.66 & 0.02 &   3.02 &    0.38 & \ldots &  49&67108868& 62 &     384 \\ 
\ldots &\ldots &\ldots &\ldots &\ldots &\ldots &\ldots &\ldots &\ldots \\
\end{tabular}
\caption{Sample of the model-dependent measurements for the \SerExp{} fit described in Table~\ref{tab:modtab2}. The full table has 39 columns and is available in machine-readable format on-line. Similar tables are also available for the \Dev{}, \Ser{}, and \DevExp{} fits.}
\end{table}

\end{document}